\documentclass[11pt]{article}
\usepackage[margin=1in]{geometry}
\usepackage{amsmath}
\usepackage{graphicx}
\usepackage{enumerate}
\usepackage{enumitem}
\usepackage{natbib}
\usepackage{url} % not crucial - just used below for the URL 

\usepackage{algorithmic} %For computer algorithm code in LATEX
\usepackage{algorithm} %For algorithm box around the algorithmic
\usepackage{setspace} %For setting spacing (e.g., double spacing, use \doublespacing)
\usepackage{rotating} %For rotating tables (provides sidewaysfigure and sidewaystable)

\usepackage{amsfonts, amssymb} %Either provides Real number and other font'related stuff.
\usepackage{amsthm}

\usepackage{epstopdf} %for adding eps in pdfLatex	
\usepackage{hyperref} %For adding hyperlinks in documents
\usepackage{bm} %Bolds everything when uses \bm as a command
\usepackage{bbm}
\usepackage{multirow} %Have multiple rows in a table
\usepackage{latexsym} %Latex-based math symbols. Generally load this in conjunction with amssymb
\usepackage{rotating} %arbitrary rotate any object
\usepackage{titlesec} %centering section headings
\usepackage{caption} %captioning for tables
\usepackage{mathtools} %ceiling/floor commands
\usepackage{color} %Provides color for text.
\usepackage{diagbox}
\usepackage{pict2e}
\usepackage{etoolbox}
\usepackage[english]{babel}
\usepackage{empheq}

\usepackage{cases}
\usepackage{tikz}
\usetikzlibrary{arrows, chains, positioning, quotes, shapes.geometric, arrows.meta}

\usepackage{lato}

\newtheorem{theorem}{Theorem}[section]

\newtheorem{condition}{Condition}[section]

\theoremstyle{definition}

\newtheorem{assumption}{Assumption}[section]
\newtheorem{assumptionNew}{Assumption}[section]
\newtheorem{example}{Example}[section]

\theoremstyle{remark}
\newtheorem{remark}{Remark}[section]

\theoremstyle{definition}

\newtheorem{GREG}{\textit{Regularity Condition}}

\definecolor{mycolor}{RGB}{0,200,200} 

\newcommand{\indep}{\,\rotatebox[origin=c]{90}{$\models$}\,}
\newcommand{\nindep}{\not \hspace{-0.125cm} \rotatebox[origin=c]{90}{$\models$}\,}

\DeclareMathOperator*{\argmin}{arg\,min}

\DeclareMathOperator*{\plim}{plim}
\newcommand{\EXP}{\text{E}}
\newcommand{\VAR}{\text{Var}}
\newcommand{\AVER}{\mathbbm{P}}

\newcommand{\cond}{\, \big| \,}
\newcommand{\Cond}{\, \Big| \,}

\newcommand{\con}{ ; }
\newcommand{\R}{\mathbbm{R}}
\newcommand{\ind}{\mathbbm{1}}
\newcommand{\T}{^\intercal}
\newcommand{\sT}{^{*\intercal}}

\newcommand{\bX}{\bm{X}}
\newcommand{\bD}{\bm{D}}
\newcommand{\bx}{\bm{x}}
\newcommand{\bW}{\bm{W}}
\newcommand{\bw}{\bm{w}}
\newcommand{\bG}{\bm{G}}
\newcommand{\bg}{\bm{g}}
\newcommand{\bh}{\bm{\phi}}

\newcommand{\bO}{\bm{O}}

\newcommand{\bmu}{\bm{\mu}}
\newcommand{\blambda}{\bm{\lambda}}
\newcommand{\bgamma}{\bm{\gamma}}
\newcommand{\Beta}{\bm{\eta}}
\newcommand{\bdelta}{\bm{\delta}}
\newcommand{\btheta}{\bm{\theta}}
\newcommand{\bbeta}{\bm{\beta}}
\newcommand{\bZ}{\bm{Z}}

\newcommand{\bphi}{\bm{\phi}}

\newcommand{\potY}[2]{Y_{#1}^{(#2)}}
\newcommand{\potW}[2]{W_{#1}^{(#2)}}

\newcommand{\HL}[1]{\hyperlink{(#1)}{(#1)}}
\newcommand{\HT}[1]{\hypertarget{(#1)}{(#1)}}

\newcommand{\PI}{\text{PSC}}

\newcommand{\pre}{\text{pre}}
\newcommand{\post}{\text{post}}

\definecolor{light-gray}{gray}{0.7}

\hypersetup{
colorlinks=true,
linkcolor=blue,
filecolor=blue,
urlcolor=blue,
citecolor=blue
}

\graphicspath{ {plot/} }% plot path

\definecolor{red1}{RGB}{255,64,64}
\definecolor{blue1}{RGB}{128,255,255}
\definecolor{green1}{RGB}{0,205,0}

\definecolor{red1}{RGB}{255,204,204}
\definecolor{blue1}{RGB}{204,204,255}
\definecolor{light-gray}{gray}{0.7}

\begin{document}

\setlength{\abovedisplayskip}{8pt}
\setlength{\belowdisplayskip}{8pt}
\setlength{\abovedisplayshortskip}{8pt}
\setlength{\belowdisplayshortskip}{8pt}

\title{\vspace*{-1cm}Single Proxy Synthetic Control}
 \author{
  Chan Park$^{1}$ and Eric J. Tchetgen Tchetgen$^{2}$ 
  \\[0.2cm]
  {\small 1: Department of Statistics, University of Illinois, Urbana-Champaign}\\
  {\small 2: Department of Statistics and Data Science, The Wharton School, University of Pennsylvania}\\
    }
 \date{}
  \maketitle
\begin{abstract}
Synthetic control methods are widely used to estimate the treatment effect on a single treated unit in time-series settings. A common approach to estimate synthetic control weights is to regress the treated unit's pre-treatment outcome and covariates' time series measurements on those of untreated units via ordinary least squares. However, this approach can perform poorly if the pre-treatment fit is not near perfect, whether the weights are normalized or not. In this paper, we introduce a single proxy synthetic control approach, which views the outcomes of untreated units as proxies of the treatment-free potential outcome of the treated unit, a perspective we leverage to construct a valid synthetic control. Under this framework, we establish an alternative identification strategy and corresponding estimation methods for synthetic controls and the treatment effect on the treated unit. Notably, unlike existing proximal synthetic control methods, which require two types of proxies for identification, ours relies on a single type of proxy, thus facilitating its practical relevance. Additionally, we adapt a conformal inference approach to perform inference about the treatment effect, obviating the need for a large number of post-treatment observations. Lastly, our framework can accommodate time-varying covariates and nonlinear models. We demonstrate the proposed approach in a simulation study and a real-world application.

\end{abstract}
\noindent%

\newpage

\section{Introduction}

Synthetic control methods have grown popular for estimating the treatment effect of an intervention in settings where a single unit is treated and pre- and post-treatment time series data are available on the treated unit and a heterogeneous pool of untreated control units \citep{Abadie2003, Abadie2010}. In the absence of a natural control unit, the main idea of the approach hinges upon constructing a so-called synthetic control, corresponding to a certain weighted average of control units' outcomes (and potentially covariates), obtained by matching the outcome time series of the treated unit to the weighted average in the pre-intervention period, to the extent empirically feasible. The resulting synthetic control is then used to forecast the treatment-free potential outcome of the treated unit in the post-treatment period, therefore delivering an estimate of the treatment effect by comparing the treated unit’s outcome to the synthetic control forecast. 

There is a fast-growing literature concerned with developing and improving approaches to constructing synthetic control weights. Following \citet{Abadie2010}, a common approach is to use ordinary (or weighted) least squares by regressing the pre-treatment outcome and available covariates of the treated unit on those of control units, typically restricting the weights to be nonnegative and sum to one; see Section \ref{sec:Review} for a more detailed discussion. Despite intuitive appeal and simplicity, the performance of the standard synthetic control approach may break down in settings where the pre-treatment synthetic control match to the treated unit's outcomes is short of perfect; an eventuality \citet{Abadie2010} warns against. In order to improve the performance of the synthetic control approach in the event of an imperfect pre-treatment match, recent papers have considered alternative formulations of the synthetic control framework. For example, \citet{GSC2017, Amjad2018, ASCM2021, FermanPinto2021, Ferman2021, Shi2023SC} rely on variants of a so-called interactive fixed effects model (IFEM; \citet{Bai2009}). In particular, the latter three papers specify a linear latent factor potential outcome model with an exogenous, common set of latent factors with corresponding unit-specific factor loadings. Under this linear factor model, a key identification condition is that the factor loading of the treated unit lies in the vector space spanned by factor loadings of donor units, and thus, there exists a linear combination of the latter that matches the former exactly. Using the corresponding matching weights, one can therefore construct an unbiased synthetic control of the treated unit's potential outcome which, under certain conditions, can be used to mimic the treated unit's outcome in the post-treatment period, had the intervention been withheld. At their core, these methods substitute the requirement of a perfect pre-treatment match of the outcome of the treated unit and the synthetic control (an empirically testable assumption) with finding a match for the treated unit's factor loadings in the linear span of the donors' factor loadings (an empirically untestable assumption). Despite the growing interest in synthetic control methods, limited research has gone beyond the IFEM or its nonparametric generalizations \citep{Qiu2022, Shi2023SC}; one notable exception is \citet{Blei2022} where the units' outcomes are viewed as averages of more granular study units, allowing for the construction of a synthetic control under specific restrictions on the model of granular study units' outcomes.

In this work, we consider an alternative theoretical framework to formalize the synthetic control approach which obviates a specification of an IFEM. Specifically, we propose to view the synthetic control model from a measurement error perspective, whereby donor units' outcomes stand as error-prone proxy measurements of the treated unit's treatment-free potential outcome. In this framework, a synthetic control outcome can be obtained via a simple form of calibration, say a linear combination of donor units, so that on average, it matches the treated unit's outcome in the pre-treatment period. Whereas the standard IFEM views the treated and control units' outcomes as proxies of latent factors, our approach views donor units' outcomes as direct proxies of the treated unit's treatment-free potential outcome. Thus, the proposed framework shares similarity with the recent proximal synthetic control framework of \citet{Shi2023SC}, which also formalizes donor outcomes as so-called outcome proxies. However, a major distinction is that the latter requires an additional group of proxies (so-called treatment proxies) to identify synthetic control weights; in contrast, our proposed approach relies on a single type of proxies, given by donor units and obviates the need to evoke existence of latent factors. 

Interestingly, similar to the connection between the proximal synthetic control approach of \citet{Shi2023SC} and proximal causal inference for independent and identically distributed (i.i.d.) data \citep{Miao2018, TT2020_Intro}, the proposed synthetic control framework is likewise inspired by the control outcome calibration approach \citep{TT2013_COCA} and its recent generalization to a so-called single proxy control framework \citep{SPC2024} both of which were proposed for i.i.d. samples subject to an endogenous treatment assignment mechanism. Therefore, we aptly refer to our approach as single proxy synthetic control (SPSC) approach. Despite this connection, the synthetic control generalization presents several new challenges related to (i) only observing a single treated unit, and therefore treatment assignment is implicitly conditioned on, and (ii) having access to pre-and post-treatment time series data for a heterogeneous pool of untreated donor units, none of which can serve as a natural control; and (iii) serial correlation and heteroskedasticity due to the time series nature of the data. We tackle each of challenges (i)-(iii) in turn and develop a general framework for single proxy control in a synthetic control setting. The proposed method is implemented in an R package available at \url{https://github.com/qkrcks0218/SPSC}.

% The rest of the paper is organized as follows. In Section \ref{sec:Setup}, we introduce notations and review synthetic control methods. In Section \ref{sec:SPSC}, we establish the identification of a synthetic control under the SPSC framework. In turn, we provide the estimation strategy for the average treatment effect on the treated using the generalized method of moments. Moreover, we apply a recent development in the conformal inference approach to our framework to construct pointwise confidence intervals of the treatment effect. As an extension, we extend our framework to nonparametric settings and discuss how to incorporate covariates in the framework. In Sections \ref{sec:Sim} and \ref{sec:Data}, we perform simulation studies investigating the performance of the proposed method, and illustrate the method by revisiting a financial data regarding the 1907 Panic, respectively. Lastly, in Section \ref{sec:Conclusion}, we provide concluding remarks. More detailed discussions and proofs are relegated to Supplementary Material.

\section{Setup And Review of Existing Synthetic Control Frameworks}	\label{sec:Setup}

\subsection{Setup}

Let us consider a setting where $N+1$ units are observed over $T$ time periods. Units and time periods are indexed by $i \in \{0,1,\ldots,N\}$ and $t \in \{1,\ldots,T\}$, respectively. Following the standard synthetic control setting, we suppose that only the first unit with index $i=0$ is treated, whereas the latter $N$ units with index $i \in \{1,\ldots,N\}$ are untreated control units; these untreated control units are also referred to as donors. Consider a binary treatment indicator $A_{t}$ which encodes whether time $t$ is in the pre-treatment period, in which case $A_{t}=0$ for $t\in \{1,\ldots,T_0\}$, or the post-treatment period, in which case $A_{t}=1$ for $t \in \{T_0+1,\ldots,T\}$, respectively. Thus, $T_0$ is the number of pre-treatment periods and $T_1=T - T_0$ is the number of post-treatment periods. Unless otherwise stated, we assume that $N$ is fixed and $T_0$ and $T_1$ are large with similar order of magnitude. Let $Y_{t}$ and $W_{it}$ denote observed outcomes of the treated unit and the $i$th control unit, respectively, for $i \in \{1,\ldots,N\}$. We define $\bW_t = (W_{1t},\ldots,W_{Nt} )\T \in \R^N$ as the $N$-dimensional vector of the untreated units' outcome at time $t$. We define $\bO_t = (Y_t,\bW_t\T,A_t)$ as the observed data at time $t$. Let $\potY{t}{a}$ and $\potW{it}{a}$ denote the potential outcomes of the treated and $i$th control units, respectively, which one would have observed had, possibly contrary to fact, the treatment been set to $A_t=a$ at time $t$. 

For illustrative purposes, we will consider the following two examples throughout:
\begin{example} \label{example-1}
\citet{Abadie2010} investigated the effects of Proposition 99, a tobacco control program implemented in California in 1988, on cigarette sales in the state. Their empirical analysis considered annual cigarette sales data from California and from $N=38$ other states, corresponding to $Y_t$ and $\bW_t$, respectively. The potential outcome $\potY{t}{0}$ represents California's cigarette sales had Proposition 99 not been implemented. The data covered the period from 1970 to 2000, resulting in $T_0=29$  pre-treatment and $T_1=12$ post-treatment time periods. 
\end{example}

\begin{example} \label{example-2} In Section \ref{sec:Data}, we revisited the analysis by  \citet{Dataset1907} to study the effects of the Panic of 1907 \citep{Moen1992} on the average log stock prices of two trust companies (Knickerbocker and Trust Company of America) that were hypothesized to have been impacted by the Panic. For comparison, a selection of $N=49$ trust companies conjectured to be immune to the Panic served as potential control units. The log stock price of these trust companies defines $Y_t$ and $\bW_t$, respectively. The potential outcome $\potY{t}{0}$ represents the average log prices of Knickerbocker and Trust Company of America had the Panic of 1907 not occurred. The tri-weekly panel data consists of $T_0=217$ pre-treatment and $T_1=167$ post-treatment time periods, respectively.
\end{example}

\hspace{\parindent}Throughout, let $\ind(\mathcal{E})$ denote the indicator function of an event $\mathcal{E}$, i.e., $\ind(\mathcal{E})=1$ if $\mathcal{E}$ is satisfied and $\ind(\mathcal{E})=0$ otherwise. Let $\R$ be the set of real numbers. Let $V_1 \indep V_2 \cond V_3$ denote that $V_1$ and $V_2$ are conditionally independent given $V_3$. Conversely, we use $V_1 \nindep V_2 \cond V_3$ to denote that $V_1$ and $V_2$ are conditionally dependent given $V_3$. Let $0_{p \times d}$, $1_{p \times d}$,  and $I_{p \times p}$ denote the $(p \times d)$-dimensional zero matrix, $(p \times d)$-dimensional matrix with ones,  and $(p \times p)$-dimensional identity matrix, respectively. 

\subsection{Review of Existing Synthetic Control Framework}		\label{sec:Review}

A common target estimand in the synthetic control setting is the average treatment effect on the treated unit (ATT) at time $t$ in the post-treatment periods, i.e., 
\begin{align*}
\tau_t^* 
=
\EXP \big\{ \potY{t}{1} - \potY{t}{0} \big\}
\ , 
\quad \quad
t \in \{T_0+1,\ldots,T\} \ . 
\end{align*}
Note that, by definition, $\potY{t}{1} - \potY{t}{0} = \tau_t^* + \nu_t$ for $t \in \{T_0+1,\ldots,T\}$ where $\nu_t$ is a mean-zero idiosyncratic residual error and, therefore, $\tau_t^*$ may be viewed as a deterministic function of time capturing the expected effect of the treatment experienced by the treated unit if one were to average over the residual $\nu_t$. In Section \ref{sec:Conformal}, we describe an approach for constructing prediction intervals for $\potY{t}{1}-\potY{t}{0}$ by appropriately accounting for the idiosyncratic error term $\nu_t$. To proceed, we make the consistency assumption:
\begin{assumption}[Consistency]	\label{assumption:consistency}
$Y_t = \potY{t}{A_t}$ almost surely and $W_{it} = \potW{it}{A_t}$ almost surely for all $i \in \{1,\ldots,N\}$ and $t\in \{1,\ldots,T\}$.
\end{assumption}
\noindent Additionally, we assume no interference, i.e., the treatment has no causal effect on control units.
\begin{assumption}[No Interference on Control Units]
\label{assumption:noitf}
$\potW{it}{0}=\potW{it}{1}$ almost surely for all $i\in \{1,\ldots,N\}$ and $t \in \{1,\ldots,T\}$.
\end{assumption}
\noindent 
In the context of Example \ref{example-1}, Assumption \ref{assumption:noitf} means that Proposition 99 does not have a causal effect on other states' cigarette sales; a similar interpretation applies to Example \ref{example-2}. 

Under Assumptions \ref{assumption:consistency} and \ref{assumption:noitf}, we have the following result almost surely for $t\in \{1,\ldots,T\}$:
\begin{align*}
&
Y_t 
=
\potY{t}{0} (1-A_t)
+
\potY{t}{1} A_t
\ , 
&&
W_{it} 
=
\potW{it}{0}
=
\potW{it}{1}
\ , \quad 
i \in \{1,\ldots,N\}
\ .
\end{align*}
Therefore, for the post-treatment period, $\potY{t}{1}$ matches the observed outcome $Y_t$ while $\potY{t}{0}$ is unobserved, implying that an additional assumption is required to establish identification of the ATT. 

In the classical synthetic control setting, a further assumption relates the observed outcomes of the untreated units with the treatment-free potential outcome of the treated unit. Specifically, following \citet{Abadie2010} and \citet{FermanPinto2021}, suppose that units' outcomes are generated from the following IFEM \citep{Bai2009} for $t \in \{1,\ldots,T\}$:
\begin{align}	\label{eq-IFEM}
Y_{t}
&
=
\text{\makebox[1.25cm]{$ \tau_{t}^* A_t + $}}
\bmu_{0} \T
\blambda_t
+
e_{0t}
\ ,
&&
\EXP \big( e_{0t} \cond \blambda_t ) = 0
\nonumber
\\
W_{it}
&
=
\text{\makebox[1.25cm]{}}
\bmu_{i} \T
\blambda_t
+
e_{it}
\ , 
&&
\EXP \big( e_{it} \cond \blambda_t ) = 0
\ , 
&&
i\in \{1,\ldots,N\} \ .
\end{align}
Here, $\tau_t^*$ is the fixed, non-random treatment effect at time $t$, $\blambda_t \in \R^r$ is a random $r$-dimensional vector of latent factors which are known a priori to causally impact the treated and donor units, despite being unobserved, and can potentially be nonstationary over time, $\bmu_i \in \R^r$ is a time-fixed $r$-dimensional vector of unit-specific factor loadings, and $e_{it}$ is a random error. For identification, it is typically assumed that the number of latent factors $r$ is no larger than the number of donor units $N$ and the pre-treatment period length $T_0$. 
% Under the interactive fixed effect model \eqref{eq-IFEM}, control units' outcomes can be viewed as proxies of interactive effects, $\bmu_i\T \blambda_t$, subject to measurement error $\epsilon_{it}$. The treated unit's outcome is modeled in a similar manner, with the additional consideration of the treatment effect over the post-treatment period. In addition, the treatment status of the treated unit at each time point depends on the interactive effects $\bmu_i\T \blambda_t$ through the latent factors $\blambda_t$. Therefore, interactive effects are unmeasured confounders of the causal relationship between treatment and the treated unit's outcome where confounding arises to the extent that the factor loadings of the treated unit differ from those of untreated donor units, i.e., $\bmu_i \neq \bmu_0$ for $i\in \{1,\ldots,N\}$. 
Combined with Assumptions \ref{assumption:consistency} and \ref{assumption:noitf}, the IFEM \eqref{eq-IFEM} implies $\potY{t}{0}
=
\bmu_{0} \T
\blambda_t
+
e_{0t}$ and $\potY{t}{1}
=
\tau_{t}^* A_t
+
\potY{t}{0}$ where the ATT is represented as $\tau_t^* = \potY{t}{1}-\potY{t}{0}$ for $t\in \{T_0+1,\ldots,T$\}; note that $\potY{t}{1}-\potY{t}{0}$ is non-random under model \eqref{eq-IFEM}. In addition, if there were a donor whose factor loading matched that of the treated unit, i.e., $\bm{\mu}_i=\bm{\mu}_0$ for some $i\in \{1,\ldots,N\}$, then $W_{it}$ would be unbiased for $\potY{t}{0}$ and, therefore, $\potY{t}{1} - W_{it}$ would be unbiased for the ATT. This suggests that confounding bias of the treatment effect on the treated unit's outcome reflects the extent to which donors' factor loadings differ from the treated unit’s.

% Then, interactive effects, which are linear combinations of the latent factors weighted by the units' factor loading, are used as key objects to model the data generating mechanism in the synthetic control framework. Specifically, donor units' outcomes can be viewed as proxies of their interactive effects subject to measurement error. The treated unit's outcome is modeled in a similar manner, with the additional consideration of the treatment effect over the post-treatment period; see equation \eqref{eq-IFEM}. In addition, the treatment status of the treated unit at each time point depends on the interactive effects through the latent factors (which are time-specific). Combined, the interactive effects are unmeasured confounders of the causal relationship between treatment and the treated unit's outcome where confounding arises to the extent that the factor loadings of the treated unit differ from those of untreated donor units, i.e., $\bmu_i \neq \bmu_0$ for $i\in \{1,\ldots,N\}$.

Next, following \citet{FermanPinto2021} and \citet{Shi2023SC}, suppose that a set of weights $\bgamma^{\dagger} = ( \gamma_{1}^{\dagger},\ldots,\gamma_{N}^{\dagger} ) \T$ satisfies
\begin{align}		\label{eq-ExistSC}
\bmu_0
=
\sum_{i =1}^{N} \gamma_{i}^{\dagger} \bmu_i \ .
\end{align}
Equations \eqref{eq-IFEM} and \eqref{eq-ExistSC} imply that there exists a synthetic control $ \bW_{t}\T \bgamma^\dagger = \sum_{i=1}^{N} \gamma_i^\dagger W_{it}$ satisfying 
\begin{align}		\label{eq-SC Equation}
\potY{t}{0}
=
\bW_{t} \T \bgamma^\dagger
+ 
e_{0t} 
-
\sum_{i=1}^{N} \gamma_i^{\dagger} e_{it} 
\ , 
\quad \quad
t\in \{1,\ldots, T\}
\ .
\end{align}
In the context of Example \ref{example-1}, equation \eqref{eq-SC Equation} means that:
\begin{align} \label{example-1-Usual}
\begin{array}{l}	
\textit{The counterfactual measurement of cigarette sales for California}\\
\textit{had, contrary to fact, Proposition 99 not been implemented}\\
\textit{is an error-prone weighted average of cigarette sales in the other 38 states.}
\end{array}
\end{align}
A similar interpretation holds for Example \ref{example-2}.  Therefore, $	\tau_t^*
=
\EXP \big\{ \potY{t}{1} - \bW_{t} \T \bgamma^\dagger \big\}$ for $t\in \{T_0+1,\ldots,T\}$, i.e., $Y_t - \bW_{t}\T \bgamma^\dagger$ is unbiased for the ATT. Unfortunately, it is impossible to obtain $\bgamma^{\dagger}$ from equation \eqref{eq-ExistSC} because the factor loadings $\bmu_i$ are unknown. Importantly, the synthetic control weights satisfying \eqref{eq-ExistSC} naturally accommodate an imperfect pre-treatment fit as shown in \eqref{eq-SC Equation}, i.e., the synthetic control can significantly deviate from the observed pre-treatment fit, however, the corresponding error is mean zero.

Based on \eqref{eq-SC Equation}, one may consider estimating $\bgamma^\dagger$ via penalized least squares minimization, say:
\begin{align}
\label{eq-OLS}
\widehat{\bgamma}_{\text{PLS}}
=
\argmin_{\bgamma}
\bigg\{
\frac{1}{T_0}
\sum_{t=1}^{T_0}
\big(
Y_t - \bW_{t}\T \bgamma
\big)^2
+
\mathcal{R}( \bgamma )
\bigg\} \ ,
\end{align}
where $\mathcal{R} (\bgamma)$ is a penalty which constraints $\bgamma$. For instance, \citet{Abadie2010} restricts the weight to lie within a simplex, meaning that they are non-negative and sum to one, \citet{Doudchenko2016} uses elastic-net penalization, and \citet{Robbins2017} uses entropy penalization. In words, $\widehat{\bgamma}_{\text{PLS}}$ is obtained by fitting a possibly constrained ordinary least squares (OLS) regression of $Y_t$ on $W_{it}$. Importantly, without penalization, the moment restriction solving \eqref{eq-OLS} reduces to $ \EXP \{ \Psi_{\text{OLS}} (\bO_t \con \bgamma) \} = 0$ for $t\in \{ 1, \ldots, T_0\}$ 
% \begin{align*}
% \EXP \big\{ \Psi_{\text{OLS}} (\bO_t \con \bgamma) \big\} = 0 \ , \ t \in \{ 1, \ldots, T_0\} \ , 
% \end{align*}
where $\Psi_{\text{OLS}} (\bO_t \con \bgamma)
=
\bW_{t}
\big( Y_t - \bW_{t}\T \bgamma \big)$ are standard least squares normal equations. 

However, as discussed in \citet{FermanPinto2021} and \citet{Shi2023SC}, the OLS weights obtained from \eqref{eq-OLS} are generally inconsistent under \eqref{eq-ExistSC} as $T_0$ tends to infinity, which can result in biased estimation of the treatment effect unless $e_{it}$ is exactly zero for all $i$ and $t$; see Supplementary Material \ref{sec:supp:OLS} for details. We remark that this result does not conflict with \citet{Abadie2010} because their synthetic control weights are assumed to satisfy a perfect pre-treatment fit; specifically, there exist values $\bgamma^{\#} = (\gamma_{1}^{\#},\ldots,\gamma_{N}^{\#}) \T$ satisfying
\begin{align}		\label{eq-ExistSC-Abadie}
\potY{t}{0}
=
\bW_{t}\T \bgamma^{\#}
\ , 
\quad \quad 
t\in \{1,\ldots,T_0\} \ .
\end{align}
In the context of Example \ref{example-1}, equation \eqref{eq-ExistSC-Abadie} means that:
\begin{align} \label{example-1-Abadie}
\begin{array}{l}	
\textit{The counterfactual measurement of cigarette sales for California}\\
\textit{had, contrary to fact, Proposition 99 not been implemented}\\ 
\textit{is equal to weighted averages of cigarette sales in the other 38 states.}
\end{array}
\end{align}
Example \ref{example-2} follows a similar interpretation. 
Note that \eqref{eq-ExistSC-Abadie} is distinct from condition \eqref{eq-ExistSC} of  \citet{FermanPinto2021} and \citet{Shi2023SC}, as reflected in their interpretations \eqref{example-1-Usual} and \eqref{example-1-Abadie}. Moreover, as discussed in \citet{FermanPinto2021}, \eqref{eq-ExistSC-Abadie} can be expected to hold approximately under \eqref{eq-ExistSC} when the variance of the error $e_{it}$ in \eqref{eq-IFEM} becomes negligible as $T_0$ becomes large. Specifically, in a noiseless setting where $e_{it}=0$ almost surely for all $i \in \{0,1,\ldots,N\}$, \eqref{eq-IFEM} and \eqref{eq-ExistSC} imply \eqref{eq-ExistSC-Abadie} because \eqref{eq-SC Equation} becomes equivalent to \eqref{eq-ExistSC-Abadie}; see \citet{Abadie2010} for related results, and Sections 1 and 3.1 of \citet{FermanPinto2021}, and Section 2 \citet{Shi2023SC} for detailed discussions.

Recently, \citet{Shi2023SC} introduced a proximal causal inference framework for synthetic controls. Specifically, they assume that they have also observed proxy variables $\bZ_{t} = (Z_{1t},\ldots,Z_{Mt})\T$ a priori known to satisfy the following condition in the pre-treatment period:
\begin{align}		\label{eq-Shi-proxy}
\bZ_{t} \indep
\big( Y_t, \bW_{t} \big)
\cond \blambda_t 
\ , 
\quad \quad
t\in \{1,\ldots,T_0\} \ .
\end{align}
A reasonable candidate for $\bZ_{t}$ maybe the outcome of units excluded from the donor pool; see \citet{Shi2023SC} for alternative choices of proxies. Then, under Assumptions \ref{assumption:consistency} and \ref{assumption:noitf}, the IFEM \eqref{eq-IFEM}, condition \eqref{eq-ExistSC}, and the existence of proxies satisfying \eqref{eq-Shi-proxy}, the synthetic control weights $\bgamma^\dagger$ in \eqref{eq-ExistSC} satisfy $\EXP ( Y_t - \bW_{ t}\T \bgamma^\dagger \cond \bZ_{ t} ) = 0$ and $ \EXP \{ \Psi_{\PI}(\bO_t \con \bgamma^\dagger) \} = 0$ for $t\in \{1,\ldots,T_0\}$  where $\Psi_{\PI}(\bO_t \con \bgamma) = \bg  ( \bZ_{ t} ) \big( Y_t - \bW_{ t}\T \bgamma \big)$; here, $\bg$ is a user-specified function of $\bZ_{t}$ with $\text{dim}(\bg) \geq d$. Based on this second result, one can estimate the synthetic control weights as the solution to the generalized method of moments (GMM) \citep{GMM1982}, i.e., $\widehat{\bgamma}_{\PI} 
= (\widehat{\gamma}_{\PI,1}, \ldots, \widehat{\gamma}_{\PI,N}) \T$ is the minimizer of $T_0^{-1} \sum_{t=1}^{T_0}
\big\{ \Psi_{\PI}(\bO_t \con \bgamma) \big\}\T 
\widehat{\Omega}
\big\{ \Psi_{\PI}(\bO_t \con \bgamma) \big\} $ where $\widehat{\Omega}$ is a user-specified symmetric and positive-definite weight matrix. Importantly, in contrast to the OLS-based estimator $\widehat{\bgamma}_{\text{PLS}}$ in \eqref{eq-OLS}, the proximal estimator $\widehat{\bgamma}_{\PI}$ is consistent for $\bgamma^\dagger$. Under certain regularity conditions, \citet{Shi2023SC} established that the resulting GMM estimator of the ATT is consistent and asymptotically normal. For instance, in the special case of constant ATT, i.e., $\tau_t^* = \tau^*$ for all $t\in \{T_0+1,\ldots,T\}$, the estimator $T_1^{-1} \sum_{t=T_0+1}^{T} (Y_t - \bW_{ t}\T \widehat{\bgamma}_{\PI})$ is consistent for $\tau^*$; see Section 3.2 of \citet{Shi2023SC} for details.

\section{Single Proxy Synthetic Control Approach}		\label{sec:SPSC}

\subsection{Assumptions} \label{sec:SPSC:Assumption}

In this section, we provide a novel synthetic control approach which obviates the need for an IFEM, and, in fact, does not necessarily postulate the existence of a latent factor $\blambda_t$. At its core, the approach views the outcomes of the untreated units $W_{it}$ as proxies for the treatment-free potential outcome of the treated unit $\potY{t}{0}$, which is formally stated as follows:
\begin{assumption}[Proxy] \label{assumption:valid proxy}
There exists a function $h^*: \R^N \rightarrow \R$ satisfying 
\begin{align*}
&
h^*( \bW_{t} )  \nindep \potY{t}{0} 
\ , \quad \quad
t\in \{1, \ldots, T_0\} \ . 
\end{align*}
\end{assumption}
Assumption \ref{assumption:valid proxy} encodes that a function of the untreated units' outcomes $\bW_{t}$ is associated with and, therefore, predictive of $\potY{t}{0}$ at time $t\in \{1,\ldots,T_0\}$. In terms of Example \ref{example-1}, Assumption \ref{assumption:valid proxy} means that there exists a function of 38 states' cigarette sales that is associated with cigarette sales in counterfactual California where Proposition 99 was not implemented; a similar interpretation also applies to Example \ref{example-2}. Note that Assumption \ref{assumption:valid proxy} allows for the existence of irrelevant donors among the donor pool, i.e., some untreated units can be independent of $\potY{t}{0}$ as long as the remaining untreated units are associated with the latter. Additionally, we make the following assumption for $h^*$:
\begin{assumption}[Existence of a Synthetic Control Bridge Function] \label{assumption:SC}
For all $t\in \{1,\ldots,T\}$, there exists a synthetic control bridge function $h^*: \R^N \rightarrow \R$ satisfying
\begin{align} \label{eq-SC Counterfactual Eq}
\potY{t}{0} = \EXP \big\{
h^*(\bW_{ t})
\cond
\potY{t}{0}
\big\} \text{ almost surely}.
\end{align}
\end{assumption}
\noindent Assumption \ref{assumption:SC} is the key identification assumption of the SPSC framework. It posits the existence of a synthetic control $h^*(\bW_{ t})$ that is conditionally unbiased for $\potY{t}{0}$. In words, there exists a function of donors $h^*$, possibly nonlinear, whose conditional expectation given $\potY{t}{0}$ recovers $\potY{t}{0}$; the function $h^*$ is a kind of bridge functions \citep{Miao2018, TT2020_Intro}, and we aptly refer to $h^*$ as a \textit{synthetic control bridge function} in this paper. The synthetic control bridge function $h^*$ is a solution to the Fredholm integral equation of the first kind \eqref{eq-SC Counterfactual Eq}, and sufficient conditions for the existence of a solution are well-studied in previous related works developed under i.i.d. settings such as \citet{Miao2018} and \citet{Cui2023}; see Supplementary Material \ref{sec:supp:Exist h} for details. Importantly, Assumption \ref{assumption:SC} may still hold in non-i.i.d. settings, such as when $(\potY{t}{0}, \bW_t)$ is non-stationary; see Supplementary Material \ref{sec:supp:IFEM} for further details.

In particular, if $h^*$ has a linear form, say $h^*(\bW_{ t}) = \bW_{ t}\T \bgamma^*$ for some $\bgamma^* \in \R^N$, the assumption implies the following linear model with an error $\overline{e}_t$:
\begin{align}		\label{eq-MeasurementErrorModel}
\bW_{ t}\T \bgamma^*
=
\potY{t}{0} 
+
\overline{e}_{t}
\ , \quad
\EXP \big\{ \overline{e}_{t} \cond \potY{t}{0} \big\} = 0 \text{ almost surely for  all $t\in \{1,\ldots,T\}$} \ .
\end{align}
Regression model \eqref{eq-MeasurementErrorModel} essentially implies that $\potY{t}{0}$ falls in the linear span of $\EXP \big\{ \bW_{ t} \cond \potY{t}{0} \big\}$, up to a mean zero residual. Thus, Assumption \ref{assumption:SC} may be interpreted as follows for Example \ref{example-1}:
\begin{align} \label{example-1-SPSC}
\hspace*{-0.25cm}
\begin{array}{l}	
\textit{There exists a weighted average of cigarette sales for the 38 donor states}
\\
\textit{which constitutes an error-prone 
 counterfactual measurement of cigarette sales for California}\\
\textit{had, contrary to fact, Proposition 99 not been implemented.}
\end{array}
\end{align}

Assumption \ref{assumption:SC} plays an analogous role as condition \eqref{eq-ExistSC} in \citet{FermanPinto2021} and \citet{Shi2023SC} and condition \eqref{eq-ExistSC-Abadie} in \citet{Abadie2010} in that it establishes a relationship between $\potY{t}{0}$ and $\bW_t$; however, Assumption \ref{assumption:SC} is fundamentally different from these assumptions. In particular, condition \eqref{eq-ExistSC} implies that the counterfactual outcome $\potY{t}{0}$ is equal to the synthetic control $\bW_t\T \bgamma^*$ plus an error; in contrast, Assumption \ref{assumption:SC} with a linear $h^*$ implies that the synthetic control $\bW_t\T \bgamma^*$ is equal to the couterfactual outcome $\potY{t}{0}$ plus a residual error. This distinction highlights that Assumption \ref{assumption:SC} and condition \eqref{eq-ExistSC} can be viewed as reversed assumptions: they differ in which variable is treated as an error-prone version of the other. Lastly, condition \eqref{eq-ExistSC-Abadie} is a special case of the former two cases where the residual error is assumed to be exactly zero, i.e., noiseless setting. Consequently, in the pre-treatment periods, Assumption \ref{assumption:SC} is strictly weaker than condition \eqref{eq-ExistSC-Abadie} because $\overline{e}_t$ is not necessarily zero.

Unlike condition \eqref{eq-ExistSC}, Assumption \ref{assumption:SC} obviates the need for latent factors, their corresponding factor loadings, the IFEM \eqref{eq-IFEM}, or any related latent factor models. Instead, Assumption \ref{assumption:SC} simply states that it is possible to construct a function of the control units' outcomes $h^*(\bW_t)$ which is conditionally unbiased for the treatment-free potential outcome of the treated units $\potY{t}{0}$, without requiring assumptions about how these outcomes are generated.
%ETT. THIS DOES NOT MAKE SENSE TO ME, HOW CAN THE POTENTIAL OUTCOME BE EXOGENOUS??? IM TAKING THIS OUT %This perspective can be particularly useful when the counterfactual outcome $\potY{t}{0}$ is considered as an \textit{exogenous} variable generated from agnostic data generating mechanism and $\bW_t$ is viewed as an  \textit{endogenous} variables containing information of $\potY{t}{0}$, up to some error. 
From this viewpoint, $h^*$ in Assumption \ref{assumption:SC} serves as a \textit{bridge function} relating $\bW_t$ and $\potY{t}{0}$ in that $h^*(\bW_t)$ is an error-prone version of $\potY{t}{0}$. This perspective can be illustrated in Example \ref{example-1}: cigarette sales in counterfactual California, had Proposition 99 not been implemented, are viewed as a variable \textit{a priori} determined by an unknown mechanism, while cigarette sales in the other 38 states are seen as error-prone transformations of this counterfactual outcome. Then, Assumption  \ref{assumption:SC} implies that cigarette sales in counterfactual California can be recovered by aggregating these latter variables up to a mean-zero error. 

Moreover, this perspective aligns with existing statistical literature.  In particular, model \eqref{eq-MeasurementErrorModel} is reminiscent of a nonclassical measurement model \citep{Carroll2006, Freedman2008}. From a regression model perspective, the donors' outcomes $\bW_{t}$ and the treated unit's treatment-free potential outcome $\potY{t}{0}$ in model \eqref{eq-MeasurementErrorModel} can be viewed as dependent and independent variables, respectively. This may appear somewhat unconventional at first glance, as some previous synthetic control methods treat $\potY{t}{0}$ and $ \bW_{t} $ as dependent and independent variables, respectively, in estimation of synthetic control weights. To be more precise, they use equation \eqref{eq-OLS} to estimate the synthetic control weights by regressing $\potY{t}{0}$ on $\bW_{t}$ using standard ordinary (or weighted) least squares. However, as model \eqref{eq-MeasurementErrorModel} suggests, our framework is different from previous works in synthetic control and better aligned with regression calibration techniques in measurement error literature \citep{Carroll2006} in that we view the problem as the reverse regression model of $\bW_{t}$ on $\potY{t}{0}$. From this perspective, synthetic control weights $\bgamma^*$ are sought to make the weighted response $\bW_{ t}\T \bgamma^*$ as close as possible to the regressor $\potY{t}{0}$. 

To summarize, the SPSC framework differs from existing synthetic control frameworks in its identifying assumptions and interpretation of the synthetic control. Specifically, in the SPSC framework, the synthetic control is viewed as an error-prone outcome measurement (see \eqref{eq-MeasurementErrorModel}), eliminating the need for a generative model for $Y_t^{(0)}$. In contrast, existing approaches interpret the synthetic control as either the projection of the outcome onto the donor's outcome space (see \eqref{eq-SC Equation} and \eqref{eq-OLS}) or the outcome itself (see \eqref{eq-ExistSC-Abadie}). Despite these differences, both frameworks share key similarities. In both frameworks, synthetic controls are constructed by weighting donor units to optimally match the treated unit during the pre-treatment period, though the matching criteria differ, as previously noted. Furthermore, synthetic controls in both approaches serve as unbiased forecasts of the mean treatment-free potential outcome, $\EXP \big\{ \potY{t}{0} \big\}$, enabling treatment effect estimation by comparing observed outcomes $Y_t$ to the synthetic controls over the post-treatment period. Additionally, like other synthetic control methods, the SPSC framework accommodates time-varying confounders, distinguishing it from difference-in-differences approaches. Most importantly, the SPSC framework is compatible with the IFEM, as shown in the next section. Thus, while the interpretation of the synthetic control differs, most features of existing synthetic control approaches carry over to the SPSC framework.

\subsection{A Generative Model} \label{sec:IFEM discussion}

While, in principle, Assumptions \ref{assumption:valid proxy} and \ref{assumption:SC} do not require a generative model, it is instructive to consider a model compatible with these assumptions. In this vein, suppose that $\potY{t}{0}$ and $W_{it}$ are generated from the following nonparametric structural equation model \citep{Pearl1995} for $t\in \{1,\ldots,T\}$:
\begin{align} \label{eq-SPSC model}
& 
\potY{t}{0} = f_0 ( \blambda_t, e_{0t} ) \ , \quad  \quad  
W_{it} = f_i ( \blambda_t , e_{it} )
\ , \quad 
i\in \{1,\ldots,N\}
\ .
\end{align}
Here, $f_0$ and $f_i$ are structural equations for $\potY{t}{0}$ and $W_{it}$, respectively, $\blambda_t = (\lambda_{1t},\ldots,\lambda_{rt})\T$ is an $r$-dimensional latent factor, and the errors satisfy $e_{it} \indep \blambda_t$ for $i \in \{0,1,\ldots,N\}$ and $e_{0t} \nindep e_{it}$, where the latter condition further strengthens Assumption \ref{assumption:valid proxy} in the sense that $W_{it}$ is relevant for $\potY{t}{0}$ even beyond $\blambda_t$. Figure \ref{fig:DAG} provides graphical representations compatible with Assumption \ref{assumption:valid proxy} and model \eqref{eq-SPSC model}. 

\begin{figure}[!htp]
\centering  

\scalebox{0.6}{
\begin{tikzpicture}
\tikzset{line width=1.5pt, outer sep=0pt,
ell/.style={draw,fill=white, minimum size=1cm, line width=1pt},
ellD/.style={draw,fill=white, minimum size=1cm, line width=1pt, dashed},
empty/.style={fill=white},
};
\node[name=Y0, ell, ellipse] at (5,0) {$\potY{t}{0}$}  ;
\node[name=W, ell, ellipse] at (1,0) {$W_{1t}$}  ;
\node[name=Tag, empty, ellipse] at (3,4) {\Large (a) Assumption \ref{assumption:valid proxy}}  ;
\begin{scope}[>={Stealth[black]}, every edge/.style={draw=black}]
%\draw[->](a1) to (a2);

\path [<->] (Y0.north) edge[bend right=90, line width=1pt, dashed] (W.north);
\end{scope}

\tikzset{line width=1.5pt, outer sep=0pt,
ell/.style={draw,fill=white, minimum size=1cm, line width=1pt},
ellD/.style={draw,fill=white, minimum size=1cm, line width=1pt, dashed},
empty/.style={fill=white},
};
\node[name=Y0, ell, ellipse] at (6+8,0) {$\potY{t}{0}$}  ;
\node[name=eY, ell, ellipse] at (4+8,0) {$e_{0t}$}  ;
\node[name=U, ell, ellipse] at (3+8,2.5) {$\lambda_t$}  ;
\node[name=eW, ell, ellipse] at (2+8,0) {$e_{1t}$}  ;
\node[name=W, ell, ellipse] at (0+8,0) {$W_{1t}$}  ;
\node[name=Tag, empty, ellipse] at (3+8,4) {\Large (b) Model \eqref{eq-SPSC model} with correlated $e_t$}  ;
\begin{scope}[>={Stealth[black]}, every edge/.style={draw=black}]
%\draw[->](a1) to (a2);

\path [->] (U.east) edge[bend left=30, line width=1pt] (Y0.north);
\path [->] (U.west) edge[bend right=30, line width=1pt] (W.north);
\path [<->] (eY.north) edge[bend right=75, line width=1pt, dashed] (eW.north);
\path [->] (eY.east) edge[line width=1pt] (Y0.west);
\path [->] (eW.west) edge[line width=1pt] (W.east);
\end{scope}

\tikzset{line width=1.5pt, outer sep=0pt,
ell/.style={draw,fill=white, minimum size=1cm, line width=1pt},
ellD/.style={draw,fill=white, minimum size=1cm, line width=1pt, dashed},
empty/.style={fill=white},
};
\node[name=Y0, ell, ellipse] at (6+17,0) {$\potY{t}{0}$}  ;
\node[name=eY, ell, ellipse] at (4+17,0) {$e_{0t}$}  ;
\node[name=U, ell, ellipse] at (3+17,2.5) {$\lambda_t$}  ;
\node[name=eW, ell, ellipse] at (2+17,0) {$e_{1t}$}  ;
\node[name=W, ell, ellipse] at (0+17,0) {$W_{1t}$}  ;
\node[name=Tag, empty, ellipse] at (3+17,4) {\Large (c) Model \eqref{eq-SPSC model} with independent $e_t$}  ;
\begin{scope}[>={Stealth[black]}, every edge/.style={draw=black}]
%\draw[->](a1) to (a2);

\path [->] (U.east) edge[bend left=30, line width=1pt] (Y0.north);
\path [->] (U.west) edge[bend right=30, line width=1pt] (W.north);
\path [->] (eY.east) edge[line width=1pt] (Y0.west);
\path [->] (eW.west) edge[line width=1pt] (W.east);
\end{scope}

\end{tikzpicture} }
\caption{\small Graphical illustrations for (a) Assumption \ref{assumption:valid proxy}, (b) model \eqref{eq-SPSC model} with correlated errors, and (c) model \eqref{eq-SPSC model} with independent errors. The dashed bow arcs depict the association between two variables. For illustration, we consider $N=1$.}
\label{fig:DAG}
\end{figure}

Under model \eqref{eq-SPSC model}, $\potY{t}{0}$ is determined by $\blambda_t$ and $e_{0t}$. Given this relationship, it is natural to consider a sufficient condition of Assumption \ref{assumption:SC} characterized in terms of $\blambda_t$ and $e_{0t}$, say:
\begin{condition} \label{assumption:SC-structural}
For all $t\in \{1,\ldots,T\}$, there exists a function $h^*: \R^N \rightarrow \R$ that satisfies $\potY{t}{0} = f_0(\blambda_t , e_{0t}) = \EXP \big\{
h^*(\bW_{ t})
\cond
\blambda_t, e_{0t}
\big\}$ almost surely.
\end{condition}
\noindent 
Condition \ref{assumption:SC-structural} is a sufficient condition for Assumption \ref{assumption:SC} because, under Condition \ref{assumption:SC-structural}, we obtain $ \EXP \big\{ h^*(\bW_{ t}) \cond \potY{t}{0} \big\} = \EXP \big[ \EXP \big\{ h^*(\bW_{ t}) \cond \blambda_t, e_{0t} \big\} \cond \potY{t}{0} \big] = \potY{t}{0}$. 

Under model \eqref{eq-SPSC model} and Condition \ref{assumption:SC-structural}, consider the special case where $f_i$ is the IFEM \eqref{eq-IFEM}:
\begin{align}
& f_i (\blambda_t, e_{it}) = \bm{\mu}_{i}\T \blambda_t + e_{it} = \sum_{\ell=1}^{r} \mu_{\ell i} \lambda_{\ell t} + e_{it}
\ , \
&&
\EXP\big( e_{it} \big) 
= 0
\ , \
\EXP \big( e_{it} \cond e_{0t} \big) = \omega_{i} e_{0t}  
\ , \
i \in \{ 0, 1,\ldots,N  \} \ .
\label{eq-IFEM2}
\end{align}
Here, $\omega_i$ is a regression coefficient obtained from regressing the $i$th donor's error $e_{it}$ on the treated unit's error $e_{it}$. We remark that $\omega_0 = 1$ and $\omega_{i} \neq 0$ for some $i \in \{1,\ldots,N\}$, encoding $e_{0t} \nindep e_{it}$. Under the IFEM, Condition \ref{assumption:SC-structural} holds with $h^*(\bW_{ t}) = \bW_{ t}\T \bgamma^*$  if $\bgamma^* = (\gamma_{1}^*,\ldots,\gamma_{N}^*)\T$ solves the following linear system:
\begin{align}
\potY{t}{0} 
&
=
\sum_{i=1}^{N}
\gamma_i^*
\EXP \big( W_{it} \cond \blambda_t, e_{0t} \big)
\quad 
\Leftrightarrow
\quad
\begin{bmatrix}
\mu_{10}
\\
\vdots 
\\
\mu_{r0}
\\
\omega_{0}
\end{bmatrix}
=
\underbrace{
\begin{bmatrix}
\mu_{1 1} & \mu_{1 2} & \cdots  & \mu_{1 N}
\\
\vdots & \vdots & \ddots & \vdots 
\\
\mu_{r 1} & \mu_{r 2} & \cdots  & \mu_{r N}
\\
\omega_{1} & \omega_{2} & \cdots & \omega_{N}
\end{bmatrix}
}_{=\mathcal{A}}
\begin{bmatrix}
\gamma_{1}^*
\\
\gamma_{2}^*
\\
\vdots 
\\
\gamma_{N}^*
\end{bmatrix}  \ .
\label{eq-LinearModel-System}
\end{align}
A sufficient condition for the existence of the weight $\bgamma^*$ is that the matrix $\mathcal{A}$ is of full row rank, which is satisfied under the following sufficient (but not necessary) conditions: (i) $r < N$, i.e., the number of donors $N$ is greater than the number of latent factors, and (ii) the factor loadings $\bm{\mu}_i$ are linearly independent. If the matrix $\mathcal{A}$ is square and invertible, $\bgamma^*$ is uniquely determined. This observation informs that a linear synthetic control satisfying Condition \ref{assumption:SC-structural}, and thus Assumption \ref{assumption:SC}, is likely to exist when the errors are correlated and there are sufficient number of donors, regardless of the distribution of the latent factors and errors.

Since equation \eqref{eq-LinearModel-System} is based on the IFEM, it has interesting connections with previous works that also rely on this model. In order to elucidate these connections, we consider the following alternative representation of equation \eqref{eq-LinearModel-System}:
\begin{align} \label{eq-SingleProximal}
\widetilde{\bmu}_0 = \sum_{i=1}^{N} \gamma_i^* \widetilde{\bmu}_i   \ ,
\end{align}
where $\widetilde{\bmu}_i = \big( \mu_{1i},\ldots,\mu_{ri}, \omega_i \big)\T$ for $i \in \{0,1,\ldots,N\}$. 
As the expression itself indicates, condition \eqref{eq-SingleProximal} is similar to condition \eqref{eq-ExistSC}, a condition used in \citet{FermanPinto2021} and \citet{Shi2023SC}, but there is a notable difference between \eqref{eq-ExistSC} and \eqref{eq-SingleProximal} in how they handle errors $e_{it}$. Specifically, in condition \eqref{eq-SingleProximal}, one can address the residual errors $e_{it}$ by accommodating the regression coefficients $\omega_{i}$ as a component of the unit-specific factor loadings $\widetilde{\bmu}_i$. In contrast, condition \eqref{eq-ExistSC} does not account for these errors. Consequently, \eqref{eq-SingleProximal} implies \eqref{eq-ExistSC} because $\bm{\mu}_i$ is a subvector of $\widetilde{\bmu}_i$, indicating that \eqref{eq-SingleProximal} is a stronger condition than \eqref{eq-ExistSC}. However, as stated in Theorem \ref{thm:SC} in Section \ref{sec:SPSC:id}, it is crucial to note that this stronger condition is offset by not requiring an additional condition for establishing identification of the synthetic control weight $\bgamma^*$. In other words, condition \eqref{eq-SingleProximal} alone is sufficient for identification of $\bgamma^*$. On the other hand, condition \eqref{eq-ExistSC} fails to do so, necessitating additional assumptions for identification of $\bgamma^*$, as exemplified by \citet{FermanPinto2021} and \citet{Shi2023SC}. Specifically, \citet{FermanPinto2021} requires either (i) $\VAR(e_{it})=0$ for all $i \in \{0,1,\ldots,N\}$, meaning a noiseless setting, or (ii) $\bgamma^*$ is a minimizer of $\mathcal{V}(\bgamma) = \EXP \{ (e_{0t} - \sum_{i=1}^{N} \gamma_{i} e_{it} )^2 \big\}$, the variance of a linear combination of error terms appearing in \eqref{eq-SC Equation}; see Propositions 1 and 2 of \citet{FermanPinto2021} for details. Interestingly, under (i), all $\omega_{it}$ can be taken as zero, and \eqref{eq-SingleProximal} becomes equivalent to \eqref{eq-ExistSC}, the assumption made by \citet{FermanPinto2021} and \citet{Shi2023SC}. Lastly, in the degenerate case where $\potY{t}{0}$ and $\bW_{t}$ share the same error, i.e., $e_{0t}=e_{1t}=\cdots=e_{N t}$ almost surely, condition \eqref{eq-SingleProximal} implies the perfect fit condition, i.e., condition \eqref{eq-ExistSC-Abadie}, in which case the unconstrained OLS weights \eqref{eq-OLS} are consistent as $T_0$ tends to infinity. 

While the IFEM with correlated errors in \eqref{eq-IFEM2} is useful for motivating the SPSC framework, the standard IFEM typically assumes no correlation among errors, i.e., $\omega_{i}=0$ for all $i\in \{1,\ldots,N\}$ in \eqref{eq-IFEM2}. When the errors are uncorrelated, the solution to equation \eqref{eq-LinearModel-System} may not exist, implying that no linear single proxy synthetic control bridge function satisfies Condition \ref{assumption:SC-structural}. This may suggest that the SPSC framework may not be compatible with a standard IFEM. However, a linear synthetic control bridge function satisfying Assumption \ref{assumption:SC} may still exist under the IFEM with uncorrelated errors, while Condition \ref{assumption:SC-structural} is violated; this is because Condition \ref{assumption:SC-structural} is not a necessary condition of Assumption \ref{assumption:SC}. With additional assumptions regarding the latent factors and errors, it is possible to conceive of a reasonable scenario where the SPSC framework remains valid within the standard IFEM. For instance, if $\blambda_t$ and $\bm{e}_t = (e_{0t}, e_{1t}, \ldots, e_{Nt})\T$ follow multivariate normal distributions with homoskedastic variances, specifically $\blambda_t \sim N_r (\bm{\nu}_t, \Sigma_{\lambda})$ and $\bm{e}_t \sim N_{N+1}(0_{(N+1)\times1}, \Sigma_{e})$, then a linear synthetic control satisfying Assumption \ref{assumption:SC} exists even when $\Sigma_{e}$ is a diagonal matrix; see Supplementary Material \ref{sec:supp:IFEM} for details.  In essence, such circumstances may arise because, despite the uncorrelated errors, $\bW_{t}$ and $\potY{t}{0}$ remain associated through the latent factors $\blambda_t$, allowing for the possibility of a linear single proxy synthetic control to exist; see Figure \ref{fig:DAG} (c) for a graphical illustration. In summary, while uncorrelated errors may undermine the plausibility of the SPSC framework, it can still be valid if certain conditions on $(\blambda_t, \bm{e}_t)$ are met such as the normality assumption.

\subsection{Identification of the Synthetic Control and the Treatment Effect} \label{sec:SPSC:id}

As a direct consequence of Assumptions \ref{assumption:consistency}, \ref{assumption:noitf}, \ref{assumption:valid proxy}, and \ref{assumption:SC}, the synthetic control bridge function $h^*$ can be represented as a solution to the moment equation given in the following result:
\begin{theorem}	\label{thm:SC}
Under Assumptions \ref{assumption:consistency}, \ref{assumption:noitf}, \ref{assumption:valid proxy}, and \ref{assumption:SC}, the synthetic control bridge function $h^*$ satisfy $\EXP \big\{ h^* ( \bW_{ t} ) \cond Y_t \big\} = Y_t $ almost surely for $t\in \{1,\ldots,T_0\}$.
\end{theorem}
The proof of the Theorem, as well as all other proofs, are provided in Supplementary Material \ref{sec:supp:proof}. Theorem \ref{thm:SC} motivates our approach for estimating the synthetic control bridge function $h^*$, as it only involves the observed data. Another consequence of Assumptions \ref{assumption:consistency}, \ref{assumption:noitf}, \ref{assumption:valid proxy}, and \ref{assumption:SC}, is that, as formalized in Theorem \ref{thm:ATT} below, the synthetic control bridge function $h^*(\bW_{ t})$ can be used to identify $\tau_t^*$:
\begin{theorem}	\label{thm:ATT}
Under Assumptions \ref{assumption:consistency}, \ref{assumption:noitf}, \ref{assumption:valid proxy}, and \ref{assumption:SC}, we have that $\EXP \big\{ \potY{t}{0} \big\} = \EXP \big\{ h^*(\bW_{ t}) \big\}$ for any $t\in \{1,\ldots,T\}$. Additionally, the ATT is identified as $	\tau_t^*
=
\EXP
\big\{
Y_t - h^*(\bW_{ t})
\big\}$ for $t\in \{T_0+1,\ldots,T\}$.
\end{theorem}
Theorem \ref{thm:ATT} provides a theoretical basis for the use of the synthetic control method to estimate the ATT. Specifically, following \citet{Abadie2003} and \citet{Shi2023SC}, we use $Y_t - h^*(\bW_{ t})$ in a standard time series regression where the ATT is identified as the deterministic component of the decomposition $Y_t - h^*(\bW_{ t}) = \tau_t^* + \epsilon_t$, with $\epsilon_t$ representing a mean-zero error. The following Sections elaborate on this approach, first describing how the identification result leads to an estimator of the synthetic control.

To facilitate the exposition, hereafter in the main text, we restrict attention to inference under a linear bridge function, i.e., $h^*(\bW_{ t} ) = \bW_{ t} \T \bgamma^*$, while allowing for the possibility for $\bgamma^*$ not to be unique. In Supplementary Material \ref{sec:supp:nonparametric full}, we present the more general case where $h^*$ is nonparametric.

\subsection{Estimation and Inference of the Treatment Effect Under a Linear Bridge Function}		\label{sec:Estimation}

We first discuss estimation of the synthetic control weights $\bgamma^*$. We consider the following time-invariant estimating function for the pre-treatment periods:
\begin{align}	\label{eq-GMM-moment-old}
\Phi_{\pre} (\bO_t \con \bgamma)
=
\bh (Y_t)
\big( Y_t - \bW_{ t} \T \bgamma \big)
\ , \quad \quad
t \in \{1,\ldots,T_0\} \ .
\end{align}
Here, $\bh: \R \rightarrow \R^p$ is a $p$-dimensional user-specified function of the treated unit's outcome. Theorem \ref{thm:SC} implies that the estimating function $\Phi_{\pre}$ satisfies $\EXP \{ \Phi_{\pre} (\bO_t \con \bgamma^*) \} = 0$ for $t\in \{1,\ldots,T_0\}$, indicating that the estimating function $\Phi_{\pre}$ can be used to obtain an estimator of $\bgamma^*$. An important remark on $\bh$ is that the dimension of $\bh$ can be smaller than the number of donors, i.e., $p < N$. Therefore, $\bh$ can be specified as a simple function, e.g., $\bh(y) = y$.

It is instructive to note that solving the estimating equation $\EXP \big\{ \Phi_{\pre}(\bO_t \con \bgamma) \big\}=0$ has a close connection to performing an instrumental variable regression. To illustrate this, consider a simple setting where $\bh(y) = y$ and $N = 1$, along with an alternative form of model \eqref{eq-MeasurementErrorModel} for $t \in \{ 1, \ldots, T_0\}$:
\begin{align}	\label{eq-IV interpretation}
&
W_t \gamma^*
=
Y_t
+
\overline{e}_{t}
&&
\Leftrightarrow
&&
Y_t
=
W_t \gamma^*
-
\overline{e}_{t}
\ , \quad
\EXP \big( \overline{e}_{t} \cond Y_t \big) = 0 \text{ almost surely}.
\end{align}
One might attempt to interpret the model on the right-hand side as a standard regression model, treating $Y_t$ as the response variable and $W_t$ as the explanatory variable. However, such an interpretation would not be correct, as the error term $-\overline{e}_t$ is orthogonal to the response variable $Y_t$. Instead, the right-hand side model exhibits the following properties: (i) the error term $-\overline{e}_{t}$ is correlated with $W_t$ (as induced from the left-hand side model), making $W_t$ an endogenous explanatory variable on the right-hand side model; (ii) the error $-\overline{e}_{t}$ is orthogonal to $Y_t$; and (iii) $Y_t$ is correlated with $W_t$ under Assumption \ref{assumption:valid proxy}. Thus, $Y_t$ can serve as an instrumental variable for $W_t$, allowing for an instrumental variable regression estimator, where $Y_t$ and $W_t$ are used as the instrument and the endogenous explanatory variable, respectively. This estimator is given by $\widehat{\gamma}_{\text{IV}} =  \big( T_0^{-1} \sum_{t=1}^{T_0}  Y_t W_t \big)^{-1} \big( T_0^{-1} \sum_{t=1}^{T_0}  Y_t^2 \big)$. Notably, $\widehat{\gamma}_{\text{IV}}$ is consistent for $\gamma^* =  \big\{ T_0^{-1} \sum_{t=1}^{T_0} \EXP ( Y_t W_t ) \big\}^{-1} \big\{ T_0^{-1} \sum_{t=1}^{T_0} \EXP ( Y_t^2) \big\}$ under some conditions, which is the solution to the estimating equation $\EXP \big\{ \Phi_{\pre} (\bO_t \con \bgamma) \big\} = 0$.  The case for a general $\bh$ and multiple donors can be understood in a similar manner, with the main difference being the use of multiple instrumental variables $\bh(Y_t) \in \R^p$ and multiple explanatory variables $\bW_t \in \R^N$.

The choice of $\bh$ affects the efficiency of the corresponding estimator of $\bgamma^*$ and the treatment effect parameter $\bbeta^*$, which we later define in this section; see Section 2 of \citet{Donald2009} for a similar discussion. Therefore, one could theoretically select the optimal $\bh$ from a set of candidates that minimizes the asymptotic variance of the estimators, thereby maximizing efficiency. For example, $\bh$ can be selected from basis functions such as polynomials up to the $p$th power, where $p$ is determined to minimize the asymptotic variance of the estimators of $(\bgamma^*, \bbeta^*)$; other examples of basis functions include truncated polynomial bases, Fourier basis functions, splines, or wavelets such as the Haar basis; see \citet{XiaohongChen2007} and references therein for more details on how to choose the optimal $\bh$ over a basis function space. However, selecting the optimal $\bh$ can be computationally intensive, and despite this burden, it may yield only marginal gains in efficiency. From a practical standpoint, we use a simple specification for $\bh$, namely the identity function $\bh(y)=y$, leading to $p=1$. In the simulation studies and data analysis, this simple choice of $\bh$ performs well and produces reasonable results compared to competing methods in settings we consider, although we cannot guarantee this to be the case in all settings one might face in practice.

A time-invariant specification of $\bh$ may sometimes lead to poorly behaved estimates of synthetic control weights, particularly in scenarios where the outcomes exhibit nonstationary behavior. To address this, the estimating function can be adapted to accommodate secular trends as follows:
\begin{align}	\label{eq-GMM-moment}
\Psi_{\pre} (\bO_t \con \Beta,  \bgamma)
=
\begin{bmatrix}
\bD_t \big( Y_t - \bD_t\T \Beta \big)
\\
\bg(t, Y_t \con \Beta) \big( Y_t - \bW_t \T \bgamma \big)
\end{bmatrix}
\ , &&
\bg(t,y \con \Beta)
=
\begin{bmatrix}
\bD_t \\ \bh(y - \bD_t\T \Beta)
\end{bmatrix}
\ , &&
t \in \{ 1,\ldots,T_0\} \ .
\end{align}
Here, $\bD_t \in \R^{d}$ is a $d$-dimensional vector of basis functions to de-trend nonstationary behaviors of the outcomes. We assume that $\bD_t$ is selected such that there exists a unique vector $\Beta^*$ satisfying $\EXP \big\{ \potY{t}{0} \big\} = \bD_{t}\T \Beta^*$ for $t \in \{1,\ldots,T_{0}\}$, meaning that the time trend of $\potY{t}{0}$ over the pre-treatment period is correctly specified by the regression model spanned by $\bD_{t}$. The selection of $\bD_t$ can be evaluated by examining the residuals from regressing $Y_t$ on $\bD_{t}$ over the pre-treatment period. For instance, to account for a linear trend, one might select $\bD_t = (1, t/T_0)\T$, where these terms account for an intercept and the drift of a nonstationary process. Alternatively, one could choose $\bD_t = \mathcal{B}_d(t)$, the $d$-dimensional cubic B-spline function, to capture nonlinear trends. While $\bD_{t}$ could also be specified as a dummy vector---allowing each component of $\Beta^*$ to represent time fixed effects for each pre-treatment period---this approach may result in an inconsistent ATT estimator. To ensure valid inference of the ATT while reducing the risk of misspecification, we recommend using a cubic B-spline basis of small to moderate dimension. In both our simulation studies and real-world analysis (Sections \ref{sec:Sim} and \ref{sec:Data}), we used a 6-dimensional cubic B-spline basis, which demonstrated reasonable performance.

The function $\bg: [0,\infty) \otimes \R \rightarrow \R^{d+p}$ can be seen as a basis function for both outcome and time period. Note that the dimension of $\bg$ may be smaller than $N$, which may arise from simple specifications of $\bD_t$ and $\bh$. For instance, we specify $\bh(y)=y$ and $\bD_t=(1,t/T_0)\T$, the dimension of $\bg$ is then equal to three, which may be substantially smaller than the number of untreated units $N$. 

The time-varying estimating function $\Psi_{\pre}$ satisfies $\EXP \{ \Psi_{\pre} (\bO_t \con \Beta^* , \bgamma^*) \} = 0$ for $t\in \{1,\ldots,T_0\}$ under Assumptions \ref{assumption:consistency}, \ref{assumption:noitf}, \ref{assumption:valid proxy}, and \ref{assumption:SC}. This ensures that an estimator of $\bgamma^*$ can be obtained by using the time-varying estimating function $\Psi_{\pre}$ rather than the time-invariant estimating function $\Phi_{\pre}$. In fact, incorporating the time-varying term can potentially enhance the finite sample performance of the proposed estimator in the presence of nonstationary behavior. For instance, under the IFEM \eqref{eq-IFEM2}, we show that incorporating time-varying components reduces the bias of the estimator of $\bgamma^*$ when the latent factor $\blambda_t$ exhibits a secular trend; see Supplementary Material \ref{sec:supp:DT} for this result. Moreover, simulation studies in Section \ref{sec:Sim} suggest that including time-varying components can help reduce bias in the presence of a time trend. Therefore, in the remainder of the paper, we use the time-varying estimating function $\Psi_{\pre}$ unless stated otherwise.

An estimator of $\bgamma^*$ can in principle be obtained based on the empirical counterpart of the moment condition. Because the dimension of $\Psi_{\pre}$ is allowed to be smaller than the dimension of $(\Beta^*, \bgamma^*)$,  standard GMM theory \citep{GMM1982} does not readily apply; typically, standard GMM requires the number of moment equations to be greater or equal to the number of unknown parameters. To regularize the problem, we include a ridge penalty term for $\bgamma^*$. Specifically, the regularized GMM estimator $\widehat{\bgamma}_{\rho}$ with regularization penalty ${\rho} \in (0,\infty)$ is defined as the solution to the following minimization problem:
\begin{align}		\label{eq-GMM-Formula}
&
\big( \widehat{\Beta}, \widehat{\bgamma}_{\rho} \big)
=
\argmin_{(\Beta,\bgamma)}
\Big[
\big\{ 
\widehat{\Psi}_{\pre} ( \Beta , \bgamma ) 
\big\} \T
\widehat{\Omega}_{\pre}
\big\{ 
\widehat{\Psi}_{\pre} ( \Beta , \bgamma ) 
\big\}
+
{\rho} \big\| \bgamma \big\|_2^2
\Big]
\ .    
\end{align}
Here, $\widehat{\Psi}_{\pre} ( \Beta , \bgamma )  = T_{0}^{-1}
\sum_{t=1}^{T_0} 
\Psi_{\pre} (\bO_t \con \Beta , \bgamma) $ is the empirical mean of the estimating function over the pre-treatment periods evaluated at $(\Beta, \bgamma)$. Also, $\widehat{\Omega}_{\pre} =  \text{diag}(I_{d \times d} , \widehat{\Omega}_{\bg} ) \in \R^{(2d+p)\times(2d+p)}$ is a user-specified symmetric positive definite block-diagonal matrix with $\widehat{\Omega}_{\bg} \in \R^{(d+p) \times (d+p)}$, which can simply be set to the identity matrix. Since the first block of $\widehat{\Omega}_{\pre}$ is the identity matrix, 
$\widehat{\Beta}$ reduces to the OLS estimator, i.e., $\widehat{\Beta}
=
\big( \sum_{t=1}^{T_0} \bD_t\bD_t\T \big)^{-1} 
\big( \sum_{t=1}^{T_0} \bD_tY_t \big) $.

Equations \eqref{eq-GMM-moment} and \eqref{eq-GMM-Formula} fortunately admit closed-form solutions. For instance, if $\widehat{\Omega}_{\pre}$ is the identity matrix, we have $\bgamma^* = \bG_{YW}^{*+}\bG_{YY}^* + \zeta_{YW}$ and $\widehat{\bgamma}_{\rho} = ( \widehat{\bG}_{YW}\T \widehat{\bG}_{YW} + {\rho} I_{N \times N} )^{-1} \widehat{\bG}_{YW}\T \widehat{\bG}_{YY}$ where 
\begin{align} \label{eq-Gyw Gyy}
&
\bG_{YW}^* = \frac{1}{T_0} \sum_{t=1}^{T_0} \EXP \big\{ \bg(t, Y_t \con \Beta^*) \bW_{ t} \T \big\} \in \R^{(d+p) \times N}
\ , 
&&
\bG_{YY}^* = \frac{1}{T_0} \sum_{t=1}^{T_0} \EXP \big\{ \bg(t, Y_t \con \Beta^*) Y_t \big\} \in \R^{d+p}
\ , \
\nonumber
\\
&
\widehat{\bG}_{YW} =
\frac{1}{T_0} 
\sum_{t=1}^{T_0} \bg(t, Y_t \con \widehat{\Beta}) \bW_{ t}\T \in \R^{(d+p) \times N}
\ , 
&&
\widehat{\bG}_{YY} = 
\frac{1}{T_0} \sum_{t=1}^{T_0} \bg(t, Y_t \con \widehat{\Beta}) Y_t \in \R^{d+p} \ .
\end{align}
Here, $M^+$ denotes the Moore-Penrose inverse of a matrix $M$, and $\zeta_{YW}$ is an arbitrary vector in the null spaces of $\bG_{YW}^*$, i.e., $\bG_{YW}^* \zeta_{YW}=0$. In general, $\bgamma^*$ may not be unique unless $\bG_{YW}^*$ is of full column rank, in which case $\bgamma^*$ is uniquely determined by $\bgamma^* = \big( \bG_{YW}\sT \bG_{YW}^* \big)^{-1} \bG_{YW}\sT \bG_{YY}^*$. However, when the number of untreated units $N$ is large, a common scenario in many synthetic control settings, the full column rank condition of $\bG_{YW}^*$ may not be met, making $\bgamma^*$ not unique.

A special instance of $\bgamma^*$ is the minimum-norm solution, denoted by $\bgamma_{0}^*= \bG_{YW}^{*+}\bG_{YY}^*$, which corresponds to $\zeta_{YW}=0$. Even if $\bgamma^*$ is not unique, $\bgamma_{0}^*$ remains unique. Moreover, under certain conditions, $\widehat{\bgamma}_{\rho}$ is consistent for $\bgamma_0^*$ as the number of pre-treatment periods $T_0$ goes to infinity and the regularization parameter ${\rho}$ decreases at a sufficiently fast rate. In other words, $\widehat{\bgamma}_\rho$ uniquely converges to $\bgamma_0^*$, allowing us to rely on standard GMM theory as if $\bgamma_0^*$ were the unique solution to the estimating equation. Consequently, we can infer the treatment effect based on the synthetic control with estimated weights $\widehat{\bgamma}_{\rho}$. 

Once the synthetic control weights are estimated, one could in principle estimate the treatment-free potential outcome and the ATT as $\widehat{Y}_{t}^{(0)} = \bW_{ t}\T \widehat{\bgamma}_{\rho}$ and $\widehat{\tau}_t = Y_t - \bW_{ t}\T \widehat{\bgamma}_{\rho}$, respectively, for $t \in \{ T_0 + 1,\ldots,T\}$. Unfortunately, without additional assumptions, it is impossible to perform inference of the ATT $\tau_t^*$ based on $\widehat{\tau}_t$ because the latter will generally fail to be consistent given that we only have access to one observation for each $t$. An alternative is to infer the random treatment effects $\xi_t^* = Y_t^{(1)} - Y_t^{(0)}$ based on pointwise prediction intervals, obviating the need for consistency of $\widehat{\tau}_t$; see Section \ref{sec:Conformal} for details. However, for the remainder of this section, we maintain our focus on inference about the ATT.

We posit a parsimonious working model for the ATT as a function of time. Specifically, we assume that the ATT follows a model indexed by a $b$-dimensional parameter $\bbeta$ via a function $\tau(\cdot \con \cdot): [0,\infty) \otimes \R^{b} \rightarrow \R$. Let $\bbeta^* \in \R^{b}$ be the true parameter satisfying $\tau_t^* 
=
\tau ( t \con \bbeta^* )$ for $t\in \{1,\ldots,T\}$. This parametrization allows us to pool information over time in the post-treatment period to infer $\bbeta^*$ and the ATT. Possible forms for $\tau(t \con \bbeta)$ are given below:
\begin{example}[\textit{Constant Effect}] 
$\tau(t \con \bbeta) = \beta $; this model is reasonable if the treatment yields an immediate, short-term effect which persists over a long period of time.		
\end{example}
\begin{example}[\textit{Linear Effect}] 
$\tau(t \con \bbeta) = \beta_0 + \beta_1 (t- T_0)_+/T_1$ where $(c)_+ = \max(c,0)$ for a constant $c$; this model is appropriate if the treatment yields a gradual, increasing effect over time.
\end{example}
\begin{example}[\textit{Nonlinear Effect}] 
This includes a quadratic model $\tau(t \con \bbeta) = \beta_0 + \beta_1 (t- T_0)_+/T_1 + \beta_2 (t - T_0)_+^2/T_1$, or an exponentially time-varying treatment model $\tau(t \con \bbeta) = \exp \big\{ \beta_0 + \beta_1 (t-T_0)_+/T_1 \big\}$, or a model spanned by nonlinear basis functions, e.g., $\tau(t \con \bbeta) = \mathcal{B}_b\T(t) \bm{\beta} $ where $\mathcal{B}_b(t)$ is the $b$-dimensional cubic B-spline function; this model is appropriate if the treatment yields a nonlinear effect over time.
\end{example}

For tractable inference, we assume that the error process is weakly independent, which is formally stated as follows:
\begin{assumption}[Weakly Dependent Error]	 		\label{assumption:weakdep}
Let $\epsilon_t = Y_t - \bW_{ t} \T \bgamma_0^* - \tau(t \con \bbeta^*)$ for $t\in \{1,\ldots,T\}$. Then, the error process $\{ \epsilon_{1},\ldots,\epsilon_{T} \}$ is weakly dependent, i.e., $\text{corr}(\epsilon_{t}, \epsilon_{t+t'})$ converges to 0 as $t' \rightarrow \pm \infty$.
\end{assumption}
\noindent Assumption \ref{assumption:weakdep} applies to many standard time series models, including autoregressive (AR) models, moving-average (MA) models, and autoregressive moving-average (ARMA) models. 

Along with these conditions, we will consider an asymptotic regime where $T_0, T_1 \rightarrow \infty$ and $T_1/T_0 \rightarrow r \in (0,\infty)$. Specifically, let $\Psi (\bO_t \con \Beta, \bgamma, \bbeta)$ be the following $(2d+p+b)$-dimensional estimating function:
\begin{align*}
% \label{eq-Moment GMM}
\Psi (\bO_t \con \Beta, \bgamma, \bbeta)
=
\begin{bmatrix}
\Psi_{\pre} (\bO_t \con \Beta , \bgamma)
\\
\Psi_{\post} (\bO_t \con \bgamma , \bbeta)
\end{bmatrix}
=
\begin{bmatrix}
(1-A_t)
\bD_t \big( Y_t - \bD_t\T \Beta \big)
\\
(1-A_t)
\bg (t,Y_t \con \Beta)
\big( Y_t - \bW_{ t} \T \bgamma \big)
\\
A_t 
\frac{\partial \tau(t \con \bbeta) }{\partial \bbeta } 
\big\{
Y_t - \bW_{ t} \T \bgamma - \tau (t \con \bbeta)
\big\}
\end{bmatrix}
\in \R^{2d+p+b} \ .
\end{align*}
Then, GMM estimators of the synthetic control weights and treatment effect parameter are obtained as the solution to the following minimization problem:
\begin{align*}
% \label{eq-GMM}
\big(
\widehat{\Beta} 
,
\widehat{\bgamma}_{\rho}
,
\widehat{\bbeta}
\big)
=
\argmin_{(\Beta,\bgamma,\bbeta)}
\Big[
\big\{ \widehat{\Psi}( \Beta, \bgamma, \bbeta) \big\} \T
\widehat{\Omega}
\big\{ \widehat{\Psi}( \Beta, \bgamma, \bbeta) \big\} 
+ \rho \big\| \bgamma \big\|_{2}^{2}
\Big]
\ ,
\end{align*}
where $	\widehat{\Psi}(\Beta, \bgamma, \bbeta)
= T^{-1}
\sum_{t=1}^{T} \Psi (\bO_t \con \Beta, \bgamma, \bbeta)$ is the empirical mean of the estimating function and $\widehat{\Omega} \in \R^{(2d+p+b) \times (2d+p+b)}$ a user-specified symmetric positive definite block-diagonal matrix as $\widehat{\Omega} = \text{diag}(\widehat{\Omega}_{\pre} , \widehat{\Omega}_{\post} )$; for simplicity, $\widehat{\Omega}$ can be chosen as the identity matrix.

Under our assumptions, the following result establishes that $(\widehat{\Beta}, \widehat{\bgamma}_{\rho}, \widehat{\bbeta})$ is asymptotically normal when the number of time periods goes to infinity and the regularization parameter diminishes at $o(N^{-1/2})$ rate:
\begin{theorem}	\label{thm:AN}
Suppose that Assumptions \ref{assumption:consistency}, \ref{assumption:noitf}, \ref{assumption:valid proxy}, \ref{assumption:SC}, \ref{assumption:weakdep}, and regularity conditions in Supplementary Material \ref{sec:supp:AN} hold. Then, as $T \rightarrow \infty$ and $\rho=o(N^{-1/2})$, we have
\begin{align*}
\sqrt{ T }
\left\{ 
\begin{pmatrix}
\widehat{\Beta} 
\\
\widehat{\bgamma}_{\rho}
\\
\widehat{\bbeta}
\end{pmatrix}
-
\begin{pmatrix}
\Beta^*
\\
\bgamma_{0}^*
\\
\bbeta^*
\end{pmatrix}
\right\}
\text{ converges in distribution to }
N \big( 0, \Sigma^*  \big) \ ,
\end{align*}
where $\Sigma^* = \Sigma_1^* \Sigma_2^* \Sigma_1\sT$ is given by
\begin{align*}
&
\Sigma_1^* 
=
\bigg[ \Omega^{*1/2}
\lim_{T \rightarrow \infty} 
\frac{\partial \EXP \big\{ \widehat{\Psi} (\Beta,\bgamma,\bbeta) \big\} }{ \partial (\Beta,\bgamma,\bbeta)\T }
\bigg|_{ \Beta=\Beta^*, \bgamma=\bgamma_0^* , \bbeta=\bbeta^* }
\bigg]^{+} 
\Omega^{*1/2}
\ , 
&&
\Sigma_2^* 
=
\lim_{T \rightarrow \infty} 
\VAR \Big\{ \sqrt{T} \cdot  \widehat{\Psi} (\Beta^*,\bgamma_0^*, \bbeta^*) \Big\} \ .
\end{align*} 
Here, $\Omega^{*1/2}$ is a symmetric positive-definite matrix satisfying $\big( \Omega^{*1/2} \big)^2 = \lim_{T \rightarrow \infty} \widehat{\Omega}$.
\end{theorem}
\noindent Note that $\Sigma^*$ is rank-deficient if the dimension of $\bm{g}$ is smaller than $N$. In this case, the asymptotic distribution is a degenerate normal distribution.  However, this degeneracy only impacts the synthetic control weight estimator $\widehat{\bgamma}_{\rho}$. Therefore, the asymptotic variance of $\widehat{\bbeta}$ remains full rank, even in this case, ensuring that inference regarding $\bbeta^*$ remains valid. For inference about $\bbeta^*$, we propose to use the $(b \times b)$-dimensional bottom-right submatrix of $\widehat{\Sigma}
=
\widehat{\Sigma}_1 
\widehat{\Sigma}_2
\widehat{\Sigma}_1 \T$, which is associated with $\widehat{\bbeta}$. Here, $\widehat{\Sigma}_1$ is defined by
\begin{align*}
\widehat{\Sigma}_1
=
\Big\{
\widehat{\mathcal{G}}\T \widehat{\Omega} \widehat{\mathcal{G}} 
+ 
\text{diag} \big( 0_{d \times d} , \rho \cdot I_{N \times N}, 0_{b \times b} \big)
\Big\}^{-1}
\Big(
\widehat{\mathcal{G}}\T \widehat{\Omega} 
\Big)
\ , \quad 
\widehat{\mathcal{G}} = \frac{\partial \widehat{\Psi} (\Beta,\bgamma,\bbeta) }{ \partial (\Beta,\bgamma,\bbeta)\T }
\bigg|_{ \Beta=\widehat{\Beta}, \bgamma=\widehat{\bgamma}_{\rho}, \bbeta=\widehat{\bbeta} } \ .
\end{align*} 
For $\widehat{\Sigma}_2$, we use a heteroskedasticity and autocorrelation consistent estimator \citep{NW1987, Andrews1991} given the time series nature of the observed sample; see Supplementary Material \ref{sec:supp:HAC} for details. Alternatively, one could implement the block bootstrap; see Supplementary Material \ref{sec:supp:BB} for details. 

Lastly, while Theorem \ref{thm:AN} specifies the required rate for the regularization parameter $\rho$ in relation to $T$, it is still necessary to select a specific value $\rho$ for the given data at hand. In practice, we select $\rho$ using cross-validation; for further details, see Supplementary Material \ref{sec:supp:CV}. In addition, one may have access to exogenous covariates that may be leveraged to improve efficiency. In Supplementary Material \ref{sec:Cov}, we provide details on how to incorporate measured covariates in the SPSC framework. 

\subsection{Conformal Inference of the Treatment Effect} \label{sec:Conformal}

Key limitations of the methodology proposed in the previous Section include (i) a parsimonious model choice for $\tau_t = \tau(t \con \bbeta)$ may be mis-specified and (ii) it potentially requires $T_0$ and $T_1$ both be large in order to rely upon a law of large numbers and central limit theorem for valid asymptotic inference,  so that our large sample analysis can reliably be used to quantify uncertainty associated with the estimated parameters. These limitations may be prohibitive in real-world applications with limited post-treatment follow-up data available. In order to address this specific challenge, previous works such as \citet{Cattaneo2021, Chernozhukov2021} developed prediction intervals to assess statistical uncertainty, obviating the need to specify a model for the treatment effect or large $T_1$. We focus on the conformal inference approach proposed by \citet{Chernozhukov2021} due to its ready adaptation to the SPSC framework. The key idea of the approach is to construct pointwise prediction intervals for the random treatment effects $\xi_t^* = \potY{t}{1} - \potY{t}{0}$ for $t\in \{T_0+1,\ldots,T\}$ by inverting permutation tests about certain null hypotheses concerning $\xi_t^*$. One crucial requirement for the approach is the existence of an unbiased predictor for $\potY{t}{0}$ for $t\in \{1,\ldots, T\}$. In the context of SPSC, the synthetic control $\bW_{ t} \T \bgamma_0^*$ is an unbiased predictor for $\potY{t}{0}$ as established in Theorem \ref{thm:ATT}, and, consequently, their approach readily applies. In what follows, we present the approach in detail. 

Consider an asymptotic regime whereby $T_0$ goes to infinity while $T_1$ is fixed. Let $s \in \{T_0+1,\ldots,T\}$ be a post-treatment period for which one aims to construct a prediction interval for the treatment effect; without loss of generality, we take $s=T_0+1$. The null hypothesis of interest can be expressed as $H_{0, T_0+1} : \xi_{T_0+1}^* = \xi_{0,{T_0+1}}$, where $\xi_{0,{T_0+1}}$ represents a hypothesized treatment effect value. Under $H_{0,T_0+1}$, the treatment-free potential outcome at time ${T_0+1}$ can be identified as $\potY{{T_0+1}}{0} = Y_{T_0+1} - \xi_{0,{T_0+1}}$ and, consequently, pre-treatment outcomes ${ Y_1, \ldots, Y_{T_0} }$ may in fact be supplemented with $Y_{T_0+1} - \xi_{0,{T_0+1}}$  to estimate the synthetic control weights. We may then redefine the pre-treatment estimating function $\Psi_{\pre}$ in equation \eqref{eq-GMM-moment} as follows:
\begin{align*}
\Psi_{\pre} (\bO_t \con \Beta, \bgamma, \xi_{0,{T_0+1}} )
=
\begin{bmatrix}
\bD_t \big( Y_t - A_t \xi_{0,T_0+1}  - \bD_t\T \Beta \big)
\\
\bg \big( t, Y_t - A_t \xi_{0,T_0+1} \con \Beta \big)  \big( Y_t - A_t\xi_{0,T_0+1}  - \bW_t \T \bgamma \big)
\end{bmatrix} 
\ , 
\quad t\in \{1,\ldots,T_0+1\} \ .
\end{align*}
At the minimum-norm synthetic control weights $\bgamma_0^*$, the redefined estimating function is mean-zero for $t \in \{ 1,\ldots, {T_0+1}\}$ under $H_{0,T_0+1}$. Therefore, a GMM estimator $\widehat{\bgamma}(\xi_{0,T_0+1})$ can be obtained by solving the following minimization problem, which is similar to \eqref{eq-GMM-Formula}: 
\begin{align*}
\big( \widehat{\Beta} (\xi_{0,T_0+1}), \widehat{\bgamma}_{\rho} (\xi_{0,T_0+1}) \big)
=
\argmin_{(\Beta,\bgamma)}
\Big[
\big\{ 
\widehat{\Psi}_{\pre} ( \Beta , \bgamma, \xi_{0,T_0+1} ) 
\big\} \T
\widehat{\Omega}_{\pre}
\big\{ 
\widehat{\Psi}_{\pre} ( \Beta , \bgamma, \xi_{0,T_0+1} ) 
\big\}
+
{\rho} \big\| \bgamma \big\|_2^2
\Big]
\ , \ 
\end{align*} 
where $\widehat{\Psi}_{\pre} ( \Beta, \bgamma, \xi_{0,T_0+1} ) 
= 
(T_0+1)^{-1}
\sum_{t=1}^{T_0+1} 
\Psi_{\pre} (\bO_t \con \Beta, \bgamma, \xi_{0,T_0+1})$ and $\widehat{\Omega}_{\pre}$ is the weight matrix used in \eqref{eq-GMM-Formula}. We may then compute residuals $\widehat{\nu}_{t} (\xi_{0,T_0+1}) = Y_t - A_t \xi_{0,{T_0+1}} - \bW_{ t} \T \widehat{\bgamma}_{\rho}(\xi_{0,{T_0+1}}) $, and use these residuals to obtain a p-value for testing the null hypothesis as follows:
\begin{align*}
p_{T_0+1}(\xi_{0, T_0+1}) 
=
\frac{1}{T_0+1	}
\sum_{t=1}^{T_0+1} \ind \Big\{ \big| \widehat{\nu}_{t} (\xi_{0, T_0+1}) \big| \geq \big| \widehat{\nu}_{T_0+1} (\xi_{0, T_0+1}) \big| \Big\}
\ .
\end{align*}
In words, the p-value is the proportion of residuals of 
magnitudes no smaller than the post-treatment residual. Under $H_{0, T_0+1}$ and regularity conditions including that the error $\nu_t = Y_t - \xi_t^* - \bW_{ t} \T \bgamma_0^*$ is stationary and weakly dependent, the p-value is approximately unbiased, i.e., $\Pr \{ p_{T_0+1}(\xi_{0,T_0+1}) \leq \alpha \} = \alpha + o(1)$ as $T_0 \rightarrow \infty$ for a user-specified confidence level $\alpha \in (0,1)$; we refer the readers to Theorem 1 of \citet{Chernozhukov2021} for technical details. Therefore, an approximate $100(1-\alpha)$\% prediction interval for $\xi_{t}^*$ can be constructed by inverting the hypothesis test based on $p_{T_0+1}(\xi_{0, T_0+1})$. This prediction interval is formally defined as
$\mathcal{C}_{T_0+1} (1-\alpha) =
\big\{
\xi
\cond		
p_{T_0+1}(\xi) > \alpha
\big\}$ and can be found via a grid-search. 
% We remark that the proposed conformal inference approach is also valid for constructing a confidence interval for the ATT by replacing $\xi_t^*$ with $\tau_t^*$. 

\section{Simulation}		\label{sec:Sim}

We conducted a simulation study to evaluate the finite sample performance of the proposed estimator under a variety of conditions. Based on the IFEM in \eqref{eq-IFEM}, we considered the following data generating mechanisms with pre- and post-treatment periods of length $T_0 = T_1 \in \{ 50, 100, 250, 500 \}$ and donor pools of size $N = 16$. 

First, for each $t \in \{1,\ldots,T\}$, we generated 4-dimensional latent factors $\blambda_t = (\lambda_{1t},\cdots,\lambda_{4t})\T$ from $N(\bm{\nu}_t,0.25 \cdot I_{4\times 4})$, with $\blambda_t$ independent across time periods. For the mean vector $\bm{\nu}_t=(\nu_{1t},\cdots,\nu_{4t})\T$, we considered the following two specifications for $j\in \{1,\ldots,4\}$:
\begin{align*}
& (\textit{No trend}): \quad \nu_{jt} = 0; 	\quad 
&& (\textit{Linear trend}): \quad \nu_{jt} = t/T_0;	\quad  
\end{align*} 

The latent factor loadings $\bm{\mu}_{i}$ for $i \in \{1,\ldots,16\}$, i.e., latent factor loadings of untreated units,  were specified as follows:
\begin{align*}
\mathfrak{M}
=
\begin{bmatrix}
\bm{\mu}_1 & \cdots & \bm{\mu}_{16}
\end{bmatrix}
=
\begin{bmatrix}
2 & 1.75 & 1.5 & 1.25 & 1 & 0.75 & 0.5 & 0.25 & 0_{1 \times 8} \\
0.8 & 0.8 & 0.6 & 0.6 & 0.4 & 0.4 & 0.2 & 0.2 & 0_{1 \times 8} \\
0 & 0 & 0 & 0 & 0 & 0 & 0 & 0 & 1_{1 \times 8} \\
0 & 0 & 0 & 0 & 0 & 0 & 0 & 0 &  0.5 \cdot 1_{1 \times 8} 
\end{bmatrix}
\in \R^{4 \times 16} \ .
\end{align*}
The latent factor loading $\bm{\mu}_{0}$, i.e., latent factor loading of the treated unit, was specified from either one of the followings:
\begin{align*}
& (\textit{Simplex}): && \bm{\mu}_{0} = (1.125,0.5,0,0)\T = \frac{1}{8} \sum_{i=1}^{8}  \bm{\mu}_{i} ; \quad \quad \quad 
&& (\textit{Non-simplex}): && \bm{\mu}_{0} = (2,1.5,0,0)\T \ .
\end{align*}
Note that condition \eqref{eq-ExistSC} is satisfied with $\bgamma^\dagger = \mathfrak{M}^+ \bm{\mu}_0$, although this vector is not the unique solution. Also, when $\bm{\mu}_0$ is chosen as \textit{(Simplex)}, $\bgamma^\dagger$ lies within a 16-dimensional simplex, thus satisfying the restriction of \citet{Abadie2010}. In contrast, when $\bm{\mu}_0$ is chosen as \textit{(Non-implex)}, $\bgamma^\dagger$ does not belong to this simplex. 

The errors $\bm{e}_t = (e_{0t} ,  e_{1t} , \cdots,  e_{16 t})\T$ were generated independently across time periods from $\bm{e}_t \sim N \big( 0_{16 \times 1} , 0.25 \cdot \text{diag} (\Sigma_{e}, I_{8 \times 8}) \big)$ where $\Sigma_e \in \R^{9 \times 9}$ were chosen from one of the following three matrices with the corresponding $\omega_i$ values in \eqref{eq-LinearModel-System}:
\begin{align*}
&
(\textit{Independent errors}):
&&
\Sigma_{e} = I_{9 \times 9}   \ ;
&&
\omega_0 = 1 , \ 
\omega_1=\cdots=\omega_{16}=0     \ ;
\\
&
(\textit{Correlated errors}):
&&
\Sigma_{e} = 
0.1 \cdot I_{9 \times 9} + 0.9  \cdot 1_{9 \times 9}    \ ;
&&
\omega_0 = 1 , \ 
\omega_1=\cdots=\omega_{8}=0.9 , \
\omega_9=\cdots=\omega_{16}=0    \ ;
\\
&
(\textit{No $Y$ error}):
&&
\Sigma_{e} = \text{diag}(0,I_{8 \times 8})   \ ;
&&
\omega_0 =\cdots=\omega_{16}=0 \ .
\end{align*} 
Under (\textit{Correlated errors}) and (\textit{No $Y$ error}), equation \eqref{eq-LinearModel-System} admit solutions, thus satisfying Condition \ref{assumption:SC-structural} and Assumption \ref{assumption:SC}. In contrast, under (\textit{Independent errors}), equation \eqref{eq-LinearModel-System} does not have a solution, violating of Condition \ref{assumption:SC-structural}. Nonetheless, as we discussed in Section \ref{sec:IFEM discussion}, it is still possible to find a synthetic control bridge function satisfying Assumption \ref{assumption:SC}; see Supplementary Material \ref{sec:supp:IFEM} for details.

With these generated variables, $\potY{t}{0}$ and $W_{it}$ at $t\in \{1,\ldots,T\}$ were generated as $\potY{t}{0} = \bm{\mu}_0\T \bm{\lambda}_t + e_{0t}$ and $W_{it} = \bm{\mu}_i\T \bm{\lambda}_t + e_{it}$ for $i\in \{1,\ldots,N\}$, respectively. Note that the latter eight untreated units $(W_{9t},\cdots,W_{16t})\T$ were independent of $\potY{t}{0}$, which resulted in multiple synthetic control bridge functions satisfying Assumption \ref{assumption:SC}. The potential outcomes under treatment at $t\in \{1,\ldots,T\}$ were generated as $\potY{t}{1} = \potY{t}{0} + 3 A_t + \epsilon_{t}$ where $\epsilon_t$ were generated independently across time periods from $N(0,0.25)$. Note that the ATT was $\tau_t^* = 3$ for $t\in \{T_0+1,\ldots,T\}$.

Using the simulated data, we estimated the aggregated ATT over the post-treatment periods, i.e., $T_1^{-1} \sum_{t=T_0+1}^{T} \tau_t^*$, based on the following six estimators. First, we obtained two ATT estimators based on the proposed SPSC approaches both time-invariant pre-treatment estimating function $\Phi_{\pre}$ in \eqref{eq-GMM-moment-old} and time-varying pre-treatment estimating function $\Psi_{\pre}$ in \eqref{eq-GMM-moment} where the function $\bh$ was chosen as $\bh(y)=y$. We specified the vector $\bD_t$ in $\Psi_{\pre}$ as $\bD_t=\mathcal{B}_6(t)$, the 6-dimensional cubic B-spline function, to adjust a potential time trend. These estimators are referred to as \textit{SPSC-NoDT} and \textit{SPSC-DT}, respectively. For comparison, we also considered two OLS-based ATT estimators based on \eqref{eq-OLS}. In the first OLS-based ATT estimator, we place no regularization on the weight; in the second OLS-based ATT estimator, we followed \citet{Abadie2010} to restrict the weight to be non-negative and its values must add up to one. These estimators are referred to as \textit{OLS-NoReg} and \textit{OLS-Standard}, respectively. Lastly, we implemented two recently developed synthetic control methods by \citet{ASCM2021} and \citet{Cattaneo2021}, which are referred to as \textit{ASC} and \textit{SCPI}, respectively. In our analysis, we implemented the OLS-Standard, ASC, and SCPI estimators using \texttt{synth} \citep{Synth2011}, \texttt{augsynth} \citep{ASCM2023package}, and \texttt{scpi} \citep{scpiPackage2023} R-packages, respectively. Unfortunately, these three packages do not appear to provide readily available standard errors, so the standard errors and empirical coverage rates of these methods are not reported. We repeated the simulation 500 times.

To simplify the discussion, we present only the results under the (\textit{Non-simplex}) case for $\bm{\mu}_0$. The results for the (\textit{Simplex}) case are included in Supplementary Material \ref{sec:supp:Simulation}. Figure \ref{fig:Sim:Constant} summarizes the empirical distribution of the estimators graphically. First, when $\blambda_t$ does not have a trend, all estimators exhibit negligible bias for the ATT regardless of error specifications.  Second, when $\blambda_t$ has a linear trend case, we find that the four estimators from OLS, ASC, and SCPI approaches are biased for the ATT. Although the SPSC-NoDT estimator outperforms these four estimators, it still exhibits non-negligible bias when the errors are independent. In contrast, the SPSC-DT estimator little bias for all error specifications. Note that 95\% Monte Carlo confidence interval for SPSC estimators shrinks as the number of time periods increases, which is consistent with the results established in Section \ref{sec:SPSC}. 

\begin{figure}[!htb]
\vspace*{-0.1cm}
\centering
\includegraphics[width=1\textwidth]{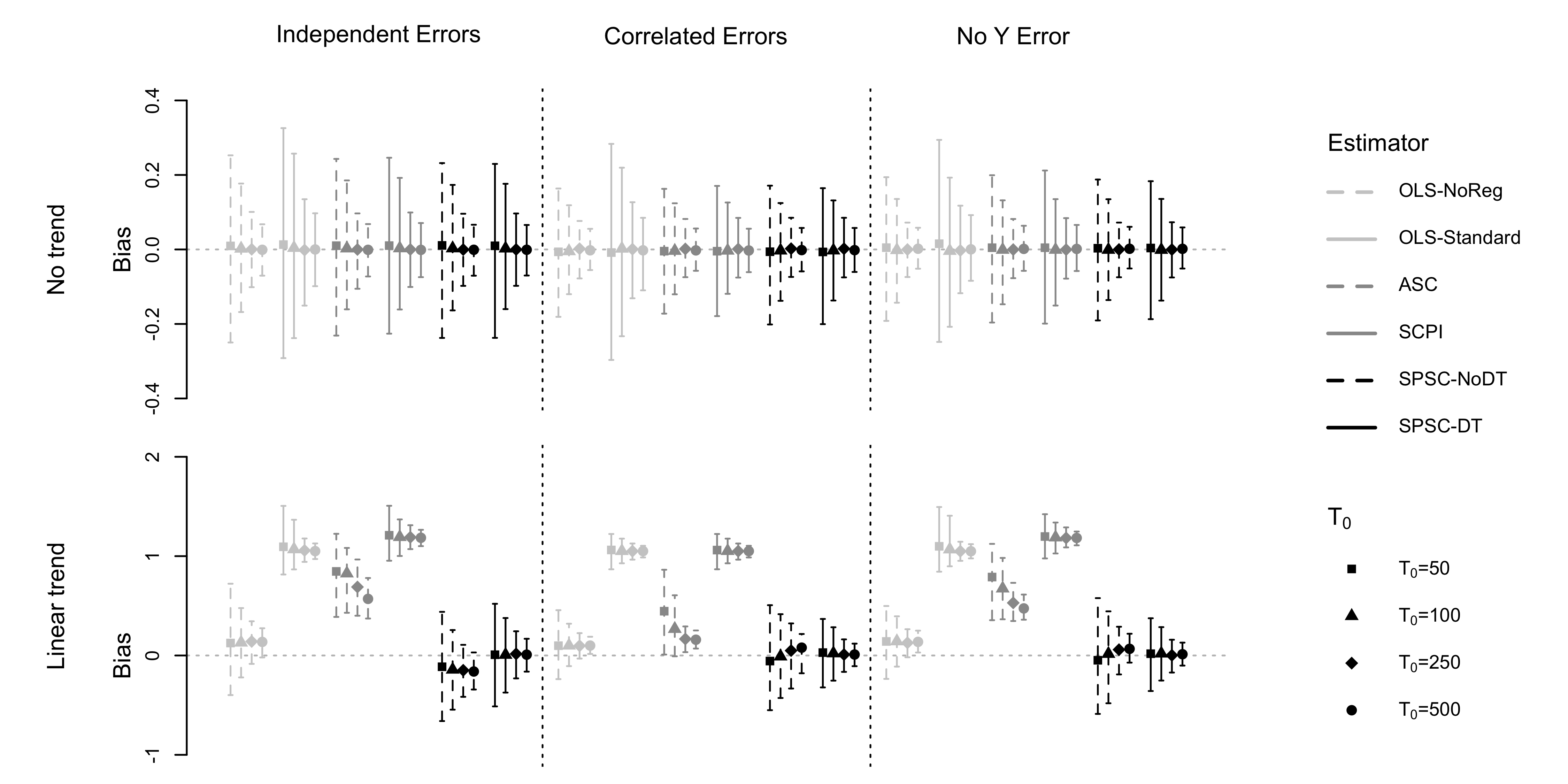}
\vspace*{-0.6cm}
\caption{Empirical Distributions of the Estimators. The top and bottom plots show results when $\blambda_t$ has no trend and a linear trend, respectively. The left, middle, and right panels show results under (\textit{Independent errors}), (\textit{Correlated errors}), and (\textit{No $Y$ error}) for $\bm{e}_t$, respectively. The vertical segments represent 95\% Monte Carlo confidence interval for each estimator.
The dots represent the empirical mean of 500 estimates.
The colors (light gray, gray, and black) and line types (solid and dashed) encode a corresponding estimator, and the shape of the dots encode the length of the pre-treatment period, respectively. The $y$-axis represents the magnitude of bias.}
\label{fig:Sim:Constant} 
\end{figure}

Table \ref{tab:Sim:ATT} provides more detailed summary statistics when errors $\bm{e}_t$ were generated from the (\textit{Independent errors}) case, a common assumption that the standard IFEM make. The results for the other two error specifications are reported in Supplementary Material \ref{sec:supp:Simulation}.  We remark that the OLS-Standard, ASC, and SCPI approaches do not provide a standard error or 95\% confidence interval for the ATT. First, when $\blambda_t$ has no trend, all estimators exhibit negligible bias and achieve the nominal coverage rate, provided that confidence intervals are available. Second, when $\blambda_t$ has a linear trend, the performance of the estimators differs in terms of both mean squared error and coverage rate. We find that the SPSC-DT estimator attains the smallest mean squared error compared to the other estimators, including the SPSC-NoDT estimator. Regarding the coverage rate, confidence intervals based on the OLS-NoReg and SPSC-NoDT estimators fail to attain the nominal coverage rate, especially when $T_0$ and $T_1$ are large due to non-diminishing bias. In contrast, confidence intervals based on the SPSC-DT estimator attain the nominal coverage rate. This demonstrates that accounting for time-varying components in the pre-treatment estimating estimation can significantly improve the performance of the SPSC estimators and is, in fact, necessary to conduct valid inference.

\begin{table}[!htp]
\renewcommand{\arraystretch}{1.1} \centering
\scriptsize
\setlength{\tabcolsep}{2pt} 
\begin{tabular}{c|c|cc|cc|cc|cc|cc|cc|}
\hline
\multicolumn{1}{|c|}{\multirow{3}{*}{$\blambda_t$}} &  \multicolumn{1}{c|}{\multirow{3}{*}{\normalfont Statistics}} & \multicolumn{12}{c|}{\normalfont Estimators and $T_0$} \\ \cline{3-14} 

\multicolumn{1}{|c|}{} & \multicolumn{1}{c|}{}  & \multicolumn{2}{c|}{\normalfont OLS-NoReg} & \multicolumn{2}{c|}{\normalfont OLS-Standard} & \multicolumn{2}{c|}{\normalfont ASC}       & \multicolumn{2}{c|}{\normalfont SCPI}      & \multicolumn{2}{c|}{\normalfont SPSC-NoDT} & \multicolumn{2}{c|}{\normalfont SPSC-DT}   \\ \cline{3-14} 
\multicolumn{1}{|c|}{} & \multicolumn{1}{c|}{}  & \multicolumn{1}{c|}{\normalfont 100} & \multicolumn{1}{c|}{\normalfont 500} & \multicolumn{1}{c|}{\normalfont 100} & \multicolumn{1}{c|}{\normalfont 500}& \multicolumn{1}{c|}{\normalfont 100} & \multicolumn{1}{c|}{\normalfont 500}& \multicolumn{1}{c|}{\normalfont 100} & \multicolumn{1}{c|}{\normalfont 500}& \multicolumn{1}{c|}{\normalfont 100} & \multicolumn{1}{c|}{\normalfont 500}& \multicolumn{1}{c|}{\normalfont 100} & \multicolumn{1}{c|}{\normalfont 500}  \\ \hline

\multicolumn{1}{|c|}{\multirow{7}{*}{$\!\!\!\begin{array}{c}\text{\normalfont \scriptsize No trend}\end{array}\!\!\!$}} & \multicolumn{1}{c|}{\normalfont \scriptsize Bias $(\times10)$} & \multicolumn{1}{c|}{\normalfont \scriptsize 0.02} & {\normalfont \scriptsize -0.01} & \multicolumn{1}{c|}{\normalfont \scriptsize 0.04} & {\normalfont \scriptsize 0.00} & \multicolumn{1}{c|}{\normalfont \scriptsize 0.03} & {\normalfont \scriptsize -0.01} & \multicolumn{1}{c|}{\normalfont \scriptsize 0.03} & {\normalfont \scriptsize -0.01} & \multicolumn{1}{c|}{\normalfont \scriptsize 0.03} & {\normalfont \scriptsize -0.01} & \multicolumn{1}{c|}{\normalfont \scriptsize 0.03} & \multicolumn{1}{c|}{\normalfont \scriptsize -0.01}\\ \cline{2-14}
\multicolumn{1}{|c|}{} & \multicolumn{1}{c|}{\normalfont \scriptsize ASE $(\times10)$} & \multicolumn{1}{c|}{\normalfont \scriptsize 0.89} & {\normalfont \scriptsize 0.37} & \multicolumn{1}{c|}{\normalfont \scriptsize -} & {\normalfont \scriptsize -} & \multicolumn{1}{c|}{\normalfont \scriptsize -} & {\normalfont \scriptsize -} & \multicolumn{1}{c|}{\normalfont \scriptsize -} & {\normalfont \scriptsize -} & \multicolumn{1}{c|}{\normalfont \scriptsize 0.84} & {\normalfont \scriptsize 0.37} & \multicolumn{1}{c|}{\normalfont \scriptsize 0.84} & \multicolumn{1}{c|}{\normalfont \scriptsize 0.37}\\ \cline{2-14}
\multicolumn{1}{|c|}{} & \multicolumn{1}{c|}{\normalfont \scriptsize BSE $(\times10)$} & \multicolumn{1}{c|}{\normalfont \scriptsize 0.97} & {\normalfont \scriptsize 0.38} & \multicolumn{1}{c|}{\normalfont \scriptsize -} & {\normalfont \scriptsize -} & \multicolumn{1}{c|}{\normalfont \scriptsize -} & {\normalfont \scriptsize -} & \multicolumn{1}{c|}{\normalfont \scriptsize -} & {\normalfont \scriptsize -} & \multicolumn{1}{c|}{\normalfont \scriptsize 0.84} & {\normalfont \scriptsize 0.37} & \multicolumn{1}{c|}{\normalfont \scriptsize 0.84} & \multicolumn{1}{c|}{\normalfont \scriptsize 0.37}\\ \cline{2-14}
\multicolumn{1}{|c|}{} & \multicolumn{1}{c|}{\normalfont \scriptsize ESE $(\times10)$} & \multicolumn{1}{c|}{\normalfont \scriptsize 0.92} & {\normalfont \scriptsize 0.37} & \multicolumn{1}{c|}{\normalfont \scriptsize 1.25} & {\normalfont \scriptsize 0.51} & \multicolumn{1}{c|}{\normalfont \scriptsize 0.90} & {\normalfont \scriptsize 0.39} & \multicolumn{1}{c|}{\normalfont \scriptsize 0.92} & {\normalfont \scriptsize 0.39} & \multicolumn{1}{c|}{\normalfont \scriptsize 0.89} & {\normalfont \scriptsize 0.37} & \multicolumn{1}{c|}{\normalfont \scriptsize 0.89} & \multicolumn{1}{c|}{\normalfont \scriptsize 0.37}\\ \cline{2-14}
\multicolumn{1}{|c|}{} & \multicolumn{1}{c|}{\normalfont \scriptsize MSE $(\times100)$} & \multicolumn{1}{c|}{\normalfont \scriptsize 0.85} & {\normalfont \scriptsize 0.14} & \multicolumn{1}{c|}{\normalfont \scriptsize 1.56} & {\normalfont \scriptsize 0.26} & \multicolumn{1}{c|}{\normalfont \scriptsize 0.82} & {\normalfont \scriptsize 0.15} & \multicolumn{1}{c|}{\normalfont \scriptsize 0.84} & {\normalfont \scriptsize 0.15} & \multicolumn{1}{c|}{\normalfont \scriptsize 0.79} & {\normalfont \scriptsize 0.14} & \multicolumn{1}{c|}{\normalfont \scriptsize 0.79} & \multicolumn{1}{c|}{\normalfont \scriptsize 0.14}\\ \cline{2-14}
\multicolumn{1}{|c|}{} & \multicolumn{1}{c|}{\normalfont \scriptsize Coverage (ASE)} & \multicolumn{1}{c|}{\normalfont \scriptsize 0.95} & {\normalfont \scriptsize 0.96} & \multicolumn{1}{c|}{\normalfont \scriptsize -} & {\normalfont \scriptsize -} & \multicolumn{1}{c|}{\normalfont \scriptsize -} & {\normalfont \scriptsize -} & \multicolumn{1}{c|}{\normalfont \scriptsize -} & {\normalfont \scriptsize -} & \multicolumn{1}{c|}{\normalfont \scriptsize 0.93} & {\normalfont \scriptsize 0.96} & \multicolumn{1}{c|}{\normalfont \scriptsize 0.93} & \multicolumn{1}{c|}{\normalfont \scriptsize 0.96}\\ \cline{2-14}
\multicolumn{1}{|c|}{} & \multicolumn{1}{c|}{\normalfont \scriptsize Coverage (BSE)} & \multicolumn{1}{c|}{\normalfont \scriptsize 0.96} & {\normalfont \scriptsize 0.96} & \multicolumn{1}{c|}{\normalfont \scriptsize -} & {\normalfont \scriptsize -} & \multicolumn{1}{c|}{\normalfont \scriptsize -} & {\normalfont \scriptsize -} & \multicolumn{1}{c|}{\normalfont \scriptsize -} & {\normalfont \scriptsize -} & \multicolumn{1}{c|}{\normalfont \scriptsize 0.93} & {\normalfont \scriptsize 0.96} & \multicolumn{1}{c|}{\normalfont \scriptsize 0.93} & \multicolumn{1}{c|}{\normalfont \scriptsize 0.96}\\ \hline
\multicolumn{1}{|c|}{\multirow{7}{*}{$\!\!\!\begin{array}{c}\text{\normalfont \scriptsize Linear trend}\end{array}\!\!\!$}} & \multicolumn{1}{c|}{\normalfont \scriptsize Bias $(\times10)$} & \multicolumn{1}{c|}{\normalfont \scriptsize 1.30} & {\normalfont \scriptsize 1.37} & \multicolumn{1}{c|}{\normalfont \scriptsize 10.70} & {\normalfont \scriptsize 10.51} & \multicolumn{1}{c|}{\normalfont \scriptsize 8.26} & {\normalfont \scriptsize 5.70} & \multicolumn{1}{c|}{\normalfont \scriptsize 11.93} & {\normalfont \scriptsize 11.85} & \multicolumn{1}{c|}{\normalfont \scriptsize -1.42} & {\normalfont \scriptsize -1.61} & \multicolumn{1}{c|}{\normalfont \scriptsize 0.07} & \multicolumn{1}{c|}{\normalfont \scriptsize 0.08}\\ \cline{2-14}
\multicolumn{1}{|c|}{} & \multicolumn{1}{c|}{\normalfont \scriptsize ASE $(\times10)$} & \multicolumn{1}{c|}{\normalfont \scriptsize 1.76} & {\normalfont \scriptsize 0.79} & \multicolumn{1}{c|}{\normalfont \scriptsize -} & {\normalfont \scriptsize -} & \multicolumn{1}{c|}{\normalfont \scriptsize -} & {\normalfont \scriptsize -} & \multicolumn{1}{c|}{\normalfont \scriptsize -} & {\normalfont \scriptsize -} & \multicolumn{1}{c|}{\normalfont \scriptsize 1.85} & {\normalfont \scriptsize 0.84} & \multicolumn{1}{c|}{\normalfont \scriptsize 1.79} & \multicolumn{1}{c|}{\normalfont \scriptsize 0.81}\\ \cline{2-14}
\multicolumn{1}{|c|}{} & \multicolumn{1}{c|}{\normalfont \scriptsize BSE $(\times10)$} & \multicolumn{1}{c|}{\normalfont \scriptsize 2.02} & {\normalfont \scriptsize 0.82} & \multicolumn{1}{c|}{\normalfont \scriptsize -} & {\normalfont \scriptsize -} & \multicolumn{1}{c|}{\normalfont \scriptsize -} & {\normalfont \scriptsize -} & \multicolumn{1}{c|}{\normalfont \scriptsize -} & {\normalfont \scriptsize -} & \multicolumn{1}{c|}{\normalfont \scriptsize 2.24} & {\normalfont \scriptsize 0.99} & \multicolumn{1}{c|}{\normalfont \scriptsize 1.94} & \multicolumn{1}{c|}{\normalfont \scriptsize 0.86}\\ \cline{2-14}
\multicolumn{1}{|c|}{} & \multicolumn{1}{c|}{\normalfont \scriptsize ESE $(\times10)$} & \multicolumn{1}{c|}{\normalfont \scriptsize 1.82} & {\normalfont \scriptsize 0.77} & \multicolumn{1}{c|}{\normalfont \scriptsize 1.16} & {\normalfont \scriptsize 0.41} & \multicolumn{1}{c|}{\normalfont \scriptsize 1.70} & {\normalfont \scriptsize 0.95} & \multicolumn{1}{c|}{\normalfont \scriptsize 0.96} & {\normalfont \scriptsize 0.43} & \multicolumn{1}{c|}{\normalfont \scriptsize 2.08} & {\normalfont \scriptsize 0.97} & \multicolumn{1}{c|}{\normalfont \scriptsize 1.94} & \multicolumn{1}{c|}{\normalfont \scriptsize 0.85}\\ \cline{2-14}
\multicolumn{1}{|c|}{} & \multicolumn{1}{c|}{\normalfont \scriptsize MSE $(\times100)$} & \multicolumn{1}{c|}{\normalfont \scriptsize 5.01} & {\normalfont \scriptsize 2.45} & \multicolumn{1}{c|}{\normalfont \scriptsize 115.72} & {\normalfont \scriptsize 110.54} & \multicolumn{1}{c|}{\normalfont \scriptsize 71.16} & {\normalfont \scriptsize 33.37} & \multicolumn{1}{c|}{\normalfont \scriptsize 143.18} & {\normalfont \scriptsize 140.58} & \multicolumn{1}{c|}{\normalfont \scriptsize 6.36} & {\normalfont \scriptsize 3.54} & \multicolumn{1}{c|}{\normalfont \scriptsize 3.75} & \multicolumn{1}{c|}{\normalfont \scriptsize 0.72}\\ \cline{2-14}
\multicolumn{1}{|c|}{} & \multicolumn{1}{c|}{\normalfont \scriptsize Coverage (ASE)} & \multicolumn{1}{c|}{\normalfont \scriptsize 0.87} & {\normalfont \scriptsize 0.58} & \multicolumn{1}{c|}{\normalfont \scriptsize -} & {\normalfont \scriptsize -} & \multicolumn{1}{c|}{\normalfont \scriptsize -} & {\normalfont \scriptsize -} & \multicolumn{1}{c|}{\normalfont \scriptsize -} & {\normalfont \scriptsize -} & \multicolumn{1}{c|}{\normalfont \scriptsize 0.84} & {\normalfont \scriptsize 0.49} & \multicolumn{1}{c|}{\normalfont \scriptsize 0.93} & \multicolumn{1}{c|}{\normalfont \scriptsize 0.94}\\ \cline{2-14}
\multicolumn{1}{|c|}{} & \multicolumn{1}{c|}{\normalfont \scriptsize Coverage (BSE)} & \multicolumn{1}{c|}{\normalfont \scriptsize 0.93} & {\normalfont \scriptsize 0.61} & \multicolumn{1}{c|}{\normalfont \scriptsize -} & {\normalfont \scriptsize -} & \multicolumn{1}{c|}{\normalfont \scriptsize -} & {\normalfont \scriptsize -} & \multicolumn{1}{c|}{\normalfont \scriptsize -} & {\normalfont \scriptsize -} & \multicolumn{1}{c|}{\normalfont \scriptsize 0.91} & {\normalfont \scriptsize 0.60} & \multicolumn{1}{c|}{\normalfont \scriptsize 0.94} & \multicolumn{1}{c|}{\normalfont \scriptsize 0.96}\\ \hline
                                                                                                                                         
\end{tabular}
\vspace*{0.2cm}
\caption{Summary Statistics of Estimation Results Under Independent Errors. 
Bias row gives the empirical bias of 500 estimates. 
ASE row gives the asymptotic standard error obtained from the sandwich estimator of the GMM. BSE row shows the bootstrap standard error obtained from the approach in Supplementary Material \ref{sec:supp:BB}.
ESE row gives the standard deviation of 500 estimates.
MSE row gives the mean squared error of 500 estimates. 
Coverage (ASE) and Coverage (BSE) rows give the empirical coverage rate of 95\% confidence intervals based on the ASE and BSE, respectively.
Bias, standard errors, and mean squared error are scaled by factors of 10, 10, and 100, respectively.
}
\label{tab:Sim:ATT}  
\end{table} 

Next, we evaluated the finite sample performance of the conformal inference approach in Section \ref{sec:Conformal}. As competing methods, we considered the ASC, SCPI, and two SPSC estimators. For each simulated data set, we obtained pointwise 95\% pointwise prediction intervals for the treatment effect $\xi_t^* = \tau_t^* + \epsilon_{t}$ at 10 post-treatment times $t \in \{ T_0 + 0.1 T_1, \ldots T_0 + 0.9 T_1 , T_0+T_1 \} $, using the proposed conformal inference approach for the SPSC estimators along with the ASC and SCPI approaches. We then evaluated the empirical coverage rates of these pointwise prediction intervals based on 500 simulation repetitions, i.e., the proportion of Monte Carlo samples where $\xi_{t}^*$ is contained in 95\% pointwise prediction intervals. 

Table \ref{tab:Table3} gives the empirical coverage rates for each simulated scenario. Surprisingly, the ASC and SPCI approaches fail to attain the nominal coverage rate; we believe this failure originates from the simulation setting where $\bm{\mu}_0$ lies outside the simplex, i.e., (\textit{Non-simplex}). These methods perform particularly poorly when $\blambda_t$ follows a linear trend and the number of time periods is large (i.e., $T_0=500$). In contrast, regardless whether $\blambda_t$ has a trend or not, both SPSC estimators attains the desired nominal coverage rate, aligning closely with theoretical expectations.

\begin{table}[!htp] 
\renewcommand{\arraystretch}{1.1} \centering 
\setlength{\tabcolsep}{4pt} 
\begin{tabular}{|c|c|cc|cc|cc|cc|}
\hline
\multicolumn{1}{|c|}{\multirow{3}{*}{$\bm{\lambda}_t$}} & \multicolumn{1}{c|}{\multirow{3}{*}{$\bm{e}_t$}} & 
\multicolumn{8}{c|}{\normalfont Estimators and $T_0$}
\\ \cline{3-10} 

\multicolumn{1}{|c|}{} & \multicolumn{1}{c|}{} & \multicolumn{2}{c|}{\normalfont ASC}       & \multicolumn{2}{c|}{\normalfont SCPI}      & \multicolumn{2}{c|}{\normalfont SPSC-NoDT} & \multicolumn{2}{c|}{\normalfont SPSC-DT}   \\ \cline{3-10}

\multicolumn{1}{|c|}{} & \multicolumn{1}{c|}{}   & \multicolumn{1}{c|}{\normalfont 100} & \multicolumn{1}{c|}{\normalfont 500} & \multicolumn{1}{c|}{\normalfont 100} & \multicolumn{1}{c|}{\normalfont 500}& \multicolumn{1}{c|}{\normalfont 100} & \multicolumn{1}{c|}{\normalfont 500}& \multicolumn{1}{c|}{\normalfont 100} & \multicolumn{1}{c|}{\normalfont 500} \\ \hline

\multicolumn{1}{|c|}{\multirow{3}{*}{$\!\!\!\begin{array}{c}\text{\normalfont \scriptsize No trend}\end{array}\!\!\!$}} & \multicolumn{1}{c|}{\normalfont \scriptsize Independent errors} & \multicolumn{1}{c|}{\normalfont \scriptsize 0.933} & \multicolumn{1}{c|}{\normalfont \scriptsize 0.925} & \multicolumn{1}{c|}{\normalfont \scriptsize 0.935} & \multicolumn{1}{c|}{\normalfont \scriptsize 0.912} & \multicolumn{1}{c|}{\normalfont \scriptsize 0.959} & \multicolumn{1}{c|}{\normalfont \scriptsize 0.948} & \multicolumn{1}{c|}{\normalfont \scriptsize 0.961} & \multicolumn{1}{c|}{\normalfont \scriptsize 0.948}\\ \cline{2-10}
\multicolumn{1}{|c|}{} & \multicolumn{1}{c|}{\normalfont \scriptsize Correlated errors} & \multicolumn{1}{c|}{\normalfont \scriptsize 0.904} & \multicolumn{1}{c|}{\normalfont \scriptsize 0.906} & \multicolumn{1}{c|}{\normalfont \scriptsize 0.925} & \multicolumn{1}{c|}{\normalfont \scriptsize 0.913} & \multicolumn{1}{c|}{\normalfont \scriptsize 0.962} & \multicolumn{1}{c|}{\normalfont \scriptsize 0.952} & \multicolumn{1}{c|}{\normalfont \scriptsize 0.963} & \multicolumn{1}{c|}{\normalfont \scriptsize 0.953}\\ \cline{2-10}
\multicolumn{1}{|c|}{} & \multicolumn{1}{c|}{\normalfont \scriptsize No $Y$ error} & \multicolumn{1}{c|}{\normalfont \scriptsize 0.922} & \multicolumn{1}{c|}{\normalfont \scriptsize 0.910} & \multicolumn{1}{c|}{\normalfont \scriptsize 0.938} & \multicolumn{1}{c|}{\normalfont \scriptsize 0.925} & \multicolumn{1}{c|}{\normalfont \scriptsize 0.964} & \multicolumn{1}{c|}{\normalfont \scriptsize 0.951} & \multicolumn{1}{c|}{\normalfont \scriptsize 0.964} & \multicolumn{1}{c|}{\normalfont \scriptsize 0.953}\\ \hline
\multicolumn{1}{|c|}{\multirow{3}{*}{$\!\!\!\begin{array}{c}\text{\normalfont \scriptsize Linear trend}\end{array}\!\!\!$}} & \multicolumn{1}{c|}{\normalfont \scriptsize Independent errors} & \multicolumn{1}{c|}{\normalfont \scriptsize 0.795} & \multicolumn{1}{c|}{\normalfont \scriptsize 0.846} & \multicolumn{1}{c|}{\normalfont \scriptsize 0.938} & \multicolumn{1}{c|}{\normalfont \scriptsize 0.889} & \multicolumn{1}{c|}{\normalfont \scriptsize 0.959} & \multicolumn{1}{c|}{\normalfont \scriptsize 0.944} & \multicolumn{1}{c|}{\normalfont \scriptsize 0.962} & \multicolumn{1}{c|}{\normalfont \scriptsize 0.949}\\ \cline{2-10}
\multicolumn{1}{|c|}{} & \multicolumn{1}{c|}{\normalfont \scriptsize Correlated errors} & \multicolumn{1}{c|}{\normalfont \scriptsize 0.844} & \multicolumn{1}{c|}{\normalfont \scriptsize 0.883} & \multicolumn{1}{c|}{\normalfont \scriptsize 0.907} & \multicolumn{1}{c|}{\normalfont \scriptsize 0.843} & \multicolumn{1}{c|}{\normalfont \scriptsize 0.957} & \multicolumn{1}{c|}{\normalfont \scriptsize 0.944} & \multicolumn{1}{c|}{\normalfont \scriptsize 0.967} & \multicolumn{1}{c|}{\normalfont \scriptsize 0.948}\\ \cline{2-10}
\multicolumn{1}{|c|}{} & \multicolumn{1}{c|}{\normalfont \scriptsize No $Y$ error} & \multicolumn{1}{c|}{\normalfont \scriptsize 0.728} & \multicolumn{1}{c|}{\normalfont \scriptsize 0.808} & \multicolumn{1}{c|}{\normalfont \scriptsize 0.920} & \multicolumn{1}{c|}{\normalfont \scriptsize 0.852} & \multicolumn{1}{c|}{\normalfont \scriptsize 0.957} & \multicolumn{1}{c|}{\normalfont \scriptsize 0.953} & \multicolumn{1}{c|}{\normalfont \scriptsize 0.957} & \multicolumn{1}{c|}{\normalfont \scriptsize 0.946}\\ \hline

\end{tabular}
\vspace*{0.2cm}
\caption{
Empirical Coverage Rates of 95\% Pointwise Prediction Intervals. The numbers in SPSC-NoDT and SPSC-DT columns give the empirical coverage rates of 95\% pointwise prediction intervals obtained from the conformal inference approach in Section \ref{sec:Conformal}. The numbers in ASC and SCPI columns give the empirical coverage rates of 95\% pointwise prediction intervals obtained from the approaches proposed by \citet{ASCM2021} and \citet{Cattaneo2021}, respectively.}
\label{tab:Table3} 
\end{table}

In Supplementary Material \ref{sec:supp:Simulation PI}, we assess the finite sample performance of the proposed conformal inference approach based on the simulation scenario given in \citet{Cattaneo2021}, which may not be compatible with the key identifying condition, Assumption \ref{assumption:SC}, of SPSC. As expected, the approach of \citet{Cattaneo2021} performs well in this setting. Although our method without time trend adjustment (i.e., SPSC-NoDT) sometimes fails to achieve the nominal coverage rate, particularly when outcomes are non-stationary, our method with time trend adjustment (i.e., SPSC-DT) consistently attains the nominal coverage rate, provided that the basis functions for time periods are appropriately chosen. This highlights the robustness of the proposed SPSC approach and its broad applicability in synthetic control settings.

\section{Application}		\label{sec:Data}

We applied the proposed method to analyze a real-world application. In particular, we revisited the dataset analyzed in \citet{Dataset1907}, which consists of time series data of length $384$ for 59 trust companies, recorded between January 5, 1906, and December 30, 1908, with a triweekly frequency. Notably, this time period includes the Panic of 1907 \citep{Moen1992}, a financial panic that lasted for three weeks in the United States starting in mid-October, 1907. As a result of the panic, there was a significant drop in the stock market during this period. From this context, we focused on the effect of the financial panic in October 1907 on the log stock price of trust companies using $T_0=217$ pre-treatment time periods and $T_1=167$ post-treatment time periods, respectively. 

The treated unit and donors were defined as follows. According to \citet{Dataset1907}, Knickerbocker, Trust Company of America, and Lincoln were the three trust companies that were most severely affected during the panic. However, % despite the absence of the financial panic, 
Lincoln's stock price showed a strong downward trend over the pre-treatment period. Therefore, we defined the average of the log stock prices of the first two trust companies as $Y_t$, the outcome of the treated units at time $t\in \{1,\ldots,384\}$. As for potential donors, \citet{Dataset1907} identified $N=49$ trust companies that had weak financial connections with the aforementioned three severely affected trust companies. 
%However, some of these trust companies seem to violate the relevance condition \eqref{eq-relevant}. Therefore, we chose donors based on the procedure proposed in Section \ref{sec:supp:Donors} of Supplementary Material, which resulted in $d=24$ trust companies. 
Accordingly, the log stock prices of these 49 trust companies were defined as $\bW_{ t}$, the outcome of the donors. Following the simulation study, we specified the time-invariant and time-varying pre-treatment estimating functions, $\Phi_{\pre}$ and $\Psi_{\pre}$, with $\bh(y)=y$ and $\bD_t=\mathcal{B}_6(t)$, the 6-dimensional cubic B-spline function, to account for a potential time trend.

We first report the ATT estimates under a constant treatment effect model $\tau(t \con \bbeta) = \beta$. Similar to Section \ref{sec:Sim}, we compare the same six estimators: the unconstrained OLS synthetic control estimator (OLS-NoReg), the standard synthetic control approach proposed by \citet{Abadie2010} (OLS-Standard), two recent approaches by \citet{ASCM2021} (ASC) and \citet{Cattaneo2021} (SCPI), and SPSC estimators without and with time-varying terms (SPSC-NoDT, SPSC-DT). The results are summarized in Table \ref{tab:data:ATT}. Interestingly, all six estimators yield similar point estimates of the treatment effect, ranging from $-1.021$ to $-0.813$. According to the 95\% confidence intervals, three estimates uniformly reject the null hypothesis of no treatment effect across time points, suggesting that the financial panic led to a significant decrease in the average log stock price of Knickerbocker and Trust Company of America. We remark again that  the OLS-Standard, ASC, and SCPI approaches do not provide a standard error or 95\% confidence interval for the ATT. In terms of the length of the confidence interval, SPSC with the time-varying components (i.e., SPSC-DT) yields the narrowest confidence interval, followed by SPSC with no time-varying component (i.e., SPSC-NoDT), and the approach based on OLS. 

\begin{table}[!htp] 
\renewcommand{\arraystretch}{1.2} \centering
\scriptsize
\setlength{\tabcolsep}{5pt} 
\begin{tabular}{|c|c|c|c|c|c|c|c|}
\hline
\multicolumn{1}{|c|}{\normalfont Estimator} & {\normalfont  OLS-NoReg} & {\normalfont  OLS-Standard} & {\normalfont   ASC} & {\normalfont  SCPI} & {\normalfont  SPSC-NoDT } & {\normalfont   SPSC-DT} \\ \hline

\multicolumn{1}{|c|}{\normalfont Estimate} & {\normalfont -1.021}  & {\normalfont -0.873}  & {\normalfont -0.912}  & {\normalfont -0.876}  & {\normalfont -0.813}  & {\normalfont -0.816}  \\ \hline
\multicolumn{1}{|c|}{\normalfont ASE} & {\normalfont 0.139}  & {\normalfont -}  & {\normalfont -}  & {\normalfont -}  & {\normalfont 0.084}  & {\normalfont 0.066}  \\ \hline
\multicolumn{1}{|c|}{\normalfont 95\% CI} & {\normalfont (-1.295,-0.748)}  & {\normalfont -}  & {\normalfont -}  & {\normalfont -}  & {\normalfont (-0.978,-0.648)}  & {\normalfont (-0.945,-0.688)}  \\ \hline

\end{tabular} 
\vspace*{0.2cm}
\caption{Summary Statistics of the Estimation of the Average Treatment Effect on the Treated.}
\label{tab:data:ATT} 
\end{table}	

\begin{figure}[!htb]
\centering
\includegraphics[width=1\textwidth]{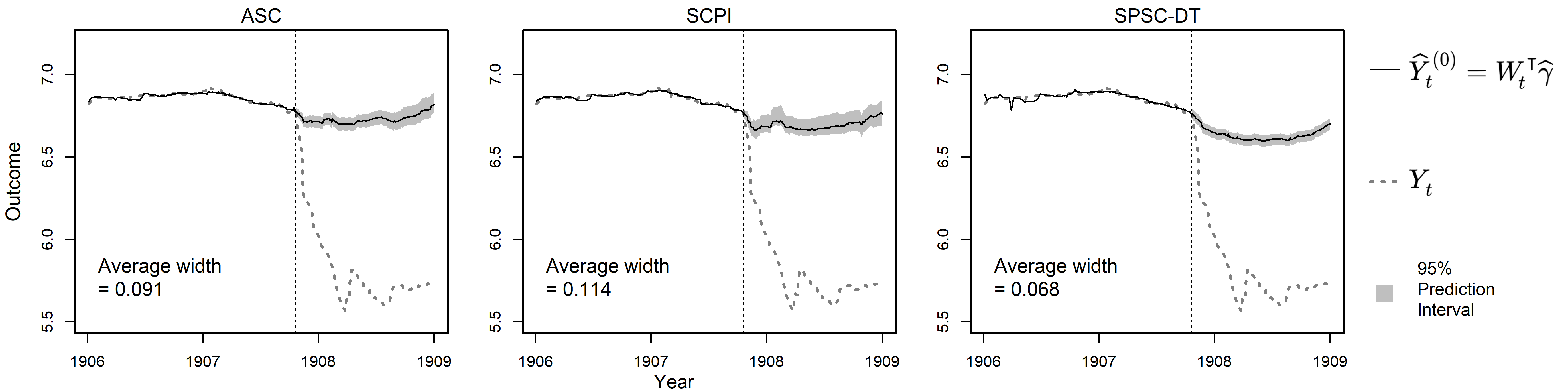}
\caption{Graphical Summaries of the 95\% Prediction Intervals over the Post-treatment Periods. 
These plots, from left to right, present the results using the approaches proposed by \citet{ASCM2021}, \citet{Cattaneo2021}, and the conformal inference approach presented in Section \ref{sec:Conformal} with the time-varying estimating function $\Psi_{\pre}$, respectively. 
The numbers show the average length of the 95\% prediction intervals over the post-treatment periods.
}
\label{fig:data:1}
\end{figure}

We also constructed the pointwise prediction intervals based on the SPSC approach with time-varying components using the conformal inference approach in Section \ref{sec:Conformal}. For comparison, we also implemented the ASC and SCPI approaches. Figure \ref{fig:data:1} provides the visual summary of the result. For the post-treatment period $t\in \{218,\ldots,384\}$, we find that $\widehat{Y}_{t}^{(0)}$, the predictive value of the treatment-free potential outcome, have similar shapes for all methods. However, 95\% pointwise prediction intervals behave differently. Specifically, we focus on the average width of the prediction intervals over the post-treatment periods. The prediction intervals from the ASC and SCPI approaches have average widths of 0.091 and 0.114, respectively; in contrast, our method with time-varying components yields prediction intervals with average widths of 0.068, over 25\% narrower than those from the competing methods; see Supplementary Material \ref{sec:supp:Data} for the distribution of the prediction interval widths across time. The comparison reveals that our method appears to produce tighter predictions of treatment effect trends. Combining results in the simulation study and the data application, we conclude that our approach appears to perform quite competitively when compared to some leading alternative methods in the literature.

Additionally, for the sake of credibility, we conducted the following additional analysis for the application; the details can be found in  Supplementary Material \ref{sec:supp:Data}. First, we studied the trend of the residuals, the difference between the observed outcome and synthetic control, over the pre-treatment time periods. We observed that the OLS-NoReg, SCPI, and SPSC-DT estimators produced residuals without a deterministic trend over time, while the other three estimators (OLS-Standard, ASC, SPSC-NoDT) showed the opposite behavior. Notably, the SPSC-DT estimator appears to satisfy the zero mean condition of Assumption \ref{assumption:SC}, whereas the SPSC-NoDT estimator seems to violate this condition due to a non-zero deterministic trend over time. This again highlights the importance of accommodating time-varying components in the SPSC estimation procedure. 

Next, we performed the following falsification study. We restricted the entire analysis to the pre-treatment period in which the causal effect is expected to be null. We artificially defined a financial panic time in late July 1907, which is roughly three months before the actual financial panic. This resulted in the lengths of the pre- and post-treatment periods equal to $T_0' = 181$ and $T_1'=36$, respectively. The proposed SPSC-NoDT and SPSC-DT estimators resulted in the placebo ATT estimates of $-0.005$ and $0.005$ with 95\% confidence intervals of $(-0.025,0.016)$ and $(-0.004,0.013)$, respectively. The placebo ATT estimate obtained from the unconstrained OLS estimator was $-0.031$ with a 95\% confidence interval of $(-0.062,0.001)$. All 95\% prediction intervals include the null, consistent with the expectation of no treatment effect in the placebo period. Lastly, the constrained OLS estimator (i.e., OLS-Standard), ASC estimator, and SCPI estimator produced placebo ATT estimates of $-0.012$, $-0.032$, and $-0.015$, respectively, which are also close to zero; however, corresponding statistical inference was not available for these estimators. Therefore, these results provide no evidence against validity of the estimators. In Supplementary Material \ref{sec:supp:Data}, we provide a trajectory of the synthetic controls along with 95\% prediction intervals under the placebo treatment. Our findings indicate that the 95\% prediction intervals from the SPSC-DT estimator support the null causal effect. However, the ASC and SCPI estimators occasionally fail to do so during certain time periods. Therefore, we conclude that the SPSC-DT estimator provides a more reliable framework for analyzing the impact of financial panic on the stock prices of the two trust companies.

%In Section \ref{sec:supp:Data} of Supplementary Material, we provide additional results when different donor pools are used. In brief, all results are similar across choices of donors. 

\vspace{-0.4cm}
\section{Concluding Remarks}	\label{sec:Conclusion}

In this paper, we propose a novel SPSC approach in which the synthetic control is defined as a linear combination of donors' outcomes whose conditional expectation matches the treatment-free potential outcome in both pre- and post-treatment periods. The model is analogous yet more general than measurement error models widely studied in standard measurement error literature. Under the framework, we establish the identification of a synthetic control, and provide an estimation strategy for the ATT. Furthermore, we introduce a method for inferring the treatment effect through pointwise prediction intervals, which remains valid even in the case of a short post-treatment period. We validate our methods through simulation studies and provide an application analyzing a real-world financial dataset related to the 1907 Panic.

We reiterate that the SPSC framework differs from existing synthetic control methods in its identifying assumptions and interpretation. It views the synthetic control as an error-prone outcome measurement, without the need for specifying a generative model for the outcome, whereas existing approaches treat it as the projection of the outcome onto the donor's outcome space or the outcome itself. Despite these differences, both frameworks construct synthetic controls by optimally weighting donor units (according to their identifying assumptions), which are then used for treatment effect estimation. Additionally, like other synthetic control methods, the SPSC framework allows for time-varying confounders, as demonstrated in the generative models in Section \ref{sec:IFEM discussion}.

While, as mentioned in Section \ref{sec:SPSC:Assumption}, the SPSC framework may be viewed as a nonstandard form of instrumental variable approach, it is important to highlight key distinctions between the proposed SPSC approach and well-known instrumental variable approaches in dynamic panel data, such as in \citet{AndersonHsiao1981} and \citet{ArellanoBond1991}. In dynamic panel data models, endogeneity arises across different time periods, with the typical assumption that there is no within-time period endogeneity. As a result, these models use lagged variables as instruments to address cross-time endogeneity. In contrast, in the SPSC framework, endogeneity occurs within each time period, without specific assumptions about cross-time endogeneity. Consequently, the instrumental variable approach in this context operates within a single time period. We remark that, like the SPSC framework, many synthetic control models are agnostic about cross-time dependence structure; for example, the cross-time dependent structure of $\blambda_t$ in the IFEM \eqref{eq-IFEM} is agnostic. A notable exception to this agnostic perspective is the ``Instrumental variable-like SC estimator'' proposed by \citet{FermanPinto2019_arxiv}, an earlier version of \citet{FermanPinto2021}, which was developed in the presence of serial correlation in $\blambda_t$. Similar to estimators used in dynamic panel data models, it employs lagged variables as instruments.

As briefly mentioned in the introduction, the proposed SPSC framework has a connection to the single proxy control framework \citep{TT2013_COCA, SPC2024} developed for i.i.d. data. In particular, \citet{SPC2024} proposed an approach that relies on a so-called outcome bridge function, which is a (potentially nonlinear) function of outcome proxies. An important property of the outcome bridge function is that it is conditionally unbiased for the treatment-free potential outcome. Therefore, the proposed SPSC approach can be viewed as an adaptation of the outcome bridge function-based single proxy control approach to the synthetic control setting, where the outcome bridge function is known a priori to be a linear function of donors' outcomes. In Supplementary Material \ref{sec:supp:nonparametric full}, we present a general SPSC framework, which is designed to accommodate nonparametric and nonlinear synthetic controls. Therefore, this framework obviates the over-reliance on a linear specification of synthetic controls in the literature and establishes a more direct connection with the outcome bridge function-based single proxy approach presented in \citet{SPC2024}. Notably, the general SPSC framework addresses underdeveloped areas of the synthetic control literature by allowing for various types of outcomes, including continuous, binary, count, or a combination of these. 

In addition to the outcome bridge function-based approach, \citet{SPC2024} introduced two other single proxy control approaches for i.i.d. sampling. One approach relies on propensity score weighting, eliminating the need for specifying an outcome bridge function. The second approach uses both the propensity score and the outcome bridge function and, more importantly, exhibits a doubly-robust property in that the treatment effect in view is identified if either propensity score or outcome bridge function, but not necessarily both, is correctly specified. Consequently, a promising direction for future research would be to develop new SPSC approaches by extending these single proxy methods to the synthetic control setting. Such new SPSC approaches can be viewed as complementing the doubly-robust proximal synthetic control approach \citep{Qiu2022}. However, such extensions pose significant challenges due to (i) a single treated unit with non-random treatment assignment, (ii) multiple heterogeneous untreated donor units; and (iii) serial correlation and heteroskedasticity due to the time series nature of the data. In particular, non-random treatment assignment undermines the conventional notion of the propensity score, rendering it undefined. Approaches for addressing these challenges and developing corresponding statistical methods will be considered elsewhere. 

\newpage

\newpage

\appendix

\section{Details of the Paper} \label{sec:supp:Detail}

\subsection{Inconsistency of the Ordinary Least Squares Estimator}	\label{sec:supp:OLS}

Following \citet{FermanPinto2021}, we provide details on why synthetic controls obtained from the ordinary least squares (OLS) may be inconsistent. For simplicity, we consider an unconstrained case, in which equation \eqref{eq-OLS} of the main paper reduces to:
\begin{align}
\tag{\ref{eq-OLS}}
\widehat{\bgamma}_{\text{OLS}}
=
\argmin_{\bgamma}
Q(\bgamma)
\ , &&
Q(\bgamma)
= 
\frac{1}{T_0}
\sum_{t=1}^{T_0}
\big(
Y_t - \bW_{ t}\T \bgamma
\big)^2
\ .
\end{align}
For a fixed $\bgamma=(\gamma_{1},\ldots,\gamma_{N}) \T$, the probability limit of $Q(\bgamma)$ as $T_0 \rightarrow \infty$  is given as follows: 
\begin{align}
\nonumber
\lim_{T_0 \rightarrow \infty}
Q(\bgamma)
&=
\lim_{T_0 \rightarrow \infty}
\frac{1}{T_0}
\sum_{t=1}^{T_0} 
\big(
Y_t - 
\bW_{ t}\T \bgamma
\big)^2
\\
\nonumber
& 
=
\lim_{T_0 \rightarrow \infty}
\frac{1}{T_0}
\sum_{t=1}^{T_0} \Big\{
\bW_{ t}\T (\bgamma^\dagger - \bgamma)
+ 
e_{0t} 
-
\sum_{i =1}^{N} \gamma_i^{\dagger} e_{it} 
\Big\}^2
\\
\nonumber
& 
=
\lim_{T_0 \rightarrow \infty}
\frac{1}{T_0}
\sum_{t=1}^{T_0} 
\bigg\{
\sum_{i =1}^{N}
(\gamma_i^\dagger - \gamma_i) \bmu_i\T 
\blambda_t
+ 
e_{0t}
-
\sum_{i=1}^{N} \gamma_i e_{it} 		
\bigg\} ^2
\\
\label{eq-problimit}
&
=
\sum_{i =1}^{N}
(\gamma_i^\dagger - \gamma_i)^2 \bmu_i\T \Lambda	\bmu_i
+
\bigg( 1+ \sum_{i =1}^{N} \gamma_i^2 \bigg) \sigma_e^2
\ .
\end{align}
where the second and third lines hold from \eqref{eq-SC Equation} and \eqref{eq-IFEM} of the main paper, respectively, which are restated below:
\begin{align}	\tag{\ref{eq-IFEM}}
Y_{t}
&
=
\text{\makebox[1.25cm]{$ \tau_{t}^* A_t + $}}
\bmu_{0} \T
\blambda_t
+
e_{0t}
\ ,
&&
\EXP \big( e_{0t} \cond \blambda_t ) = 0
\nonumber
\\
W_{it}
&
=
\text{\makebox[1.25cm]{}}
\bmu_{i} \T
\blambda_t
+
e_{it}
\ , 
&&
\EXP \big( e_{it} \cond \blambda_t ) = 0
\ , 
&&
i \in \{1,\ldots,N\} \ ,
&&
t \in \{1,\ldots,T\} \ .
\nonumber
\end{align}
and
\begin{align}		\tag{\ref{eq-SC Equation}}
\potY{t}{0}
=
\bW_{ t} \T \bgamma^\dagger
+ 
e_{0t} 
-
\sum_{i =1}^{N} \gamma_i^{\dagger} e_{it} 
\ , 
\quad \quad
t \in \{1,\ldots, T\}
\ .
\end{align}
The last line holds under the following additional assumptions on $\blambda_t$ and $e_{it}$ as $T_0 \rightarrow \infty$:
\begin{align*}
& 
\frac{1}{T_0} \sum_{t=1}^{T_0} \blambda_t = o_P(1) \ ,
&&
\frac{1}{T_0} \sum_{t=1}^{T_0} \blambda_t \blambda_t \T = \Lambda + o_P(1)
\\
&
\frac{1}{T_0} \sum_{t=1}^{T_0} {e}_{it} = o_P(1) \ ,
&&
\frac{1}{T_0} \sum_{t=1}^{T_0} {e}_{it} {e}_{jt} = \ind(i=j) \sigma_e^2 
\ , 
&&
\frac{1}{T_0} \sum_{t=1}^{T_0} e_{it} \blambda_t = o_P(1) 
\ .
\end{align*}
where $\Lambda$ is positive semidefinite. Clearly, $\gamma_i^\dagger$ is not the minimizer of \eqref{eq-problimit} unless $\sigma_e^2 = 0$, i.e., a noiseless setting. Therefore, the OLS weights defined in \eqref{eq-OLS} converge to the minimizer of $Q(\bgamma)$ as $T_0 \rightarrow \infty$, which is different from the true synthetic control weights $\bgamma^\dagger$ satisfying $\bm{\mu}_{0} = \sum_{i =1}^{N} \gamma_i^{\dagger} \bm{\mu}_i$. This implies that the OLS estimator is inconsistent for $\bgamma^\dagger$ unless $\sigma_e^2=0$.

\subsection{Choice of the Regularization Parameter $\rho$} \label{sec:supp:CV}

We choose the regularization parameter $\rho$ based on leave-one-out cross-validation; see Algorithm \ref{alg:LOOCV} below.
\begin{algorithm}[!htb]
\begin{algorithmic}[1]
\REQUIRE Length of the pre-treatment periods $T_0$
\FOR{$t \in \{1,\ldots,T_0\}$}

\STATE Let $\widehat{\bG}_{YW,(-t),\rho}$ and $\widehat{\bG}_{YY,(-t),\rho}$ be
\begin{align*}
&
\widehat{\bG}_{YW,(-t),\rho} =
\frac{1}{T_0-1} 
\sum_{s \leq T_0, s \neq t} \bg(s, Y_s \con \widehat{\Beta}) \bW_{ s}\T \in \R^{(d+p) \times d}
\ , 
\\
&
\widehat{\bG}_{YY,(-t),\rho} = 
\frac{1}{T_0-1} 
\sum_{s \leq T_0, s \neq t} \bg(s, Y_s \con \widehat{\Beta}) Y_s \in \R^{d+p} 
\end{align*}
where $\widehat{\Beta}
=
\big( \sum_{t=1}^{T_0} \bD_t\bD_t\T \big)^{-1} 
\big( \sum_{t=1}^{T_0} \bD_tY_t \big) $.

\STATE Let $\widehat{\bgamma}_{(-t),\rho}
=
\big\{ \widehat{\bG}_{YW,(-t),\rho}\T \widehat{\Omega}_{\bg} \widehat{\bG}_{YW,(-t),\rho} + \rho I_{\dim(\bg)\times\dim(\bg)} \big\}^{-1}
\big\{ \widehat{\bG}_{YW,(-t),\rho}\T \widehat{\Omega}_{\bg} \widehat{\bG}_{YY,(-t),\rho} \big\}$

\STATE Calculate the leave-one-out residual $
\widehat{e}_{t,\rho} = Y_t - \bW_{ t} \T \widehat{\bgamma}_{(-t),\rho}$

\ENDFOR

\STATE Obtain the squared error based on the leave-one-out residuals $
\widehat{MSE}_{\rho} = \sum_{t=1}^{T_0} \widehat{e}_{t,\rho}^2 / T_0$

\RETURN Obtain the optimal $\rho$ that minimizes the squared error:
\begin{align*}
\rho_{\text{opt}} = \argmin_{\rho} 
\widehat{MSE}_{\rho}
\end{align*}
\end{algorithmic}
\caption{Leave-one-out Cross-validation for Choosing the Regularization Parameter $\rho$}
\label{alg:LOOCV}
\end{algorithm}

\subsection{A Heteroskedasticity and Autocorrelation Consistent Covariance Matrix Estimator}		\label{sec:supp:HAC}

We provide details of a heteroskedasticity and autocorrelation consistent (HAC) covariance matrix estimator, which are obtained by following approaches of \citet{NW1987} and \citet{Andrews1991}. Let $\big( \widehat{\Beta}, \widehat{\bgamma}_{\rho}, \widehat{\bbeta} \big) $ and $\Psi (\bO_t \con \Beta, \bgamma , \bbeta)$ be the GMM estimators used in Theorem \ref{thm:AN} and the corresponding estimating function, respectively. Then, for a given bandwidth $\omega >0$ and a kernel function $\mathcal{K}(z)$, a heteroskedasticity and autocorrelation consistent estimator of $\Sigma_2^* =
\lim_{T \rightarrow \infty} \VAR \big\{ \sqrt{T} \cdot \widehat{\Psi}(\Beta^*, \bgamma_0^*, \bbeta^*) \big\}$ is given as 
\begin{align*}
\widehat{\Sigma}_2
=
\frac{1}{T}
\sum_{t=1}^{T}
\left[
\begin{array}{l}
\big\{
\Psi( \bO_t \con \widehat{\Beta}, \widehat{\bgamma}_{\rho} , \widehat{\bbeta} )
\big\}
\big\{
\Psi( \bO_t \con \widehat{\Beta}, \widehat{\bgamma}_{\rho} , \widehat{\bbeta} )
\big\}\T
\\
+
\sum_{s=1}^{T} 
\mathcal{K} \big( s / \omega \big)
\big\{
\Psi( \bO_t \con \widehat{\Beta}, \widehat{\bgamma}_{\rho} , \widehat{\bbeta} )
\big\}
\big\{
\Psi( \bO_{t+s} \con \widehat{\Beta}, \widehat{\bgamma}_{\rho} , \widehat{\bbeta} )
\big\}\T
\\
+
\sum_{s=1}^{T} 
\mathcal{K} \big( s / \omega \big)
\big\{
\Psi( \bO_{t+s} \con \widehat{\Beta}, \widehat{\bgamma}_{\rho} , \widehat{\bbeta} )
\big\}
\big\{
\Psi( \bO_{t} \con \widehat{\Beta}, \widehat{\bgamma}_{\rho} , \widehat{\bbeta} )
\big\}\T
\end{array}
\right] \ .
\end{align*} 

Popular choices for the kernel function are Bartlett and quadratic spectral functions, which are defined as follows:
\begin{itemize}[itemsep=0cm,leftmargin=1cm]
\item Bartlett kernel: $\mathcal{K}(z) = \big\{ 1- |z| \big\} \ind\big\{ |z| \leq 1 \big\}$
\item Quadratic spectral kernel: $\mathcal{K}(z) = \big\{ {25}/{(12 \pi^2 z^2)} \big\} \cdot \big\{ \sin (6\pi z/5) / (6\pi z/5) - \cos (6\pi z/5) \big\} $
\end{itemize}

For these two kernel functions, the bandwidth parameter $\omega$ can be chosen based on the approximation to the first-order autoregressive model; see Algorithm \ref{alg:bandwidth} for details. We use the quadratic spectral kernel function for the simulation studies and the data analysis of the main paper.

\begin{algorithm}[!htb]
\begin{algorithmic}[1]
\STATE Let $\big(  \widehat{\Beta}, \widehat{\bgamma}_{\rho} , \widehat{\bbeta} ) \big) $ and $\Psi (\bO_t \con \Beta, \bgamma , \bbeta)$ be the GMM estimators used in Theorem \ref{thm:AN} and the corresponding estimating function related to $\bbeta$, respectively. 

\FOR{$s \in \{1,\ldots,\dim( \Psi_{\post} )\}$}

\STATE Fit AR(1) model for the $s$th component of the time series $\big\{ \Psi_{\post}(\bO_t \con \widehat{\Beta}, \widehat{\bgamma}_{\rho}, \widehat{\bbeta} ) \big\}_{t \in \{1,\ldots,T\}}$. 
\STATE Let $\widehat{\kappa}_s$ and $\widehat{\sigma}_s^2$ be the estimated coefficient of the autoregressive coefficient and the estimated variance of the error from the AR(1) model above, respectively.
\ENDFOR

\STATE For Barlett and quadratic spectral kernel functions, we choose the bandwidth as
\begin{align*}
&
\omega_{\text{Bartlett}} 
=
1.1447 \big\{ \alpha_1 \cdot T \big\}^{1/3}
, 
&&
\hspace*{-0.25cm}
\alpha_1
=
\bigg\{ \sum_{s=1}^{b} \frac{ \widehat{\sigma}_s^4 }{ (1-\widehat{\kappa}_s)^4 } \bigg\}^{-1}
\bigg\{ \sum_{s=1}^{b} \frac{ 4 \widehat{\kappa}_s^2 \widehat{\sigma}_s^4 }{ (1-\widehat{\kappa}_s)^6 (1+\widehat{\kappa}_s)^2} \bigg\} 		
\\
&
\omega_{\text{QS}} 
=
1.3221 \big\{ \alpha_2 \cdot T \big\}^{1/5} 		
,
&&
\hspace*{-0.25cm}
\alpha_2
=
\bigg\{ \sum_{s=1}^{b} \frac{ \widehat{\sigma}_s^4 }{ (1-\widehat{\kappa}_s)^4 } \bigg\}^{-1}
\bigg\{ \sum_{s=1}^{b} \frac{ 4 \widehat{\kappa}_s^2 \widehat{\sigma}_s^4 }{ (1-\widehat{\kappa}_s)^8} \bigg\} \ .
\end{align*}

\RETURN Bandwidth parameters $\omega_{\text{Bartlett}}$ and $\omega_{\text{QS}}$.
\end{algorithmic}
\caption{Choice of Bandwidth Parameters for Bartlett and Quadratic Spectral Kernel Functions}
\label{alg:bandwidth}
\end{algorithm}

\subsection{Block Bootstrap} \label{sec:supp:BB}

In this section, we provide a moving block bootstrap method \citep{Kunsch1989,Liu1992} adapted to our setting. Algorithm \ref{alg:MBB} provides details of the block bootstrap implementation. We remark that other block bootstrap methods can be adopted with minor modifications; see \citet{Lahiri1999} for examples of block bootstrap methods.

\begin{algorithm}[!htb]
\begin{algorithmic}[1]
\REQUIRE Length of the block $L < T_0$, Number of bootstrap repetitions $B$

\STATE Let the pre- and post-treatment blocks be
\begin{align*}
&
B_{\pre,1} = \big\{ \bO_1,\ldots,\bO_L \big\} , 
&& \ldots \ , 
&& B_{\pre,T_0-L+1} = \big\{ \bO_{T_0-L+1},\ldots,\bO_{T_0} \big\} 	
\\
&
B_{\post,1} = \big\{ \bO_{T_0+1},\ldots, \bO_{T_0+L} \big\} , 
&& \ldots \ , 
&& B_{\post,T_0-L+1} = \big\{ \bO_{T-L+1},\ldots, \bO_{T} \big\} 	
\end{align*} 	

\FOR{$b \in \{1,\ldots,B\}$}

\STATE Randomly sample $K_{\pre} = \lceil T_0/L \rceil$ pre-treatment blocks and $K_{\post} = \lceil T_1/L \rceil$ post-treatment blocks with replacement, respectively; we denote these blocks as $\big\{ B_{\pre,1}^{(b)} ,\ldots, B_{\pre,K_{\pre}}^{(b)} \big\}$ and $\big\{ B_{\post,1}^{(b)} ,\ldots, B_{\post,K_{\post}}^{(b)} \big\}$

\STATE Choose the first $T_0$ and $T_1$ observations from the resampled blocks, i.e.,
\begin{align*}
&
\big\{ \bO_1^{(b)},\ldots,\bO_{T_0} ^{(b)} \big\}
=
\text{first $T_0$ observations of } \big\{ B_{\pre,1}^{(b)} ,\ldots, B_{\pre,K_{\pre}}^{(b)} \big\}
\\
&
\big\{ \bO_{T_0+1}^{(b)},\ldots,\bO_{T} ^{(b)} \big\}
=
\text{first $T_1$ observations of } \big\{ B_{\post,1}^{(b)} ,\ldots, B_{\post,K_{\post}}^{(b)} \big\}
\end{align*}

\STATE Calculate $\widehat{\bbeta}^{(b)}$ from the GMM in Section \ref{sec:Estimation} using $\big\{ \bO_1^{(b)},\ldots,\bO_{T_0} ^{(b)} , \bO_{T_0+1}^{(b)},\ldots,\bO_{T} ^{(b)} \big\}$. 	

\ENDFOR

\RETURN Report the variance of the bootstrap estimates $\big\{ \widehat{\bbeta}^{(1)}, \ldots, \widehat{\bbeta}^{(B)} \big\}$
\end{algorithmic}
\caption{Moving Block Bootstrap in Single Proxy Synthetic Control Framework}
\label{alg:MBB}
\end{algorithm}

The choice of block length $L$ is critical to the performance of block bootstrap methods. The optimal choice of $L$ for minimizing mean square error is known to be $O(T^{1/3})$. In the simulation studies in Section \ref{sec:Sim}, we use the bandwidth of the Bartlett kernel function $\omega_{\text{Bartlett}}$ in Algorithm \ref{alg:bandwidth} of which the rate is $O(T^{1/3})$. As discussed, this choice seems reasonable based on the simulation results reported in Section \ref{sec:supp:Simulation}. 

\subsection{Examples Where Assumption \ref{assumption:SC} Holds But Condition \ref{assumption:SC-structural} Is Violated} \label{sec:supp:IFEM}

We provide an example that violation of Condition \ref{assumption:SC-structural} does not necessarily imply violation of Assumption \ref{assumption:SC}. Consider the following interactive fixed effects model (IFEM): 
\begin{align}
&
\potY{t}{0}
= \bm{\mu}_0\T \blambda_t + e_{0t}	
\ , \quad
W_{it}
=
\bm{\mu}_i\T \blambda_t + e_{it}
\ , \quad
i \in \{1,\ldots,N\}
\ , \quad 
t \in \{1,\ldots, T\} \ ,
\nonumber
\\
& \text{where }
\quad
\blambda_t \stackrel{iid}{\sim} N (\bm{\nu}_t, \Sigma_{\rho})
\quad , \quad 
\big( 
e_{0t} \ , \ e_{1t} \ , \ \cdots \ , \ e_{Nt}
\big)\T
\stackrel{iid}{\sim}
N( 0_{(N+1) \times 1} , 
\Sigma_e ) \ .
\label{eq-IFEM-supp}
\end{align}
Note that the model allows for non-stationarity behaviors, as $\bm{\nu}_t$ can be time-varying. The model also allows for $\sigma_{ij}=0$ for $i\neq j$, i.e., independent errors, where $\sigma_{ij}$ for $i,j \in \{0,1,\ldots,N\}$ is the components of $\Sigma_e$. Therefore, for any $\bgamma \in \R^{N}$, we find
\begin{align*}
\bW_t\T \bgamma
=
\bigg( \sum_{i=1}^{N} \gamma_i \bm{\mu}_i\T \bigg) \blambda_t + \bigg( \sum_{i=1}^{N} \gamma_i e_{it} \bigg) 
\quad \Rightarrow \quad 
\EXP \big( \bW_t\T \bgamma \cond \blambda_t, e_{0t} \big)
=
\bigg( \sum_{i=1}^{N} \gamma_i \bm{\mu}_i\T \bigg) \blambda_t \ .
\end{align*}
Note that $\EXP \big( \bW_t\T \bgamma \cond \blambda_t, e_{0t} \big)$ does not depend on $e_{0t}$  for any $\bgamma$ where as $\potY{t}{0} =  	\bm{\mu}_0\T\blambda_t + e_{0t}$ depends on $e_{0t}$. Therefore, there is no $\bgamma$ satisfying Condition \ref{assumption:SC-structural} when errors are independent, unless $\sigma_{00}=\VAR(e_{0t})$ is zero.

However, one can still find $\bm{\gamma}$ such that Assumption \ref{assumption:SC} is satisfied despite independent errors, i.e., there exists a weight vector $\bgamma^*$ satisfying
\begin{align*}
\EXP \big\{ \bW_t\T \bgamma^* \cond 
\potY{t}{0} \big\}
=
\EXP \big( \bW_t\T \bgamma^* \cond \bm{\mu}_0\T \bm{\lambda}_t + e_{0t} \big) 
= 
\bm{\mu}_0\T \bm{\lambda}_t + e_{0t}
=
\potY{t}{0}
\ .
\end{align*}
Specifically, let $\bm{\gamma}^*=(\gamma_1^*,\cdots,\gamma_N^*)$ be a vector that solves the following two equations:
\begin{align}
&
\bm{\mu_0}
=
\sum_{i=1}^{N} \gamma_i^*  \bm{\mu}_i
\label{eq-IFEM-SPSC-1}
\ , \\
&
\bm{\mu}_0\T \Sigma_{\rho}
\bm{\mu}_0 + \sigma_{00}
=
\sum_{i=1}^{N} \gamma_{i}^* ( \bm{\mu}_0\T \Sigma_{\rho} \bm{\mu}_i + \sigma_{0i} )   \ .
\label{eq-IFEM-SPSC-2}
\end{align}
Again, note that $\sigma_{0i}=0$ for all $i \in \{1,\ldots,N\}$ when errors are independent.

From a property of joint normal distributions, we can represent the conditional distribution of $e_{it}$ given $e_{0t}$ as follows:
\begin{align*}
e_{it} \cond e_{0t}
\stackrel{D}{=}
\frac{\sigma_{0i}}{\sigma_{00}} e_{0t}
+ \xi_{it} \ ,
\quad
\xi_{it} \sim N \bigg( 0, \sigma_{ii}^2 - \bigg( \frac{\sigma_{0i}}{\sigma_{00}} \bigg)^2
\bigg) 
\ , 
\quad 
e_{0t} \indep \xi_{it}
\end{align*} 
Therefore, we find the joint distribution of $(\bW_t\T \bgamma^*,\potY{t}{0})$ is represented as follows:
\begin{align*}
&
\begin{pmatrix}
\bW_t\T \bgamma^*
\\
\potY{t}{0}
\end{pmatrix}  
\\
& 
\stackrel{D}{=}
\begin{pmatrix}
\sum_{i=1}^{N} \gamma_{i}^* 
(\bm{\mu}_i\T \bm{\lambda}_t +
\frac{\sigma_{0i}}{\sigma_{00}}
\epsilon_0
+
\xi_i
)
\\
\bm{\mu}_0\T \bm{\lambda}_t + \epsilon_0
\end{pmatrix} 
\\
&
\sim 
N
\left(
\begin{pmatrix}
\sum_{i=1}^{N} \gamma_{i}^* 
(\bm{\mu}_i\T \bm{\nu}_t) \\ 
\bm{\mu}_0\T \bm{\nu}_t
\end{pmatrix}
,
\begin{pmatrix}
(*) & \sum_{i=1}^{N} \gamma_{i}^* (\bm{\mu}_0\T \Sigma_{\rho} \bm{\mu}_i
+
\sigma_{0i})
\\
\sum_{i=1}^{N} \gamma_{i}^* (\bm{\mu}_0\T \Sigma_{\rho} 
\bm{\mu}_i
+
\sigma_{0i})
&
\bm{\mu}_0\T \Sigma_{\rho}
\bm{\mu}_0 + \sigma_{00}
\end{pmatrix}
\right)  ,
\end{align*}
where $(*)$ denotes a generic variance. This implies that the conditional distribution $\bW_t\T \bgamma^* \cond \potY{t}{0}$ is given by 
\begin{align}
\bW_t\T \bm{\gamma}^* \cond ( \potY{t}{0} =y )
&
=
\bW_t\T \bm{\gamma}^* \cond ( \bm{\mu}_0\T \bm{\lambda}_t + \epsilon_0 = y )
\nonumber
\\
&
\sim 
N
\Bigg(
\underbrace{
\sum_{i=1}^{N} \gamma_{i}^* \bm{\mu}_i\T \bm{\nu}_{t}
+
\frac{ \sum_{i=1}^{N} \gamma_{i}^* (\bm{\mu}_0\T \Sigma_{\rho} \bm{\mu}_i
+ \sigma_{0i})} { \bm{\mu}_0\T \Sigma_{\rho} 
\bm{\mu}_0 + \sigma_{00} }
( y - \bm{\mu}_0\T \bm{\nu}_{t})}_{=y} , (*)
\Bigg) \ ,    
\label{eq-IFEM-SPSC-3}
\end{align}
where the last line holds from \eqref{eq-IFEM-SPSC-1} and \eqref{eq-IFEM-SPSC-2}. 
Therefore, we get $ \EXP \big\{ \bW_t\T \bm{\gamma}^* \cond \potY{t}{0}\big\} = \potY{t}{0}$ almost surely, implying that Assumption \ref{assumption:SC} is satisfied with $h^*(\bW_t) = \bW_t\T \bm{\gamma}^*$. Note that the synthetic control weight vector $\bgamma^*$ exists even when errors are independent and the outcomes are non-stationary.

Assumption \ref{assumption:SC} can be satisfied for non-continuous outcomes. To illustrate this, consider the following latent variable model for count data:
\begin{align*}
&
\potY{t}{0} = \bm{\mu}_0\T \bm{\blambda}_t + V_{0t} 
\sim \text{Poisson} \bigg( \sum_{j=1}^{r} \mu_{j0} + \kappa_0 \bigg)
\ , \quad 
\\
&
W_{it} = \bm{\mu}_i\T \bm{\blambda}_t + V_{it} 
\sim \text{Poisson} \bigg( \sum_{j=1}^{r} \mu_{ji} + \kappa_i \bigg)
\ , \quad i \in \{1,\ldots,N\} \ , 
\\
&
\bm{\lambda}_t = (\lambda_{1t},\ldots,\lambda_{rt})\T \text{ where }  \lambda_{jt} \sim \text{Poisson}(\nu_{jt}) \text{ for } j \in \{1,\ldots,r\} \ , \\
&
V_{it} \sim \text{Poisson}(\kappa_{i}) \text{ for } i \in \{0,1,\ldots,N\} \ , \\
& 
\lambda_{1t} \indep \ldots \indep \lambda_{rt} \indep V_{0t} \indep V_{1t} \indep \ldots \indep V_{Nt} \ .
\end{align*}
Note that $(\potY{t}{0},\bW_{t})$ can be non-stationary because $\bm{\nu}_t=(\nu_{1t},\ldots,\nu_{rt})\T$ is allowed to be time-varying. 

Let $\bgamma^* = (\gamma_1^*,\ldots,\gamma_N^*)\T$ be a vector that satisfies $
\bm{\mu}_0 
= 
\sum_{i=1}^N \gamma_i^* \bm{\mu}_i$. Then, we find
\begin{align*}
& \bW_t\T \bgamma^*
\cond \potY{t}{0}
\\
&
\stackrel{D}{=}
\bigg( \sum_{i=1}^{N} \gamma_i^* \bm{\mu}_i\T \blambda_{t}  + \sum_{i=1}^{N} \bgamma_i^* V_{it}\bigg) 
\, \bigg| \, (  \bm{\mu}_0\T \blambda_t + V_{0t} )
\\
&
\stackrel{D}{=}
\bigg( \sum_{i=1}^{N} \gamma_i^* \bm{\mu}_i\T \blambda_{t} \bigg) 
\, \bigg| \, ( \bm{\mu}_0\T \blambda_t + V_{0t} )
\oplus \sum_{i=1}^{N} \gamma_i^* V_{it}
&&
\Leftarrow
\quad (V_{1t},\ldots,V_{Nt}) \indep (\blambda_t, V_{0t})
\\
&
\stackrel{D}{=}
\bm{\mu}_0 \blambda_{t}
\, \big| \, ( \bm{\mu}_0\T \blambda_t + V_{0t} )
\oplus 
\text{Poisson} \bigg( \sum_{i=1}^{N} \gamma_i^*\kappa_{i} \bigg)
&&
\Leftarrow
\quad 
\bm{\mu}_0=\sum_{i=1}^{N} \gamma_i^* \bm{\mu}_i
\\
&
\stackrel{D}{=}
\text{Bin}
\bigg(
\bm{\mu}_0\T \blambda_t + V_{0t} 
, 
\frac{ \sum_{j=1}^{r} \mu_{ji} }{\sum_{j=1}^{r} \mu_{ji} + \kappa_0}
\bigg)
\oplus 
\text{Poisson} \bigg( \sum_{i=1}^{N} \gamma_i^*\kappa_{i} \bigg)
&&
\Leftarrow
\quad 
\text{see \eqref{eq-Pois-Binom}}
\\
&
\stackrel{D}{=}
\text{Bin}
\bigg(
\potY{t}{0}
, 
\frac{ \sum_{j=1}^{r} \mu_{ji} }{\sum_{j=1}^{r} \mu_{ji} + \kappa_0}
\bigg)
\oplus 
\text{Poisson} \bigg( \sum_{i=1}^{N} \gamma_i^*\kappa_{i} \bigg)
\end{align*}

Note that the fourth line holds from
\begin{align}
\label{eq-Pois-Binom}
V_i \stackrel{\text{indep}}{\sim} \text{Poisson}(\mu_i) \ , \ i=1,2 \ ,
\quad \Rightarrow \quad 
V_1 \cond (V_1+V_2) \stackrel{D}{=} \text{Bin} \bigg( V_1+V_2 , \frac{\mu_1}{\mu_1+\mu_2} \bigg) \ .
\end{align}
Consequently, we find
\begin{align*}
&
\EXP \big\{ \bW_t\T \bgamma^*
\cond \potY{t}{0} \big\}
=
\frac{ \sum_{j=1}^{r} \mu_{ji} }{\sum_{j=1}^{r} \mu_{ji} + \kappa_0}
\potY{t}{0}
+
\sum_{i=1}^{N} \gamma_i^*\kappa_{i} 
\\
& \Leftrightarrow \quad 
\EXP \bigg\{ 
\underbrace{
\frac{\sum_{j=1}^{r} \mu_{ji} + \kappa_0}{ \sum_{j=1}^{r} \mu_{ji} }
( \bW_t - \bm{\kappa})\T  \bgamma^*
}_{=:h^*(\bW_t)}
\, \bigg| \, \potY{t}{0} \bigg\}
=
\potY{t}{0} \quad \Leftarrow \quad 
\bm{\kappa}=(\kappa_1,\ldots,\kappa_N)\T
\ . 
\end{align*} 
Therefore, we can find a synthetic control bridge function $h^*(\bW_t)$.
% ETT THIS IS VERY INTERESTING I WONDER HOW IT RELATES TO THE JOHANNSON PAPER I EMAILED YOU SEE HIS POISSON EXAMPLE, IF RELATED MIGHT BE GOOD TO CITE IT AS FURTHER THEORETICAL FOUNDATION FOR OUR METHODS 

\subsection{Details on Accommodating Time-varying Components}		\label{sec:supp:DT}

We provide a rationale that accommodating time-varying components can improve the performance of the SPSC approach. Specifically, we consider the IFEM \eqref{eq-IFEM-supp}, where $\EXP(\blambda_t) = \bm{\nu}_t$ is assumed to lie within the space spanned by $\bD_t$, i.e.,
\begin{align*}
\bm{\nu}_t=\EXP(\blambda_t)
=
\mathcal{E}\sT \bD_t \ , \quad \mathcal{E}^* \in \R^{d \times r} \ .
\end{align*}
This assumption is reasonable if $\bD_t$ is defined as a set of rich basis functions, such as polynomials, trigonometric functions, splines, or wavelets. 

\subsubsection{Using the Time-invariant Estimating Equation}

Suppose that the synthetic control weights are estimated by the time-invariant pre-treatment estimating equation:
\begin{align*}	
\widehat{\bgamma}_{\rho}
=
\argmin_{\bgamma}
\Bigg[ 
\bigg\|
\frac{1}{T_0}
\sum_{t=1}^{T_0}
\bh (Y_t)
\big( Y_t - \bW_{ t} \T \bgamma \big)
\bigg\|_2^2
+ \rho \big\| \bgamma \big\|_2^2
\Bigg]
\ . 
\end{align*}
For simplicity, we choose $\bh(y)=(1,y)\T$. Then, it is straightforward to show that the regularized GMM estimator of $\bgamma$ is given by
\begin{align*}
\widehat{\bgamma}_{\rho}
=
\big(
\widehat{\bG}_{YW} \T \widehat{\bG}_{YW}
+
\rho I_{N \times N} 
\big)^{-1}
\big( 
\widehat{\bG}_{YW} \T \widehat{\bG}_{YY}
\big)
\end{align*}
where 
\begin{align*}
&
\widehat{\bG}_{YW} =
\frac{1}{T_0} 
\sum_{t=1}^{T_0} \begin{pmatrix}
\bW_{ t}\T \\ Y_t \bW_{ t}\T
\end{pmatrix}   \in \R^{2 \times N}
\ , 
&&
\widehat{\bG}_{YY} = 
\frac{1}{T_0} \sum_{t=1}^{T_0} \begin{pmatrix}
Y_t \\ Y_t^2
\end{pmatrix} \in \R^{2} \ .
\end{align*}
From the law of large numbers, we find
\begin{align*}
&
\widehat{\bG}_{YW} \stackrel{P}{\rightarrow} 
{\bG}_{YW}^*
=
\begin{pmatrix}	
\overline{\bm{\nu}}\T
\mathfrak{M} 
\\
\bm{\mu}_0\T  
\Lambda
\mathfrak{M} 
+
\Sigma_{0,(-0)}
\end{pmatrix}
\ , 
\quad 
&&
\hspace*{-1cm}
\widehat{\bG}_{YY}
\stackrel{P}{\rightarrow} 
{\bG}_{YY}^*
=
\begin{pmatrix}	
\overline{\bm{\nu}}\T
\bm{\mu}_0
\\
\bm{\mu}_0\T  
\Lambda
\bm{\mu}_0
+\sigma_{00}
\end{pmatrix}
\ , 
\\
&
\mathfrak{M} =	\big[ \bm{\mu}_1 \ , \ \cdots \ , \ \bm{\mu}_{N} \big]
\in \R^{r \times N}
\ , \quad 
&&
\hspace*{-1cm}
\Sigma_{0,(-0)} = (\sigma_{01},\cdots,\sigma_{0N}) \in \R^{1 \times N}
\\
&
\overline{\bm{\nu}}
=
\lim_{T_0 \rightarrow \infty}
\frac{1}{T_0} \sum_{t=1}^{T_0} \EXP \big( \blambda_t) 
=
\lim_{T_0 \rightarrow \infty}
\frac{1}{T_0} \sum_{t=1}^{T_0} \bm{\nu}_t
\in \R^{r}
\ , \\
&
\Lambda
=
\lim_{T_0 \rightarrow \infty}
\frac{1}{T_0} \sum_{t=1}^{T_0} \EXP \big( \blambda_t \blambda_t\T \big)  
=
\Sigma_{\rho} + 
\lim_{T_0 \rightarrow \infty}
\frac{1}{T_0} \sum_{t=1}^{T_0} \bm{\nu}_t  \bm{\nu}_t\T
\in \R^{r \times r}
\ .
\end{align*}
Therefore, the limit of $\widehat{\bgamma}_{\rho}$ is $\bgamma^\dagger = \big( \bG_{YW}^* \big)^+ \bG_{YY}^*$, which is the minimum-norm solution of the equation $ \bG_{YY}^*
=
\bG_{YW}^* \bgamma $. Therefore, $\bgamma^\dagger$ satisfies
\begin{align*}
\bG_{YY}^*
=
\bG_{YW}^* \bgamma^\dagger
\quad \Leftrightarrow \quad 
\left\{
\begin{array}{l}	
\overline{\bm{\nu}}\T \bm{\mu}_0
=
\overline{\bm{\nu}}\T \mathfrak{M}
\bgamma^\dagger
=
\overline{\bm{\nu}}\T
\sum_{i=1}^{N} \bm{\mu}_i \gamma_i^\dagger
\
\\
\bm{\mu}_0\T \Lambda \bm{\mu}_0 + \sigma_{00}
=
\big\{ \bm{\mu}_0\T \Lambda \mathfrak{M} + \Sigma_{0,(-0)} \big\} \bgamma^\dagger
=	
\sum_{i=1}^{N}
\big(
\bm{\mu}_0\T
\Lambda \bm{\mu}_i + \sigma_{0i} \big) \gamma_i^\dagger
\end{array}
\right. 
\ .
\end{align*}

If  $\bm{\nu}_t=\bm{\nu}$ for all $t$, i.e., the mean of $\blambda_t$ is time-invariant, we have 
\begin{align}
&
\overline{\bm{\nu}}\T \bm{\mu}_0 
=
\overline{\bm{\nu}}\T
\sum_{i=1}^{N} \bm{\mu}_i \gamma_i^\dagger
\quad \Rightarrow \quad 
\bm{\mu}_0\T \bm{\nu}
=
\sum_{i=1}^{N} \gamma_{i}^\dagger \bm{\mu}_i\T \bm{\nu}
\label{eq-time-invariant-IFEM-1}
\end{align}	
and
\begin{align}
&
\bm{\mu}_0\T \Lambda \bm{\mu}_0 + \sigma_{00} 
=	
\sum_{i=1}^{N}
\big(
\bm{\mu}_0\T
\Lambda \bm{\mu}_i + \sigma_{0i} \big) \gamma_i^\dagger   
\nonumber
\\
&
\Rightarrow \quad
\bm{\mu}_0\T \bigg( \Sigma_{\rho} + \lim_{T_0 \rightarrow \infty} \sum_{t=1}^{T_0} \bm{\nu}_t\bm{\nu}_t\T \bigg)
\bm{\mu}_0 + \sigma_{00}
=
\sum_{i=1}^{N} \gamma_{i}^\dagger \bigg\{ \bm{\mu}_0\T \bigg( \Sigma_{\rho} + \lim_{T_0 \rightarrow \infty} \sum_{t=1}^{T_0} \bm{\nu}_t\bm{\nu}_t\T \bigg) \bm{\mu}_i + \sigma_{0i} \bigg\}
\nonumber
\\
&
\Rightarrow \quad
\bm{\mu}_0\T \big( \Sigma_{\rho} +  \bm{\nu}\bm{\nu}\T \big)
\bm{\mu}_0 + \sigma_{00}
=
\sum_{i=1}^{N} \gamma_{i}^\dagger \big\{ \bm{\mu}_0\T \big( \Sigma_{\rho} +  \bm{\nu}\bm{\nu}\T \big) \bm{\mu}_i + \sigma_{0i} \big\}
\nonumber
\\
&
\stackrel{\eqref{eq-time-invariant-IFEM-1}}{\Rightarrow} \quad
\bm{\mu}_0\T \Sigma_{\rho} 
\bm{\mu}_0 + \sigma_{00}
=
\sum_{i=1}^{N} \gamma_{i}^\dagger  \bm{\mu}_0\T \Sigma_{\rho} \bm{\mu}_i + \sigma_{0i} \ . 
\label{eq-time-invariant-IFEM-2}
\end{align}
Therefore, $\bW\T \bgamma^\dagger$ satisfies \eqref{eq-IFEM-SPSC-3}, implying that it is a valid synthetic control satisfying Assumption \ref{assumption:SC}. Therefore, $\bW_t\T \widehat{\bgamma}_{\rho}$ can be used to obtain a consistent estimate for the ATT.

However, the solution $\bgamma^\dagger$ in general does not satisfy \eqref{eq-IFEM-SPSC-1} when $\EXP(\blambda_t)=\bm{\nu}_t$ is time-varying. This is because conditions \eqref{eq-time-invariant-IFEM-1} and \eqref{eq-time-invariant-IFEM-2} are not generally satisfied. 	Therefore, $\bW\T \bgamma^\dagger$ fails to satisfy \eqref{eq-IFEM-SPSC-3}. Therefore, the synthetic control $\bW_t\T \widehat{\bgamma}_{\rho}$ converges to a invalid synthetic control that fails to satisfy Assumption \ref{assumption:SC}, leading to an inconsistent estimate for the ATT.

\subsubsection{Using the Time-varying Estimating Equation}

Suppose that the synthetic control weights are estimated by the time-varying pre-treatment estimating equation:
\begin{align*}	
\widehat{\bgamma}_{\rho}
=
\argmin_{\bgamma}
\Bigg[ 
\bigg\|
\frac{1}{T_0}
\sum_{t=1}^{T_0}
\bigg[ 
\begin{array}{c}
\bD_t
\\[-0.2cm]
\bh (Y_t - \bD_t\T \Beta^*)
\end{array}
\bigg]
\big( Y_t - \bW_{ t} \T \bgamma \big)
\bigg\|_2^2
+ \rho \big\| \bgamma \big\|_2^2
\Bigg]
\ ,
\end{align*}
where $\Beta^*$ satisfies $\EXP(Y_t) = \bm{\mu}_0\T \bm{\nu}_t = \bD_t\T \Beta^*$ for $t \in \{1,\ldots,T_0\}$. Note that $\bm{\eta}^* = \mathcal{E}^* \bm{\mu}_0$. For simplicity, we choose $\bh(y)=y$. Then, it is straightforward to show that the regularized GMM estimator of $\bgamma$ is given by
\begin{align*}
\widehat{\bgamma}_{\rho}
=
\big(
\widehat{\bG}_{YW} \T \widehat{\bG}_{YW}
+
\rho I_{N \times N} 
\big)^{-1}
\big( 
\widehat{\bG}_{YW} \T \widehat{\bG}_{YY}
\big)
\end{align*}
where 
\begin{align*}
&
\widehat{\bG}_{YW} =
\frac{1}{T_0} 
\sum_{t=1}^{T_0} \begin{pmatrix}
\bD_t \bW_{ t}\T 
\\
(Y_t - \bD_t\T \Beta^*) \bW_{ t}\T
\end{pmatrix}   \in \R^{(d+1) \times N}
\ , 
&&
\widehat{\bG}_{YY} = 
\frac{1}{T_0} \sum_{t=1}^{T_0} \begin{pmatrix}
\bD_t Y_t  
\\
(Y_t - \bD_t\T \Beta^*) Y_t
\end{pmatrix} \in \R^{d+1} \ .
\end{align*}	
Note that $Y_t - \bD_t\T \Beta^* = \bm{\mu}_0\T (\blambda_t - \bm{\nu}_t) + e_{0t}$, which results in
\begin{align*}
\EXP \big\{ \big( Y_t - \bD_t\T \Beta^* ) \bW_t\T \big\}
&
=
\EXP \big[ \big\{  \bm{\mu}_0\T (\blambda_t - \bm{\nu}_t) + e_{0t} \big\} 
\big( \blambda_t\T \mathfrak{M} + \bm{e}_t\T \big) \big]
\\
&
=
\bm{\mu}_0\T \EXP \big\{ (\blambda_t-\bm{\nu}_t) \blambda_t\T \big\} \mathfrak{M}
+ \Sigma_{0,(-0)}
\ , 
&&
\Leftarrow \quad 
\Sigma_{0,(-0)} = (\sigma_{01},\cdots,\sigma_{0N}) \in \R^{1 \times N}
\\
&
=
\bm{\mu}_0\T \Sigma_{\rho} \mathfrak{M}
+ \Sigma_{0,(-0)} \ ,
\\
\EXP \big\{ \big( Y_t - \bD_t\T \Beta^* ) Y_t \big\}
&
=
\EXP \big[ \big\{  \bm{\mu}_0\T (\blambda_t - \bm{\nu}_t) + e_{0t} \big\} 
\big( \blambda_t\T \bm{\mu}_0 + e_{0t} \big) \big]
\\
&
=
\bm{\mu}_0\T \EXP \big\{ (\blambda_t-\bm{\nu}_t) \blambda_t\T \big\} \bm{\mu}_0
+ \sigma_{00}
\\
&
=
\bm{\mu}_0\T \Sigma_{\rho} \bm{\mu}_0
+ \sigma_{00}
\ .
\end{align*}
From the law of large numbers, we find
\begin{align*}
&
\widehat{\bG}_{YW} \stackrel{P}{\rightarrow} 
{\bG}_{YW}^*
=
\begin{pmatrix}	
(\overline{\bm{D} \bm{\nu}}\T)
\mathfrak{M} 
\\
\bm{\mu}_0\T  
\Sigma_{\rho}
\mathfrak{M} 
+
\Sigma_{0,(-0)}
\end{pmatrix}
\ , 
\quad 
&&
\hspace*{-1cm}
\widehat{\bG}_{YY}
\stackrel{P}{\rightarrow} 
{\bG}_{YY}^*
=
\begin{pmatrix}	
(\overline{\bm{D} \bm{\nu}}\T)
\bm{\mu}_0
\\
\bm{\mu}_0\T  
\Sigma_{\rho}
\bm{\mu}_0
+\sigma_{00}
\end{pmatrix}
\ , 
\\
&
\mathfrak{M} =	\big[ \bm{\mu}_1 \ , \ \cdots \ , \ \bm{\mu}_{N} \big]
\in \R^{r \times N}
\ , \quad 
&&
\hspace*{-1cm}
\Sigma_{0,(-0)} = (\sigma_{01},\cdots,\sigma_{0N}) \in \R^{1 \times N}
\\
&
(\overline{\bm{D} \bm{\nu}}\T)
=
\lim_{T_0 \rightarrow \infty}
\frac{1}{T_0} \sum_{t=1}^{T_0}  
(\bm{D}_t \bm{\nu}_t\T)
\in \R^{d \times r}
\ , \quad 
&&
\hspace*{-1cm}
\overline{\bm{\nu}}
=
\lim_{T_0 \rightarrow \infty}
\frac{1}{T_0} \sum_{t=1}^{T_0} \bm{\nu}_t
\in \R^{r}
\ .
\end{align*}
Since $\bm{\nu}_t = \mathcal{E}\sT \bD_t$, we have $(\overline{\bm{D} \bm{\nu}}\T) = (\overline{\bm{D} \bm{D}}\T)
\mathcal{E}^*$ where
\begin{align*}
&
(\overline{\bm{D} \bm{D}}\T)
=
\lim_{T_0 \rightarrow \infty}
\frac{1}{T_0} \sum_{t=1}^{T_0}  
(\bm{D}_t \bm{D}_t\T) 
\in \R^{d \times d}
\ .
\end{align*}
We can re-define $\bm{D}_t$ so that $(\overline{\bm{D} \bm{D}}\T)$ is of full rank. Therefore, the limit of $\widehat{\bgamma}_{\rho}$ is $\bgamma^\dagger = \big( \bG_{YW}^* \big)^+ \bG_{YY}^*$, which is the minimum-norm solution of the equation $ \bG_{YY}^*
=
\bG_{YW}^* \bgamma $. Therefore, $\bgamma^\dagger$ satisfies
\begin{align*}
\bG_{YY}^*
=
\bG_{YW}^* \bgamma^\dagger
\quad \Leftrightarrow \quad 
\left\{
\begin{array}{l}	
(\overline{\bm{D} \bm{D}}\T)
\mathcal{E}^* \bm{\mu}_0
=	
(\overline{\bm{D} \bm{D}}\T)
\mathcal{E}^* \mathfrak{M}
\bgamma^\dagger 
\\
\bm{\mu}_0\T \Sigma_{\rho} \bm{\mu}_0 + \sigma_{00}
=
\sum_{i=1}^{N}
\big(
\bm{\mu}_0\T
\Sigma_{\rho} \bm{\mu}_i + \sigma_{0i} \big) \gamma_i^\dagger
\end{array}
\right. 
\ .
\end{align*}
Consequently, we further obtain
\begin{align}
(\overline{\bm{D} \bm{D}}\T)
\mathcal{E}^* \bm{\mu}_0
=	
(\overline{\bm{D} \bm{D}}\T)
\mathcal{E}^* \mathfrak{M}
\bgamma^\dagger
\quad 
& 
\Leftrightarrow
\quad
\mathcal{E}^* \bm{\mu}_0
=
\sum_{i=1}^{N} \mathcal{E}^* \bm{\mu}_i \gamma_i^\dagger
\nonumber
\\
&
\Rightarrow \quad 
\bD_t\T
\mathcal{E}^* \bm{\mu}_0
=
\sum_{i=1}^{N} \bD_t\T \mathcal{E}^* \bm{\mu}_i
\nonumber
\\
&
\Leftrightarrow \quad 
\bm{\nu}_t\T \bm{\mu}_0
=
\sum_{i=1}^{N} \bm{\nu}_t\T \bm{\mu}_i \gamma_i^\dagger
\label{eq-time-invariant-IFEM-3}
\end{align}	
and
\begin{align}
&
\bm{\mu}_0\T \Sigma_{\rho} \bm{\mu}_0 + \sigma_{00}
=
\sum_{i=1}^{N}
\big(
\bm{\mu}_0\T
\Sigma_{\rho} \bm{\mu}_i + \sigma_{0i} \big) \gamma_i^\dagger \ .
\label{eq-time-invariant-IFEM-4}
\end{align}
Therefore, $\bW\T \bgamma^\dagger$ satisfies \eqref{eq-IFEM-SPSC-3}, implying that $\bW\T \bgamma^\dagger$ is a valid synthetic control satisfying Assumption \ref{assumption:SC}. Therefore, $\bW_t\T \widehat{\bgamma}_{\rho}$ can be used to obtain a consistent estimate for the ATT.

\subsection{Extension: Covariate Adjustment}		\label{sec:Cov}

In practice, a rich collection of measured exogenous covariates may be available. One may want to incorporate these covariates in the synthetic control analysis because using these covariates may improve efficiency. In this Section, we provide details on the SPSC framework by incorporating measured covariates. Specifically, we denote $q$-dimensional measured exogenous covariates for unit $i  \in \{ 0,\ldots,N\}$ at time $t \in \{1,\ldots,T\}$ as $\bX_{it} \in \R^{q}$; we remind the readers that $i=0$ is the treated unit and $i \in \{1,\ldots,N\}$ are the untreated units. Let $\bX_{t} = ( \bX_{1 t}\T,\ldots,\bX_{N t}\T)\T \in \R^{Nq}$ be the collection of all measured covariates of donors at time $t$. To account for covariates, we modify Assumptions \ref{assumption:valid proxy} and \ref{assumption:SC} as follows:

\begin{assumption}[Proxy \& Existence of a Synthetic Control Bridge Function] \label{assumption:valid proxy Cov}
There exists a function $h: \R^{N(q+1)} \rightarrow \R$ satisfying 
\begin{align}
&
h^*(\bW_{ t}, \bX_{0t}, \bX_{ t} ) \nindep \potY{t}{0} \cond ( \bX_{0t}, \bX_t )
\ , 
&&
t \in \{1, \ldots, T_0\} \ ,
\\
& 
\potY{t}{0} = \EXP \big\{
h^*(\bW_{ t}, \bX_{0t}, \bX_{t} )
\cond
\potY{t}{0}, \bX_{0t}, \bX_{t}
\big\} \ \ \text{almost surely} 
\ , 
&&
t \in \{1, \ldots, T\} \ .
\label{eq-relevant Cov}
\end{align}
\end{assumption} 

\begin{theorem}	\label{thm:Extension NP Cov}

Suppose that Assumptions \ref{assumption:consistency}, \ref{assumption:noitf}, \ref{assumption:valid proxy Cov} are satisfied. Then, for $t \in \{1,\ldots,T_0\}$, the synthetic control bridge function $h^*$ satisfies 
\begin{align}
\EXP \big\{ Y_t - h^*(\bW_{ t},\bX_{0t}, \bX_{t} ) \cond Y_t, \bX_{0t}, \bX_{t} \big\} = 0 \ \text{ almost surely } \ .
\label{eq-Fredholm}
\end{align}
Moreover, for $t \in \{1,\ldots,T\}$, we have 
\begin{align*}
\EXP \big\{ \potY{t}{0} - h^*(\bW_{ t},\bX_{0t}, \bX_{t} ) \big\} = 0 \ .
\end{align*}
Lastly, the ATT at time $t  \in \{ T_0+1,\ldots,T\}$ is identified as 
\begin{align*}
\tau_t^*
=
\EXP
\big\{
Y_t - h^*(\bW_{ t}, \bX_{0t},\bX_{t} ) 
\big\} \ .
\end{align*}
\end{theorem}

Leveraging the result of the Theorem, estimation and inference of the ATT with covariate adjustment can be established, which is a straightforward extension of Section \ref{sec:Estimation}. Consider that the bridge function is linear as follows:
\begin{align*}
h^*(\bW_{ t}, \bX_{0t}, \bX_{t} )
=
\bW_{ t}\T \bgamma^*
-
\bX_{0 t} \T \bdelta_{0}^*
+
\bX_{ t} \T \bdelta^* \ .
\end{align*}
Following \ref{sec:Estimation}, we define the following estimating function:
\begin{align}			\label{eq-Moment-Cov}
&
\Psi_{\text{Cov}}( \bO_t \con \Beta, \bgamma, \bdelta_0, \bdelta, \bbeta )
\\
&
=
\begin{bmatrix}
(1-A_t)
\bD_t \big( Y_t - \bX_{0t}\T \bdelta_{0} - \bD_t\T \Beta \big)
\\
(1-A_t)
\bg (t,Y_t, \bX_{0t}, \bX_{t} \con \Beta, \bdelta_0, \bdelta)
\big\{ \big( Y_t - \bX_{0t}\T \bdelta_{0} \big)  - \big( \bW_{ t} \T \bgamma - \bX_{t}\T \bdelta  \big) \big\}
\\
A_t 
\frac{\partial \tau(t \con \bbeta) }{\partial \bbeta }  
\big\{ \big( Y_t - \bX_{0t}\T \bdelta_{0} \big)  - \big( \bW_{ t} \T \bgamma - \bX_{t}\T \bdelta  \big)  - \tau (t \con \bbeta)
\big\}
\end{bmatrix}
\in \R^{2d+p+q+b} \ ,
\nonumber
\end{align}
where $\bO_t = (Y_t,\bW_{ t}, \bX_{0t}, \bX_{t}, A_t)$ is the collection of the observed data at time $t$, $\bg(\cdot)$ is a $(d+p+q)$-dimensional user-specified function of $(t, Y_t,\bX_{0t},\bX_{ t})$, and $\tau(t \con \bbeta)$ is a user-specified treatment effect function. A example for $\bg(\cdot)$ includes:
\begin{align*}
\bg (t,y, \bx_0, \bx \con \Beta, \bdelta_0, \bdelta)
=
\begin{bmatrix}
\bD_t 
\\ \bh(y - \bx_0\T \bdelta_0 - \bD_t\T \Beta)
\\ \bx_{0t}
\\ \bx_{t}
\end{bmatrix}
\in \R^{d+p+(N+1)q} \ .
\end{align*}
We allow $\dim(\bg)$ to be smaller than $\dim(\Beta,\bgamma,\bbeta,\bdelta_0,\bdelta)$. Let $\bgamma_0^*$ be the minimum-norm solution for $\EXP \big\{ \Psi_{\text{Cov}} (\bO_t \con  \Beta^*, \bgamma, \bdelta_0^*, \bdelta^*, \bbeta^* ) \big\} = 0$.

We assume that the treatment effect function is chosen so that the associated error process is weakly dependent:
\begin{assumption}[Weakly Dependent Error in the Presence of Covariates] \label{assumption:weakdep Cov}
Let $\epsilon_t$ be $\epsilon_t = \big(
Y_t
-
\bX_{0 t} \T \bdelta_{0}^*
\big)
-
\big(
\bW_{ t}\T \bgamma_{0}^*
-
\bX_{ t} \T \bdelta^*
\big) - \tau (t \con \bbeta^*) $. Then, the error process $\big\{ \epsilon_1,\ldots,\epsilon_{T} \big\}$ satisfies Assumption \ref{assumption:weakdep}, i.e., $\text{corr}(\epsilon_{t}, \epsilon_{t+t'})$ converges to 0 as $t' \rightarrow \pm \infty$.
\end{assumption}	

We then establish the asymptotic normality of the regularized GMM estimators $(\widehat{\Beta}, \widehat{\bgamma}_{\rho}, \widehat{\bdelta}_0, \widehat{\bdelta}, \widehat{\bbeta} ) $; see the formal statement below:
\begin{theorem}	\label{thm:AN Cov}
Suppose that Assumptions \ref{assumption:consistency}, \ref{assumption:noitf}, \ref{assumption:valid proxy Cov}, and \ref{assumption:weakdep Cov} hold, $(\Beta^*,\bdelta_0^*,\bdelta^*,\bbeta^*)$ are unique, and Regularity Conditions in Section \ref{sec:supp:AN} hold. Let $( \widehat{\Beta},\widehat{\bgamma}_{\rho}, \widehat{\bdelta}_0, \widehat{\bdelta}, \widehat{\bbeta} ) $ be the regularized GMM estimators where the estimating function \eqref{eq-Moment-Cov} is used, i.e., 
\begin{align*}
\big(
\widehat{\Beta}
,
\widehat{\bgamma}_{\rho}
, 
\widehat{\bdelta}_0
,
\widehat{\bdelta} 
,
\widehat{\bbeta}
\big)
=
\argmin_{(\Beta,\bgamma,\bdelta_0,\bdelta,\bbeta)}
\Big[
\big\{ \widehat{\Psi}_{\text{Cov}}( \Beta,\bgamma, \bdelta_0, \bdelta, \bbeta) \big\} \T
\widehat{\Omega}_{\text{Cov}}
\big\{ \widehat{\Psi}_{\text{Cov}}( \Beta,\bgamma, \bdelta_0, \bdelta, \bbeta) \big\} 
+ \rho \big\| \bgamma \big\|^2
\Big]
\ ,
\end{align*}
where $	\widehat{\Psi}_{\text{Cov}}(\Beta,\bgamma, \bdelta_0, \bdelta, \bbeta)
= T^{-1}
\sum_{t=1}^{T} \Psi_{\text{Cov}} (\bO_t \con \bgamma, \bdelta_0, \bdelta, \bbeta)$ is the empirical mean of the estimating function and $\widehat{\Omega}_{\text{Cov}} = \text{diag}(I_{d \times d}, \widehat{\Omega}_{\bg}, \widehat{\Omega}_{\post})$ is a user-specified symmetric, block-diagonal positive definite matrix. Then, as $T \rightarrow \infty$, we have
\begin{align*}
\sqrt{ T }
\left\{
\begin{pmatrix}
\widehat{\Beta}
\\
\widehat{\bgamma}_{\rho}
\\
\widehat{\bdelta}_0
\\
\widehat{\bdelta} 
\\
\widehat{\bbeta}
\end{pmatrix}
-
\begin{pmatrix}
\Beta^*
\\
\bgamma_0^*
\\
\bdelta_0^*
\\
\bdelta^*
\\
\bbeta^* 
\end{pmatrix}
\right\}
\text{ converges in distribution to }
N \big( 0, \Sigma_{\text{Cov},1}^* \Sigma_{\text{Cov},2}^* \Sigma_{\text{Cov},1}\sT \big) \ ,
\end{align*}
where
\begin{align*}
&
\Sigma_{\text{Cov},1}^* 
=
\bigg[ \Omega^{*1/2}
\lim_{T \rightarrow \infty} 
\frac{\partial \EXP \big\{ \widehat{\Psi} (\Beta,\bgamma,\bdelta_0,\bdelta,\bbeta) \big\} }{ \partial (\Beta,\bgamma,\bdelta_0,\bdelta,\bbeta)\T }
\bigg|_{ \Beta=\Beta^*, \bgamma=\bgamma_0^* , \bbeta=\bbeta^* }
\bigg]^{+} 
\Omega^{*1/2}
\ , 
\\
&
\Sigma_{\text{Cov},2}^* 
=
\lim_{T \rightarrow \infty} 
\VAR \Big\{ \sqrt{T} \cdot  \widehat{\Psi} (\Beta^*,\bgamma_0^*, \bbeta^*) \Big\} \ .
\end{align*} 
Here, $\Omega^{*1/2}$ is a symmetric positive-definite matrix satisfying $\big( \Omega^{*1/2} \big)^2 = \lim_{T \rightarrow \infty} \widehat{\Omega}$.
\end{theorem}
Estimators of $\Sigma_{\text{Cov},1}^*$ and $\Sigma_{\text{Cov},2}^*$ can be similarly defined as in Section \ref{sec:Estimation}, thus we omit the details here.

\subsection{Additional Simulation Studies} \label{sec:supp:Simulation}

We restate the data generating process of the simulation studies in Section \ref{sec:Sim}. The length of pre- and post-treatment periods were given by $T_0 = T_1 \in \{ 50, 100, 250, 500 \}$ and the number of donors were given by $N = 16$.

First, for each $t \in \{1,\ldots,T\}$, we independently generated 4-dimensional latent factors $\blambda_t = (\lambda_{1t},\cdots,\lambda_{4t})\T$ from $N(\bm{\nu}_t,0.25\cdot I_{4\times 4})$, with $\blambda_t$ being independent across time periods. For the mean vector $\bm{\nu}_t=(\nu_{1t},\cdots,\nu_{4t})\T$, we considered the following four specifications for $j \in \{1,\ldots,4\}$:
\begin{align*}
& (\textit{No trend with no intercept}): && \nu_{jt} = 0; 	\quad 
&& (\textit{No trend with intercept}): && \nu_{jt} = 1;
\\
& (\textit{Linear trend with no intercept}): &&\nu_{jt} = t/T_0;	\quad 
&& (\textit{Linear trend with intercept}): && \nu_{jt} = 1+t/T_0 \ .
\end{align*} 

The latent factor loadings $\bm{\mu}_{i}$ for $i \in \{1,\ldots,16\}$, i.e., latent factor loadings of untreated units,  were specified as follows:
\begin{align*}
\mathfrak{M}
=
\begin{bmatrix}
\bm{\mu}_1 & \cdots & \bm{\mu}_{16}
\end{bmatrix}
=
\begin{bmatrix}
2 & 1.75 & 1.5 & 1.25 & 1 & 0.75 & 0.5 & 0.25 & 0_{1 \times 8} \\
0.8 & 0.8 & 0.6 & 0.6 & 0.4 & 0.4 & 0.2 & 0.2 & 0_{1 \times 8} \\
0 & 0 & 0 & 0 & 0 & 0 & 0 & 0 & 1_{1 \times 8} \\
0 & 0 & 0 & 0 & 0 & 0 & 0 & 0 &  0.5 \cdot 1_{1 \times 8} 
\end{bmatrix}
\in \R^{4 \times 16} \ .
\end{align*}
The latent factor loading $\bm{\mu}_{0}$, i.e., latent factor loading of the treated unit, was specified from either one of the followings:
\begin{align*}
& (\textit{Simplex}): && \bm{\mu}_{0} = (1.125,0.5,0,0)\T = \sum_{i=1}^{8}  \bm{\mu}_{i}/8 ; \quad \quad \quad 
&& (\textit{Non-simplex}): && \bm{\mu}_{0} = (2,1.5,0,0)\T \ .
\end{align*}

The errors $\bm{e}_t = (e_{0t} ,  e_{1t} , \cdots,  e_{16 t})\T$ were generated independently across time periods from $\bm{e}_t \sim N \big( 0_{16 \times 1} , 0.25 \cdot \text{diag} (\Sigma_{e}, I_{8 \times 8}) \big)$ where $\Sigma_e \in \R^{9 \times 9}$ were chosen from one of the following three matrices with the corresponding $\omega_i$ values in \eqref{eq-LinearModel-System}:
\begin{align*}
&
(\textit{Independent errors}):
&&
\Sigma_{e} = I_{9 \times 9}   \ ;
\\
&
&&
\omega_0 = 1 , \ 
\omega_1=\cdots=\omega_{16}=0     \ ;
\\
&
(\textit{Correlated errors}):
&&
\Sigma_{e} = 
0.1 \cdot I_{9 \times 9} + 0.9  \cdot 1_{9 \times 9}    \ ;
\\
&
&&
\omega_0 = 1 , \ 
\omega_1=\cdots=\omega_{8}=0.9 , \
\omega_9=\cdots=\omega_{16}=0    \ ;
\\
&
(\textit{No $Y$ error}):
&&
\Sigma_{e} = \text{diag}(0,I_{8 \times 8})   \ ;
\\
&
&&
\omega_0 =\cdots=\omega_{16}=0 \ .
\end{align*} 

With these generated variables, $\potY{t}{a}$ and $W_{it}$ at $t \in \{1,\ldots,T\}$ were generated as 
\begin{align*}
\potY{t}{0} 
&
= \bm{\mu}_0\T \bm{\lambda}_t + e_{0t}
\ , \quad 
&&
&&
\potY{t}{1} 
= \potY{t}{0} + 3 A_t + \epsilon_{t} \ , 
&&
\epsilon_{t} \stackrel{iid}{\sim} N(0,0.25) \ , 
\\
W_{it} 
&
= \bm{\mu}_i\T \bm{\lambda}_t + e_{it}
\ , 
&&
i \in \{1,\ldots,N\} \ .
\end{align*} 

We review the six approaches employed for the analysis:
\begin{itemize}[leftmargin=1cm,itemsep=0cm,parsep=0cm]
\item[](\textit{OLS-NoReg}) 
OLS-based approach based on \eqref{eq-ExistSC-Abadie} with no regularization.
\item[](\textit{OLS-Standard}) 
The standard synthetic control approach proposed by \citet{Abadie2010}; we used \texttt{synth} R-package \citep{Synth2011}.
\item[] (\textit{ASC}) 
The augmented synthetic control approach proposed by \citet{ASCM2021}; we used \texttt{augsynth} \citep{ASCM2023package} R-package.
\item[] (\textit{SCPI}) 
The synthetic control prediction interval approach proposed by \citet{Cattaneo2021}; we used \texttt{scpi} \citep{scpiPackage2023} R-package. 
\item[] (\textit{SPSC-NoDT}) 
The single proxy synthetic control approach with no time-varying component; we used \texttt{SPSC} \citep{SPSC2024package} R-package.
\item[] (\textit{SPSC-DT}) 
The single proxy synthetic control approach with time-varying components \texttt{SPSC} \citep{SPSC2024package} R-package.
\end{itemize}
For the two SPSC estimators, we set $\bh(y) = y$ and, for SPSC-DT, we additionally set $\bD_t=\mathcal{B}_6(t)$, 6-dimensional cubic B-spline bases functions. 

\begin{figure}[!htb]
\centering
\includegraphics[width= \textwidth]{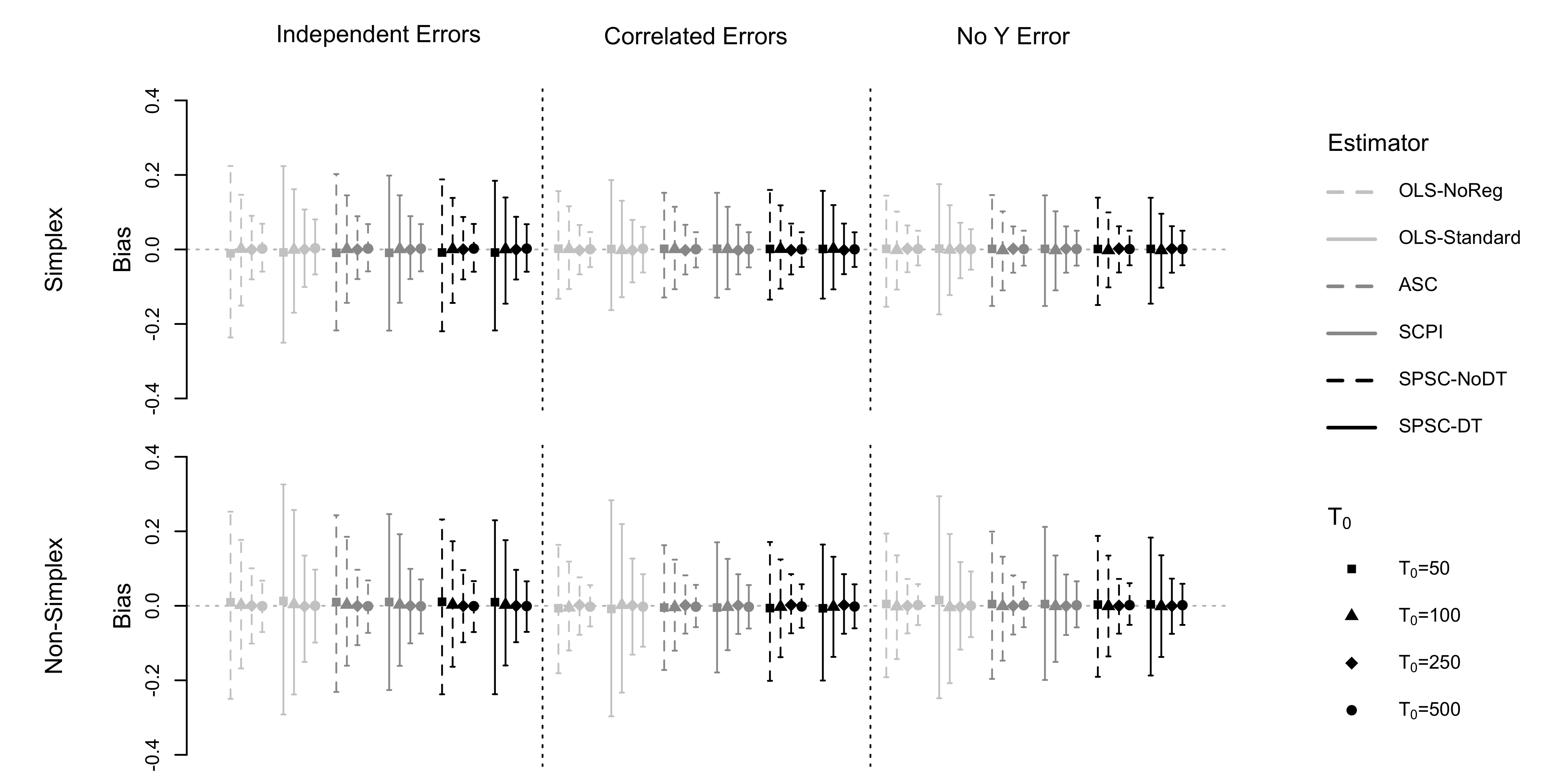}%
\caption{A Graphical Summary of Empirical Distributions of the Estimates for the ATT under (No trend with no intercept)}
\label{Fig:Supp:Sim:L0}
\end{figure}

\begin{figure}[!htb]
\centering
\includegraphics[width= \textwidth]{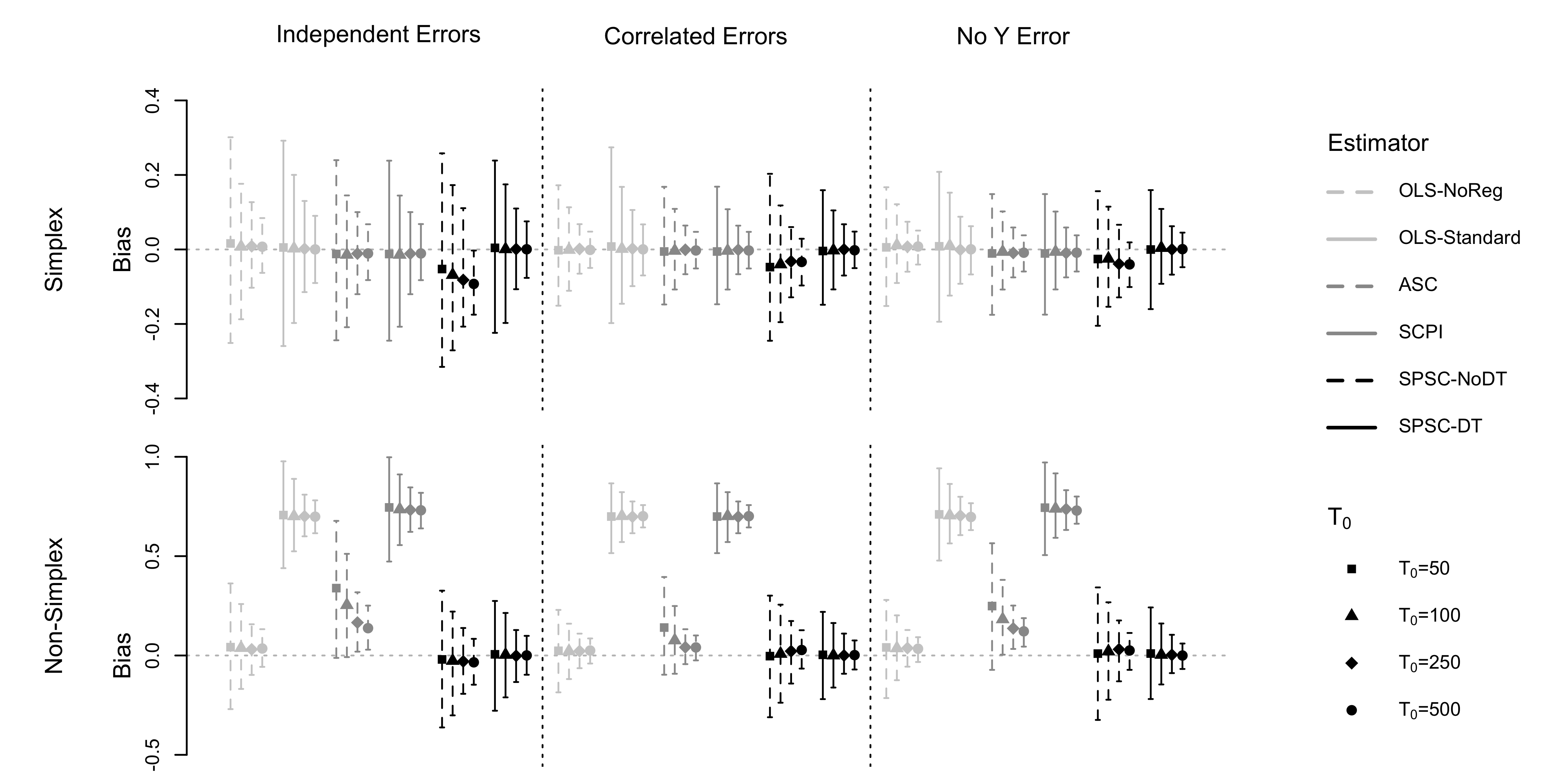}%
\caption{A Graphical Summary of Empirical Distributions of the Estimates for the ATT under (No trend with intercept)}
\label{Fig:Supp:Sim:L1}
\end{figure}

\begin{figure}[!htb]
\centering
\includegraphics[width= \textwidth]{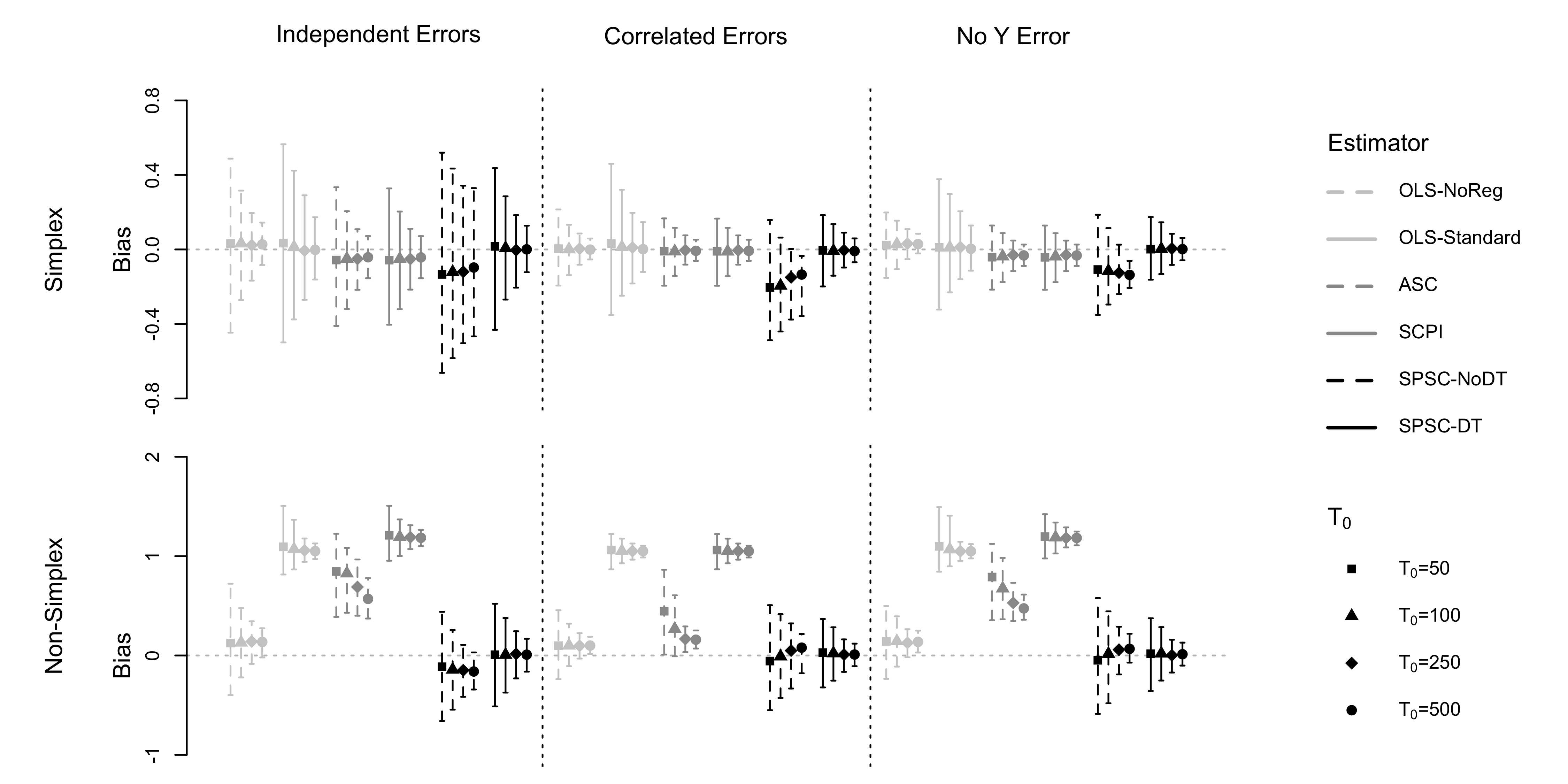}%
\caption{A Graphical Summary of Empirical Distributions of the Estimates for the ATT under (Linear trend with no intercept)}
\label{Fig:Supp:Sim:L2}
\end{figure}

\begin{figure}[!htb]
\centering
\includegraphics[width= \textwidth]{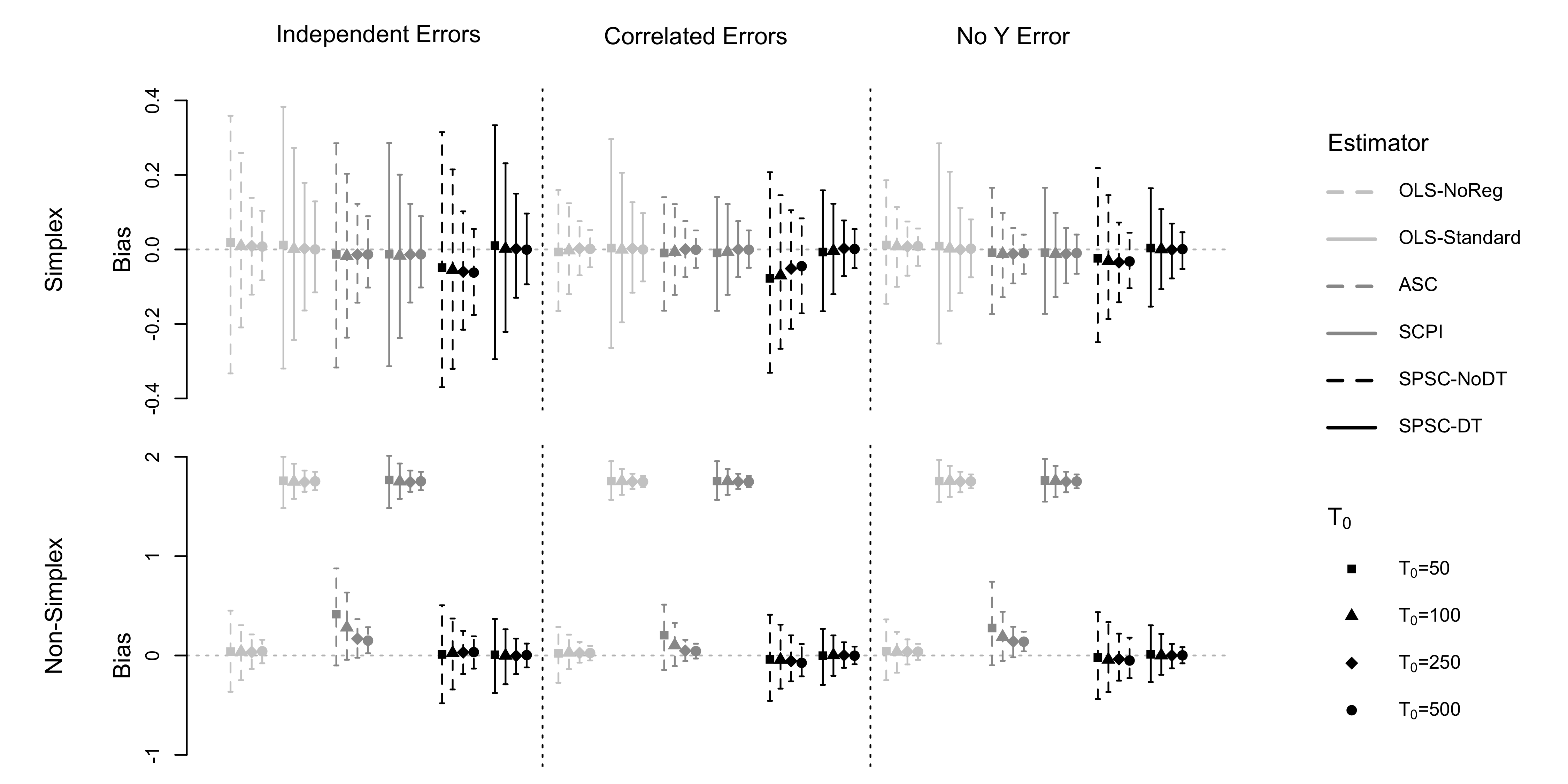}%
\caption{A Graphical Summary of Empirical Distributions of the Estimates for the ATT under (Linear trend with intercept)}
\label{Fig:Supp:Sim:L3}
\end{figure}

We first estimated the ATT $\tau_t^*=3$. Figures \ref{Fig:Supp:Sim:L0}-\ref{Fig:Supp:Sim:L3} summarize the empirical distribution of the estimators graphically. Each figure is drawn in the following format:
\begin{itemize}[leftmargin=1cm,itemsep=0cm,parsep=0cm]
\item The vertical segments represent 95\% Monte Carlo confidence interval for each estimator obtained from 500 estimates.
\item The dots represent the empirical mean of 500 estimates.
\item The colors (light gray, gray, and black) and line types (solid and dashed) encode a corresponding estimator
\item The shape of the dots encode the length of the pre-treatment period, respectively. 
\item The $y$-axis represents the magnitude of bias.
\end{itemize}
We find that the results are similar to those in Figure \ref{fig:Sim:Constant} of the main paper. When $\bm{\mu}_0$ is generated from (Non-simplex) and $\blambda_t$ has non-zero mean, we find the OLS-Standard, ASC, and SCPI estimators appear to have significant magnitudes of biases even under a large sample size.

\newpage

\newpage

\newpage

In Tables \ref{Tab:Supp:Sim:T0} and \ref{Tab:Supp:Sim:T1}, we first present numerical summaries of the simulation studies considered in the main paper. Each table is written in the following format:
\begin{itemize}[leftmargin=1cm,itemsep=0cm,parsep=0cm]
\item Bias row shows the empirical bias of 500 estimates;
\item ASE row shows the asymptotic standard error obtained from the sandwich variance estimator;
\item BSE row shows the bootstrap standard error obtained from the approach in Section \ref{sec:supp:BB} of the Supplementary Material;
\item ESE row shows the standard deviation of 500 estimates;
\item MSE row shows the mean squared error of 500 estimates;
\item Cover (ASE) and Cover (BSE) show the empirical coverage rates of 95\% confidence intervals based on the asymptotic and bootstrap standard errors, respectively;
\item Bias, standard errors, and mean squared error are scaled by factors of 10, 10, and 100, respectively, for readability.	
\end{itemize}
We remark that the results in Tables \ref{Tab:Supp:Sim:T0}-\ref{Tab:Supp:Sim:T3} are similar to those in Table \ref{tab:Sim:ATT} of the main text. In particular, when $\bm{\mu}_0$ is generated from (Non-simplex) and $\blambda_t$ has a non-zero mean, we find the OLS-Standard, ASC, and SCPI estimators yield significant magnitudes of biases even under a large sample size. Moreover, these biases are not negligible compared to the magnitude of the empirical standard errors. Consequently, we can deduce that the failure of the ASC and SCPI approaches to attain the nominal coverage rate can be attributed to the non-diminishing bias.

\newpage

\begin{table}[!htp]
\renewcommand{\arraystretch}{1.3} \centering
\scriptsize
\setlength{\tabcolsep}{1pt}

\begin{tabular}{|c|c|c|c|cccccccccccc|}
\hline
\multirow{3}{*}{$\bm{\lambda}_t$}                                                                   & \multirow{3}{*}{$\bm{\mu}_0$} & \multirow{3}{*}{$\bm{e}_t$}                                                    & \multirow{3}{*}{Statistics} & \multicolumn{12}{c|}{Estimators and $T_0$}                                                                                                                                                                                                                                                                                                       \\ \cline{5-16}

&                                &                                                                                &                             & \multicolumn{2}{c|}{OLS-NoReg}                            & \multicolumn{2}{c|}{OLS-Standard}                         & \multicolumn{2}{c|}{ASC}                                  & \multicolumn{2}{c|}{SCPI}                                 & \multicolumn{2}{c|}{SPSC-NoDT}                            & \multicolumn{2}{c|}{SPSC-DT}         \\ \cline{5-16} 
&                                &                                                                                &                             & \multicolumn{1}{c|}{100}    & \multicolumn{1}{c|}{500}    & \multicolumn{1}{c|}{100}    & \multicolumn{1}{c|}{500}    & \multicolumn{1}{c|}{100}    & \multicolumn{1}{c|}{500}    & \multicolumn{1}{c|}{100}    & \multicolumn{1}{c|}{500}    & \multicolumn{1}{c|}{100}    & \multicolumn{1}{c|}{500}    & \multicolumn{1}{c|}{100}    & 500    \\ \hline

\multirow{42}{*}{\begin{tabular}[c]{@{}c@{}}No\\ trend \\ with \\ no \\ intercept\end{tabular}} & \multirow{21}{*}{Simplex}      & \multirow{7}{*}{\begin{tabular}[c]{@{}c@{}}Independent \\ errors\end{tabular}} & Bias $(\times 10)$          & \multicolumn{1}{c|}{0.015}  & \multicolumn{1}{c|}{0.021}  & \multicolumn{1}{c|}{-0.008} & \multicolumn{1}{c|}{0.033}  & \multicolumn{1}{c|}{0.009}  & \multicolumn{1}{c|}{0.023}  & \multicolumn{1}{c|}{0.008}  & \multicolumn{1}{c|}{0.023}  & \multicolumn{1}{c|}{0.007}  & \multicolumn{1}{c|}{0.022}  & \multicolumn{1}{c|}{0.006}  & 0.022  \\ \cline{4-16} 
&                                &                                                                                & ASE $(\times 10)$           & \multicolumn{1}{c|}{0.779}  & \multicolumn{1}{c|}{0.329}  & \multicolumn{1}{c|}{-}      & \multicolumn{1}{c|}{-}      & \multicolumn{1}{c|}{-}      & \multicolumn{1}{c|}{-}      & \multicolumn{1}{c|}{-}      & \multicolumn{1}{c|}{-}      & \multicolumn{1}{c|}{0.727}  & \multicolumn{1}{c|}{0.325}  & \multicolumn{1}{c|}{0.728}  & 0.325  \\ \cline{4-16} 
&                                &                                                                                & BSE $(\times 10)$           & \multicolumn{1}{c|}{0.838}  & \multicolumn{1}{c|}{0.338}  & \multicolumn{1}{c|}{-}      & \multicolumn{1}{c|}{-}      & \multicolumn{1}{c|}{-}      & \multicolumn{1}{c|}{-}      & \multicolumn{1}{c|}{-}      & \multicolumn{1}{c|}{-}      & \multicolumn{1}{c|}{0.728}  & \multicolumn{1}{c|}{0.325}  & \multicolumn{1}{c|}{0.730}  & 0.324  \\ \cline{4-16} 
&                                &                                                                                & ESE $(\times 10)$           & \multicolumn{1}{c|}{0.730}  & \multicolumn{1}{c|}{0.319}  & \multicolumn{1}{c|}{0.836}  & \multicolumn{1}{c|}{0.377}  & \multicolumn{1}{c|}{0.722}  & \multicolumn{1}{c|}{0.319}  & \multicolumn{1}{c|}{0.722}  & \multicolumn{1}{c|}{0.318}  & \multicolumn{1}{c|}{0.716}  & \multicolumn{1}{c|}{0.316}  & \multicolumn{1}{c|}{0.714}  & 0.316  \\ \cline{4-16} 
&                                &                                                                                & MSE $(\times 100)$          & \multicolumn{1}{c|}{0.532}  & \multicolumn{1}{c|}{0.102}  & \multicolumn{1}{c|}{0.697}  & \multicolumn{1}{c|}{0.143}  & \multicolumn{1}{c|}{0.520}  & \multicolumn{1}{c|}{0.102}  & \multicolumn{1}{c|}{0.520}  & \multicolumn{1}{c|}{0.102}  & \multicolumn{1}{c|}{0.512}  & \multicolumn{1}{c|}{0.100}  & \multicolumn{1}{c|}{0.509}  & 0.100  \\ \cline{4-16} 
&                                &                                                                                & Coverage (ASE)              & \multicolumn{1}{c|}{0.954}  & \multicolumn{1}{c|}{0.952}  & \multicolumn{1}{c|}{-}      & \multicolumn{1}{c|}{-}      & \multicolumn{1}{c|}{-}      & \multicolumn{1}{c|}{-}      & \multicolumn{1}{c|}{-}      & \multicolumn{1}{c|}{-}      & \multicolumn{1}{c|}{0.950}  & \multicolumn{1}{c|}{0.954}  & \multicolumn{1}{c|}{0.948}  & 0.954  \\ \cline{4-16} 
&                                &                                                                                & Coverage (BSE)              & \multicolumn{1}{c|}{0.966}  & \multicolumn{1}{c|}{0.954}  & \multicolumn{1}{c|}{-}      & \multicolumn{1}{c|}{-}      & \multicolumn{1}{c|}{-}      & \multicolumn{1}{c|}{-}      & \multicolumn{1}{c|}{-}      & \multicolumn{1}{c|}{-}      & \multicolumn{1}{c|}{0.946}  & \multicolumn{1}{c|}{0.950}  & \multicolumn{1}{c|}{0.946}  & 0.948  \\ \cline{3-16} 
&                                & \multirow{7}{*}{\begin{tabular}[c]{@{}c@{}}Correlated \\ errors\end{tabular}}  & Bias $(\times 10)$          & \multicolumn{1}{c|}{0.010}  & \multicolumn{1}{c|}{0.006}  & \multicolumn{1}{c|}{-0.018} & \multicolumn{1}{c|}{0.022}  & \multicolumn{1}{c|}{0.010}  & \multicolumn{1}{c|}{0.006}  & \multicolumn{1}{c|}{0.009}  & \multicolumn{1}{c|}{0.006}  & \multicolumn{1}{c|}{0.014}  & \multicolumn{1}{c|}{0.004}  & \multicolumn{1}{c|}{0.015}  & 0.004  \\ \cline{4-16} 
&                                &                                                                                & ASE $(\times 10)$           & \multicolumn{1}{c|}{0.534}  & \multicolumn{1}{c|}{0.237}  & \multicolumn{1}{c|}{-}      & \multicolumn{1}{c|}{-}      & \multicolumn{1}{c|}{-}      & \multicolumn{1}{c|}{-}      & \multicolumn{1}{c|}{-}      & \multicolumn{1}{c|}{-}      & \multicolumn{1}{c|}{0.533}  & \multicolumn{1}{c|}{0.238}  & \multicolumn{1}{c|}{0.533}  & 0.238  \\ \cline{4-16} 
&                                &                                                                                & BSE $(\times 10)$           & \multicolumn{1}{c|}{0.551}  & \multicolumn{1}{c|}{0.241}  & \multicolumn{1}{c|}{-}      & \multicolumn{1}{c|}{-}      & \multicolumn{1}{c|}{-}      & \multicolumn{1}{c|}{-}      & \multicolumn{1}{c|}{-}      & \multicolumn{1}{c|}{-}      & \multicolumn{1}{c|}{0.529}  & \multicolumn{1}{c|}{0.237}  & \multicolumn{1}{c|}{0.529}  & 0.237  \\ \cline{4-16} 
&                                &                                                                                & ESE $(\times 10)$           & \multicolumn{1}{c|}{0.568}  & \multicolumn{1}{c|}{0.242}  & \multicolumn{1}{c|}{0.688}  & \multicolumn{1}{c|}{0.298}  & \multicolumn{1}{c|}{0.565}  & \multicolumn{1}{c|}{0.243}  & \multicolumn{1}{c|}{0.565}  & \multicolumn{1}{c|}{0.243}  & \multicolumn{1}{c|}{0.565}  & \multicolumn{1}{c|}{0.243}  & \multicolumn{1}{c|}{0.564}  & 0.243  \\ \cline{4-16} 
&                                &                                                                                & MSE $(\times 100)$          & \multicolumn{1}{c|}{0.322}  & \multicolumn{1}{c|}{0.059}  & \multicolumn{1}{c|}{0.473}  & \multicolumn{1}{c|}{0.089}  & \multicolumn{1}{c|}{0.319}  & \multicolumn{1}{c|}{0.059}  & \multicolumn{1}{c|}{0.319}  & \multicolumn{1}{c|}{0.059}  & \multicolumn{1}{c|}{0.318}  & \multicolumn{1}{c|}{0.059}  & \multicolumn{1}{c|}{0.318}  & 0.059  \\ \cline{4-16} 
&                                &                                                                                & Coverage (ASE)              & \multicolumn{1}{c|}{0.940}  & \multicolumn{1}{c|}{0.944}  & \multicolumn{1}{c|}{-}      & \multicolumn{1}{c|}{-}      & \multicolumn{1}{c|}{-}      & \multicolumn{1}{c|}{-}      & \multicolumn{1}{c|}{-}      & \multicolumn{1}{c|}{-}      & \multicolumn{1}{c|}{0.934}  & \multicolumn{1}{c|}{0.954}  & \multicolumn{1}{c|}{0.936}  & 0.954  \\ \cline{4-16} 
&                                &                                                                                & Coverage (BSE)              & \multicolumn{1}{c|}{0.940}  & \multicolumn{1}{c|}{0.952}  & \multicolumn{1}{c|}{-}      & \multicolumn{1}{c|}{-}      & \multicolumn{1}{c|}{-}      & \multicolumn{1}{c|}{-}      & \multicolumn{1}{c|}{-}      & \multicolumn{1}{c|}{-}      & \multicolumn{1}{c|}{0.922}  & \multicolumn{1}{c|}{0.946}  & \multicolumn{1}{c|}{0.936}  & 0.946  \\ \cline{3-16} 
&                                & \multirow{7}{*}{\begin{tabular}[c]{@{}c@{}}No \\ $Y$ error\end{tabular}}       & Bias $(\times 10)$          & \multicolumn{1}{c|}{-0.016} & \multicolumn{1}{c|}{0.012}  & \multicolumn{1}{c|}{-0.008} & \multicolumn{1}{c|}{0.007}  & \multicolumn{1}{c|}{-0.021} & \multicolumn{1}{c|}{0.011}  & \multicolumn{1}{c|}{-0.021} & \multicolumn{1}{c|}{0.011}  & \multicolumn{1}{c|}{-0.020} & \multicolumn{1}{c|}{0.011}  & \multicolumn{1}{c|}{-0.020} & 0.011  \\ \cline{4-16} 
&                                &                                                                                & ASE $(\times 10)$           & \multicolumn{1}{c|}{0.527}  & \multicolumn{1}{c|}{0.234}  & \multicolumn{1}{c|}{-}      & \multicolumn{1}{c|}{-}      & \multicolumn{1}{c|}{-}      & \multicolumn{1}{c|}{-}      & \multicolumn{1}{c|}{-}      & \multicolumn{1}{c|}{-}      & \multicolumn{1}{c|}{0.525}  & \multicolumn{1}{c|}{0.234}  & \multicolumn{1}{c|}{0.525}  & 0.234  \\ \cline{4-16} 
&                                &                                                                                & BSE $(\times 10)$           & \multicolumn{1}{c|}{0.542}  & \multicolumn{1}{c|}{0.238}  & \multicolumn{1}{c|}{-}      & \multicolumn{1}{c|}{-}      & \multicolumn{1}{c|}{-}      & \multicolumn{1}{c|}{-}      & \multicolumn{1}{c|}{-}      & \multicolumn{1}{c|}{-}      & \multicolumn{1}{c|}{0.523}  & \multicolumn{1}{c|}{0.233}  & \multicolumn{1}{c|}{0.522}  & 0.233  \\ \cline{4-16} 
&                                &                                                                                & ESE $(\times 10)$           & \multicolumn{1}{c|}{0.541}  & \multicolumn{1}{c|}{0.231}  & \multicolumn{1}{c|}{0.626}  & \multicolumn{1}{c|}{0.275}  & \multicolumn{1}{c|}{0.539}  & \multicolumn{1}{c|}{0.229}  & \multicolumn{1}{c|}{0.539}  & \multicolumn{1}{c|}{0.229}  & \multicolumn{1}{c|}{0.536}  & \multicolumn{1}{c|}{0.231}  & \multicolumn{1}{c|}{0.536}  & 0.231  \\ \cline{4-16} 
&                                &                                                                                & MSE $(\times 100)$          & \multicolumn{1}{c|}{0.292}  & \multicolumn{1}{c|}{0.053}  & \multicolumn{1}{c|}{0.391}  & \multicolumn{1}{c|}{0.076}  & \multicolumn{1}{c|}{0.290}  & \multicolumn{1}{c|}{0.052}  & \multicolumn{1}{c|}{0.290}  & \multicolumn{1}{c|}{0.052}  & \multicolumn{1}{c|}{0.287}  & \multicolumn{1}{c|}{0.053}  & \multicolumn{1}{c|}{0.287}  & 0.053  \\ \cline{4-16} 
&                                &                                                                                & Coverage (ASE)              & \multicolumn{1}{c|}{0.944}  & \multicolumn{1}{c|}{0.954}  & \multicolumn{1}{c|}{-}      & \multicolumn{1}{c|}{-}      & \multicolumn{1}{c|}{-}      & \multicolumn{1}{c|}{-}      & \multicolumn{1}{c|}{-}      & \multicolumn{1}{c|}{-}      & \multicolumn{1}{c|}{0.948}  & \multicolumn{1}{c|}{0.954}  & \multicolumn{1}{c|}{0.948}  & 0.952  \\ \cline{4-16} 
&                                &                                                                                & Coverage (BSE)              & \multicolumn{1}{c|}{0.946}  & \multicolumn{1}{c|}{0.956}  & \multicolumn{1}{c|}{-}      & \multicolumn{1}{c|}{-}      & \multicolumn{1}{c|}{-}      & \multicolumn{1}{c|}{-}      & \multicolumn{1}{c|}{-}      & \multicolumn{1}{c|}{-}      & \multicolumn{1}{c|}{0.946}  & \multicolumn{1}{c|}{0.948}  & \multicolumn{1}{c|}{0.948}  & 0.950  \\ \cline{2-16} 
& \multirow{21}{*}{Non-simplex}  & \multirow{7}{*}{\begin{tabular}[c]{@{}c@{}}Independent \\ errors\end{tabular}} & Bias $(\times 10)$          & \multicolumn{1}{c|}{0.022}  & \multicolumn{1}{c|}{-0.009} & \multicolumn{1}{c|}{0.042}  & \multicolumn{1}{c|}{0.000}  & \multicolumn{1}{c|}{0.028}  & \multicolumn{1}{c|}{-0.010} & \multicolumn{1}{c|}{0.032}  & \multicolumn{1}{c|}{-0.011} & \multicolumn{1}{c|}{0.031}  & \multicolumn{1}{c|}{-0.009} & \multicolumn{1}{c|}{0.027}  & -0.008 \\ \cline{4-16} 
&                                &                                                                                & ASE $(\times 10)$           & \multicolumn{1}{c|}{0.895}  & \multicolumn{1}{c|}{0.373}  & \multicolumn{1}{c|}{-}      & \multicolumn{1}{c|}{-}      & \multicolumn{1}{c|}{-}      & \multicolumn{1}{c|}{-}      & \multicolumn{1}{c|}{-}      & \multicolumn{1}{c|}{-}      & \multicolumn{1}{c|}{0.837}  & \multicolumn{1}{c|}{0.373}  & \multicolumn{1}{c|}{0.837}  & 0.372  \\ \cline{4-16} 
&                                &                                                                                & BSE $(\times 10)$           & \multicolumn{1}{c|}{0.970}  & \multicolumn{1}{c|}{0.383}  & \multicolumn{1}{c|}{-}      & \multicolumn{1}{c|}{-}      & \multicolumn{1}{c|}{-}      & \multicolumn{1}{c|}{-}      & \multicolumn{1}{c|}{-}      & \multicolumn{1}{c|}{-}      & \multicolumn{1}{c|}{0.845}  & \multicolumn{1}{c|}{0.374}  & \multicolumn{1}{c|}{0.839}  & 0.372  \\ \cline{4-16} 
&                                &                                                                                & ESE $(\times 10)$           & \multicolumn{1}{c|}{0.920}  & \multicolumn{1}{c|}{0.373}  & \multicolumn{1}{c|}{1.248}  & \multicolumn{1}{c|}{0.506}  & \multicolumn{1}{c|}{0.905}  & \multicolumn{1}{c|}{0.387}  & \multicolumn{1}{c|}{0.918}  & \multicolumn{1}{c|}{0.392}  & \multicolumn{1}{c|}{0.888}  & \multicolumn{1}{c|}{0.371}  & \multicolumn{1}{c|}{0.889}  & 0.371  \\ \cline{4-16} 
&                                &                                                                                & MSE $(\times 100)$          & \multicolumn{1}{c|}{0.845}  & \multicolumn{1}{c|}{0.139}  & \multicolumn{1}{c|}{1.556}  & \multicolumn{1}{c|}{0.256}  & \multicolumn{1}{c|}{0.817}  & \multicolumn{1}{c|}{0.150}  & \multicolumn{1}{c|}{0.841}  & \multicolumn{1}{c|}{0.154}  & \multicolumn{1}{c|}{0.788}  & \multicolumn{1}{c|}{0.137}  & \multicolumn{1}{c|}{0.789}  & 0.138  \\ \cline{4-16} 
&                                &                                                                                & Coverage (ASE)              & \multicolumn{1}{c|}{0.950}  & \multicolumn{1}{c|}{0.958}  & \multicolumn{1}{c|}{-}      & \multicolumn{1}{c|}{-}      & \multicolumn{1}{c|}{-}      & \multicolumn{1}{c|}{-}      & \multicolumn{1}{c|}{-}      & \multicolumn{1}{c|}{-}      & \multicolumn{1}{c|}{0.932}  & \multicolumn{1}{c|}{0.956}  & \multicolumn{1}{c|}{0.934}  & 0.956  \\ \cline{4-16} 
&                                &                                                                                & Coverage (BSE)              & \multicolumn{1}{c|}{0.962}  & \multicolumn{1}{c|}{0.962}  & \multicolumn{1}{c|}{-}      & \multicolumn{1}{c|}{-}      & \multicolumn{1}{c|}{-}      & \multicolumn{1}{c|}{-}      & \multicolumn{1}{c|}{-}      & \multicolumn{1}{c|}{-}      & \multicolumn{1}{c|}{0.934}  & \multicolumn{1}{c|}{0.958}  & \multicolumn{1}{c|}{0.930}  & 0.960  \\ \cline{3-16} 
&                                & \multirow{7}{*}{\begin{tabular}[c]{@{}c@{}}Correlated\\ errors\end{tabular}}   & Bias $(\times 10)$          & \multicolumn{1}{c|}{-0.046} & \multicolumn{1}{c|}{-0.023} & \multicolumn{1}{c|}{0.022}  & \multicolumn{1}{c|}{-0.024} & \multicolumn{1}{c|}{-0.037} & \multicolumn{1}{c|}{-0.025} & \multicolumn{1}{c|}{-0.031} & \multicolumn{1}{c|}{-0.028} & \multicolumn{1}{c|}{-0.028} & \multicolumn{1}{c|}{-0.017} & \multicolumn{1}{c|}{-0.026} & -0.019 \\ \cline{4-16} 
&                                &                                                                                & ASE $(\times 10)$           & \multicolumn{1}{c|}{0.630}  & \multicolumn{1}{c|}{0.274}  & \multicolumn{1}{c|}{-}      & \multicolumn{1}{c|}{-}      & \multicolumn{1}{c|}{-}      & \multicolumn{1}{c|}{-}      & \multicolumn{1}{c|}{-}      & \multicolumn{1}{c|}{-}      & \multicolumn{1}{c|}{0.660}  & \multicolumn{1}{c|}{0.295}  & \multicolumn{1}{c|}{0.667}  & 0.297  \\ \cline{4-16} 
&                                &                                                                                & BSE $(\times 10)$           & \multicolumn{1}{c|}{0.668}  & \multicolumn{1}{c|}{0.280}  & \multicolumn{1}{c|}{-}      & \multicolumn{1}{c|}{-}      & \multicolumn{1}{c|}{-}      & \multicolumn{1}{c|}{-}      & \multicolumn{1}{c|}{-}      & \multicolumn{1}{c|}{-}      & \multicolumn{1}{c|}{0.665}  & \multicolumn{1}{c|}{0.295}  & \multicolumn{1}{c|}{0.679}  & 0.300  \\ \cline{4-16} 
&                                &                                                                                & ESE $(\times 10)$           & \multicolumn{1}{c|}{0.619}  & \multicolumn{1}{c|}{0.278}  & \multicolumn{1}{c|}{1.049}  & \multicolumn{1}{c|}{0.497}  & \multicolumn{1}{c|}{0.630}  & \multicolumn{1}{c|}{0.277}  & \multicolumn{1}{c|}{0.644}  & \multicolumn{1}{c|}{0.284}  & \multicolumn{1}{c|}{0.670}  & \multicolumn{1}{c|}{0.293}  & \multicolumn{1}{c|}{0.666}  & 0.296  \\ \cline{4-16} 
&                                &                                                                                & MSE $(\times 100)$          & \multicolumn{1}{c|}{0.385}  & \multicolumn{1}{c|}{0.077}  & \multicolumn{1}{c|}{1.098}  & \multicolumn{1}{c|}{0.247}  & \multicolumn{1}{c|}{0.398}  & \multicolumn{1}{c|}{0.077}  & \multicolumn{1}{c|}{0.415}  & \multicolumn{1}{c|}{0.081}  & \multicolumn{1}{c|}{0.449}  & \multicolumn{1}{c|}{0.086}  & \multicolumn{1}{c|}{0.443}  & 0.088  \\ \cline{4-16} 
&                                &                                                                                & Coverage (ASE)              & \multicolumn{1}{c|}{0.948}  & \multicolumn{1}{c|}{0.942}  & \multicolumn{1}{c|}{-}      & \multicolumn{1}{c|}{-}      & \multicolumn{1}{c|}{-}      & \multicolumn{1}{c|}{-}      & \multicolumn{1}{c|}{-}      & \multicolumn{1}{c|}{-}      & \multicolumn{1}{c|}{0.938}  & \multicolumn{1}{c|}{0.944}  & \multicolumn{1}{c|}{0.936}  & 0.940  \\ \cline{4-16} 
&                                &                                                                                & Coverage (BSE)              & \multicolumn{1}{c|}{0.960}  & \multicolumn{1}{c|}{0.948}  & \multicolumn{1}{c|}{-}      & \multicolumn{1}{c|}{-}      & \multicolumn{1}{c|}{-}      & \multicolumn{1}{c|}{-}      & \multicolumn{1}{c|}{-}      & \multicolumn{1}{c|}{-}      & \multicolumn{1}{c|}{0.940}  & \multicolumn{1}{c|}{0.948}  & \multicolumn{1}{c|}{0.936}  & 0.946  \\ \cline{3-16} 
&                                & \multirow{7}{*}{\begin{tabular}[c]{@{}c@{}}No\\ $Y$ error\end{tabular}}        & Bias $(\times 10)$          & \multicolumn{1}{c|}{-0.017} & \multicolumn{1}{c|}{0.019}  & \multicolumn{1}{c|}{-0.038} & \multicolumn{1}{c|}{0.003}  & \multicolumn{1}{c|}{-0.009} & \multicolumn{1}{c|}{0.016}  & \multicolumn{1}{c|}{-0.009} & \multicolumn{1}{c|}{0.015}  & \multicolumn{1}{c|}{-0.009} & \multicolumn{1}{c|}{0.018}  & \multicolumn{1}{c|}{-0.011} & 0.017  \\ \cline{4-16} 
&                                &                                                                                & ASE $(\times 10)$           & \multicolumn{1}{c|}{0.693}  & \multicolumn{1}{c|}{0.293}  & \multicolumn{1}{c|}{-}      & \multicolumn{1}{c|}{-}      & \multicolumn{1}{c|}{-}      & \multicolumn{1}{c|}{-}      & \multicolumn{1}{c|}{-}      & \multicolumn{1}{c|}{-}      & \multicolumn{1}{c|}{0.665}  & \multicolumn{1}{c|}{0.293}  & \multicolumn{1}{c|}{0.665}  & 0.293  \\ \cline{4-16} 
&                                &                                                                                & BSE $(\times 10)$           & \multicolumn{1}{c|}{0.737}  & \multicolumn{1}{c|}{0.301}  & \multicolumn{1}{c|}{-}      & \multicolumn{1}{c|}{-}      & \multicolumn{1}{c|}{-}      & \multicolumn{1}{c|}{-}      & \multicolumn{1}{c|}{-}      & \multicolumn{1}{c|}{-}      & \multicolumn{1}{c|}{0.673}  & \multicolumn{1}{c|}{0.292}  & \multicolumn{1}{c|}{0.670}  & 0.292  \\ \cline{4-16} 
&                                &                                                                                & ESE $(\times 10)$           & \multicolumn{1}{c|}{0.712}  & \multicolumn{1}{c|}{0.290}  & \multicolumn{1}{c|}{1.023}  & \multicolumn{1}{c|}{0.437}  & \multicolumn{1}{c|}{0.717}  & \multicolumn{1}{c|}{0.303}  & \multicolumn{1}{c|}{0.730}  & \multicolumn{1}{c|}{0.311}  & \multicolumn{1}{c|}{0.695}  & \multicolumn{1}{c|}{0.291}  & \multicolumn{1}{c|}{0.699}  & 0.291  \\ \cline{4-16} 
&                                &                                                                                & MSE $(\times 100)$          & \multicolumn{1}{c|}{0.506}  & \multicolumn{1}{c|}{0.084}  & \multicolumn{1}{c|}{1.045}  & \multicolumn{1}{c|}{0.191}  & \multicolumn{1}{c|}{0.513}  & \multicolumn{1}{c|}{0.092}  & \multicolumn{1}{c|}{0.532}  & \multicolumn{1}{c|}{0.097}  & \multicolumn{1}{c|}{0.482}  & \multicolumn{1}{c|}{0.085}  & \multicolumn{1}{c|}{0.488}  & 0.085  \\ \cline{4-16} 
&                                &                                                                                & Coverage (ASE)              & \multicolumn{1}{c|}{0.936}  & \multicolumn{1}{c|}{0.954}  & \multicolumn{1}{c|}{-}      & \multicolumn{1}{c|}{-}      & \multicolumn{1}{c|}{-}      & \multicolumn{1}{c|}{-}      & \multicolumn{1}{c|}{-}      & \multicolumn{1}{c|}{-}      & \multicolumn{1}{c|}{0.936}  & \multicolumn{1}{c|}{0.946}  & \multicolumn{1}{c|}{0.932}  & 0.948  \\ \cline{4-16} 
&                                &                                                                                & Coverage (BSE)              & \multicolumn{1}{c|}{0.952}  & \multicolumn{1}{c|}{0.958}  & \multicolumn{1}{c|}{-}      & \multicolumn{1}{c|}{-}      & \multicolumn{1}{c|}{-}      & \multicolumn{1}{c|}{-}      & \multicolumn{1}{c|}{-}      & \multicolumn{1}{c|}{-}      & \multicolumn{1}{c|}{0.936}  & \multicolumn{1}{c|}{0.948}  & \multicolumn{1}{c|}{0.930}  & 0.948  \\ \hline
\end{tabular}

\caption{Summary Statistics of the Estimation Results Under (No trend with no intercept) for $\blambda_t$.} 
\label{Tab:Supp:Sim:T0}

\end{table}

\newpage

\begin{table}[!htp]
\renewcommand{\arraystretch}{1.3} \centering
\scriptsize
\setlength{\tabcolsep}{1pt}

\begin{tabular}{|c|c|c|c|cccccccccccc|}
\hline
\multirow{3}{*}{$\bm{\lambda}_t$}                                                                   & \multirow{3}{*}{$\bm{\mu}_0$} & \multirow{3}{*}{$\bm{e}_t$}                                                    & \multirow{3}{*}{Statistics} & \multicolumn{12}{c|}{Estimators and $T_0$}                                                                                                                                                                                                                                                                                                       \\ \cline{5-16}

&                                &                                                                                &                             & \multicolumn{2}{c|}{OLS-NoReg}                            & \multicolumn{2}{c|}{OLS-Standard}                         & \multicolumn{2}{c|}{ASC}                                  & \multicolumn{2}{c|}{SCPI}                                 & \multicolumn{2}{c|}{SPSC-NoDT}                            & \multicolumn{2}{c|}{SPSC-DT}         \\ \cline{5-16} 
&                                &                                                                                &                             & \multicolumn{1}{c|}{100}    & \multicolumn{1}{c|}{500}    & \multicolumn{1}{c|}{100}    & \multicolumn{1}{c|}{500}    & \multicolumn{1}{c|}{100}    & \multicolumn{1}{c|}{500}    & \multicolumn{1}{c|}{100}    & \multicolumn{1}{c|}{500}    & \multicolumn{1}{c|}{100}    & \multicolumn{1}{c|}{500}    & \multicolumn{1}{c|}{100}    & 500    \\ \hline

\multirow{42}{*}{\begin{tabular}[c]{@{}c@{}}No\\ trend \\ with \\ intercept\end{tabular}} & \multirow{21}{*}{Simplex}      & \multirow{7}{*}{\begin{tabular}[c]{@{}c@{}}Independent \\ errors\end{tabular}} & Bias $(\times 10)$          & \multicolumn{1}{c|}{0.075}  & \multicolumn{1}{c|}{0.073}  & \multicolumn{1}{c|}{0.026}  & \multicolumn{1}{c|}{0.001}  & \multicolumn{1}{c|}{-0.141} & \multicolumn{1}{c|}{-0.107} & \multicolumn{1}{c|}{-0.142} & \multicolumn{1}{c|}{-0.107} & \multicolumn{1}{c|}{-0.683} & \multicolumn{1}{c|}{-0.925} & \multicolumn{1}{c|}{0.008}  & 0.005  \\ \cline{4-16} 
&                                &                                                                                & ASE $(\times 10)$           & \multicolumn{1}{c|}{0.916}  & \multicolumn{1}{c|}{0.399}  & \multicolumn{1}{c|}{-}      & \multicolumn{1}{c|}{-}      & \multicolumn{1}{c|}{-}      & \multicolumn{1}{c|}{-}      & \multicolumn{1}{c|}{-}      & \multicolumn{1}{c|}{-}      & \multicolumn{1}{c|}{0.924}  & \multicolumn{1}{c|}{0.418}  & \multicolumn{1}{c|}{0.886}  & 0.399  \\ \cline{4-16} 
&                                &                                                                                & BSE $(\times 10)$           & \multicolumn{1}{c|}{1.002}  & \multicolumn{1}{c|}{0.411}  & \multicolumn{1}{c|}{-}      & \multicolumn{1}{c|}{-}      & \multicolumn{1}{c|}{-}      & \multicolumn{1}{c|}{-}      & \multicolumn{1}{c|}{-}      & \multicolumn{1}{c|}{-}      & \multicolumn{1}{c|}{0.982}  & \multicolumn{1}{c|}{0.443}  & \multicolumn{1}{c|}{0.898}  & 0.400  \\ \cline{4-16} 
&                                &                                                                                & ESE $(\times 10)$           & \multicolumn{1}{c|}{0.930}  & \multicolumn{1}{c|}{0.390}  & \multicolumn{1}{c|}{1.022}  & \multicolumn{1}{c|}{0.467}  & \multicolumn{1}{c|}{0.909}  & \multicolumn{1}{c|}{0.387}  & \multicolumn{1}{c|}{0.907}  & \multicolumn{1}{c|}{0.388}  & \multicolumn{1}{c|}{1.115}  & \multicolumn{1}{c|}{0.466}  & \multicolumn{1}{c|}{0.921}  & 0.390  \\ \cline{4-16} 
&                                &                                                                                & MSE $(\times 100)$          & \multicolumn{1}{c|}{0.868}  & \multicolumn{1}{c|}{0.157}  & \multicolumn{1}{c|}{1.042}  & \multicolumn{1}{c|}{0.217}  & \multicolumn{1}{c|}{0.844}  & \multicolumn{1}{c|}{0.161}  & \multicolumn{1}{c|}{0.841}  & \multicolumn{1}{c|}{0.162}  & \multicolumn{1}{c|}{1.708}  & \multicolumn{1}{c|}{1.072}  & \multicolumn{1}{c|}{0.847}  & 0.152  \\ \cline{4-16} 
&                                &                                                                                & Coverage (ASE)              & \multicolumn{1}{c|}{0.944}  & \multicolumn{1}{c|}{0.940}  & \multicolumn{1}{c|}{-}      & \multicolumn{1}{c|}{-}      & \multicolumn{1}{c|}{-}      & \multicolumn{1}{c|}{-}      & \multicolumn{1}{c|}{-}      & \multicolumn{1}{c|}{-}      & \multicolumn{1}{c|}{0.842}  & \multicolumn{1}{c|}{0.400}  & \multicolumn{1}{c|}{0.938}  & 0.956  \\ \cline{4-16} 
&                                &                                                                                & Coverage (BSE)              & \multicolumn{1}{c|}{0.960}  & \multicolumn{1}{c|}{0.954}  & \multicolumn{1}{c|}{-}      & \multicolumn{1}{c|}{-}      & \multicolumn{1}{c|}{-}      & \multicolumn{1}{c|}{-}      & \multicolumn{1}{c|}{-}      & \multicolumn{1}{c|}{-}      & \multicolumn{1}{c|}{0.854}  & \multicolumn{1}{c|}{0.436}  & \multicolumn{1}{c|}{0.948}  & 0.960  \\ \cline{3-16} 
&                                & \multirow{7}{*}{\begin{tabular}[c]{@{}c@{}}Correlated \\ errors\end{tabular}}  & Bias $(\times 10)$          & \multicolumn{1}{c|}{-0.010} & \multicolumn{1}{c|}{-0.012} & \multicolumn{1}{c|}{0.016}  & \multicolumn{1}{c|}{0.003}  & \multicolumn{1}{c|}{-0.034} & \multicolumn{1}{c|}{-0.027} & \multicolumn{1}{c|}{-0.035} & \multicolumn{1}{c|}{-0.027} & \multicolumn{1}{c|}{-0.399} & \multicolumn{1}{c|}{-0.335} & \multicolumn{1}{c|}{-0.026} & -0.022 \\ \cline{4-16} 
&                                &                                                                                & ASE $(\times 10)$           & \multicolumn{1}{c|}{0.554}  & \multicolumn{1}{c|}{0.248}  & \multicolumn{1}{c|}{-}      & \multicolumn{1}{c|}{-}      & \multicolumn{1}{c|}{-}      & \multicolumn{1}{c|}{-}      & \multicolumn{1}{c|}{-}      & \multicolumn{1}{c|}{-}      & \multicolumn{1}{c|}{0.742}  & \multicolumn{1}{c|}{0.334}  & \multicolumn{1}{c|}{0.556}  & 0.251  \\ \cline{4-16} 
&                                &                                                                                & BSE $(\times 10)$           & \multicolumn{1}{c|}{0.577}  & \multicolumn{1}{c|}{0.252}  & \multicolumn{1}{c|}{-}      & \multicolumn{1}{c|}{-}      & \multicolumn{1}{c|}{-}      & \multicolumn{1}{c|}{-}      & \multicolumn{1}{c|}{-}      & \multicolumn{1}{c|}{-}      & \multicolumn{1}{c|}{0.763}  & \multicolumn{1}{c|}{0.345}  & \multicolumn{1}{c|}{0.554}  & 0.251  \\ \cline{4-16} 
&                                &                                                                                & ESE $(\times 10)$           & \multicolumn{1}{c|}{0.557}  & \multicolumn{1}{c|}{0.237}  & \multicolumn{1}{c|}{0.795}  & \multicolumn{1}{c|}{0.358}  & \multicolumn{1}{c|}{0.552}  & \multicolumn{1}{c|}{0.237}  & \multicolumn{1}{c|}{0.552}  & \multicolumn{1}{c|}{0.237}  & \multicolumn{1}{c|}{0.783}  & \multicolumn{1}{c|}{0.328}  & \multicolumn{1}{c|}{0.558}  & 0.240  \\ \cline{4-16} 
&                                &                                                                                & MSE $(\times 100)$          & \multicolumn{1}{c|}{0.310}  & \multicolumn{1}{c|}{0.056}  & \multicolumn{1}{c|}{0.631}  & \multicolumn{1}{c|}{0.128}  & \multicolumn{1}{c|}{0.306}  & \multicolumn{1}{c|}{0.057}  & \multicolumn{1}{c|}{0.305}  & \multicolumn{1}{c|}{0.057}  & \multicolumn{1}{c|}{0.772}  & \multicolumn{1}{c|}{0.220}  & \multicolumn{1}{c|}{0.311}  & 0.058  \\ \cline{4-16} 
&                                &                                                                                & Coverage (ASE)              & \multicolumn{1}{c|}{0.950}  & \multicolumn{1}{c|}{0.950}  & \multicolumn{1}{c|}{-}      & \multicolumn{1}{c|}{-}      & \multicolumn{1}{c|}{-}      & \multicolumn{1}{c|}{-}      & \multicolumn{1}{c|}{-}      & \multicolumn{1}{c|}{-}      & \multicolumn{1}{c|}{0.902}  & \multicolumn{1}{c|}{0.824}  & \multicolumn{1}{c|}{0.954}  & 0.944  \\ \cline{4-16} 
&                                &                                                                                & Coverage (BSE)              & \multicolumn{1}{c|}{0.958}  & \multicolumn{1}{c|}{0.954}  & \multicolumn{1}{c|}{-}      & \multicolumn{1}{c|}{-}      & \multicolumn{1}{c|}{-}      & \multicolumn{1}{c|}{-}      & \multicolumn{1}{c|}{-}      & \multicolumn{1}{c|}{-}      & \multicolumn{1}{c|}{0.898}  & \multicolumn{1}{c|}{0.842}  & \multicolumn{1}{c|}{0.950}  & 0.950  \\ \cline{3-16} 
&                                & \multirow{7}{*}{\begin{tabular}[c]{@{}c@{}}No \\ $Y$ error\end{tabular}}       & Bias $(\times 10)$          & \multicolumn{1}{c|}{0.113}  & \multicolumn{1}{c|}{0.079}  & \multicolumn{1}{c|}{0.098}  & \multicolumn{1}{c|}{0.004}  & \multicolumn{1}{c|}{-0.063} & \multicolumn{1}{c|}{-0.087} & \multicolumn{1}{c|}{-0.064} & \multicolumn{1}{c|}{-0.087} & \multicolumn{1}{c|}{-0.247} & \multicolumn{1}{c|}{-0.401} & \multicolumn{1}{c|}{0.042}  & 0.008  \\ \cline{4-16} 
&                                &                                                                                & ASE $(\times 10)$           & \multicolumn{1}{c|}{0.544}  & \multicolumn{1}{c|}{0.244}  & \multicolumn{1}{c|}{-}      & \multicolumn{1}{c|}{-}      & \multicolumn{1}{c|}{-}      & \multicolumn{1}{c|}{-}      & \multicolumn{1}{c|}{-}      & \multicolumn{1}{c|}{-}      & \multicolumn{1}{c|}{0.679}  & \multicolumn{1}{c|}{0.305}  & \multicolumn{1}{c|}{0.544}  & 0.244  \\ \cline{4-16} 
&                                &                                                                                & BSE $(\times 10)$           & \multicolumn{1}{c|}{0.566}  & \multicolumn{1}{c|}{0.248}  & \multicolumn{1}{c|}{-}      & \multicolumn{1}{c|}{-}      & \multicolumn{1}{c|}{-}      & \multicolumn{1}{c|}{-}      & \multicolumn{1}{c|}{-}      & \multicolumn{1}{c|}{-}      & \multicolumn{1}{c|}{0.678}  & \multicolumn{1}{c|}{0.305}  & \multicolumn{1}{c|}{0.542}  & 0.244  \\ \cline{4-16} 
&                                &                                                                                & ESE $(\times 10)$           & \multicolumn{1}{c|}{0.543}  & \multicolumn{1}{c|}{0.249}  & \multicolumn{1}{c|}{0.732}  & \multicolumn{1}{c|}{0.344}  & \multicolumn{1}{c|}{0.547}  & \multicolumn{1}{c|}{0.254}  & \multicolumn{1}{c|}{0.547}  & \multicolumn{1}{c|}{0.254}  & \multicolumn{1}{c|}{0.705}  & \multicolumn{1}{c|}{0.314}  & \multicolumn{1}{c|}{0.539}  & 0.249  \\ \cline{4-16} 
&                                &                                                                                & MSE $(\times 100)$          & \multicolumn{1}{c|}{0.307}  & \multicolumn{1}{c|}{0.068}  & \multicolumn{1}{c|}{0.545}  & \multicolumn{1}{c|}{0.118}  & \multicolumn{1}{c|}{0.302}  & \multicolumn{1}{c|}{0.072}  & \multicolumn{1}{c|}{0.302}  & \multicolumn{1}{c|}{0.072}  & \multicolumn{1}{c|}{0.557}  & \multicolumn{1}{c|}{0.259}  & \multicolumn{1}{c|}{0.292}  & 0.062  \\ \cline{4-16} 
&                                &                                                                                & Coverage (ASE)              & \multicolumn{1}{c|}{0.942}  & \multicolumn{1}{c|}{0.944}  & \multicolumn{1}{c|}{-}      & \multicolumn{1}{c|}{-}      & \multicolumn{1}{c|}{-}      & \multicolumn{1}{c|}{-}      & \multicolumn{1}{c|}{-}      & \multicolumn{1}{c|}{-}      & \multicolumn{1}{c|}{0.932}  & \multicolumn{1}{c|}{0.732}  & \multicolumn{1}{c|}{0.954}  & 0.956  \\ \cline{4-16} 
&                                &                                                                                & Coverage (BSE)              & \multicolumn{1}{c|}{0.950}  & \multicolumn{1}{c|}{0.948}  & \multicolumn{1}{c|}{-}      & \multicolumn{1}{c|}{-}      & \multicolumn{1}{c|}{-}      & \multicolumn{1}{c|}{-}      & \multicolumn{1}{c|}{-}      & \multicolumn{1}{c|}{-}      & \multicolumn{1}{c|}{0.928}  & \multicolumn{1}{c|}{0.736}  & \multicolumn{1}{c|}{0.952}  & 0.958  \\ \cline{2-16} 
& \multirow{21}{*}{Non-simplex}  & \multirow{7}{*}{\begin{tabular}[c]{@{}c@{}}Independent \\ errors\end{tabular}} & Bias $(\times 10)$          & \multicolumn{1}{c|}{0.381}  & \multicolumn{1}{c|}{0.351}  & \multicolumn{1}{c|}{7.010}  & \multicolumn{1}{c|}{6.987}  & \multicolumn{1}{c|}{2.540}  & \multicolumn{1}{c|}{1.375}  & \multicolumn{1}{c|}{7.358}  & \multicolumn{1}{c|}{7.308}  & \multicolumn{1}{c|}{-0.267} & \multicolumn{1}{c|}{-0.344} & \multicolumn{1}{c|}{0.036}  & 0.004  \\ \cline{4-16} 
&                                &                                                                                & ASE $(\times 10)$           & \multicolumn{1}{c|}{1.104}  & \multicolumn{1}{c|}{0.474}  & \multicolumn{1}{c|}{-}      & \multicolumn{1}{c|}{-}      & \multicolumn{1}{c|}{-}      & \multicolumn{1}{c|}{-}      & \multicolumn{1}{c|}{-}      & \multicolumn{1}{c|}{-}      & \multicolumn{1}{c|}{1.282}  & \multicolumn{1}{c|}{0.575}  & \multicolumn{1}{c|}{1.058}  & 0.476  \\ \cline{4-16} 
&                                &                                                                                & BSE $(\times 10)$           & \multicolumn{1}{c|}{1.223}  & \multicolumn{1}{c|}{0.489}  & \multicolumn{1}{c|}{-}      & \multicolumn{1}{c|}{-}      & \multicolumn{1}{c|}{-}      & \multicolumn{1}{c|}{-}      & \multicolumn{1}{c|}{-}      & \multicolumn{1}{c|}{-}      & \multicolumn{1}{c|}{1.321}  & \multicolumn{1}{c|}{0.596}  & \multicolumn{1}{c|}{1.086}  & 0.483  \\ \cline{4-16} 
&                                &                                                                                & ESE $(\times 10)$           & \multicolumn{1}{c|}{1.141}  & \multicolumn{1}{c|}{0.474}  & \multicolumn{1}{c|}{0.939}  & \multicolumn{1}{c|}{0.438}  & \multicolumn{1}{c|}{1.357}  & \multicolumn{1}{c|}{0.564}  & \multicolumn{1}{c|}{0.935}  & \multicolumn{1}{c|}{0.448}  & \multicolumn{1}{c|}{1.302}  & \multicolumn{1}{c|}{0.571}  & \multicolumn{1}{c|}{1.090}  & 0.485  \\ \cline{4-16} 
&                                &                                                                                & MSE $(\times 100)$          & \multicolumn{1}{c|}{1.444}  & \multicolumn{1}{c|}{0.347}  & \multicolumn{1}{c|}{50.020} & \multicolumn{1}{c|}{49.005} & \multicolumn{1}{c|}{8.289}  & \multicolumn{1}{c|}{2.208}  & \multicolumn{1}{c|}{55.011} & \multicolumn{1}{c|}{53.601} & \multicolumn{1}{c|}{1.762}  & \multicolumn{1}{c|}{0.443}  & \multicolumn{1}{c|}{1.187}  & 0.235  \\ \cline{4-16} 
&                                &                                                                                & Coverage (ASE)              & \multicolumn{1}{c|}{0.920}  & \multicolumn{1}{c|}{0.884}  & \multicolumn{1}{c|}{-}      & \multicolumn{1}{c|}{-}      & \multicolumn{1}{c|}{-}      & \multicolumn{1}{c|}{-}      & \multicolumn{1}{c|}{-}      & \multicolumn{1}{c|}{-}      & \multicolumn{1}{c|}{0.950}  & \multicolumn{1}{c|}{0.914}  & \multicolumn{1}{c|}{0.942}  & 0.930  \\ \cline{4-16} 
&                                &                                                                                & Coverage (BSE)              & \multicolumn{1}{c|}{0.948}  & \multicolumn{1}{c|}{0.894}  & \multicolumn{1}{c|}{-}      & \multicolumn{1}{c|}{-}      & \multicolumn{1}{c|}{-}      & \multicolumn{1}{c|}{-}      & \multicolumn{1}{c|}{-}      & \multicolumn{1}{c|}{-}      & \multicolumn{1}{c|}{0.946}  & \multicolumn{1}{c|}{0.922}  & \multicolumn{1}{c|}{0.948}  & 0.936  \\ \cline{3-16} 
&                                & \multirow{7}{*}{\begin{tabular}[c]{@{}c@{}}Correlated\\ errors\end{tabular}}   & Bias $(\times 10)$          & \multicolumn{1}{c|}{0.218}  & \multicolumn{1}{c|}{0.249}  & \multicolumn{1}{c|}{7.019}  & \multicolumn{1}{c|}{7.012}  & \multicolumn{1}{c|}{0.763}  & \multicolumn{1}{c|}{0.406}  & \multicolumn{1}{c|}{7.018}  & \multicolumn{1}{c|}{7.011}  & \multicolumn{1}{c|}{0.085}  & \multicolumn{1}{c|}{0.276}  & \multicolumn{1}{c|}{0.004}  & 0.020  \\ \cline{4-16} 
&                                &                                                                                & ASE $(\times 10)$           & \multicolumn{1}{c|}{0.728}  & \multicolumn{1}{c|}{0.317}  & \multicolumn{1}{c|}{-}      & \multicolumn{1}{c|}{-}      & \multicolumn{1}{c|}{-}      & \multicolumn{1}{c|}{-}      & \multicolumn{1}{c|}{-}      & \multicolumn{1}{c|}{-}      & \multicolumn{1}{c|}{1.084}  & \multicolumn{1}{c|}{0.485}  & \multicolumn{1}{c|}{0.827}  & 0.372  \\ \cline{4-16} 
&                                &                                                                                & BSE $(\times 10)$           & \multicolumn{1}{c|}{0.784}  & \multicolumn{1}{c|}{0.325}  & \multicolumn{1}{c|}{-}      & \multicolumn{1}{c|}{-}      & \multicolumn{1}{c|}{-}      & \multicolumn{1}{c|}{-}      & \multicolumn{1}{c|}{-}      & \multicolumn{1}{c|}{-}      & \multicolumn{1}{c|}{1.098}  & \multicolumn{1}{c|}{0.495}  & \multicolumn{1}{c|}{0.849}  & 0.384  \\ \cline{4-16} 
&                                &                                                                                & ESE $(\times 10)$           & \multicolumn{1}{c|}{0.749}  & \multicolumn{1}{c|}{0.321}  & \multicolumn{1}{c|}{0.680}  & \multicolumn{1}{c|}{0.286}  & \multicolumn{1}{c|}{0.880}  & \multicolumn{1}{c|}{0.324}  & \multicolumn{1}{c|}{0.680}  & \multicolumn{1}{c|}{0.286}  & \multicolumn{1}{c|}{1.237}  & \multicolumn{1}{c|}{0.477}  & \multicolumn{1}{c|}{0.816}  & 0.362  \\ \cline{4-16} 
&                                &                                                                                & MSE $(\times 100)$          & \multicolumn{1}{c|}{0.607}  & \multicolumn{1}{c|}{0.165}  & \multicolumn{1}{c|}{49.721} & \multicolumn{1}{c|}{49.244} & \multicolumn{1}{c|}{1.354}  & \multicolumn{1}{c|}{0.270}  & \multicolumn{1}{c|}{49.718} & \multicolumn{1}{c|}{49.241} & \multicolumn{1}{c|}{1.535}  & \multicolumn{1}{c|}{0.303}  & \multicolumn{1}{c|}{0.665}  & 0.131  \\ \cline{4-16} 
&                                &                                                                                & Coverage (ASE)              & \multicolumn{1}{c|}{0.944}  & \multicolumn{1}{c|}{0.854}  & \multicolumn{1}{c|}{-}      & \multicolumn{1}{c|}{-}      & \multicolumn{1}{c|}{-}      & \multicolumn{1}{c|}{-}      & \multicolumn{1}{c|}{-}      & \multicolumn{1}{c|}{-}      & \multicolumn{1}{c|}{0.892}  & \multicolumn{1}{c|}{0.926}  & \multicolumn{1}{c|}{0.956}  & 0.948  \\ \cline{4-16} 
&                                &                                                                                & Coverage (BSE)              & \multicolumn{1}{c|}{0.958}  & \multicolumn{1}{c|}{0.874}  & \multicolumn{1}{c|}{-}      & \multicolumn{1}{c|}{-}      & \multicolumn{1}{c|}{-}      & \multicolumn{1}{c|}{-}      & \multicolumn{1}{c|}{-}      & \multicolumn{1}{c|}{-}      & \multicolumn{1}{c|}{0.896}  & \multicolumn{1}{c|}{0.940}  & \multicolumn{1}{c|}{0.954}  & 0.962  \\ \cline{3-16} 
&                                & \multirow{7}{*}{\begin{tabular}[c]{@{}c@{}}No\\ $Y$ error\end{tabular}}        & Bias $(\times 10)$          & \multicolumn{1}{c|}{0.367}  & \multicolumn{1}{c|}{0.340}  & \multicolumn{1}{c|}{7.051}  & \multicolumn{1}{c|}{6.971}  & \multicolumn{1}{c|}{1.826}  & \multicolumn{1}{c|}{1.216}  & \multicolumn{1}{c|}{7.391}  & \multicolumn{1}{c|}{7.294}  & \multicolumn{1}{c|}{0.205}  & \multicolumn{1}{c|}{0.251}  & \multicolumn{1}{c|}{0.036}  & -0.002 \\ \cline{4-16} 
&                                &                                                                                & ASE $(\times 10)$           & \multicolumn{1}{c|}{0.815}  & \multicolumn{1}{c|}{0.353}  & \multicolumn{1}{c|}{-}      & \multicolumn{1}{c|}{-}      & \multicolumn{1}{c|}{-}      & \multicolumn{1}{c|}{-}      & \multicolumn{1}{c|}{-}      & \multicolumn{1}{c|}{-}      & \multicolumn{1}{c|}{1.103}  & \multicolumn{1}{c|}{0.495}  & \multicolumn{1}{c|}{0.797}  & 0.356  \\ \cline{4-16} 
&                                &                                                                                & BSE $(\times 10)$           & \multicolumn{1}{c|}{0.886}  & \multicolumn{1}{c|}{0.363}  & \multicolumn{1}{c|}{-}      & \multicolumn{1}{c|}{-}      & \multicolumn{1}{c|}{-}      & \multicolumn{1}{c|}{-}      & \multicolumn{1}{c|}{-}      & \multicolumn{1}{c|}{-}      & \multicolumn{1}{c|}{1.122}  & \multicolumn{1}{c|}{0.509}  & \multicolumn{1}{c|}{0.813}  & 0.361  \\ \cline{4-16} 
&                                &                                                                                & ESE $(\times 10)$           & \multicolumn{1}{c|}{0.840}  & \multicolumn{1}{c|}{0.326}  & \multicolumn{1}{c|}{0.788}  & \multicolumn{1}{c|}{0.345}  & \multicolumn{1}{c|}{0.975}  & \multicolumn{1}{c|}{0.387}  & \multicolumn{1}{c|}{0.797}  & \multicolumn{1}{c|}{0.350}  & \multicolumn{1}{c|}{1.244}  & \multicolumn{1}{c|}{0.484}  & \multicolumn{1}{c|}{0.811}  & 0.331  \\ \cline{4-16} 
&                                &                                                                                & MSE $(\times 100)$          & \multicolumn{1}{c|}{0.839}  & \multicolumn{1}{c|}{0.222}  & \multicolumn{1}{c|}{50.338} & \multicolumn{1}{c|}{48.716} & \multicolumn{1}{c|}{4.284}  & \multicolumn{1}{c|}{1.627}  & \multicolumn{1}{c|}{55.260} & \multicolumn{1}{c|}{53.321} & \multicolumn{1}{c|}{1.586}  & \multicolumn{1}{c|}{0.297}  & \multicolumn{1}{c|}{0.658}  & 0.109  \\ \cline{4-16} 
&                                &                                                                                & Coverage (ASE)              & \multicolumn{1}{c|}{0.922}  & \multicolumn{1}{c|}{0.856}  & \multicolumn{1}{c|}{-}      & \multicolumn{1}{c|}{-}      & \multicolumn{1}{c|}{-}      & \multicolumn{1}{c|}{-}      & \multicolumn{1}{c|}{-}      & \multicolumn{1}{c|}{-}      & \multicolumn{1}{c|}{0.916}  & \multicolumn{1}{c|}{0.930}  & \multicolumn{1}{c|}{0.942}  & 0.966  \\ \cline{4-16} 
&                                &                                                                                & Coverage (BSE)              & \multicolumn{1}{c|}{0.936}  & \multicolumn{1}{c|}{0.868}  & \multicolumn{1}{c|}{-}      & \multicolumn{1}{c|}{-}      & \multicolumn{1}{c|}{-}      & \multicolumn{1}{c|}{-}      & \multicolumn{1}{c|}{-}      & \multicolumn{1}{c|}{-}      & \multicolumn{1}{c|}{0.920}  & \multicolumn{1}{c|}{0.948}  & \multicolumn{1}{c|}{0.956}  & 0.968  \\ \hline
\end{tabular} 

\caption{Summary Statistics of the Estimation Results Under (No trend with intercept) for $\blambda_t$.} 
\label{Tab:Supp:Sim:T1}

\end{table}

\newpage

\begin{table}[!htp]
\renewcommand{\arraystretch}{1.3} \centering
\scriptsize
\setlength{\tabcolsep}{1pt}

\begin{tabular}{|c|c|c|c|cccccccccccc|}
\hline
\multirow{3}{*}{$\bm{\lambda}_t$}                                                                   & \multirow{3}{*}{$\bm{\mu}_0$} & \multirow{3}{*}{$\bm{e}_t$}                                                    & \multirow{3}{*}{Statistics} & \multicolumn{12}{c|}{Estimators and $T_0$}                                                                                                                                                                                                                                                                                                       \\ \cline{5-16}

&                                &                                                                                &                             & \multicolumn{2}{c|}{OLS-NoReg}                            & \multicolumn{2}{c|}{OLS-Standard}                         & \multicolumn{2}{c|}{ASC}                                  & \multicolumn{2}{c|}{SCPI}                                 & \multicolumn{2}{c|}{SPSC-NoDT}                            & \multicolumn{2}{c|}{SPSC-DT}         \\ \cline{5-16} 
&                                &                                                                                &                             & \multicolumn{1}{c|}{100}    & \multicolumn{1}{c|}{500}    & \multicolumn{1}{c|}{100}    & \multicolumn{1}{c|}{500}    & \multicolumn{1}{c|}{100}    & \multicolumn{1}{c|}{500}    & \multicolumn{1}{c|}{100}    & \multicolumn{1}{c|}{500}    & \multicolumn{1}{c|}{100}    & \multicolumn{1}{c|}{500}    & \multicolumn{1}{c|}{100}    & 500    \\ \hline

\multirow{42}{*}{\begin{tabular}[c]{@{}c@{}}Linear\\ trend \\ with \\ no\\ intercept\end{tabular}} & \multirow{21}{*}{Simplex}      & \multirow{7}{*}{\begin{tabular}[c]{@{}c@{}}Independent \\ errors\end{tabular}} & Bias $(\times 10)$          & \multicolumn{1}{c|}{0.323} & \multicolumn{1}{c|}{0.274}  & \multicolumn{1}{c|}{0.127}   & \multicolumn{1}{c|}{-0.021}  & \multicolumn{1}{c|}{-0.498} & \multicolumn{1}{c|}{-0.420} & \multicolumn{1}{c|}{-0.507}  & \multicolumn{1}{c|}{-0.426}  & \multicolumn{1}{c|}{-1.212} & \multicolumn{1}{c|}{-0.971} & \multicolumn{1}{c|}{0.071}  & 0.006  \\ \cline{4-16} 
&                                &                                                                                & ASE $(\times 10)$           & \multicolumn{1}{c|}{1.425} & \multicolumn{1}{c|}{0.636}  & \multicolumn{1}{c|}{-}       & \multicolumn{1}{c|}{-}       & \multicolumn{1}{c|}{-}      & \multicolumn{1}{c|}{-}      & \multicolumn{1}{c|}{-}       & \multicolumn{1}{c|}{-}       & \multicolumn{1}{c|}{1.225}  & \multicolumn{1}{c|}{0.552}  & \multicolumn{1}{c|}{1.429}  & 0.649  \\ \cline{4-16} 
&                                &                                                                                & BSE $(\times 10)$           & \multicolumn{1}{c|}{1.628} & \multicolumn{1}{c|}{0.658}  & \multicolumn{1}{c|}{-}       & \multicolumn{1}{c|}{-}       & \multicolumn{1}{c|}{-}      & \multicolumn{1}{c|}{-}      & \multicolumn{1}{c|}{-}       & \multicolumn{1}{c|}{-}       & \multicolumn{1}{c|}{1.794}  & \multicolumn{1}{c|}{0.825}  & \multicolumn{1}{c|}{1.534}  & 0.683  \\ \cline{4-16} 
&                                &                                                                                & ESE $(\times 10)$           & \multicolumn{1}{c|}{1.482} & \multicolumn{1}{c|}{0.598}  & \multicolumn{1}{c|}{2.061}   & \multicolumn{1}{c|}{0.900}   & \multicolumn{1}{c|}{1.361}  & \multicolumn{1}{c|}{0.582}  & \multicolumn{1}{c|}{1.355}   & \multicolumn{1}{c|}{0.582}   & \multicolumn{1}{c|}{2.885}  & \multicolumn{1}{c|}{2.871}  & \multicolumn{1}{c|}{1.518}  & 0.645  \\ \cline{4-16} 
&                                &                                                                                & MSE $(\times 100)$          & \multicolumn{1}{c|}{2.298} & \multicolumn{1}{c|}{0.432}  & \multicolumn{1}{c|}{4.254}   & \multicolumn{1}{c|}{0.808}   & \multicolumn{1}{c|}{2.096}  & \multicolumn{1}{c|}{0.515}  & \multicolumn{1}{c|}{2.089}   & \multicolumn{1}{c|}{0.520}   & \multicolumn{1}{c|}{9.779}  & \multicolumn{1}{c|}{9.168}  & \multicolumn{1}{c|}{2.304}  & 0.415  \\ \cline{4-16} 
&                                &                                                                                & Coverage (ASE)              & \multicolumn{1}{c|}{0.926} & \multicolumn{1}{c|}{0.946}  & \multicolumn{1}{c|}{-}       & \multicolumn{1}{c|}{-}       & \multicolumn{1}{c|}{-}      & \multicolumn{1}{c|}{-}      & \multicolumn{1}{c|}{-}       & \multicolumn{1}{c|}{-}       & \multicolumn{1}{c|}{0.438}  & \multicolumn{1}{c|}{0.034}  & \multicolumn{1}{c|}{0.946}  & 0.952  \\ \cline{4-16} 
&                                &                                                                                & Coverage (BSE)              & \multicolumn{1}{c|}{0.960} & \multicolumn{1}{c|}{0.954}  & \multicolumn{1}{c|}{-}       & \multicolumn{1}{c|}{-}       & \multicolumn{1}{c|}{-}      & \multicolumn{1}{c|}{-}      & \multicolumn{1}{c|}{-}       & \multicolumn{1}{c|}{-}       & \multicolumn{1}{c|}{0.668}  & \multicolumn{1}{c|}{0.156}  & \multicolumn{1}{c|}{0.962}  & 0.968  \\ \cline{3-16} 
&                                & \multirow{7}{*}{\begin{tabular}[c]{@{}c@{}}Correlated \\ errors\end{tabular}}  & Bias $(\times 10)$          & \multicolumn{1}{c|}{0.006} & \multicolumn{1}{c|}{-0.009} & \multicolumn{1}{c|}{0.131}   & \multicolumn{1}{c|}{0.017}   & \multicolumn{1}{c|}{-0.105} & \multicolumn{1}{c|}{-0.074} & \multicolumn{1}{c|}{-0.110}  & \multicolumn{1}{c|}{-0.076}  & \multicolumn{1}{c|}{-1.945} & \multicolumn{1}{c|}{-1.347} & \multicolumn{1}{c|}{-0.079} & -0.088 \\ \cline{4-16} 
&                                &                                                                                & ASE $(\times 10)$           & \multicolumn{1}{c|}{0.654} & \multicolumn{1}{c|}{0.294}  & \multicolumn{1}{c|}{-}       & \multicolumn{1}{c|}{-}       & \multicolumn{1}{c|}{-}      & \multicolumn{1}{c|}{-}      & \multicolumn{1}{c|}{-}       & \multicolumn{1}{c|}{-}       & \multicolumn{1}{c|}{0.902}  & \multicolumn{1}{c|}{0.405}  & \multicolumn{1}{c|}{0.662}  & 0.301  \\ \cline{4-16} 
&                                &                                                                                & BSE $(\times 10)$           & \multicolumn{1}{c|}{0.703} & \multicolumn{1}{c|}{0.302}  & \multicolumn{1}{c|}{-}       & \multicolumn{1}{c|}{-}       & \multicolumn{1}{c|}{-}      & \multicolumn{1}{c|}{-}      & \multicolumn{1}{c|}{-}       & \multicolumn{1}{c|}{-}       & \multicolumn{1}{c|}{0.978}  & \multicolumn{1}{c|}{0.468}  & \multicolumn{1}{c|}{0.711}  & 0.320  \\ \cline{4-16} 
&                                &                                                                                & ESE $(\times 10)$           & \multicolumn{1}{c|}{0.701} & \multicolumn{1}{c|}{0.289}  & \multicolumn{1}{c|}{1.468}   & \multicolumn{1}{c|}{0.697}   & \multicolumn{1}{c|}{0.663}  & \multicolumn{1}{c|}{0.285}  & \multicolumn{1}{c|}{0.662}   & \multicolumn{1}{c|}{0.285}   & \multicolumn{1}{c|}{1.352}  & \multicolumn{1}{c|}{0.686}  & \multicolumn{1}{c|}{0.704}  & 0.332  \\ \cline{4-16} 
&                                &                                                                                & MSE $(\times 100)$          & \multicolumn{1}{c|}{0.491} & \multicolumn{1}{c|}{0.084}  & \multicolumn{1}{c|}{2.167}   & \multicolumn{1}{c|}{0.486}   & \multicolumn{1}{c|}{0.450}  & \multicolumn{1}{c|}{0.087}  & \multicolumn{1}{c|}{0.450}   & \multicolumn{1}{c|}{0.087}   & \multicolumn{1}{c|}{5.606}  & \multicolumn{1}{c|}{2.284}  & \multicolumn{1}{c|}{0.500}  & 0.118  \\ \cline{4-16} 
&                                &                                                                                & Coverage (ASE)              & \multicolumn{1}{c|}{0.922} & \multicolumn{1}{c|}{0.950}  & \multicolumn{1}{c|}{-}       & \multicolumn{1}{c|}{-}       & \multicolumn{1}{c|}{-}      & \multicolumn{1}{c|}{-}      & \multicolumn{1}{c|}{-}       & \multicolumn{1}{c|}{-}       & \multicolumn{1}{c|}{0.470}  & \multicolumn{1}{c|}{0.176}  & \multicolumn{1}{c|}{0.922}  & 0.926  \\ \cline{4-16} 
&                                &                                                                                & Coverage (BSE)              & \multicolumn{1}{c|}{0.948} & \multicolumn{1}{c|}{0.958}  & \multicolumn{1}{c|}{-}       & \multicolumn{1}{c|}{-}       & \multicolumn{1}{c|}{-}      & \multicolumn{1}{c|}{-}      & \multicolumn{1}{c|}{-}       & \multicolumn{1}{c|}{-}       & \multicolumn{1}{c|}{0.518}  & \multicolumn{1}{c|}{0.260}  & \multicolumn{1}{c|}{0.948}  & 0.942  \\ \cline{3-16} 
&                                & \multirow{7}{*}{\begin{tabular}[c]{@{}c@{}}No \\ $Y$ error\end{tabular}}       & Bias $(\times 10)$          & \multicolumn{1}{c|}{0.280} & \multicolumn{1}{c|}{0.288}  & \multicolumn{1}{c|}{0.105}   & \multicolumn{1}{c|}{0.036}   & \multicolumn{1}{c|}{-0.373} & \multicolumn{1}{c|}{-0.321} & \multicolumn{1}{c|}{-0.376}  & \multicolumn{1}{c|}{-0.321}  & \multicolumn{1}{c|}{-1.154} & \multicolumn{1}{c|}{-1.364} & \multicolumn{1}{c|}{0.028}  & 0.016  \\ \cline{4-16} 
&                                &                                                                                & ASE $(\times 10)$           & \multicolumn{1}{c|}{0.635} & \multicolumn{1}{c|}{0.285}  & \multicolumn{1}{c|}{-}       & \multicolumn{1}{c|}{-}       & \multicolumn{1}{c|}{-}      & \multicolumn{1}{c|}{-}      & \multicolumn{1}{c|}{-}       & \multicolumn{1}{c|}{-}       & \multicolumn{1}{c|}{0.858}  & \multicolumn{1}{c|}{0.388}  & \multicolumn{1}{c|}{0.642}  & 0.288  \\ \cline{4-16} 
&                                &                                                                                & BSE $(\times 10)$           & \multicolumn{1}{c|}{0.683} & \multicolumn{1}{c|}{0.292}  & \multicolumn{1}{c|}{-}       & \multicolumn{1}{c|}{-}       & \multicolumn{1}{c|}{-}      & \multicolumn{1}{c|}{-}      & \multicolumn{1}{c|}{-}       & \multicolumn{1}{c|}{-}       & \multicolumn{1}{c|}{0.851}  & \multicolumn{1}{c|}{0.374}  & \multicolumn{1}{c|}{0.664}  & 0.299  \\ \cline{4-16} 
&                                &                                                                                & ESE $(\times 10)$           & \multicolumn{1}{c|}{0.671} & \multicolumn{1}{c|}{0.286}  & \multicolumn{1}{c|}{1.391}   & \multicolumn{1}{c|}{0.633}   & \multicolumn{1}{c|}{0.662}  & \multicolumn{1}{c|}{0.293}  & \multicolumn{1}{c|}{0.663}   & \multicolumn{1}{c|}{0.294}   & \multicolumn{1}{c|}{1.040}  & \multicolumn{1}{c|}{0.383}  & \multicolumn{1}{c|}{0.705}  & 0.309  \\ \cline{4-16} 
&                                &                                                                                & MSE $(\times 100)$          & \multicolumn{1}{c|}{0.528} & \multicolumn{1}{c|}{0.165}  & \multicolumn{1}{c|}{1.941}   & \multicolumn{1}{c|}{0.401}   & \multicolumn{1}{c|}{0.577}  & \multicolumn{1}{c|}{0.189}  & \multicolumn{1}{c|}{0.580}   & \multicolumn{1}{c|}{0.189}   & \multicolumn{1}{c|}{2.411}  & \multicolumn{1}{c|}{2.006}  & \multicolumn{1}{c|}{0.497}  & 0.095  \\ \cline{4-16} 
&                                &                                                                                & Coverage (ASE)              & \multicolumn{1}{c|}{0.906} & \multicolumn{1}{c|}{0.820}  & \multicolumn{1}{c|}{-}       & \multicolumn{1}{c|}{-}       & \multicolumn{1}{c|}{-}      & \multicolumn{1}{c|}{-}      & \multicolumn{1}{c|}{-}       & \multicolumn{1}{c|}{-}       & \multicolumn{1}{c|}{0.648}  & \multicolumn{1}{c|}{0.062}  & \multicolumn{1}{c|}{0.920}  & 0.930  \\ \cline{4-16} 
&                                &                                                                                & Coverage (BSE)              & \multicolumn{1}{c|}{0.930} & \multicolumn{1}{c|}{0.832}  & \multicolumn{1}{c|}{-}       & \multicolumn{1}{c|}{-}       & \multicolumn{1}{c|}{-}      & \multicolumn{1}{c|}{-}      & \multicolumn{1}{c|}{-}       & \multicolumn{1}{c|}{-}       & \multicolumn{1}{c|}{0.642}  & \multicolumn{1}{c|}{0.054}  & \multicolumn{1}{c|}{0.926}  & 0.940  \\ \cline{2-16} 
& \multirow{21}{*}{Non-simplex}  & \multirow{7}{*}{\begin{tabular}[c]{@{}c@{}}Independent \\ errors\end{tabular}} & Bias $(\times 10)$          & \multicolumn{1}{c|}{1.302} & \multicolumn{1}{c|}{1.366}  & \multicolumn{1}{c|}{10.695}  & \multicolumn{1}{c|}{10.506}  & \multicolumn{1}{c|}{8.262}  & \multicolumn{1}{c|}{5.699}  & \multicolumn{1}{c|}{11.927}  & \multicolumn{1}{c|}{11.849}  & \multicolumn{1}{c|}{-1.421} & \multicolumn{1}{c|}{-1.612} & \multicolumn{1}{c|}{0.069}  & 0.083  \\ \cline{4-16} 
&                                &                                                                                & ASE $(\times 10)$           & \multicolumn{1}{c|}{1.760} & \multicolumn{1}{c|}{0.789}  & \multicolumn{1}{c|}{-}       & \multicolumn{1}{c|}{-}       & \multicolumn{1}{c|}{-}      & \multicolumn{1}{c|}{-}      & \multicolumn{1}{c|}{-}       & \multicolumn{1}{c|}{-}       & \multicolumn{1}{c|}{1.851}  & \multicolumn{1}{c|}{0.836}  & \multicolumn{1}{c|}{1.795}  & 0.814  \\ \cline{4-16} 
&                                &                                                                                & BSE $(\times 10)$           & \multicolumn{1}{c|}{2.021} & \multicolumn{1}{c|}{0.816}  & \multicolumn{1}{c|}{-}       & \multicolumn{1}{c|}{-}       & \multicolumn{1}{c|}{-}      & \multicolumn{1}{c|}{-}      & \multicolumn{1}{c|}{-}       & \multicolumn{1}{c|}{-}       & \multicolumn{1}{c|}{2.242}  & \multicolumn{1}{c|}{0.991}  & \multicolumn{1}{c|}{1.941}  & 0.865  \\ \cline{4-16} 
&                                &                                                                                & ESE $(\times 10)$           & \multicolumn{1}{c|}{1.823} & \multicolumn{1}{c|}{0.766}  & \multicolumn{1}{c|}{1.159}   & \multicolumn{1}{c|}{0.406}   & \multicolumn{1}{c|}{1.702}  & \multicolumn{1}{c|}{0.949}  & \multicolumn{1}{c|}{0.962}   & \multicolumn{1}{c|}{0.433}   & \multicolumn{1}{c|}{2.084}  & \multicolumn{1}{c|}{0.973}  & \multicolumn{1}{c|}{1.938}  & 0.847  \\ \cline{4-16} 
&                                &                                                                                & MSE $(\times 100)$          & \multicolumn{1}{c|}{5.013} & \multicolumn{1}{c|}{2.451}  & \multicolumn{1}{c|}{115.724} & \multicolumn{1}{c|}{110.539} & \multicolumn{1}{c|}{71.157} & \multicolumn{1}{c|}{33.373} & \multicolumn{1}{c|}{143.178} & \multicolumn{1}{c|}{140.580} & \multicolumn{1}{c|}{6.357}  & \multicolumn{1}{c|}{3.543}  & \multicolumn{1}{c|}{3.753}  & 0.723  \\ \cline{4-16} 
&                                &                                                                                & Coverage (ASE)              & \multicolumn{1}{c|}{0.872} & \multicolumn{1}{c|}{0.582}  & \multicolumn{1}{c|}{-}       & \multicolumn{1}{c|}{-}       & \multicolumn{1}{c|}{-}      & \multicolumn{1}{c|}{-}      & \multicolumn{1}{c|}{-}       & \multicolumn{1}{c|}{-}       & \multicolumn{1}{c|}{0.840}  & \multicolumn{1}{c|}{0.486}  & \multicolumn{1}{c|}{0.928}  & 0.940  \\ \cline{4-16} 
&                                &                                                                                & Coverage (BSE)              & \multicolumn{1}{c|}{0.934} & \multicolumn{1}{c|}{0.610}  & \multicolumn{1}{c|}{-}       & \multicolumn{1}{c|}{-}       & \multicolumn{1}{c|}{-}      & \multicolumn{1}{c|}{-}      & \multicolumn{1}{c|}{-}       & \multicolumn{1}{c|}{-}       & \multicolumn{1}{c|}{0.914}  & \multicolumn{1}{c|}{0.604}  & \multicolumn{1}{c|}{0.942}  & 0.958  \\ \cline{3-16} 
&                                & \multirow{7}{*}{\begin{tabular}[c]{@{}c@{}}Correlated\\ errors\end{tabular}}   & Bias $(\times 10)$          & \multicolumn{1}{c|}{1.002} & \multicolumn{1}{c|}{0.998}  & \multicolumn{1}{c|}{10.518}  & \multicolumn{1}{c|}{10.508}  & \multicolumn{1}{c|}{2.675}  & \multicolumn{1}{c|}{1.589}  & \multicolumn{1}{c|}{10.521}  & \multicolumn{1}{c|}{10.507}  & \multicolumn{1}{c|}{-0.096} & \multicolumn{1}{c|}{0.789}  & \multicolumn{1}{c|}{0.219}  & 0.098  \\ \cline{4-16} 
&                                &                                                                                & ASE $(\times 10)$           & \multicolumn{1}{c|}{1.033} & \multicolumn{1}{c|}{0.464}  & \multicolumn{1}{c|}{-}       & \multicolumn{1}{c|}{-}       & \multicolumn{1}{c|}{-}      & \multicolumn{1}{c|}{-}      & \multicolumn{1}{c|}{-}       & \multicolumn{1}{c|}{-}       & \multicolumn{1}{c|}{1.484}  & \multicolumn{1}{c|}{0.660}  & \multicolumn{1}{c|}{1.330}  & 0.614  \\ \cline{4-16} 
&                                &                                                                                & BSE $(\times 10)$           & \multicolumn{1}{c|}{1.162} & \multicolumn{1}{c|}{0.477}  & \multicolumn{1}{c|}{-}       & \multicolumn{1}{c|}{-}       & \multicolumn{1}{c|}{-}      & \multicolumn{1}{c|}{-}      & \multicolumn{1}{c|}{-}       & \multicolumn{1}{c|}{-}       & \multicolumn{1}{c|}{1.655}  & \multicolumn{1}{c|}{0.766}  & \multicolumn{1}{c|}{1.424}  & 0.650  \\ \cline{4-16} 
&                                &                                                                                & ESE $(\times 10)$           & \multicolumn{1}{c|}{1.088} & \multicolumn{1}{c|}{0.438}  & \multicolumn{1}{c|}{0.642}   & \multicolumn{1}{c|}{0.286}   & \multicolumn{1}{c|}{1.538}  & \multicolumn{1}{c|}{0.520}  & \multicolumn{1}{c|}{0.644}   & \multicolumn{1}{c|}{0.286}   & \multicolumn{1}{c|}{2.158}  & \multicolumn{1}{c|}{0.877}  & \multicolumn{1}{c|}{1.345}  & 0.584  \\ \cline{4-16} 
&                                &                                                                                & MSE $(\times 100)$          & \multicolumn{1}{c|}{2.185} & \multicolumn{1}{c|}{1.187}  & \multicolumn{1}{c|}{111.044} & \multicolumn{1}{c|}{110.490} & \multicolumn{1}{c|}{9.519}  & \multicolumn{1}{c|}{2.794}  & \multicolumn{1}{c|}{111.110} & \multicolumn{1}{c|}{110.482} & \multicolumn{1}{c|}{4.656}  & \multicolumn{1}{c|}{1.391}  & \multicolumn{1}{c|}{1.853}  & 0.350  \\ \cline{4-16} 
&                                &                                                                                & Coverage (ASE)              & \multicolumn{1}{c|}{0.824} & \multicolumn{1}{c|}{0.434}  & \multicolumn{1}{c|}{-}       & \multicolumn{1}{c|}{-}       & \multicolumn{1}{c|}{-}      & \multicolumn{1}{c|}{-}      & \multicolumn{1}{c|}{-}       & \multicolumn{1}{c|}{-}       & \multicolumn{1}{c|}{0.796}  & \multicolumn{1}{c|}{0.716}  & \multicolumn{1}{c|}{0.934}  & 0.960  \\ \cline{4-16} 
&                                &                                                                                & Coverage (BSE)              & \multicolumn{1}{c|}{0.864} & \multicolumn{1}{c|}{0.440}  & \multicolumn{1}{c|}{-}       & \multicolumn{1}{c|}{-}       & \multicolumn{1}{c|}{-}      & \multicolumn{1}{c|}{-}      & \multicolumn{1}{c|}{-}       & \multicolumn{1}{c|}{-}       & \multicolumn{1}{c|}{0.822}  & \multicolumn{1}{c|}{0.792}  & \multicolumn{1}{c|}{0.948}  & 0.968  \\ \cline{3-16} 
&                                & \multirow{7}{*}{\begin{tabular}[c]{@{}c@{}}No\\ $Y$ error\end{tabular}}        & Bias $(\times 10)$          & \multicolumn{1}{c|}{1.430} & \multicolumn{1}{c|}{1.373}  & \multicolumn{1}{c|}{10.681}  & \multicolumn{1}{c|}{10.492}  & \multicolumn{1}{c|}{6.747}  & \multicolumn{1}{c|}{4.740}  & \multicolumn{1}{c|}{11.887}  & \multicolumn{1}{c|}{11.833}  & \multicolumn{1}{c|}{0.163}  & \multicolumn{1}{c|}{0.670}  & \multicolumn{1}{c|}{0.199}  & 0.127  \\ \cline{4-16} 
&                                &                                                                                & ASE $(\times 10)$           & \multicolumn{1}{c|}{1.221} & \multicolumn{1}{c|}{0.544}  & \multicolumn{1}{c|}{-}       & \multicolumn{1}{c|}{-}       & \multicolumn{1}{c|}{-}      & \multicolumn{1}{c|}{-}      & \multicolumn{1}{c|}{-}       & \multicolumn{1}{c|}{-}       & \multicolumn{1}{c|}{1.562}  & \multicolumn{1}{c|}{0.696}  & \multicolumn{1}{c|}{1.253}  & 0.563  \\ \cline{4-16} 
&                                &                                                                                & BSE $(\times 10)$           & \multicolumn{1}{c|}{1.383} & \multicolumn{1}{c|}{0.561}  & \multicolumn{1}{c|}{-}       & \multicolumn{1}{c|}{-}       & \multicolumn{1}{c|}{-}      & \multicolumn{1}{c|}{-}      & \multicolumn{1}{c|}{-}       & \multicolumn{1}{c|}{-}       & \multicolumn{1}{c|}{1.830}  & \multicolumn{1}{c|}{0.841}  & \multicolumn{1}{c|}{1.353}  & 0.598  \\ \cline{4-16} 
&                                &                                                                                & ESE $(\times 10)$           & \multicolumn{1}{c|}{1.274} & \multicolumn{1}{c|}{0.560}  & \multicolumn{1}{c|}{1.095}   & \multicolumn{1}{c|}{0.368}   & \multicolumn{1}{c|}{1.671}  & \multicolumn{1}{c|}{0.649}  & \multicolumn{1}{c|}{0.805}   & \multicolumn{1}{c|}{0.345}   & \multicolumn{1}{c|}{2.253}  & \multicolumn{1}{c|}{0.788}  & \multicolumn{1}{c|}{1.306}  & 0.613  \\ \cline{4-16} 
&                                &                                                                                & MSE $(\times 100)$          & \multicolumn{1}{c|}{3.663} & \multicolumn{1}{c|}{2.197}  & \multicolumn{1}{c|}{115.286} & \multicolumn{1}{c|}{110.212} & \multicolumn{1}{c|}{48.307} & \multicolumn{1}{c|}{22.885} & \multicolumn{1}{c|}{141.959} & \multicolumn{1}{c|}{140.136} & \multicolumn{1}{c|}{5.093}  & \multicolumn{1}{c|}{1.069}  & \multicolumn{1}{c|}{1.742}  & 0.391  \\ \cline{4-16} 
&                                &                                                                                & Coverage (ASE)              & \multicolumn{1}{c|}{0.758} & \multicolumn{1}{c|}{0.300}  & \multicolumn{1}{c|}{-}       & \multicolumn{1}{c|}{-}       & \multicolumn{1}{c|}{-}      & \multicolumn{1}{c|}{-}      & \multicolumn{1}{c|}{-}       & \multicolumn{1}{c|}{-}       & \multicolumn{1}{c|}{0.832}  & \multicolumn{1}{c|}{0.822}  & \multicolumn{1}{c|}{0.930}  & 0.918  \\ \cline{4-16} 
&                                &                                                                                & Coverage (BSE)              & \multicolumn{1}{c|}{0.820} & \multicolumn{1}{c|}{0.328}  & \multicolumn{1}{c|}{-}       & \multicolumn{1}{c|}{-}       & \multicolumn{1}{c|}{-}      & \multicolumn{1}{c|}{-}      & \multicolumn{1}{c|}{-}       & \multicolumn{1}{c|}{-}       & \multicolumn{1}{c|}{0.880}  & \multicolumn{1}{c|}{0.898}  & \multicolumn{1}{c|}{0.942}  & 0.928  \\ \hline
\end{tabular}

\caption{Summary Statistics of the Estimation Results Under (Linear trend with no intercept) for $\blambda_t$.} 
\label{Tab:Supp:Sim:T2}

\end{table}

\newpage

\begin{table}[!htp]
\renewcommand{\arraystretch}{1.3} \centering
\scriptsize
\setlength{\tabcolsep}{1pt}

\begin{tabular}{|c|c|c|c|cccccccccccc|}
\hline
\multirow{3}{*}{$\bm{\lambda}_t$}                                                                   & \multirow{3}{*}{$\bm{\mu}_0$} & \multirow{3}{*}{$\bm{e}_t$}                                                    & \multirow{3}{*}{Statistics} & \multicolumn{12}{c|}{Estimators and $T_0$}                                                                                                                                                                                                                                                                                                       \\ \cline{5-16}

&                                &                                                                                &                             & \multicolumn{2}{c|}{OLS-NoReg}                            & \multicolumn{2}{c|}{OLS-Standard}                         & \multicolumn{2}{c|}{ASC}                                  & \multicolumn{2}{c|}{SCPI}                                 & \multicolumn{2}{c|}{SPSC-NoDT}                            & \multicolumn{2}{c|}{SPSC-DT}         \\ \cline{5-16} 
&                                &                                                                                &                             & \multicolumn{1}{c|}{100}    & \multicolumn{1}{c|}{500}    & \multicolumn{1}{c|}{100}    & \multicolumn{1}{c|}{500}    & \multicolumn{1}{c|}{100}    & \multicolumn{1}{c|}{500}    & \multicolumn{1}{c|}{100}    & \multicolumn{1}{c|}{500}    & \multicolumn{1}{c|}{100}    & \multicolumn{1}{c|}{500}    & \multicolumn{1}{c|}{100}    & 500    \\ \hline

\multirow{42}{*}{\begin{tabular}[c]{@{}c@{}}Linear\\ trend \\ with\\ intercept\end{tabular}} & \multirow{21}{*}{Simplex}      & \multirow{7}{*}{\begin{tabular}[c]{@{}c@{}}Independent \\ errors\end{tabular}} & Bias $(\times 10)$          & \multicolumn{1}{c|}{0.091}  & \multicolumn{1}{c|}{0.079} & \multicolumn{1}{c|}{0.012}   & \multicolumn{1}{c|}{0.001}   & \multicolumn{1}{c|}{-0.173} & \multicolumn{1}{c|}{-0.134} & \multicolumn{1}{c|}{-0.169}  & \multicolumn{1}{c|}{-0.133}  & \multicolumn{1}{c|}{-0.547} & \multicolumn{1}{c|}{-0.621} & \multicolumn{1}{c|}{0.020}  & -0.005 \\ \cline{4-16} 
&                                &                                                                                & ASE $(\times 10)$           & \multicolumn{1}{c|}{1.131}  & \multicolumn{1}{c|}{0.498} & \multicolumn{1}{c|}{-}       & \multicolumn{1}{c|}{-}       & \multicolumn{1}{c|}{-}      & \multicolumn{1}{c|}{-}      & \multicolumn{1}{c|}{-}       & \multicolumn{1}{c|}{-}       & \multicolumn{1}{c|}{1.183}  & \multicolumn{1}{c|}{0.535}  & \multicolumn{1}{c|}{1.107}  & 0.502  \\ \cline{4-16} 
&                                &                                                                                & BSE $(\times 10)$           & \multicolumn{1}{c|}{1.261}  & \multicolumn{1}{c|}{0.514} & \multicolumn{1}{c|}{-}       & \multicolumn{1}{c|}{-}       & \multicolumn{1}{c|}{-}      & \multicolumn{1}{c|}{-}      & \multicolumn{1}{c|}{-}       & \multicolumn{1}{c|}{-}       & \multicolumn{1}{c|}{1.269}  & \multicolumn{1}{c|}{0.576}  & \multicolumn{1}{c|}{1.134}  & 0.506  \\ \cline{4-16} 
&                                &                                                                                & ESE $(\times 10)$           & \multicolumn{1}{c|}{1.198}  & \multicolumn{1}{c|}{0.494} & \multicolumn{1}{c|}{1.345}   & \multicolumn{1}{c|}{0.606}   & \multicolumn{1}{c|}{1.152}  & \multicolumn{1}{c|}{0.495}  & \multicolumn{1}{c|}{1.152}   & \multicolumn{1}{c|}{0.495}   & \multicolumn{1}{c|}{1.341}  & \multicolumn{1}{c|}{0.591}  & \multicolumn{1}{c|}{1.147}  & 0.507  \\ \cline{4-16} 
&                                &                                                                                & MSE $(\times 100)$          & \multicolumn{1}{c|}{1.441}  & \multicolumn{1}{c|}{0.249} & \multicolumn{1}{c|}{1.806}   & \multicolumn{1}{c|}{0.367}   & \multicolumn{1}{c|}{1.355}  & \multicolumn{1}{c|}{0.263}  & \multicolumn{1}{c|}{1.353}   & \multicolumn{1}{c|}{0.262}   & \multicolumn{1}{c|}{2.095}  & \multicolumn{1}{c|}{0.734}  & \multicolumn{1}{c|}{1.313}  & 0.257  \\ \cline{4-16} 
&                                &                                                                                & Coverage (ASE)              & \multicolumn{1}{c|}{0.928}  & \multicolumn{1}{c|}{0.942} & \multicolumn{1}{c|}{-}       & \multicolumn{1}{c|}{-}       & \multicolumn{1}{c|}{-}      & \multicolumn{1}{c|}{-}      & \multicolumn{1}{c|}{-}       & \multicolumn{1}{c|}{-}       & \multicolumn{1}{c|}{0.886}  & \multicolumn{1}{c|}{0.758}  & \multicolumn{1}{c|}{0.930}  & 0.958  \\ \cline{4-16} 
&                                &                                                                                & Coverage (BSE)              & \multicolumn{1}{c|}{0.948}  & \multicolumn{1}{c|}{0.958} & \multicolumn{1}{c|}{-}       & \multicolumn{1}{c|}{-}       & \multicolumn{1}{c|}{-}      & \multicolumn{1}{c|}{-}      & \multicolumn{1}{c|}{-}       & \multicolumn{1}{c|}{-}       & \multicolumn{1}{c|}{0.908}  & \multicolumn{1}{c|}{0.798}  & \multicolumn{1}{c|}{0.938}  & 0.950  \\ \cline{3-16} 
&                                & \multirow{7}{*}{\begin{tabular}[c]{@{}c@{}}Correlated \\ errors\end{tabular}}  & Bias $(\times 10)$          & \multicolumn{1}{c|}{-0.026} & \multicolumn{1}{c|}{0.015} & \multicolumn{1}{c|}{-0.002}  & \multicolumn{1}{c|}{-0.001}  & \multicolumn{1}{c|}{-0.066} & \multicolumn{1}{c|}{-0.006} & \multicolumn{1}{c|}{-0.064}  & \multicolumn{1}{c|}{-0.005}  & \multicolumn{1}{c|}{-0.697} & \multicolumn{1}{c|}{-0.450} & \multicolumn{1}{c|}{-0.035} & 0.012  \\ \cline{4-16} 
&                                &                                                                                & ASE $(\times 10)$           & \multicolumn{1}{c|}{0.594}  & \multicolumn{1}{c|}{0.265} & \multicolumn{1}{c|}{-}       & \multicolumn{1}{c|}{-}       & \multicolumn{1}{c|}{-}      & \multicolumn{1}{c|}{-}      & \multicolumn{1}{c|}{-}       & \multicolumn{1}{c|}{-}       & \multicolumn{1}{c|}{0.944}  & \multicolumn{1}{c|}{0.424}  & \multicolumn{1}{c|}{0.600}  & 0.271  \\ \cline{4-16} 
&                                &                                                                                & BSE $(\times 10)$           & \multicolumn{1}{c|}{0.626}  & \multicolumn{1}{c|}{0.271} & \multicolumn{1}{c|}{-}       & \multicolumn{1}{c|}{-}       & \multicolumn{1}{c|}{-}      & \multicolumn{1}{c|}{-}      & \multicolumn{1}{c|}{-}       & \multicolumn{1}{c|}{-}       & \multicolumn{1}{c|}{0.969}  & \multicolumn{1}{c|}{0.441}  & \multicolumn{1}{c|}{0.604}  & 0.272  \\ \cline{4-16} 
&                                &                                                                                & ESE $(\times 10)$           & \multicolumn{1}{c|}{0.612}  & \multicolumn{1}{c|}{0.259} & \multicolumn{1}{c|}{1.028}   & \multicolumn{1}{c|}{0.467}   & \multicolumn{1}{c|}{0.594}  & \multicolumn{1}{c|}{0.258}  & \multicolumn{1}{c|}{0.594}   & \multicolumn{1}{c|}{0.258}   & \multicolumn{1}{c|}{1.115}  & \multicolumn{1}{c|}{0.723}  & \multicolumn{1}{c|}{0.619}  & 0.265  \\ \cline{4-16} 
&                                &                                                                                & MSE $(\times 100)$          & \multicolumn{1}{c|}{0.374}  & \multicolumn{1}{c|}{0.067} & \multicolumn{1}{c|}{1.054}   & \multicolumn{1}{c|}{0.218}   & \multicolumn{1}{c|}{0.357}  & \multicolumn{1}{c|}{0.066}  & \multicolumn{1}{c|}{0.357}   & \multicolumn{1}{c|}{0.066}   & \multicolumn{1}{c|}{1.726}  & \multicolumn{1}{c|}{0.724}  & \multicolumn{1}{c|}{0.384}  & 0.070  \\ \cline{4-16} 
&                                &                                                                                & Coverage (ASE)              & \multicolumn{1}{c|}{0.948}  & \multicolumn{1}{c|}{0.954} & \multicolumn{1}{c|}{-}       & \multicolumn{1}{c|}{-}       & \multicolumn{1}{c|}{-}      & \multicolumn{1}{c|}{-}      & \multicolumn{1}{c|}{-}       & \multicolumn{1}{c|}{-}       & \multicolumn{1}{c|}{0.830}  & \multicolumn{1}{c|}{0.626}  & \multicolumn{1}{c|}{0.936}  & 0.954  \\ \cline{4-16} 
&                                &                                                                                & Coverage (BSE)              & \multicolumn{1}{c|}{0.956}  & \multicolumn{1}{c|}{0.954} & \multicolumn{1}{c|}{-}       & \multicolumn{1}{c|}{-}       & \multicolumn{1}{c|}{-}      & \multicolumn{1}{c|}{-}      & \multicolumn{1}{c|}{-}       & \multicolumn{1}{c|}{-}       & \multicolumn{1}{c|}{0.828}  & \multicolumn{1}{c|}{0.642}  & \multicolumn{1}{c|}{0.940}  & 0.950  \\ \cline{3-16} 
&                                & \multirow{7}{*}{\begin{tabular}[c]{@{}c@{}}No \\ $Y$ error\end{tabular}}       & Bias $(\times 10)$          & \multicolumn{1}{c|}{0.085}  & \multicolumn{1}{c|}{0.089} & \multicolumn{1}{c|}{0.025}   & \multicolumn{1}{c|}{0.021}   & \multicolumn{1}{c|}{-0.121} & \multicolumn{1}{c|}{-0.098} & \multicolumn{1}{c|}{-0.120}  & \multicolumn{1}{c|}{-0.099}  & \multicolumn{1}{c|}{-0.309} & \multicolumn{1}{c|}{-0.322} & \multicolumn{1}{c|}{0.002}  & 0.007  \\ \cline{4-16} 
&                                &                                                                                & ASE $(\times 10)$           & \multicolumn{1}{c|}{0.580}  & \multicolumn{1}{c|}{0.260} & \multicolumn{1}{c|}{-}       & \multicolumn{1}{c|}{-}       & \multicolumn{1}{c|}{-}      & \multicolumn{1}{c|}{-}      & \multicolumn{1}{c|}{-}       & \multicolumn{1}{c|}{-}       & \multicolumn{1}{c|}{0.826}  & \multicolumn{1}{c|}{0.372}  & \multicolumn{1}{c|}{0.582}  & 0.261  \\ \cline{4-16} 
&                                &                                                                                & BSE $(\times 10)$           & \multicolumn{1}{c|}{0.611}  & \multicolumn{1}{c|}{0.265} & \multicolumn{1}{c|}{-}       & \multicolumn{1}{c|}{-}       & \multicolumn{1}{c|}{-}      & \multicolumn{1}{c|}{-}      & \multicolumn{1}{c|}{-}       & \multicolumn{1}{c|}{-}       & \multicolumn{1}{c|}{0.830}  & \multicolumn{1}{c|}{0.373}  & \multicolumn{1}{c|}{0.587}  & 0.261  \\ \cline{4-16} 
&                                &                                                                                & ESE $(\times 10)$           & \multicolumn{1}{c|}{0.583}  & \multicolumn{1}{c|}{0.266} & \multicolumn{1}{c|}{0.925}   & \multicolumn{1}{c|}{0.419}   & \multicolumn{1}{c|}{0.603}  & \multicolumn{1}{c|}{0.273}  & \multicolumn{1}{c|}{0.603}   & \multicolumn{1}{c|}{0.273}   & \multicolumn{1}{c|}{0.844}  & \multicolumn{1}{c|}{0.388}  & \multicolumn{1}{c|}{0.578}  & 0.266  \\ \cline{4-16} 
&                                &                                                                                & MSE $(\times 100)$          & \multicolumn{1}{c|}{0.347}  & \multicolumn{1}{c|}{0.078} & \multicolumn{1}{c|}{0.855}   & \multicolumn{1}{c|}{0.176}   & \multicolumn{1}{c|}{0.378}  & \multicolumn{1}{c|}{0.084}  & \multicolumn{1}{c|}{0.377}   & \multicolumn{1}{c|}{0.084}   & \multicolumn{1}{c|}{0.806}  & \multicolumn{1}{c|}{0.254}  & \multicolumn{1}{c|}{0.333}  & 0.070  \\ \cline{4-16} 
&                                &                                                                                & Coverage (ASE)              & \multicolumn{1}{c|}{0.952}  & \multicolumn{1}{c|}{0.946} & \multicolumn{1}{c|}{-}       & \multicolumn{1}{c|}{-}       & \multicolumn{1}{c|}{-}      & \multicolumn{1}{c|}{-}      & \multicolumn{1}{c|}{-}       & \multicolumn{1}{c|}{-}       & \multicolumn{1}{c|}{0.924}  & \multicolumn{1}{c|}{0.858}  & \multicolumn{1}{c|}{0.956}  & 0.956  \\ \cline{4-16} 
&                                &                                                                                & Coverage (BSE)              & \multicolumn{1}{c|}{0.970}  & \multicolumn{1}{c|}{0.952} & \multicolumn{1}{c|}{-}       & \multicolumn{1}{c|}{-}       & \multicolumn{1}{c|}{-}      & \multicolumn{1}{c|}{-}      & \multicolumn{1}{c|}{-}       & \multicolumn{1}{c|}{-}       & \multicolumn{1}{c|}{0.922}  & \multicolumn{1}{c|}{0.850}  & \multicolumn{1}{c|}{0.954}  & 0.942  \\ \cline{2-16} 
& \multirow{21}{*}{Non-simplex}  & \multirow{7}{*}{\begin{tabular}[c]{@{}c@{}}Independent \\ errors\end{tabular}} & Bias $(\times 10)$          & \multicolumn{1}{c|}{0.383}  & \multicolumn{1}{c|}{0.424} & \multicolumn{1}{c|}{17.500}  & \multicolumn{1}{c|}{17.531}  & \multicolumn{1}{c|}{2.818}  & \multicolumn{1}{c|}{1.497}  & \multicolumn{1}{c|}{17.517}  & \multicolumn{1}{c|}{17.530}  & \multicolumn{1}{c|}{0.256}  & \multicolumn{1}{c|}{0.339}  & \multicolumn{1}{c|}{-0.009} & 0.040  \\ \cline{4-16} 
&                                &                                                                                & ASE $(\times 10)$           & \multicolumn{1}{c|}{1.394}  & \multicolumn{1}{c|}{0.608} & \multicolumn{1}{c|}{-}       & \multicolumn{1}{c|}{-}       & \multicolumn{1}{c|}{-}      & \multicolumn{1}{c|}{-}      & \multicolumn{1}{c|}{-}       & \multicolumn{1}{c|}{-}       & \multicolumn{1}{c|}{1.724}  & \multicolumn{1}{c|}{0.775}  & \multicolumn{1}{c|}{1.362}  & 0.613  \\ \cline{4-16} 
&                                &                                                                                & BSE $(\times 10)$           & \multicolumn{1}{c|}{1.571}  & \multicolumn{1}{c|}{0.629} & \multicolumn{1}{c|}{-}       & \multicolumn{1}{c|}{-}       & \multicolumn{1}{c|}{-}      & \multicolumn{1}{c|}{-}      & \multicolumn{1}{c|}{-}       & \multicolumn{1}{c|}{-}       & \multicolumn{1}{c|}{1.798}  & \multicolumn{1}{c|}{0.810}  & \multicolumn{1}{c|}{1.416}  & 0.621  \\ \cline{4-16} 
&                                &                                                                                & ESE $(\times 10)$           & \multicolumn{1}{c|}{1.470}  & \multicolumn{1}{c|}{0.627} & \multicolumn{1}{c|}{0.909}   & \multicolumn{1}{c|}{0.450}   & \multicolumn{1}{c|}{1.798}  & \multicolumn{1}{c|}{0.708}  & \multicolumn{1}{c|}{0.911}   & \multicolumn{1}{c|}{0.450}   & \multicolumn{1}{c|}{1.789}  & \multicolumn{1}{c|}{0.799}  & \multicolumn{1}{c|}{1.399}  & 0.638  \\ \cline{4-16} 
&                                &                                                                                & MSE $(\times 100)$          & \multicolumn{1}{c|}{2.303}  & \multicolumn{1}{c|}{0.572} & \multicolumn{1}{c|}{307.074} & \multicolumn{1}{c|}{307.536} & \multicolumn{1}{c|}{11.169} & \multicolumn{1}{c|}{2.743}  & \multicolumn{1}{c|}{307.689} & \multicolumn{1}{c|}{307.513} & \multicolumn{1}{c|}{3.260}  & \multicolumn{1}{c|}{0.751}  & \multicolumn{1}{c|}{1.953}  & 0.407  \\ \cline{4-16} 
&                                &                                                                                & Coverage (ASE)              & \multicolumn{1}{c|}{0.926}  & \multicolumn{1}{c|}{0.886} & \multicolumn{1}{c|}{-}       & \multicolumn{1}{c|}{-}       & \multicolumn{1}{c|}{-}      & \multicolumn{1}{c|}{-}      & \multicolumn{1}{c|}{-}       & \multicolumn{1}{c|}{-}       & \multicolumn{1}{c|}{0.934}  & \multicolumn{1}{c|}{0.928}  & \multicolumn{1}{c|}{0.934}  & 0.954  \\ \cline{4-16} 
&                                &                                                                                & Coverage (BSE)              & \multicolumn{1}{c|}{0.962}  & \multicolumn{1}{c|}{0.894} & \multicolumn{1}{c|}{-}       & \multicolumn{1}{c|}{-}       & \multicolumn{1}{c|}{-}      & \multicolumn{1}{c|}{-}      & \multicolumn{1}{c|}{-}       & \multicolumn{1}{c|}{-}       & \multicolumn{1}{c|}{0.938}  & \multicolumn{1}{c|}{0.934}  & \multicolumn{1}{c|}{0.946}  & 0.952  \\ \cline{3-16} 
&                                & \multirow{7}{*}{\begin{tabular}[c]{@{}c@{}}Correlated\\ errors\end{tabular}}   & Bias $(\times 10)$          & \multicolumn{1}{c|}{0.287}  & \multicolumn{1}{c|}{0.247} & \multicolumn{1}{c|}{17.548}  & \multicolumn{1}{c|}{17.488}  & \multicolumn{1}{c|}{1.028}  & \multicolumn{1}{c|}{0.440}  & \multicolumn{1}{c|}{17.547}  & \multicolumn{1}{c|}{17.487}  & \multicolumn{1}{c|}{-0.427} & \multicolumn{1}{c|}{-0.738} & \multicolumn{1}{c|}{0.023}  & -0.018 \\ \cline{4-16} 
&                                &                                                                                & ASE $(\times 10)$           & \multicolumn{1}{c|}{0.862}  & \multicolumn{1}{c|}{0.378} & \multicolumn{1}{c|}{-}       & \multicolumn{1}{c|}{-}       & \multicolumn{1}{c|}{-}      & \multicolumn{1}{c|}{-}      & \multicolumn{1}{c|}{-}       & \multicolumn{1}{c|}{-}       & \multicolumn{1}{c|}{1.472}  & \multicolumn{1}{c|}{0.662}  & \multicolumn{1}{c|}{1.020}  & 0.459  \\ \cline{4-16} 
&                                &                                                                                & BSE $(\times 10)$           & \multicolumn{1}{c|}{0.950}  & \multicolumn{1}{c|}{0.388} & \multicolumn{1}{c|}{-}       & \multicolumn{1}{c|}{-}       & \multicolumn{1}{c|}{-}      & \multicolumn{1}{c|}{-}      & \multicolumn{1}{c|}{-}       & \multicolumn{1}{c|}{-}       & \multicolumn{1}{c|}{1.475}  & \multicolumn{1}{c|}{0.665}  & \multicolumn{1}{c|}{1.067}  & 0.478  \\ \cline{4-16} 
&                                &                                                                                & ESE $(\times 10)$           & \multicolumn{1}{c|}{0.906}  & \multicolumn{1}{c|}{0.382} & \multicolumn{1}{c|}{0.661}   & \multicolumn{1}{c|}{0.300}   & \multicolumn{1}{c|}{1.138}  & \multicolumn{1}{c|}{0.394}  & \multicolumn{1}{c|}{0.661}   & \multicolumn{1}{c|}{0.300}   & \multicolumn{1}{c|}{1.696}  & \multicolumn{1}{c|}{0.806}  & \multicolumn{1}{c|}{1.012}  & 0.455  \\ \cline{4-16} 
&                                &                                                                                & MSE $(\times 100)$          & \multicolumn{1}{c|}{0.902}  & \multicolumn{1}{c|}{0.207} & \multicolumn{1}{c|}{308.369} & \multicolumn{1}{c|}{305.912} & \multicolumn{1}{c|}{2.349}  & \multicolumn{1}{c|}{0.349}  & \multicolumn{1}{c|}{308.345} & \multicolumn{1}{c|}{305.890} & \multicolumn{1}{c|}{3.052}  & \multicolumn{1}{c|}{1.193}  & \multicolumn{1}{c|}{1.022}  & 0.207  \\ \cline{4-16} 
&                                &                                                                                & Coverage (ASE)              & \multicolumn{1}{c|}{0.924}  & \multicolumn{1}{c|}{0.882} & \multicolumn{1}{c|}{-}       & \multicolumn{1}{c|}{-}       & \multicolumn{1}{c|}{-}      & \multicolumn{1}{c|}{-}      & \multicolumn{1}{c|}{-}       & \multicolumn{1}{c|}{-}       & \multicolumn{1}{c|}{0.902}  & \multicolumn{1}{c|}{0.734}  & \multicolumn{1}{c|}{0.934}  & 0.946  \\ \cline{4-16} 
&                                &                                                                                & Coverage (BSE)              & \multicolumn{1}{c|}{0.948}  & \multicolumn{1}{c|}{0.888} & \multicolumn{1}{c|}{-}       & \multicolumn{1}{c|}{-}       & \multicolumn{1}{c|}{-}      & \multicolumn{1}{c|}{-}      & \multicolumn{1}{c|}{-}       & \multicolumn{1}{c|}{-}       & \multicolumn{1}{c|}{0.906}  & \multicolumn{1}{c|}{0.736}  & \multicolumn{1}{c|}{0.940}  & 0.954  \\ \cline{3-16} 
&                                & \multirow{7}{*}{\begin{tabular}[c]{@{}c@{}}No\\ $Y$ error\end{tabular}}        & Bias $(\times 10)$          & \multicolumn{1}{c|}{0.385}  & \multicolumn{1}{c|}{0.388} & \multicolumn{1}{c|}{17.573}  & \multicolumn{1}{c|}{17.520}  & \multicolumn{1}{c|}{1.884}  & \multicolumn{1}{c|}{1.388}  & \multicolumn{1}{c|}{17.575}  & \multicolumn{1}{c|}{17.519}  & \multicolumn{1}{c|}{-0.421} & \multicolumn{1}{c|}{-0.506} & \multicolumn{1}{c|}{0.010}  & 0.014  \\ \cline{4-16} 
&                                &                                                                                & ASE $(\times 10)$           & \multicolumn{1}{c|}{0.983}  & \multicolumn{1}{c|}{0.435} & \multicolumn{1}{c|}{-}       & \multicolumn{1}{c|}{-}       & \multicolumn{1}{c|}{-}      & \multicolumn{1}{c|}{-}      & \multicolumn{1}{c|}{-}       & \multicolumn{1}{c|}{-}       & \multicolumn{1}{c|}{1.494}  & \multicolumn{1}{c|}{0.674}  & \multicolumn{1}{c|}{0.979}  & 0.439  \\ \cline{4-16} 
&                                &                                                                                & BSE $(\times 10)$           & \multicolumn{1}{c|}{1.092}  & \multicolumn{1}{c|}{0.449} & \multicolumn{1}{c|}{-}       & \multicolumn{1}{c|}{-}       & \multicolumn{1}{c|}{-}      & \multicolumn{1}{c|}{-}      & \multicolumn{1}{c|}{-}       & \multicolumn{1}{c|}{-}       & \multicolumn{1}{c|}{1.515}  & \multicolumn{1}{c|}{0.678}  & \multicolumn{1}{c|}{1.012}  & 0.446  \\ \cline{4-16} 
&                                &                                                                                & ESE $(\times 10)$           & \multicolumn{1}{c|}{1.091}  & \multicolumn{1}{c|}{0.432} & \multicolumn{1}{c|}{0.797}   & \multicolumn{1}{c|}{0.360}   & \multicolumn{1}{c|}{1.316}  & \multicolumn{1}{c|}{0.508}  & \multicolumn{1}{c|}{0.798}   & \multicolumn{1}{c|}{0.360}   & \multicolumn{1}{c|}{1.800}  & \multicolumn{1}{c|}{1.006}  & \multicolumn{1}{c|}{1.080}  & 0.439  \\ \cline{4-16} 
&                                &                                                                                & MSE $(\times 100)$          & \multicolumn{1}{c|}{1.336}  & \multicolumn{1}{c|}{0.337} & \multicolumn{1}{c|}{309.455} & \multicolumn{1}{c|}{307.081} & \multicolumn{1}{c|}{5.278}  & \multicolumn{1}{c|}{2.185}  & \multicolumn{1}{c|}{309.528} & \multicolumn{1}{c|}{307.059} & \multicolumn{1}{c|}{3.412}  & \multicolumn{1}{c|}{1.265}  & \multicolumn{1}{c|}{1.164}  & 0.192  \\ \cline{4-16} 
&                                &                                                                                & Coverage (ASE)              & \multicolumn{1}{c|}{0.900}  & \multicolumn{1}{c|}{0.846} & \multicolumn{1}{c|}{-}       & \multicolumn{1}{c|}{-}       & \multicolumn{1}{c|}{-}      & \multicolumn{1}{c|}{-}      & \multicolumn{1}{c|}{-}       & \multicolumn{1}{c|}{-}       & \multicolumn{1}{c|}{0.880}  & \multicolumn{1}{c|}{0.750}  & \multicolumn{1}{c|}{0.924}  & 0.960  \\ \cline{4-16} 
&                                &                                                                                & Coverage (BSE)              & \multicolumn{1}{c|}{0.932}  & \multicolumn{1}{c|}{0.866} & \multicolumn{1}{c|}{-}       & \multicolumn{1}{c|}{-}       & \multicolumn{1}{c|}{-}      & \multicolumn{1}{c|}{-}      & \multicolumn{1}{c|}{-}       & \multicolumn{1}{c|}{-}       & \multicolumn{1}{c|}{0.882}  & \multicolumn{1}{c|}{0.744}  & \multicolumn{1}{c|}{0.936}  & 0.962  \\ \hline
\end{tabular} 

\caption{Summary Statistics of the Estimation Results Under (Linear trend with intercept) for $\blambda_t$.} 
\label{Tab:Supp:Sim:T3}

\end{table}

\newpage

We report the performance of conformal inference in Section \ref{sec:Conformal} of the main paper under the simulation scenarios in Section \ref{sec:Sim} of the main paper. First, we obtain the pointwise 95\% pointwise prediction interval for the random treatment effect at 10 post-treatment times $t \in \mathcal{T} = \{ T_0 + 0.1 T_1, T_0 + 0.2 T_1, \ldots T_0 + 0.9 T_1 , T\}$, which are $\xi_{t}^* = 3 + \epsilon_{t}$.

As competing methods, we construct 95\% pointwise prediction intervals using the ASC and SCPI approaches. For the ASC approach, we use the conformal approach to construct prediction intervals, which is the default option employed in \texttt{augsynth} package. For the SCPI approach, we use the prediction interval estimating out-of-sample uncertainty with sub-Gaussian bounds, which is stored in \texttt{CI.all.gaussian} object of a \texttt{scpi} output; see below for an example R-code:
\begin{itemize}[leftmargin=2cm,itemsep=0cm]
\item[] \texttt{scpi.est $\leftarrow$ scpi::scpi(SCD) \# SCD is a scdata object}
\item[] \texttt{scpi.PI \ $\leftarrow$ scpi.est\$inference.results\$CI.all.gaussian}
\end{itemize}
For each simulation repetition and each method, we calculate $\ind \big(\xi_{t} ^* \in \mathcal{C}_{t} \big)$ where $\mathcal{C}_{t}$ is a 95\% prediction interval obtained from each method, i.e., the indicator of whether a 95\% prediction interval at $t$ obtained from each method includes the random treatment effect. Ideally, the average of these indicators across simulation repetitions (i.e., the empirical coverage rate of 95\% prediction intervals) should be close to the nominal coverage rate of 0.95.

Table \ref{tab:supp:Table3} shows the empirical coverage rates obtained from 500 repetitions for each simulation scenario. We find that the conformal inference approach for the SPSC achieves the nominal coverage rate across all simulation scenarios in general. However, we find that the ASC approach fails to do so for all reported scenarios. Likewise, the SCPI approach appears to struggle to attain the desired nominal coverage rate, especially when the latent factor $\blambda_t$ has a trend. Next, we calculate the average length of the 95\% prediction intervals for $t \in \mathcal{T}$ across 500 repetitions.   The table also shows the average lengths of the prediction intervals. 
We find that the length of the prediction intervals decreases as the length of the pre-treatment periods increases. 
Notably, the SPSC-DT approach consistently yields prediction intervals that are not only comparable in length but often the shortest when compared to other estimators across all scenarios. In particular, the SPSC-DT estimator outperforms the others in terms of coverage and length, particularly when a linear trend is present. These findings demonstrate that the proposed conformal inference method for the SPSC framework is robust and broadly applicable, regardless of the data generating process.

\newpage

\begin{table}[!htp] 
\renewcommand{\arraystretch}{1.2} \centering
\scriptsize
\setlength{\tabcolsep}{7pt} 

\begin{tabular}{|c|c|c|c|cc|cc|cc|cc|}
\hline
\multirow{3}{*}{$\blambda_t$}                                                                                        & \multirow{3}{*}{$\bm{\mu}_0$} & \multirow{3}{*}{$\bm{e}_t$}                                                        & \multirow{3}{*}{Statistics} & \multicolumn{8}{c|}{Estimators and $T_0$}   \\ \cline{5-12} 
&                                &                                                                                    &                             & \multicolumn{2}{c|}{ASC}           & \multicolumn{2}{c|}{SCPI}          & \multicolumn{2}{c|}{SPSC-NoDT}     & \multicolumn{2}{c|}{SPSC-DT}       \\ \cline{5-12} 
&                                &                                                                                    &                             & \multicolumn{1}{c|}{100}   & 500   & \multicolumn{1}{c|}{100}   & 500   & \multicolumn{1}{c|}{100}   & 500   & \multicolumn{1}{c|}{100}   & 500   \\ \hline
\multirow{12}{*}{\begin{tabular}[c]{@{}c@{}}No\\      trend\\      with\\      no\\      intercept\end{tabular}}     & \multirow{6}{*}{Simplex}       & \multirow{2}{*}{\begin{tabular}[c]{@{}c@{}}Independent\\      errors\end{tabular}} & Coverage                    & \multicolumn{1}{c|}{0.925} & 0.918 & \multicolumn{1}{c|}{0.907} & 0.939 & \multicolumn{1}{c|}{0.963} & 0.954 & \multicolumn{1}{c|}{0.963} & 0.954 \\ \cline{4-12} 
&                                &                                                                                    & Length                      & \multicolumn{1}{c|}{2.009} & 1.850 & \multicolumn{1}{c|}{2.034} & 2.036 & \multicolumn{1}{c|}{2.249} & 2.055 & \multicolumn{1}{c|}{2.244} & 2.054 \\ \cline{3-12} 
&                                & \multirow{2}{*}{\begin{tabular}[c]{@{}c@{}}Correlated\\      errors\end{tabular}}  & Coverage                    & \multicolumn{1}{c|}{0.691} & 0.567 & \multicolumn{1}{c|}{0.972} & 0.954 & \multicolumn{1}{c|}{0.959} & 0.950 & \multicolumn{1}{c|}{0.960} & 0.950 \\ \cline{4-12} 
&                                &                                                                                    & Length                      & \multicolumn{1}{c|}{0.423} & 0.283 & \multicolumn{1}{c|}{0.872} & 0.706 & \multicolumn{1}{c|}{0.755} & 0.696 & \multicolumn{1}{c|}{0.766} & 0.696 \\ \cline{3-12} 
&                                & \multirow{2}{*}{\begin{tabular}[c]{@{}c@{}}No\\      $Y$ error\end{tabular}}       & Coverage                    & \multicolumn{1}{c|}{0.714} & 0.581 & \multicolumn{1}{c|}{0.986} & 0.961 & \multicolumn{1}{c|}{0.959} & 0.948 & \multicolumn{1}{c|}{0.960} & 0.948 \\ \cline{4-12} 
&                                &                                                                                    & Length                      & \multicolumn{1}{c|}{0.465} & 0.305 & \multicolumn{1}{c|}{0.951} & 0.680 & \multicolumn{1}{c|}{0.675} & 0.610 & \multicolumn{1}{c|}{0.681} & 0.610 \\ \cline{2-12} 
& \multirow{6}{*}{Non-simplex}   & \multirow{2}{*}{\begin{tabular}[c]{@{}c@{}}Independent\\      errors\end{tabular}} & Coverage                    & \multicolumn{1}{c|}{0.933} & 0.925 & \multicolumn{1}{c|}{0.935} & 0.912 & \multicolumn{1}{c|}{0.959} & 0.948 & \multicolumn{1}{c|}{0.961} & 0.948 \\ \cline{4-12} 
&                                &                                                                                    & Length                      & \multicolumn{1}{c|}{2.597} & 2.459 & \multicolumn{1}{c|}{2.848} & 2.407 & \multicolumn{1}{c|}{2.839} & 2.556 & \multicolumn{1}{c|}{2.824} & 2.554 \\ \cline{3-12} 
&                                & \multirow{2}{*}{\begin{tabular}[c]{@{}c@{}}Correlated\\      errors\end{tabular}}  & Coverage                    & \multicolumn{1}{c|}{0.904} & 0.906 & \multicolumn{1}{c|}{0.925} & 0.913 & \multicolumn{1}{c|}{0.962} & 0.952 & \multicolumn{1}{c|}{0.963} & 0.953 \\ \cline{4-12} 
&                                &                                                                                    & Length                      & \multicolumn{1}{c|}{1.326} & 1.229 & \multicolumn{1}{c|}{1.440} & 1.255 & \multicolumn{1}{c|}{1.859} & 1.701 & \multicolumn{1}{c|}{1.935} & 1.699 \\ \cline{3-12} 
&                                & \multirow{2}{*}{\begin{tabular}[c]{@{}c@{}}No\\      $Y$ error\end{tabular}}       & Coverage                    & \multicolumn{1}{c|}{0.922} & 0.910 & \multicolumn{1}{c|}{0.938} & 0.925 & \multicolumn{1}{c|}{0.964} & 0.951 & \multicolumn{1}{c|}{0.964} & 0.953 \\ \cline{4-12} 
&                                &                                                                                    & Length                      & \multicolumn{1}{c|}{1.704} & 1.608 & \multicolumn{1}{c|}{1.978} & 1.648 & \multicolumn{1}{c|}{1.852} & 1.668 & \multicolumn{1}{c|}{1.856} & 1.670 \\ \hline
\multirow{12}{*}{\begin{tabular}[c]{@{}c@{}}No\\      trend\\      with\\      intercept\end{tabular}}               & \multirow{6}{*}{Simplex}       & \multirow{2}{*}{\begin{tabular}[c]{@{}c@{}}Independent\\      errors\end{tabular}} & Coverage                    & \multicolumn{1}{c|}{0.921} & 0.917 & \multicolumn{1}{c|}{0.911} & 0.940 & \multicolumn{1}{c|}{0.962} & 0.953 & \multicolumn{1}{c|}{0.959} & 0.951 \\ \cline{4-12} 
&                                &                                                                                    & Length                      & \multicolumn{1}{c|}{1.998} & 1.861 & \multicolumn{1}{c|}{2.068} & 2.069 & \multicolumn{1}{c|}{2.599} & 2.410 & \multicolumn{1}{c|}{2.245} & 2.061 \\ \cline{3-12} 
&                                & \multirow{2}{*}{\begin{tabular}[c]{@{}c@{}}Correlated\\      errors\end{tabular}}  & Coverage                    & \multicolumn{1}{c|}{0.674} & 0.580 & \multicolumn{1}{c|}{0.969} & 0.958 & \multicolumn{1}{c|}{0.962} & 0.955 & \multicolumn{1}{c|}{0.961} & 0.948 \\ \cline{4-12} 
&                                &                                                                                    & Length                      & \multicolumn{1}{c|}{0.410} & 0.286 & \multicolumn{1}{c|}{0.868} & 0.704 & \multicolumn{1}{c|}{1.782} & 1.653 & \multicolumn{1}{c|}{0.762} & 0.700 \\ \cline{3-12} 
&                                & \multirow{2}{*}{\begin{tabular}[c]{@{}c@{}}No\\      $Y$ error\end{tabular}}       & Coverage                    & \multicolumn{1}{c|}{0.735} & 0.603 & \multicolumn{1}{c|}{0.985} & 0.954 & \multicolumn{1}{c|}{0.953} & 0.954 & \multicolumn{1}{c|}{0.968} & 0.946 \\ \cline{4-12} 
&                                &                                                                                    & Length                      & \multicolumn{1}{c|}{0.489} & 0.324 & \multicolumn{1}{c|}{0.935} & 0.666 & \multicolumn{1}{c|}{1.469} & 1.366 & \multicolumn{1}{c|}{0.693} & 0.618 \\ \cline{2-12} 
& \multirow{6}{*}{Non-simplex}   & \multirow{2}{*}{\begin{tabular}[c]{@{}c@{}}Independent\\      errors\end{tabular}} & Coverage                    & \multicolumn{1}{c|}{0.927} & 0.921 & \multicolumn{1}{c|}{0.898} & 0.877 & \multicolumn{1}{c|}{0.960} & 0.947 & \multicolumn{1}{c|}{0.958} & 0.946 \\ \cline{4-12} 
&                                &                                                                                    & Length                      & \multicolumn{1}{c|}{3.064} & 2.746 & \multicolumn{1}{c|}{2.741} & 2.410 & \multicolumn{1}{c|}{3.778} & 3.511 & \multicolumn{1}{c|}{2.887} & 2.638 \\ \cline{3-12} 
&                                & \multirow{2}{*}{\begin{tabular}[c]{@{}c@{}}Correlated\\      errors\end{tabular}}  & Coverage                    & \multicolumn{1}{c|}{0.911} & 0.907 & \multicolumn{1}{c|}{0.893} & 0.883 & \multicolumn{1}{c|}{0.965} & 0.953 & \multicolumn{1}{c|}{0.967} & 0.953 \\ \cline{4-12} 
&                                &                                                                                    & Length                      & \multicolumn{1}{c|}{1.435} & 1.266 & \multicolumn{1}{c|}{1.420} & 1.278 & \multicolumn{1}{c|}{3.043} & 2.831 & \multicolumn{1}{c|}{2.181} & 1.813 \\ \cline{3-12} 
&                                & \multirow{2}{*}{\begin{tabular}[c]{@{}c@{}}No\\      $Y$ error\end{tabular}}       & Coverage                    & \multicolumn{1}{c|}{0.917} & 0.918 & \multicolumn{1}{c|}{0.908} & 0.890 & \multicolumn{1}{c|}{0.964} & 0.947 & \multicolumn{1}{c|}{0.967} & 0.950 \\ \cline{4-12} 
&                                &                                                                                    & Length                      & \multicolumn{1}{c|}{2.198} & 1.958 & \multicolumn{1}{c|}{2.046} & 1.751 & \multicolumn{1}{c|}{3.130} & 2.901 & \multicolumn{1}{c|}{1.968} & 1.744 \\ \hline
\multirow{12}{*}{\begin{tabular}[c]{@{}c@{}}Linear\\      trend\\      with\\      no\\      intercept\end{tabular}} & \multirow{6}{*}{Simplex}       & \multirow{2}{*}{\begin{tabular}[c]{@{}c@{}}Independent\\      errors\end{tabular}} & Coverage                    & \multicolumn{1}{c|}{0.934} & 0.915 & \multicolumn{1}{c|}{0.926} & 0.951 & \multicolumn{1}{c|}{0.950} & 0.928 & \multicolumn{1}{c|}{0.962} & 0.941 \\ \cline{4-12} 
&                                &                                                                                    & Length                      & \multicolumn{1}{c|}{2.100} & 1.879 & \multicolumn{1}{c|}{2.372} & 2.252 & \multicolumn{1}{c|}{2.675} & 2.376 & \multicolumn{1}{c|}{2.288} & 2.074 \\ \cline{3-12} 
&                                & \multirow{2}{*}{\begin{tabular}[c]{@{}c@{}}Correlated\\      errors\end{tabular}}  & Coverage                    & \multicolumn{1}{c|}{0.744} & 0.593 & \multicolumn{1}{c|}{0.985} & 0.969 & \multicolumn{1}{c|}{0.919} & 0.935 & \multicolumn{1}{c|}{0.959} & 0.949 \\ \cline{4-12} 
&                                &                                                                                    & Length                      & \multicolumn{1}{c|}{0.490} & 0.300 & \multicolumn{1}{c|}{1.028} & 0.767 & \multicolumn{1}{c|}{1.603} & 1.421 & \multicolumn{1}{c|}{0.774} & 0.700 \\ \cline{3-12} 
&                                & \multirow{2}{*}{\begin{tabular}[c]{@{}c@{}}No\\      $Y$ error\end{tabular}}       & Coverage                    & \multicolumn{1}{c|}{0.780} & 0.620 & \multicolumn{1}{c|}{0.991} & 0.973 & \multicolumn{1}{c|}{0.941} & 0.930 & \multicolumn{1}{c|}{0.961} & 0.955 \\ \cline{4-12} 
&                                &                                                                                    & Length                      & \multicolumn{1}{c|}{0.553} & 0.353 & \multicolumn{1}{c|}{1.063} & 0.725 & \multicolumn{1}{c|}{1.392} & 1.225 & \multicolumn{1}{c|}{0.704} & 0.620 \\ \cline{2-12} 
& \multirow{6}{*}{Non-simplex}   & \multirow{2}{*}{\begin{tabular}[c]{@{}c@{}}Independent\\      errors\end{tabular}} & Coverage                    & \multicolumn{1}{c|}{0.795} & 0.846 & \multicolumn{1}{c|}{0.938} & 0.889 & \multicolumn{1}{c|}{0.959} & 0.944 & \multicolumn{1}{c|}{0.962} & 0.949 \\ \cline{4-12} 
&                                &                                                                                    & Length                      & \multicolumn{1}{c|}{2.969} & 2.719 & \multicolumn{1}{c|}{3.223} & 2.646 & \multicolumn{1}{c|}{3.755} & 3.336 & \multicolumn{1}{c|}{2.937} & 2.668 \\ \cline{3-12} 
&                                & \multirow{2}{*}{\begin{tabular}[c]{@{}c@{}}Correlated\\      errors\end{tabular}}  & Coverage                    & \multicolumn{1}{c|}{0.844} & 0.883 & \multicolumn{1}{c|}{0.907} & 0.843 & \multicolumn{1}{c|}{0.957} & 0.944 & \multicolumn{1}{c|}{0.967} & 0.948 \\ \cline{4-12} 
&                                &                                                                                    & Length                      & \multicolumn{1}{c|}{1.486} & 1.299 & \multicolumn{1}{c|}{1.678} & 1.432 & \multicolumn{1}{c|}{2.883} & 2.534 & \multicolumn{1}{c|}{2.207} & 1.820 \\ \cline{3-12} 
&                                & \multirow{2}{*}{\begin{tabular}[c]{@{}c@{}}No\\      $Y$ error\end{tabular}}       & Coverage                    & \multicolumn{1}{c|}{0.728} & 0.808 & \multicolumn{1}{c|}{0.920} & 0.852 & \multicolumn{1}{c|}{0.957} & 0.953 & \multicolumn{1}{c|}{0.957} & 0.946 \\ \cline{4-12} 
&                                &                                                                                    & Length                      & \multicolumn{1}{c|}{2.113} & 1.908 & \multicolumn{1}{c|}{2.384} & 1.897 & \multicolumn{1}{c|}{3.039} & 2.676 & \multicolumn{1}{c|}{1.976} & 1.756 \\ \hline
\multirow{12}{*}{\begin{tabular}[c]{@{}c@{}}Linear\\      trend\\      with\\      intercept\end{tabular}}           & \multirow{6}{*}{Simplex}       & \multirow{2}{*}{\begin{tabular}[c]{@{}c@{}}Independent\\      errors\end{tabular}} & Coverage                    & \multicolumn{1}{c|}{0.926} & 0.917 & \multicolumn{1}{c|}{0.917} & 0.949 & \multicolumn{1}{c|}{0.959} & 0.951 & \multicolumn{1}{c|}{0.962} & 0.949 \\ \cline{4-12} 
&                                &                                                                                    & Length                      & \multicolumn{1}{c|}{2.034} & 1.860 & \multicolumn{1}{c|}{2.281} & 2.168 & \multicolumn{1}{c|}{2.633} & 2.443 & \multicolumn{1}{c|}{2.278} & 2.069 \\ \cline{3-12} 
&                                & \multirow{2}{*}{\begin{tabular}[c]{@{}c@{}}Correlated\\      errors\end{tabular}}  & Coverage                    & \multicolumn{1}{c|}{0.693} & 0.580 & \multicolumn{1}{c|}{0.979} & 0.962 & \multicolumn{1}{c|}{0.961} & 0.948 & \multicolumn{1}{c|}{0.960} & 0.948 \\ \cline{4-12} 
&                                &                                                                                    & Length                      & \multicolumn{1}{c|}{0.436} & 0.288 & \multicolumn{1}{c|}{0.953} & 0.735 & \multicolumn{1}{c|}{1.928} & 1.767 & \multicolumn{1}{c|}{0.761} & 0.702 \\ \cline{3-12} 
&                                & \multirow{2}{*}{\begin{tabular}[c]{@{}c@{}}No\\      $Y$ error\end{tabular}}       & Coverage                    & \multicolumn{1}{c|}{0.755} & 0.606 & \multicolumn{1}{c|}{0.987} & 0.965 & \multicolumn{1}{c|}{0.955} & 0.952 & \multicolumn{1}{c|}{0.966} & 0.955 \\ \cline{4-12} 
&                                &                                                                                    & Length                      & \multicolumn{1}{c|}{0.516} & 0.335 & \multicolumn{1}{c|}{0.993} & 0.694 & \multicolumn{1}{c|}{1.555} & 1.424 & \multicolumn{1}{c|}{0.703} & 0.621 \\ \cline{2-12} 
& \multirow{6}{*}{Non-simplex}   & \multirow{2}{*}{\begin{tabular}[c]{@{}c@{}}Independent\\      errors\end{tabular}} & Coverage                    & \multicolumn{1}{c|}{0.921} & 0.919 & \multicolumn{1}{c|}{0.882} & 0.825 & \multicolumn{1}{c|}{0.958} & 0.950 & \multicolumn{1}{c|}{0.962} & 0.951 \\ \cline{4-12} 
&                                &                                                                                    & Length                      & \multicolumn{1}{c|}{3.119} & 2.769 & \multicolumn{1}{c|}{3.118} & 2.686 & \multicolumn{1}{c|}{3.945} & 3.607 & \multicolumn{1}{c|}{2.947} & 2.647 \\ \cline{3-12} 
&                                & \multirow{2}{*}{\begin{tabular}[c]{@{}c@{}}Correlated\\      errors\end{tabular}}  & Coverage                    & \multicolumn{1}{c|}{0.899} & 0.902 & \multicolumn{1}{c|}{0.874} & 0.817 & \multicolumn{1}{c|}{0.960} & 0.949 & \multicolumn{1}{c|}{0.966} & 0.952 \\ \cline{4-12} 
&                                &                                                                                    & Length                      & \multicolumn{1}{c|}{1.467} & 1.275 & \multicolumn{1}{c|}{1.584} & 1.352 & \multicolumn{1}{c|}{3.237} & 2.978 & \multicolumn{1}{c|}{2.245} & 1.814 \\ \cline{3-12} 
&                                & \multirow{2}{*}{\begin{tabular}[c]{@{}c@{}}No\\      $Y$ error\end{tabular}}       & Coverage                    & \multicolumn{1}{c|}{0.916} & 0.907 & \multicolumn{1}{c|}{0.841} & 0.770 & \multicolumn{1}{c|}{0.959} & 0.946 & \multicolumn{1}{c|}{0.967} & 0.949 \\ \cline{4-12} 
&                                &                                                                                    & Length                      & \multicolumn{1}{c|}{2.276} & 1.994 & \multicolumn{1}{c|}{2.430} & 2.121 & \multicolumn{1}{c|}{3.284} & 3.032 & \multicolumn{1}{c|}{2.010} & 1.747 \\ \hline
\end{tabular}

\caption{
Empirical Coverage Rates of 95\% Pointwise Prediction Intervals.}
\label{tab:supp:Table3}

\end{table}

\newpage

\begin{figure}[!htb]
\centering
\includegraphics[width=1\textwidth]{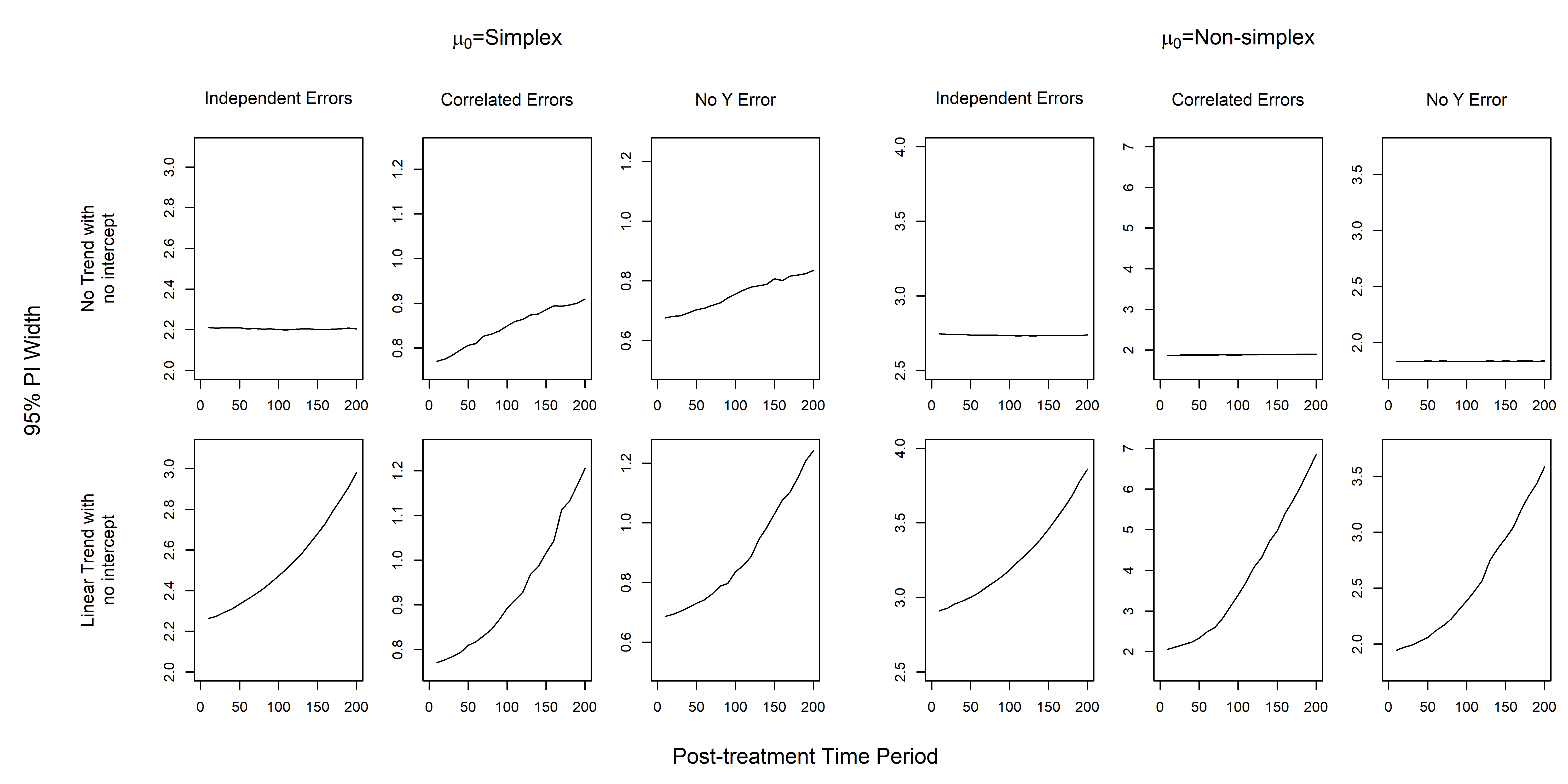}
\caption{95\% prediction interval widths over $T_1=400$ post-treatment time periods.} 

\label{fig:supp:PI width wider}

\end{figure}

Lastly, we study the width of 95\% prediction intervals obtained from the SPSC-DT estimator. We use the same simulation setup as before, with the modification that $T_0=100$ and $T_1=400$, focusing only on cases where $\blambda_t$ does not have an intercept. We calculate the 95\% prediction intervals at $t \in \{120, 140, \ldots, 480, 500\}$ and compute the average prediction interval width over 500 simulation repetitions. As shown in Figure \ref{fig:supp:PI width wider}, the prediction intervals may or may not widen as the post-treatment period progresses. Specifically, the width of the SPSC prediction intervals remains stable or shows only a slight increase when there is no systematic drift in the outcomes, but it tends to widen when such drift is present. Note that such systematic drift can be empirically verified by examining the trend of the estimated synthetic control, $\bW_t\T \widehat{\bgamma}_{\rho}$. 

\subsection{Simulation Studies under the Scenario Given in \citet{Cattaneo2021}} \label{sec:supp:Simulation PI}

For a fair comparison, we adopt a simulation scenario setup in \citet{Cattaneo2021}. In particular, we consider the following data generating process. First, we consider the length of the pre- and post-treatment periods as $T_0 = 100$ and $T_1=1$. Second, we choose the number of donors as $N = 10$, which are generated from the following AR(1) model:
\begin{align*}
&
W_{it} = \kappa W_{i, t-1} + \eta_{it} 
\ , \
t \in \{1,\ldots,T_0\}
\ , \ 
i \in \{1,\ldots,d\} \ .
\end{align*}
Here, the autocorrelation coefficient $\kappa$ is chosen from $\kappa \in \{0,0.5,1\}$, $\eta$ are generated from the standard normal distribution and are independent and identically distributed, and the baseline value $W_{i0}$ is set to zero. For the post-treatment period, we consider the following model for donors:
\begin{align*}
& W_{1 , T_0+1} = \kappa W_{1 T_0} + \eta_{1 , T_0+1} + \zeta \texttt{sd} ( W_{11},\ldots,W_{1 T_0} ) 
\\
&
W_{i , T_0+1} = \kappa W_{i T_0} + \eta_{i , T_0+1} \ , \quad i \in \{2,\ldots,N\} \ ,
\end{align*}
where $\zeta \in \{ -1,-0.5,0,0.5,1 \}$ parameterizes the degree of the shift in the first donor's post-treatment outcome. The treatment-free potential outcome of the treated unit is given by
\begin{align*}
\potY{t}{0} = 0.3 W_{1t} + 0.4 W_{2t} + 0.3 W_{3t} + 0.5 e_{t} \ , \ t \in \{1,\ldots, T_0+T_1\} \ ,
\end{align*}
where $e_t$ are independently generated from a standard normal distribution. We consider $\potY{t}{0} = \potY{t}{1}$, i.e., no treatment effect. We remark that $\EXP \big\{ \potY{t}{0} \cond \bW_t \big\} = 0.3 W_{1t} + 0.4 W_{2t} + 0.3 W_{3t}$ is not a valid synthetic control for the SPSC framework because $\EXP \big\{ \potY{t}{0} - \big( 0.3 W_{1t} + 0.4 W_{2t} + 0.3 W_{3t} \big) \cond \potY{t}{0} \big\} \neq 0 $, thereby violating Assumption \ref{assumption:SC}. Therefore, the proposed conformal inference approach for the SPSC framework in Section \ref{sec:Conformal} may fail in this data generating process. 

For our methods, the time-invariant and time-varying pre-treatment estimating equations are given by
\begin{align*}
&
\Phi_{\pre} (\bO_t \con \bgamma)
= 
\bh(Y_t) \big( Y_t - \bW_t \T \bgamma \big) 
\ ,
\\
&
\Psi_{\pre} (\bO_t \con \Beta,  \bgamma)
=
\begin{bmatrix}
\bD_t \big( Y_t - \bD_t\T \Beta \big)
\\
\bg(t, Y_t \con \Beta) \big( Y_t - \bW_t \T \bgamma \big)
\end{bmatrix}
=
\begin{bmatrix}
\bD_t \big( Y_t - \bD_t\T \Beta \big)
\\
\begin{bmatrix}
\bD_t \\ \bh(Y_t - \bD_t\T \Beta)
\end{bmatrix}
\big( Y_t - \bW_t \T \bgamma \big)
\end{bmatrix} \ ,
\end{align*}
where $\bh(y) = y$ and $\bD_t 	= 	\mathcal{B}_6 (t) \in \R^{6} $ is the 6-dimensional cubic B-spline bases function. 

We repeat the simulation 500 times and calculate the empirical coverage rates of 95\% confidence intervals from these repetitions for each simulation scenario. The results are presented in Table \ref{tab:supp:Table10}. First, we find that the SCPI approach achieves the nominal coverage rate across all simulation scenarios in general. However, we find that the conformal inference approach for SPSC without time-varying components (i.e., SPSC-NoDT) fails to achieve the nominal coverage rate, especially when the autocorrelation coefficient is large (i.e., $\kappa=1$). We conjecture that the undercoverage observed in these cases may be attributed to the nonstationarity of $W_{it}$ and $\potY{t}{0}$. Nevertheless, even in these challenging cases, the conformal inference approach for SPSC with time-varying components (i.e., SPSC-DT) shows significant improvement, attaining the nominal coverage rate across all considered simulation scenarios. This result further confirms that accounting for time-varying components is both useful and necessary for improving the performance of the proposed conformal inference approach in the presence of nonstationarity. Second, regarding the length of the prediction intervals, the SPSC-DT estimator produces the shortest intervals when $\kappa=0$ and $\kappa=0.5$. For $\kappa=1$, the SCPI estimator yields the shortest intervals, although the SPSC-DT estimator remains highly competitive.

\begin{table}[!htp]
\renewcommand{\arraystretch}{1.3} \centering
\scriptsize
\setlength{\tabcolsep}{3pt} 
\begin{tabular}{|c|c|ccccccccccccccc|}
\hline
\multirow{3}{*}{Statistics} & \multirow{3}{*}{Estimators} & \multicolumn{15}{c|}{$\kappa$ (second row) and $\zeta$ (third row)}                                                                                                                                                                                                                                                                                                                                                                                                             \\ \cline{3-17} 
&                             & \multicolumn{5}{c|}{0}                                                                                                                           & \multicolumn{5}{c|}{0.5}                                                                                                                       & \multicolumn{5}{c|}{1}                                                                                                         \\ \cline{3-17} 
&                             & \multicolumn{1}{c|}{-1}    & \multicolumn{1}{c|}{-0.5}  & \multicolumn{1}{c|}{0}     & \multicolumn{1}{c|}{0.5}    & \multicolumn{1}{c|}{1}      & \multicolumn{1}{c|}{-1}    & \multicolumn{1}{c|}{-0.5}  & \multicolumn{1}{c|}{0}     & \multicolumn{1}{c|}{0.5}   & \multicolumn{1}{c|}{1}     & \multicolumn{1}{c|}{-1}     & \multicolumn{1}{c|}{-0.5}   & \multicolumn{1}{c|}{0}      & \multicolumn{1}{c|}{0.5}    & 1      \\ \hline
\multirow{3}{*}{Coverage}   & SPSC-NoDT                   & \multicolumn{1}{c|}{0.958} & \multicolumn{1}{c|}{0.956} & \multicolumn{1}{c|}{0.964} & \multicolumn{1}{c|}{0.962}  & \multicolumn{1}{c|}{0.979}  & \multicolumn{1}{c|}{0.965} & \multicolumn{1}{c|}{0.957} & \multicolumn{1}{c|}{0.964} & \multicolumn{1}{c|}{0.967} & \multicolumn{1}{c|}{0.951} & \multicolumn{1}{c|}{0.860}  & \multicolumn{1}{c|}{0.860}  & \multicolumn{1}{c|}{0.851}  & \multicolumn{1}{c|}{0.847}  & 0.874  \\ \cline{2-17} 
& SPSC-DT                     & \multicolumn{1}{c|}{0.952} & \multicolumn{1}{c|}{0.954} & \multicolumn{1}{c|}{0.964} & \multicolumn{1}{c|}{0.962}  & \multicolumn{1}{c|}{0.974}  & \multicolumn{1}{c|}{0.961} & \multicolumn{1}{c|}{0.951} & \multicolumn{1}{c|}{0.952} & \multicolumn{1}{c|}{0.959} & \multicolumn{1}{c|}{0.943} & \multicolumn{1}{c|}{0.953}  & \multicolumn{1}{c|}{0.941}  & \multicolumn{1}{c|}{0.937}  & \multicolumn{1}{c|}{0.942}  & 0.947  \\ \cline{2-17} 
& SCPI                        & \multicolumn{1}{c|}{0.983} & \multicolumn{1}{c|}{0.971} & \multicolumn{1}{c|}{0.981} & \multicolumn{1}{c|}{0.978}  & \multicolumn{1}{c|}{0.986}  & \multicolumn{1}{c|}{0.980} & \multicolumn{1}{c|}{0.971} & \multicolumn{1}{c|}{0.979} & \multicolumn{1}{c|}{0.988} & \multicolumn{1}{c|}{0.976} & \multicolumn{1}{c|}{0.988}  & \multicolumn{1}{c|}{0.992}  & \multicolumn{1}{c|}{0.986}  & \multicolumn{1}{c|}{0.983}  & 0.991  \\ \hline
\multirow{3}{*}{Length}     & SPSC-NoDT                   & \multicolumn{1}{c|}{2.303} & \multicolumn{1}{c|}{2.317} & \multicolumn{1}{c|}{2.313} & \multicolumn{1}{c|}{2.321}  & \multicolumn{1}{c|}{2.325}  & \multicolumn{1}{c|}{2.445} & \multicolumn{1}{c|}{2.469} & \multicolumn{1}{c|}{2.462} & \multicolumn{1}{c|}{2.464} & \multicolumn{1}{c|}{2.449} & \multicolumn{1}{c|}{8.169}  & \multicolumn{1}{c|}{8.596}  & \multicolumn{1}{c|}{8.531}  & \multicolumn{1}{c|}{8.970}  & 8.346  \\ \cline{2-17} 
& SPSC-DT                     & \multicolumn{1}{c|}{2.283} & \multicolumn{1}{c|}{2.294} & \multicolumn{1}{c|}{2.292} & \multicolumn{1}{c|}{2.294}  & \multicolumn{1}{c|}{2.298}  & \multicolumn{1}{c|}{2.363} & \multicolumn{1}{c|}{2.381} & \multicolumn{1}{c|}{2.370} & \multicolumn{1}{c|}{2.370} & \multicolumn{1}{c|}{2.362} & \multicolumn{1}{c|}{3.547}  & \multicolumn{1}{c|}{3.576}  & \multicolumn{1}{c|}{3.553}  & \multicolumn{1}{c|}{3.508}  & 3.485  \\ \cline{2-17} 
& SCPI                        & \multicolumn{1}{c|}{2.546} & \multicolumn{1}{c|}{2.543} & \multicolumn{1}{c|}{2.582} & \multicolumn{1}{c|}{2.573}  & \multicolumn{1}{c|}{2.569}  & \multicolumn{1}{c|}{2.580} & \multicolumn{1}{c|}{2.569} & \multicolumn{1}{c|}{2.554} & \multicolumn{1}{c|}{2.588} & \multicolumn{1}{c|}{2.575} & \multicolumn{1}{c|}{2.985}  & \multicolumn{1}{c|}{2.999}  & \multicolumn{1}{c|}{2.966}  & \multicolumn{1}{c|}{2.962}  & 2.973  \\ \hline
\multirow{3}{*}{Bias}       & SPSC-NoDT                   & \multicolumn{1}{c|}{0.013} & \multicolumn{1}{c|}{0.012} & \multicolumn{1}{c|}{0.026} & \multicolumn{1}{c|}{-0.007} & \multicolumn{1}{c|}{-0.004} & \multicolumn{1}{c|}{0.047} & \multicolumn{1}{c|}{0.030} & \multicolumn{1}{c|}{0.007} & \multicolumn{1}{c|}{0.017} & \multicolumn{1}{c|}{0.005} & \multicolumn{1}{c|}{-0.064} & \multicolumn{1}{c|}{-0.042} & \multicolumn{1}{c|}{-0.063} & \multicolumn{1}{c|}{0.038}  & 0.082  \\ \cline{2-17} 
& SPSC-DT                     & \multicolumn{1}{c|}{0.015} & \multicolumn{1}{c|}{0.010} & \multicolumn{1}{c|}{0.025} & \multicolumn{1}{c|}{-0.002} & \multicolumn{1}{c|}{-0.005} & \multicolumn{1}{c|}{0.041} & \multicolumn{1}{c|}{0.028} & \multicolumn{1}{c|}{0.009} & \multicolumn{1}{c|}{0.011} & \multicolumn{1}{c|}{0.001} & \multicolumn{1}{c|}{0.014}  & \multicolumn{1}{c|}{-0.007} & \multicolumn{1}{c|}{0.004}  & \multicolumn{1}{c|}{-0.006} & -0.002 \\ \cline{2-17} 
& SCPI                        & \multicolumn{1}{c|}{0.009} & \multicolumn{1}{c|}{0.016} & \multicolumn{1}{c|}{0.026} & \multicolumn{1}{c|}{0.000}  & \multicolumn{1}{c|}{-0.005} & \multicolumn{1}{c|}{0.031} & \multicolumn{1}{c|}{0.028} & \multicolumn{1}{c|}{0.000} & \multicolumn{1}{c|}{0.001} & \multicolumn{1}{c|}{0.002} & \multicolumn{1}{c|}{0.009}  & \multicolumn{1}{c|}{-0.007} & \multicolumn{1}{c|}{0.019}  & \multicolumn{1}{c|}{-0.004} & -0.010 \\ \hline
\end{tabular}
\caption{Empirical Coverage Rates and Lengths of 95\% Pointwise Prediction Intervals.}
\label{tab:supp:Table10}
\end{table}

\subsection{Additional Results of the Data Analysis}		\label{sec:supp:Data}

\begin{figure}[!htb]
\centering
\includegraphics[width=1\textwidth]{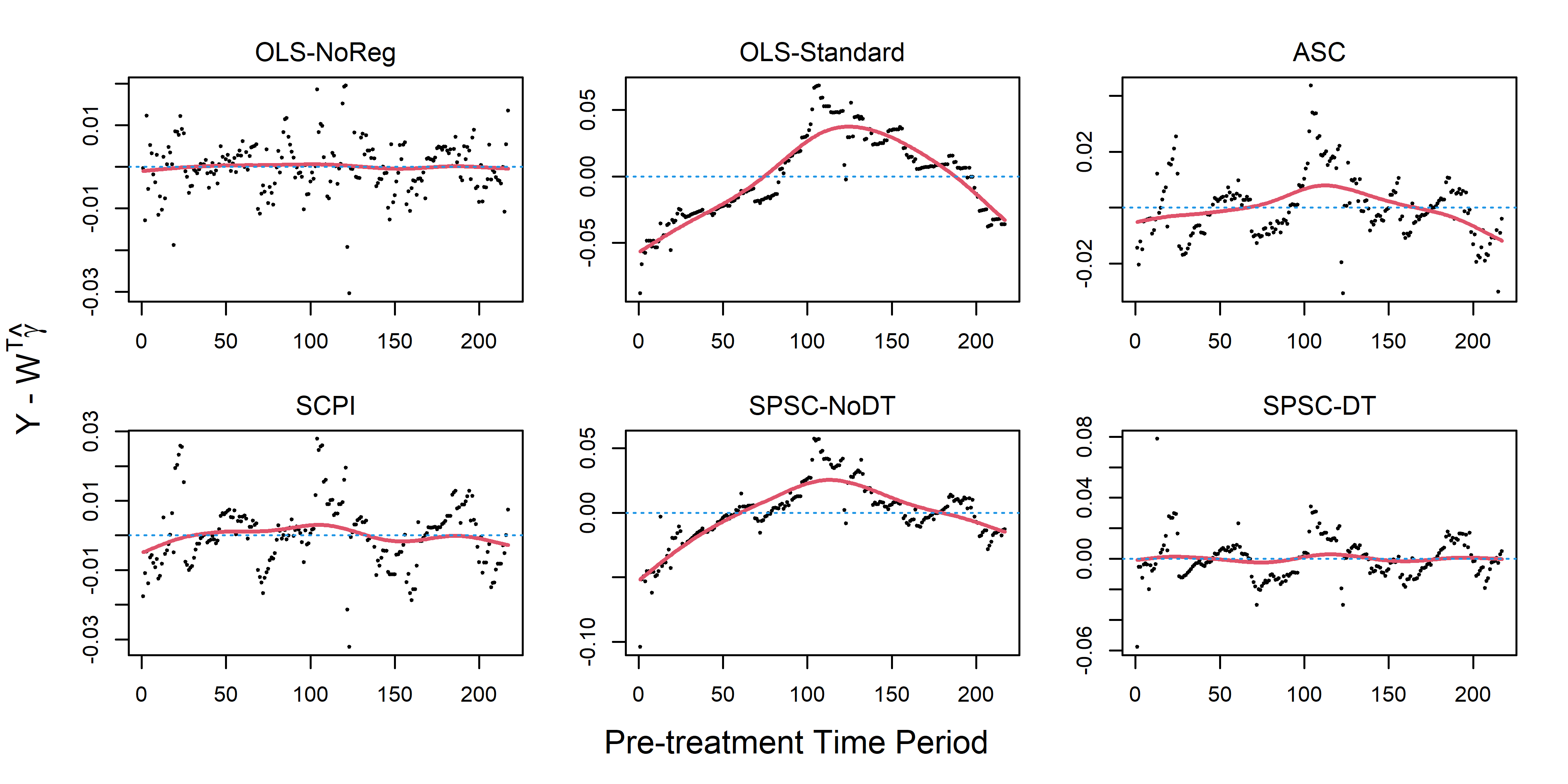}
\caption{Graphical Summaries of Residuals $Y_t - \bW_t\T \widehat{\bgamma}$ over the Pre-treatment Periods. The red solid lines depict the smoothing curve of the residuals. The blue dotted lines represent zero residual.}
\label{fig:supp:Diagnosis}
\end{figure}

In this Section, we provide additional results of the data analysis in Section \ref{sec:Data}. First,  Figure \ref{fig:supp:Diagnosis} presents graphical summaries of residuals $Y_t - \bW_t\T \widehat{\bgamma}$ over the pre-treatment periods. Note that the OLS-NoReg, SCPI, and SPSC-DT estimators produced residuals without a deterministic trend over time, while the other three estimators showed the opposite behavior. Notably, the SPSC-DT estimator appears to satisfy the zero mean condition of Assumption \ref{assumption:SC}, whereas the SPSC-NoDT estimator seems to violate this condition due to a non-zero deterministic trend over time. This again highlights the importance of accommodating time-varying components in the SPSC estimation procedure.

\begin{figure}[!htb]
\centering
\includegraphics[width=1\textwidth]{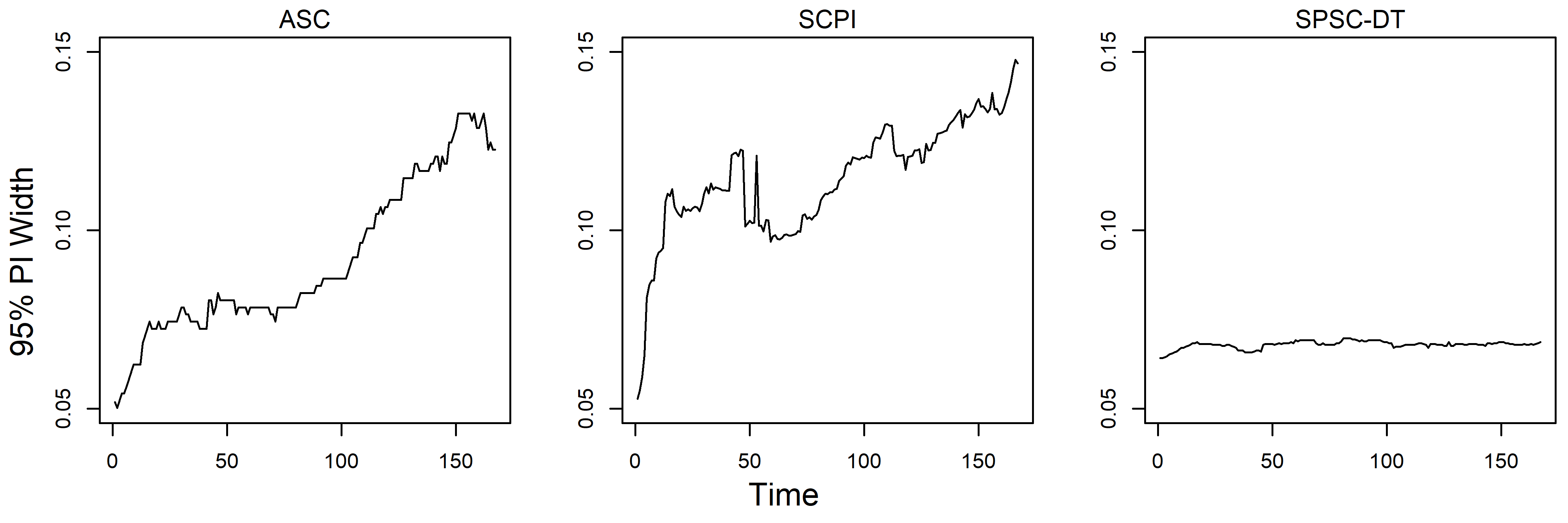}
\caption{95\% Prediction Interval Widths over $T_1=167$ Post-treatment Time Periods}
\label{fig:supp:PI width}

\end{figure}

Next, we provide the width of the 95\% prediction intervals obtained from each method in Figure \ref{fig:supp:PI width}. We remark that the SPSC-DT estimator exhibits relatively stable prediction interval widths compared to the other two methods. It is important to highlight that the stable prediction interval width of the SPSC-DT estimator is specific to this particular dataset. Depending on the underlying data-generating process, the prediction interval width may exhibit greater variability over time. For instance, in some cases illustrated in Figure \ref{fig:supp:PI width wider}, prediction intervals tens to widen  as the post-treatment period progresses. Furthermore, Figure \ref{fig:supp:PI width wider} suggests that the width of the SPSC prediction intervals remains stable or shows only a slight increase when there is no systematic drift in the outcomes, but it tends to widen when such drift is present. This systematic drift can be empirically verified by examining the trend of the estimated synthetic control, $\bW_t\T \widehat{\bgamma}_{\rho}$.  As shown in Figure \ref{fig:data:1} of the main paper, the synthetic control does not exhibit a noticeable upward or downward trend. We hypothesize that this lack of trend explains the stable prediction interval width observed in Figure  \ref{fig:supp:PI width}.

\begin{figure}[!htb]
\centering
\includegraphics[width=1\textwidth]{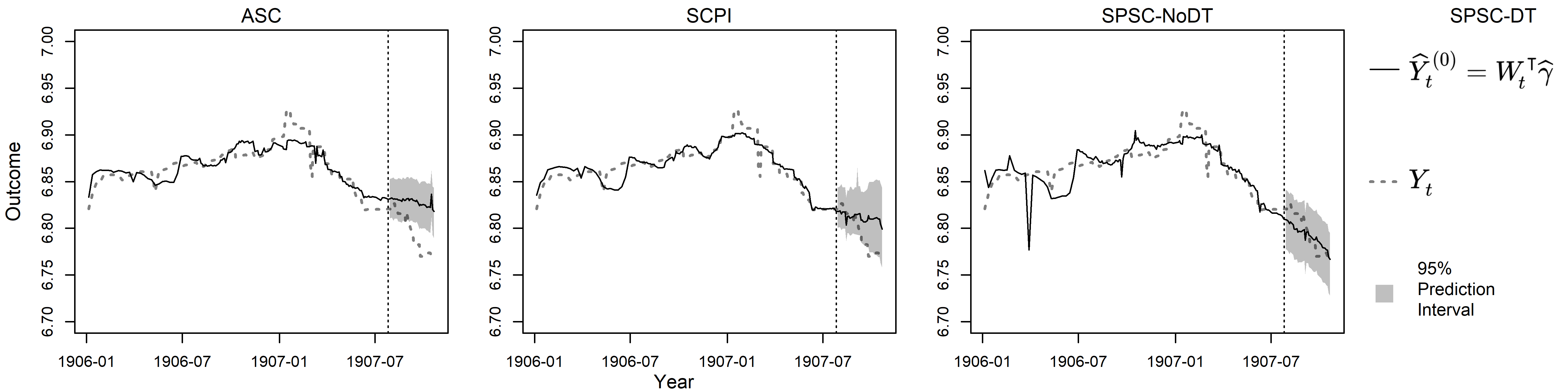}
\caption{Graphical Summaries of the 95\% Prediction Intervals over the Placebo Post-treatment Periods. 
These plots, from left to right, present the results using the approaches proposed by \citet{ASCM2021}, \citet{Cattaneo2021}, and the conformal inference approach presented in Section \ref{sec:Conformal} with time-invariant and time-varying estimating equations, respectively. 
}
\label{fig:supp:Conformal Placebo}
\end{figure}

Lastly, we provide the details of the placebo study. Figure \ref{fig:supp:Conformal Placebo} visually shows the synthetic controls under the placebo treatment. For the proposed SPSC approach, we find 95\% prediction intervals for $\potY{t}{0}$ include the true treatment-free potential outcome $\potY{t}{0}$ for all $T_1'=36$ placebo post-treatment periods. These results suggest that our SPSC approach seems reasonable for analyzing the effect of the 1907 panic on the stock price of the two trust companies. In contrast, the 95\% prediction intervals from the SCPI method cover the true treatment-free outcome for 29 placebo post-treatment periods, while the ASC estimator achieves coverage for only 17 periods.

Based on these additional analyses, we can further strengthen the causal conclusions established in the main paper especially those drawn from SPSC, i.e., the 1907 panic led to a decrease in the average log stock price of Knickerbocker and Trust Company of America. 

\newpage

\section{Nonparametric Single Proxy Synthetic Control Framework}		\label{sec:supp:nonparametric full}

\subsection{Overview} \label{sec:supp:nonparametric}

The SPSC framework can be generalized to the case in which the synthetic control is nonlinear and/or nonparametric, thus allowing the outcome to have arbitrary types such as binary, count, and continuous over a bounded interval. The estimation of inference of the synthetic control bridge function and the ATT is analogous to that established in the absence of covariates, so we suppress covariates for notational brevity. 

In Section \ref{sec:supp:Exist h}, sufficient conditions for the existence of the synthetic control bridge function $h^*$ is discussed. In Section \ref{sec:supp:Exist h}, we discuss sufficient conditions for the uniqueness of $h^*$ is discussed. Lastly, in Section \ref{sec:supp:nonparametric estimation}, we provide details about inference of the ATT without the uniqueness assumption.

\subsection{Sufficient Conditions for the Existence of the Synthetic Control Bridge Function} \label{sec:supp:Exist h}

In this Section, we provide sufficient conditions for the existence of the synthetic control bridge function $h^*$ satisfying Assumption \ref{assumption:valid proxy Cov}. We restate the assumption for readability after suppressing covariates:
\renewcommand{\theassumptionNew}{\ref{assumption:valid proxy Cov}}
\begin{assumptionNew}[Existence of Bridge Function in the Presence of Covariates] 
For all $t \in \{1,\ldots,T\}$, there exists a function $h^*: \R^{d} \rightarrow \R$ that satisfies 
\begin{align*}
\potY{t}{0} = \EXP \big\{
h^*(\bW_{ t} )
\cond
\potY{t}{0}
\big\} \ \ \text{almost surely} \ . 
\end{align*}
\end{assumptionNew}	

%Also, we focus on the case where $\potY{t}{0}$ and $\bW_t$ are stationary. In case $\potY{t}{0}$ and $\bW_t$ exhibit non-stationary behavior, say they have deterministic trends $\zeta_{Y,t}$ and $\bm{\zeta}_{W,t}$, respectively), then we may re-define $\widetilde{Y}_t^{(0)} := \potY{t}{0} - \zeta_{Y,t}$ and $\widetilde{\bW}_t\T := \bW_t - \bm{\zeta}_{W,t}$, and 

%Then, we find
%\begin{align*}
%&
%\exists \widetilde{h}^* \text{ s.t. }
%\widetilde{Y}_t^{(0)}   = \EXP \big\{
%\widetilde{h}^*( \widetilde{\bW}_t\T  )
%\cond
%\widetilde{Y}_t^{(0)} 
%\big\} 
%\\
%\Leftrightarrow
%\quad
%&
%\exists \widetilde{h}^* \text{ s.t. }
%\widetilde{Y}_t^{(0)} 
%+ \zeta_{Y,t}  = \EXP \big\{
%h^*( \widetilde{\bW}_t\T ) + \zeta_{Y,t}
%\cond
%\widetilde{Y}_t^{(0)} 
%+ \zeta_{Y,t}
%\big\} 
%\\
%\Leftrightarrow
%\quad
%&
%\exists h^* \text{ s.t. }
%Y_t^{(0)} 
%+ \zeta_{Y,t}  = \EXP \big\{
%h^*( {\bW}_t\T )
%\cond
%{Y}_t^{(0)}  
%\big\} 
%\end{align*}
%Where we defined $h^*(\bw):= \widetilde{h}^*(\bw-\bm{\zeta}_{W,t}) + \bm{\zeta}_{Y,t}$

In brief, we follow the approach in \citet{Miao2018}. The proof relies on Theorem 15.18 of \citet{Kress2014}, which is stated below for completeness.\\[0.25cm]
\noindent
\textbf{Theorem 15.18.} \citep{Kress2014}
Let $A:X \rightarrow Y$ be a compact operator with singular system $\big\{ \mu_n,\phi_n,g_n \big\}_{n \in \{1,2,\ldots\}}$. The integral equation of the first kind $A\phi = f$ is solvable if and only if 
\begin{align*}
& 1. \quad
\text{$f \in \mathcal{N}(A^{\text{adjoint}})^\perp = \big\{ f \, \big| \, A^{\text{adjoint}}(f) = 0 \big\}^{\perp}$}
\ , 
&&
2. \quad
\text{$\sum_{n=1}^{\infty} \mu_n^{-2} \big| \langle f,g_n \rangle |^2 < \infty$}
\end{align*}

To apply the Theorem, we introduce some additional notations. Let $\mathcal{L}_{W}$ and $\mathcal{L}_{ \potY{}{0} }$ be the spaces of square-integrable functions of $\bW_{ t}$ and $\potY{t}{0}$, respectively, which are equipped with the inner products $\langle h_1, h_2 \rangle_{W} = \int h_1(\bw) h_2(\bw) \, f_W (\bw) \, d\bw = \EXP \big\{ h_1(\bW_{ t}) h_2(\bW_{ t}) \big\} $ and $\langle g_1, g_2 \rangle_{\potY{}{0}} = \int g_1(y) g_2(y) \, f_{\potY{}{0}} (y) \, dy = \EXP \big\{ g_1(\potY{t}{0}) g_2(\potY{t}{0}) \big\} $, respectively. Let $\mathcal{K}: \mathcal{L}_{W} \rightarrow \mathcal{L}_{ \potY{}{0} }$ be the conditional expectation of $h(\bW_{ t} ) \in \mathcal{L}_{W}$ given $\potY{t}{0}$, i.e.,
\begin{align*}
\mathcal{K} (h) \in \mathcal{L}_{\potY{}{0}} 
\text{ satisfying }
\big( \mathcal{K}(h) \big) (y) 
=
\EXP \big\{ h(\bW_{ t}) \cond \potY{t}{0}=y \big\}
\text{ for } h \in \mathcal{L}_{W}
\end{align*}
Then, the synthetic control bridge function $h^* \in \mathcal{L}_{W}$ solves $\mathcal{K} ( h^* ) = [\text{identity map}] \in \mathcal{L}_{\potY{}{t}} $, i.e., 
\begin{align*}
\int h^*(\bw) f_{W | \potY{}{0} } (\bw \cond y ) \, d\bw = y , \ \forall y
\end{align*}

Now, we assume the following conditions:
\begin{itemize}[leftmargin=1cm, itemsep=0cm]
\item[] \HT{NPSC-1} The variables $(\potY{t}{0} , \bW_{ t})$ are stationary; 
\item[] \HT{NPSC-2} $\iint
f_{W | \potY{}{0} } (\bw \cond y )
f_{\potY{}{0} | W } (y \cond \bw )
\, d\bw \, d y
< \infty$;
\item[] \HT{NPSC-3} For $g \in \mathcal{L}_{\potY{}{0}}$, $\EXP \big\{ g(\potY{t}{0}) \cond \bW_{ t} \big\} = 0$ implies $g(\potY{t}{0})= 0$ almost surely;
\item[] \HT{NPSC-4} $\EXP \big[ \big\{ \potY{t}{0} \big\}^2 \big] < \infty$;
\item[] \HT{NPSC-5} Let the singular system of $\mathcal{K}$ be $\big\{ \mu_n,\phi_n,g_n \big\}_{n \in \{1,2,\ldots\}}$. \\Then, we have $\sum_{n=1}^{\infty} \mu_n^{-2} \big| \langle \potY{t}{0} ,g_n \rangle |^2 < \infty$.
\end{itemize}

We remark that the expectation can be defined without using $t$ under Condition \HL{NPSC-1}. First, we show that $\mathcal{K}$ is a compact operator under Condition \HL{NPSC-2}. Let $\mathcal{K}^{\text{adjoint}} : \mathcal{L}_{\potY{}{0}} \rightarrow \mathcal{L}_{W}$ be the conditional expectation of $g(\potY{t}{0}) \in \mathcal{L}_{\potY{}{0}}$ given $\bW_{ t}$, i.e., 
\begin{align*}
\mathcal{K}^{\text{adjoint}} (g) \in \mathcal{L}_{W} 
\text{ satisfying }
\big( \mathcal{K}(g) \big) (\bw) 
=
\EXP \big\{ g(\potY{t}{0}) \cond \bW_{ t}=\bw \big\}
\text{ for } g \in \mathcal{L}_{\potY{}{0}}
\end{align*}
Then, $\mathcal{K}$ and $\mathcal{K}^{\text{adjoint}}$ are the adjoint operator of each other as follows:
\begin{align*}
\langle \mathcal{K}(h) , g \rangle_{\potY{}{0}}
& =
\EXP
\big[
\EXP \big\{ h(\bW_{ t}) \cond \potY{t}{0} \big\}
g(\potY{t}{0})
\big]	
\\
& 
=
\EXP
\big[
h(\bW_{ t}) g(\potY{t}{0})
\big]	
\\
&
=
\EXP
\big[
h(\bW_{ t}) \EXP \big\{ g(\potY{t}{0}) \cond \bW_{ t} \big\}
\big]	
=
\langle h , \mathcal{K}^{\text{adjoint}}(g) \rangle_{W}
\end{align*}
Additionally, as shown in page 5659 of \citet{Carrasco2007}, $\mathcal{K}$ and $\mathcal{K}^{\text{adjoint}}$ are compact operators under Condition \HL{NPSC-2}. Moreover, by Theorem 15.16 of \citet{Kress2014}, there exists a singular value decomposition of $\mathcal{K}$ as $\big\{ \mu_n,\phi_n,g_n \big\}_{n \in \{1,2,\ldots\}}$. 

Second, we show that $\mathcal{N}(\mathcal{K}^{\text{adjoint}})^\perp = \mathcal{L}_{\potY{}{0}}$, which suffices to show $\mathcal{N}(\mathcal{K}^{\text{adjoint}}) = \big\{ 0 \big\} \subseteq \mathcal{L}_{\potY{}{0}}$. Under Condition \HL{NPSC-3}, we have 
\begin{align*}
g \in \mathcal{N}(\mathcal{K}^{\text{adjoint}})
\quad 
\Rightarrow 
\quad 
\EXP \big\{ g (\potY{t}{0}) \cond \bW_{ t} = \bw \big\}
=
0, \ \forall \bw
\quad \Rightarrow
\quad
g(\potY{t}{0}) = 0 
\end{align*}
where the first arrow is from the definition of the null space $\mathcal{N}$, and the second arrow is from Condition \HL{NPSC-3}. Therefore, any $g \in \mathcal{N}(\mathcal{K}^{\text{adjoint}})$ must satisfy $g(y) = 0 $ almost surely, i.e., $\mathcal{N}(\mathcal{K}^{\text{adjoint}})= \big\{ 0 \big\} \subseteq \mathcal{L}_{\potY{}{0}}$ almost surely. 

Third, from the definition of $\mathcal{L}_{W}$, $g(\potY{t}{0}) = \potY{t}{0} \in \mathcal{L}_{\potY{}{0}} = \mathcal{N}(\mathcal{K}^{\text{adjoint}})^\perp $ under Condition \HL{NPSC-4}.

Combining the three results, we establish that $\potY{t}{0}$ satisfies the first condition of Theorem 15.18 of \citet{Kress2014}. The second condition of the Theorem is exactly the same as Condition \HL{NPSC-5}. Therefore, we establish that the Fredholm integral equation of the first kind $\mathcal{K} ( h ) = [\text{identity map}] $ is solvable under Conditions \HL{NPSC-1}-\HL{NPSC-5}.

Note that Conditions \HL{NPSC-1} through \HL{NPSC-5} are sufficient but not necessary. In particular, it is possible to conceive of a scenario where a synthetic control bridge function exists even without the stationarity assumption \HL{NPSC-1}. However, this generally requires an additional assumption on the data generating process for $(\potY{t}{0},\bW_t)$ (e.g., IFEM) and the form of $h^*$ (e.g., linearity) to ensure stationary behavior in $\potY{t}{0}-h^*(\bW_t)$; see Section \ref{sec:supp:IFEM} for a specific example. As a result, case-specific models and assumptions are needed to account for non-stationary behavior in the outcomes. Since the purpose of this section is to demonstrate the possibility of relaxing the linearity of the synthetic control bridge function under stationarity, we do not further explore other cases without stationarity here. We intend to pursue this direction in future research.

\subsection{Uniqueness of Synthetic Control Bridge Function Under Completeness}		\label{sec:supp:NP Unique}

%In Section \ref{sec:Cov}, we showed that \eqref{eq-Fredholm} is satisfied Assumptions \ref{assumption:consistency}, \ref{assumption:noitf}, \ref{assumption:valid proxy Cov}, and \ref{assumption:valid proxy Cov}; for readability, we restate Assumption \ref{assumption:valid proxy Cov} and \eqref{eq-Fredholm} below:
%\renewcommand{\theassumptionNew}{\ref{assumption:valid proxy Cov}}
%\begin{assumptionNew}[Existence of Bridge Function in the Presence of Covariates] 
%For all $t \in \{1,\ldots,T\}$, there exists a function $h^*: \R^{d} \rightarrow \R$ that satisfies 
%\begin{align*}
%\potY{t}{0} = \EXP \big\{
%h^*(\bW_{ t}  )
%\cond
%\potY{t}{0} 
%\big\} \ \ \text{almost surely} \ . 
%\end{align*}
%\end{assumptionNew}	
%and
%\begin{align}
%\EXP \big\{ Y_t - h^*(\bW_{ t} ) \cond Y_t  \big\} = 0 \ .
%\tag{\ref{eq-Fredholm}}
%\end{align}
%
%In this section, we show the reverse is satisfied. 
%Suppose a function $h^*$ satisfies \eqref{eq-Fredholm}. Then, we obtain the following result for $t \in \{1,\ldots,T_0\}$:
%\begin{align*}
%y 
%& =
%\EXP \big\{ h^* ( \bW_{ t} ) \cond Y_t=y \big\} 
%=
%\EXP \big\{ h^* (\bW_{ t}) \cond \potY{t}{0} = y \big\} \ ,
%\end{align*} 
%where the second equality holds from Assumption \ref{assumption:consistency}. 
%Therefore, $h^*$ satisfies Assumption \ref{assumption:valid proxy Cov}. 

We provide a sufficient condition for the uniqueness of the bridge function. Consider the following completeness assumption:
\begin{assumption}[Completeness] \label{assumption-complete}
For $t \in \{1,\ldots,T\}$, suppose $\EXP \big\{ q(\bW_{ t}) \cond \potY{t}{0} \big\} = 0$ almost surely for a square integrable function $q$. Then, $q(\bW_{ t} ) = 0$ almost surely for $t \in \{1,\ldots,T\}$.
\end{assumption}
The assumption states that $\potY{t}{0}$ should be $\bW_{ t}$-relevant for all time periods in the sense that any variation in $\bW_{ t}$ is captured by variation in $\potY{t}{0}$. 

Let $h_1^*$ and $h_2^*$ be the synthetic control bridge functions satisfying Assumptions \ref{assumption:consistency}, \ref{assumption:noitf}, \ref{assumption:valid proxy Cov}, and \ref{assumption:valid proxy Cov}, and \ref{assumption-complete}. We then find $\EXP \big\{ h_1^*(\bW_{ t}) - h_2^*(\bW_{ t}) \cond \potY{t}{0} \big\} = 0$, implying $h_1^*(\bW_{ t})$ and $h_2^*(\bW_{ t}) = 0$ for all $t$, implying that a function $h^*$ satisfying \ref{assumption:valid proxy Cov} is unique. 

We remark that Assumption \ref{assumption-complete} may not be satisfied if the cardinality of the support of $\bW_{ t}$ is strictly larger than that of $\potY{t}{0}$. For instance, suppose that the outcomes are binary and two donors are available, i.e., $\bW_{ t} \in \{0,1\}^2$ and $\potY{t}{0} \in \{0,1\}$. Then, the equation in Assumption \ref{assumption-complete} reduces to
\begin{align} \label{eq-underdetermine}
\begin{bmatrix}
p_{W|Y}( 0,0 \cond 0)
&
p_{W|Y}( 0,1 \cond 0)
&
p_{W|Y}( 1,0 \cond 0)
&
p_{W|Y}( 1,1 \cond 0)
\\
p_{W|Y}( 0,0 \cond 1)
&
p_{W|Y}( 0,1 \cond 1)
&
p_{W|Y}( 1,0 \cond 1)
&
p_{W|Y}( 1,1 \cond 1)
\end{bmatrix}
\begin{bmatrix}
q(0,0) \\ q(0,1) \\ q(1,0) \\ q(1,1) 
\end{bmatrix}
=
\begin{bmatrix}
0 \\ 0
\end{bmatrix}
\end{align}
where $p_{W|Y}(a,b \cond y) = \Pr\{ \bW_{ t}=(a,b) \cond \potY{t}{0} = y \}$. Since \eqref{eq-underdetermine} is an underdetermined system, there are multiple non-zero $q$ functions satisfying \eqref{eq-underdetermine}, indicating that Assumption \ref{assumption-complete} cannot be satisfied. 

In the following section, we introduce a nonparametric SPSC framework that accommodates non-unique synthetic control bridge functions.

\subsection{Single Proxy Synthetic Control Approach without the Uniqueness Assumption}		\label{sec:supp:nonparametric estimation}

The synthetic control bridge function $h$ is defined as a function satisfying \eqref{eq-Fredholm}; we restate the equation below for readability.
\begin{align}		\label{eq-Fredholm2}
\EXP \big\{ Y_t - h^*(\bW_{ t}) \cond Y_t \big\} = 0 \text{ almost surely} \ , \quad t \in \{1,\ldots,T_0\} \ .
\tag{\ref{eq-Fredholm}}
\end{align}
We consider the case where there are multiple synthetic control bridge functions $h$ satisfying \eqref{eq-Fredholm2}. Even so, identification of the ATT established in Theorem \ref{thm:Extension NP Cov} is satisfied regardless of the choice of the bridge function. However, estimation and inference of the ATT can be complicated in the presence of multiple synthetic control bridge functions. To resolve this issue, we use approaches proposed by a series of recent works \citep{Li2023, Zhang2023}. In brief, their approaches involve the following three stages. In the first stage, we estimate a set of synthetic control bridge functions based on a sieve estimator; see Stage 1 below. In the second stage, we define a criterion function, denoted by $M$, and focus on the estimation of the minimizer of $M$, denoted by $h_0$. Then, an estimator of the ATT can be constructed based on the estimator of $h_0$; see Stage 2 below. In the third stage, we consider a de-biasing procedure for the estimator obtained in the previous stage to attain the asymptotic normality; see Stage 3 below. The following sections present details under general nonparametric settings, but the method can be applied to the parametric synthetic controls, including cases where there are multiple synthetic control weights that satisfy Assumption \ref{assumption:SC}. We have included only the essential assumptions and notations in this work to ensure clarity. We refer the readers to \citet{Li2023} and \citet{Zhang2023} for additional details.\\[0.25cm]

\noindent \textbf{Stage 1}: Estimation of the Solution Set $\mathcal{H}_0$ \\

Let $\mathcal{H}$ be a collection of user-specified smooth functions, and let $\mathcal{H}_0$ be the collection of the solutions of \eqref{eq-Fredholm2}, i.e.,
\begin{align*}
\mathcal{H}_0
=
\Big\{ h \in \mathcal{H} \, \Big| \,
Y_t = \EXP \big\{ h(\bW_{ t}) \cond Y_t \big\}, \ t \in \{1,\ldots,T_0\}
\Big\}
\end{align*}
Alternatively, we can represent $\mathcal{H}_0$ using a criterion function. Let $\mathfrak{C}: \mathcal{H} \rightarrow \R$ be a criterion function having the following form:
\begin{align*}
\mathfrak{C} (h)
=
\EXP
\Big[
\big[
Y_t
-
\EXP \big\{ h(\bW_{ t}) \cond Y_t \big \}
\big]^2
\Big] 
\ , \ t \in \{1,\ldots,T_0\}
\ .
\end{align*}
It is straightforward to check that $\mathcal{H}_0 = \big\{ h \in \mathcal{H} \cond \mathfrak{C}(h) = 0 \big\}$.

We consider a sieve approach as follows. First, we choose a sequence of approximating bases functions of $\bW_{ t}$, denoted by $\big\{ \varphi_k(\bw) \big\}_{k \in \{1,2,\ldots\}}$. For this sequence, we define an approximating function space for $\mathcal{H}$ by using the first $k_T$ bases functions, i.e.,
\begin{align*}
\mathcal{H}_T
=
\bigg\{
h \in \mathcal{H}
\, \bigg| \,
h(\bw)
=
\sum_{\ell=1}^{k_T}
b_{\ell} \varphi_{\ell}(\bw) 
\bigg\} \ ,
\end{align*}
where $k_T$ is a known parameter and $b_1,\ldots,b_{k_T}$ are unknown scalar parameters.

A sample analogue of the criterion function $\mathfrak{C}$, denoted by $\mathfrak{C}_T$, can be obtained based on the sieve approach. We choose a sequence of approximating bases functions of $\potY{t}{0}$, denoted by $ \big\{ \phi_k (y) \big\}_{k \in \{1,2,\ldots\}}$. Then, we choose the first $k_T$ bases function and construct a $k_T$-dimensional function of $y$, denoted by $\bphi (y) = \big\{ \phi_1(y),\ldots,\phi_{k_T}(y) \big\}\T$. Using the pre-treatment observations, we construct a $(T_0 \times k_T)$ matrix as follows:
\begin{align*}
\Phi_{\pre}
=
\begin{bmatrix}
\bphi \T (Y_1)
\\
\vdots
\\
\bphi \T (Y_{T_0})
\end{bmatrix} 
\in \R^{T_0 \times k_T}
\ .
\end{align*}
For a given function $h$, a sieve estimator of the conditional expectation $\EXP \big\{ h(\bW_{ t}) \cond Y_t \big\}$ for $t \in \{1,\ldots,T_0\}$ can be obtained by regressing $h(\bW_{ t})$ on $ \bphi(Y_t) $, i.e., 
\begin{align*}
\widehat{\mu}_{\pre} (y \con h)
=
\texttt{sieve}
\Big(
\EXP \big\{ h(\bW_{ t}) \cond Y_t = y \big\}
\Big)
=
\bphi \T (y)
\big( \Phi_{\pre} \T \Phi_{\pre} \big)^{-1}
\bigg\{
\sum_{t=1}^{T_0}
h(\bW_{ t}) \bphi(Y_t)
\bigg\} \ .
\end{align*}
Therefore, $\mathfrak{C}_T$ can be obtained based on a sieve estimator, i.e.,
\begin{align*}
\mathfrak{C}_T (h)
=
\frac{1}{T_0} \sum_{t=1}^{T_0} 
\Big\{
Y_t
-
\widehat{\mu}_{\pre} (Y_t \con h)
\Big\}^2
\end{align*}
The proposed estimator of $\mathcal{H}_0$ is 
\begin{align*}
\widehat{\mathcal{H}}_0 = \Big\{ h \in \mathcal{H}_T \Cond \mathfrak{C}_T(h) \leq c_T \Big\} 
\end{align*}
where $c_T$ is an appropriately chosen sequence with $c_T \rightarrow 0$ as $T \rightarrow \infty$. Under regularity conditions, we have
\begin{align*}
d_{H} \big(\widehat{\mathcal{H}}_0, {\mathcal{H}}_0 , \big\| \cdot \big\|_{\infty} \big) = o_P(1)
\end{align*}
where $d_{H} (\mathcal{H}_1,\mathcal{H}_2, \big\| \cdot \big\| )$ is the Hausdorff distance between $\mathcal{H}_1$ and $\mathcal{H}_2$ with respect to a given norm $\big\| \cdot \big\|$; see Section 3.2 of \citet{Li2023} and Section 3.2 of \citet{Zhang2023} for details. \\

\noindent \textbf{Stage 2}: A Representer-based Estimator \\

After obtaining a consistent set estimator of $\mathcal{H}_0$ (i.e., $\widehat{\mathcal{H}}_0$), we select an estimator of $h$ from $\widehat{\mathcal{H}}_0$ so that it converges to a unique element in $\mathcal{H}_0$. Specifically, we define a function $M : \mathcal{H} \rightarrow \R$ that has a unique minimum $h_0$ on $\mathcal{H}_0$. Let $M_T$ be its sample analogue, and let $\widehat{h}_0$ be the minimum of $M_T(h)$ over $\widehat{\mathcal{H}}_0$, i.e.,
\begin{align*}
\widehat{h}_0 \in \argmin_{h \in \widehat{\mathcal{H}}_0} M_T(h) \ . 
\end{align*}
To obtain a unique minimum $\widehat{h}_0$, $\mathcal{H}$ and $M$ are chosen to satisfy the following assumption:
\begin{assumption} The following conditions are satisfied:
\begin{itemize}[leftmargin=1cm, itemsep=0cm]
\item[1.] The set $\mathcal{H}$ is convex;
\item[2.] The functional $M: \mathcal{H} \rightarrow \R$ is strictly convex, and have a unique minimum at $h_0$ on $\mathcal{H}_0$;
\item[3.] The sample analogue $M_T : \mathcal{H} \rightarrow \R$ is continuous and $\sup_{h \in \mathcal{H}} \big| M_T(h) - M(h) \big| = o_P(1)$.
\end{itemize}
\end{assumption}
Possible choices for $M$ and its sample analogue $M_T$ are 
\begin{align*}
&
M(h)
=
\EXP \big[ \big\{ h(\bW_{ t} ) \big\}^2 \big]
\ , \ t \in \{1,\ldots,T_0\}
\ , 
&&
M_T (h)
=
\frac{1}{T_0}
\sum_{t=1}^{T_0}
\big\{ h(\bW_{ t} ) \big\}^2
\end{align*}
Under regularity conditions, we have $\big\| \widehat{h}_0 - h_0 \big\|_{\infty} = o_P(1)$; see Theorem 3 of \citet{Li2023} and Proposition 3.2 of \citet{Zhang2023} for details. In turn, we obtain an estimator of the ATT as $\widehat{\tau}_t = Y_t - \widehat{h}_0(\bW_{ t})$ for $t \in \{T_0+1,\ldots,T\}$ where inference based on $\widehat{\tau}_t$ can be established by the conformal inference in Section \ref{sec:Conformal} in the Supplementary Material. Alternatively, we may posit a parametric form for the ATT as $\tau_t = \tau(t \con \bbeta)$. Considering $\widehat{h}_0$ as a fixed function, an estimator of $\bbeta$ can be obtained as a solution to the following equation:
\begin{align}	
&
\widehat{\bbeta} \text{ solves }
\frac{1}{T_1}
\sum_{t=T_0+1}^{T}
{\Psi}_{\post} (\bO_t \con \bbeta, \widehat{h}_0)
= 0
\ , 
\label{eq-betahat}
\\
&
{\Psi}_{\post} (\bO_t \con \bbeta, h)
=
\frac{\partial \tau(t \con \bbeta) }{\partial \bbeta}
\Big\{ Y_t - \tau(t \con {\bbeta}) - h (\bW_{ t} ) \Big\}
\in \R^{\text{dim}(\bbeta)}
\ , \ 
t \in \{T_0+1,\ldots,T\}
\ .
\label{eq-supp-postEE}
\end{align}
To characterize the asymptotic property of $\widehat{\bbeta}$, we additionally define the following objects. Let $\langle h_1, h_2 \rangle_w$ be
\begin{align*}
\langle h_1, h_2 \rangle_w
=
\EXP \big[
\EXP \big\{ h_1(\bW_{ t}) \cond \potY{t}{0} \big\}
\EXP \big\{ h_2(\bW_{ t}) \cond \potY{t}{0} \big\} 
\big] \ , \
t \in \{T_0+1,\ldots,T\} \ ,
\end{align*}
and $\overline{\mathcal{H}}$ be the closure of the linear span of $\mathcal{H}$ under $\big\| \cdot \big\|_w$. Then, we assume the following conditions.
\begin{assumption} 	\label{assumption:supp:g}
The following conditions are satisfied:
\begin{itemize}[leftmargin=1cm, itemsep=0cm]
\item[1.] For any $h \in \overline{\mathcal{H}}$, there exists a function $g_{0,h} \in \mathcal{H}$ satisfying $\langle g_{0,h}, h \rangle_w = \EXP \big\{ h(\bW_{ t}) \big\}$ for $t \in \{T_0+1,\ldots,T\}$. 
\item[2.] There exists a projection of $\mathcal{H}$ on $\mathcal{H}_T$, denoted by $\Pi_{T}: \mathcal{H} \rightarrow \mathcal{H}_T$, which satisfies
\begin{align*}
\sup_{h \in \mathcal{H}} \big\| h - \Pi_T h \big\| = O(\eta_T) \ .
\end{align*}
where $\eta_T=o(1)$ satisfies regularity conditions; see Assumptions 7-10 of \citet{Li2023} and Assumptions 4-7 of \citet{Zhang2023} for details.
\end{itemize}

\end{assumption}
We now characterize the asymptotic representation of $T_1^{1/2}
\big( \widehat{\bbeta} - \bbeta^* \big)$ under regularity conditions including stationarity and independent errors. Applying a first-order Taylor expansion, we find
\begin{align*}
0 
&
= 
\frac{1}{T_1} \sum_{t=T_0+1}^{T} {\Psi}_{\post} (\bO_t \con \widehat{\bbeta}, \widehat{h}_0)
\\
&
=
\frac{1}{T_1} \sum_{t=T_0+1}^{T}
\Bigg\{ {\Psi}_{\post} (\bO_t \con \bbeta^*, \widehat{h}_0)
+
\frac{\partial {\Psi}_{\post} (\bO_t \con \bbeta, \widehat{h}_0) }{\partial \bbeta\T} \bigg|_{\bbeta=\bbeta^*}
\cdot \big( \widehat{\bbeta} - \bbeta^* \big)
\Bigg\}
+
o_P(1) \ .
\end{align*}
Therefore, we find that \eqref{eq-betahat} has the following asymptotic representation for $t \in \{T_0+1,\ldots,T\}$:
\begin{align}
&
\sqrt{T_1} 
\big( \widehat{\bbeta} - \bbeta^* \big)
\nonumber
\\
& 
= 
\bigg[ 
\underbrace{
\frac{1}{T_1} \sum_{t=T_0+1}^{T} 
\frac{\partial {\Psi}_{\post} (\bO_t \con \bbeta, \widehat{h}_0) }{\partial \bbeta\T} \bigg|_{\bbeta=\bbeta^*} 
}_{=: V (\bbeta^*, \widehat{h}_0) }
\bigg]^{-1}
\nonumber 
\\
& \hspace*{1cm} \times 
\bigg[ \frac{1}{\sqrt{ T_1} } \sum_{t=T_0+1}^{T} 
\frac{\partial \tau(t \con \bbeta^*)}{\partial \bbeta}
\Big\{
Y_t - \tau(t \con {\bbeta}^*) - \widehat{h}_0(\bW_{ t} )
\Big\}
\bigg]
+
o_P(1)
\nonumber
\\
& 
=
V^{-1} (\bbeta^*, \widehat{h}_0)
\bigg[ \frac{1}{\sqrt{ T_1} } \sum_{t=T_0+1}^{T} 
\frac{\partial \tau(t \con \bbeta^*)}{\partial \bbeta}
\Big\{
Y_t 
- \tau(t \con \bbeta^*)
- \widehat{h}_0(\bW_{ t} )
\Big\}
\bigg]
+
o_P(1)
\nonumber
\\
& 
=
V^{-1} (\bbeta^*, \widehat{h}_0)
\bigg[ \frac{1}{\sqrt{ T_1} } \sum_{t=T_0+1}^{T} 
\frac{\partial \tau(t \con \bbeta^*)}{\partial \bbeta}
\Big\{
Y_t 
- \tau(t \con \bbeta^*)
- h_0(\bW_{ t}) 
\Big\}
\bigg]
\label{eq-supp-Asymp1}
\\
&
\quad +
V^{-1} (\bbeta^*, \widehat{h}_0)
\bigg[ \frac{1}{\sqrt{ T_1} } \sum_{t=T_0+1}^{T} 
\frac{\partial \tau(t \con \bbeta^*)}{\partial \bbeta}
\Big[
\EXP \big\{ h_0(\bW_{ t}) - \widehat{h}_0(\bW_{ t} ) \big\} 
\Big]
\bigg]
\label{eq-supp-Asymp2}
\\
&
\quad +
V^{-1} (\bbeta^*, \widehat{h}_0)
\Bigg[ \frac{1}{\sqrt{ T_1} } \sum_{t=T_0+1}^{T} 
\frac{\partial \tau(t \con \bbeta^*)}{\partial \bbeta}
\Bigg[
\begin{array}{l}
\big\{ h_0(\bW_{ t})
- \widehat{h}_0(\bW_{ t} )
\big\}
\\
- \EXP \big\{ h_0(\bW_{ t}) - \widehat{h}_0(\bW_{ t} ) \big\}
\end{array} 
\Bigg]
\Bigg]
\label{eq-supp-Asymp3}
\\
&
\quad
+
o_P(1) 
\nonumber
\ .
\end{align}
Following Theorem 4 of \citet{Li2023} and Supplementary Material of \citet{Zhang2023}, we establish that \eqref{eq-supp-Asymp3} is $o_P(1)$. In addition, for $t \in \{T_0+1,\ldots,T\}$, the numerator of \eqref{eq-supp-Asymp2} is equal to
\begin{align}
& 
\frac{1}{\sqrt{ T_1} } \sum_{t=T_0+1}^{T} 
\frac{\partial \tau(t \con \bbeta^*)}{\partial \bbeta}
\EXP \big\{ h_0(\bW_{ t}) - \widehat{h}_0(\bW_{ t} ) \big\} 
\nonumber
\\
&=
-
\frac{1}{\sqrt{ T_1} } \sum_{t=T_0+1}^{T} 
\frac{\partial \tau(t \con \bbeta^*)}{\partial \bbeta}
\EXP \big\{ g_{0,h_0} (\bW_{ t}) \cond \potY{t}{0} \big\} 
\big\{ \potY{t}{0} - h_0(\bW_{ t} ) \big\}
\nonumber
\\
&
\quad
+
\frac{1}{\sqrt{ T_1} } \sum_{t=T_0+1}^{T} 
\frac{\partial \tau(t \con \bbeta^*)}{\partial \bbeta}
\widehat{\EXP} \big\{ \Pi_T g_{0,h_0} (\bW_{ t}) \cond \potY{t}{0} \big\} 
\big[ \potY{t}{0} - \widehat{\EXP} \big\{ \widehat{h}_0(\bW_{ t}) \cond \potY{t}{0} \big\} \big]
+
o_P(1)	
\label{eq-asymptotic rep}
\ .
\end{align}
Here, $g_{0,h}$ and its projection $\Pi_T g_{0,h}$ are chosen to satisfy Assumption \ref{assumption:supp:g}, and $\widehat{\EXP}$ is a generic estimator of the conditional expectation operator of the distribution $\bW_{ t} | \potY{t}{0}$ having a fast convergence rate; see Stage 3 below for details on how these estimators are constructed. Combining all results, we have the following result for $t \in \{T_0+1,\ldots,T\}$:
\begin{align*}
&
\sqrt{T_1} 
\big( \widehat{\bbeta} - \bbeta^* \big)
\\
&
=
V^{-1} (\bbeta^*, \widehat{h}_0)
\bigg[ \frac{1}{\sqrt{ T_1} } \sum_{t=T_0+1}^{T} 
\frac{\partial \tau(t \con \bbeta^*)}{\partial \bbeta}
\left[ 
\begin{array}{l}
Y_t 
- \tau(t \con \bbeta^*)
- h_0(\bW_{ t}) 
\\
-
\EXP \big\{ g_{0,h_0} (\bW_{ t}) \cond \potY{t}{0} \big\}
\big\{ \potY{t}{0} - h_0(\bW_{ t} ) \big\} 
\end{array}
\right]
\bigg]
\\
&
\quad
+
V^{-1} (\bbeta^*, \widehat{h}_0)
\sqrt{T_1}
r_T(\widehat{h}_0)
+
o_P(1)
\end{align*}
where
\begin{align*}
r_T(\widehat{h}_0)
= 
\frac{1}{ T_1 } \sum_{t=T_0+1}^{T} 
\frac{\partial \tau(t \con \bbeta^*)}{\partial \bbeta}
\widehat{\EXP} \big\{ \Pi_T g_{0,h_0} (\bW_{ t}) \cond \potY{t}{0} \big\} 
\big[ \potY{t}{0} - \widehat{\EXP} \big\{ \widehat{h}_0(\bW_{ t}) \cond \potY{t}{0} \big\} \big] \ .
\end{align*}

\noindent\textbf{Stage 3}: A De-biased Estimator\\

To obtain the asymptotic normality of $\widehat{\bbeta}$, we need to de-bias $\widehat{\bbeta}$ by subtracting an estimated value of $r_T(\widehat{h}_0)$. To do so, we define a new criterion function and its sample analogue for $h \in \mathcal{H}$ as follows:
\begin{align*}
&
\mathcal{R} (h)
=
\EXP \Big[ \big[ \EXP \big\{ h(\bW_{ t}) \cond \potY{t}{0} \big\} \big]^2 \Big]
-
2 \EXP \big\{ h(\bW_{ t}) \big\}
\ , \ 
t \in \{T_0+1,\ldots,T\} \ ,
\\
&
\mathcal{R}_T (h)
=
\frac{1}{T_1} \sum_{t=T_0+1}^{T}
\big[ \widehat{\EXP} \big\{ h(\bW_{ t}) \cond \potY{t}{0} \big\} \big]^2
-
\frac{2}{T_1} \sum_{t=T_0+1}^{T} h(\bW_{ t}) \ .
\end{align*}
We obtain an estimator of $\Pi_T g_{0,h_0}$, denoted by $\widehat{g}$, as 
\begin{align*}
\widehat{g} \in \argmin_{\widehat{h}_0 \in \mathcal{H} } \mathcal{R}_T(\widehat{h}_0)
\end{align*}
and the resulting estimator of $r_T(\widehat{h}_0)$ is
\begin{align*}
\widehat{r}_T^{\text{inf}}(\widehat{h}_0)
= 
\frac{1}{ T_1 } \sum_{t=T_0+1}^{T} 
\frac{\partial \tau(t \con \widehat{\bbeta})}{\partial \bbeta}
\widehat{\EXP} \big\{ \widehat{g} (\bW_{ t}) \cond \potY{t}{0} \big\} 
\big[ \potY{t}{0} - \widehat{\EXP} \big\{ \widehat{h}_0(\bW_{ t}) \cond \potY{t}{0} \big\} \big] \ , \ 
t \in \{T_0+1,\ldots,T\} \ .
\end{align*}
Unfortunately, the above estimator $	\widehat{r}_T^{\text{inf}}$ is infeasible because it involves with counterfactual outcomes. Therefore, we use $Y_t - \tau(t \con \widehat{\bbeta}) $ as realizations of the treatment-free potential outcomes $\potY{t}{0}$ and construct a $(T_1 \times k_T)$ matrix as follows:
\begin{align*}
{\Phi}_{\post}
=
\begin{bmatrix}
\bphi \T \Big( Y_{T_0+1} - \tau(T_0+1 \con \widehat{\bbeta}) \Big)
\\
\vdots
\\
\bphi \T \Big( Y_{T} - \tau(T \con \widehat{\bbeta}) \Big)
\end{bmatrix}
\in \R^{T_1 \times k_T}
\ .
\end{align*}
We consider additional sieve estimators of ${\EXP} \big\{ \widehat{g} (\bW_{ t}) \cond \potY{t}{0} \big\} $ and ${\EXP} \big\{ \widehat{h}_0(\bW_{ t}) \cond \potY{t}{0} \big\}$ for $t \in \{T_0+1,\ldots,T\}$:
\begin{align*}
\widehat{\mu}_{\post} (y \con \widehat{g})
&
=
{ \texttt{sieve} }
\Big(
\EXP \big\{ \widehat{g} (\bW_{ t}) \cond \potY{t}{0} = y \big\} 
\Big)
\\
&
=
\bphi \T (y)
\big( \Phi_{\post} \T \Phi_{\post} \big)^{-1}
\bigg\{
\sum_{t=T_0+1}^{T}
\widehat{g}(\bW_{ t} ) \bphi \Big( Y_t - \tau(t \con \widehat{\bbeta}) \Big)
\bigg\}
\\
\widehat{\mu}_{\post}(y \con \widehat{h}_0 )
&
=
{ \texttt{sieve} }
\Big(
\EXP \big\{ \widehat{h}_0(\bW_{ t}) \cond \potY{t}{0} = y \big\}
\Big)
\\
&
=
\bphi \T (y)
\big( \Phi_{\post} \T \Phi_{\post} \big)^{-1}
\bigg\{
\sum_{t=T_0+1}^{T}
\widehat{h}_0(\bW_{ t} ) \bphi \Big( Y_t - \tau(t \con \widehat{\bbeta}) \Big)
\bigg\} \ .
\end{align*}
Using these sieve estimators, we obtain a feasible estimator of $r_T(\widehat{h}_0)$ as
\begin{align*}
\widehat{r}_T (\widehat{h}_0)
= 
\frac{1}{ T_1 } \sum_{t=T_0+1}^{T} 
\frac{\partial \tau(t \con \widehat{\bbeta})}{\partial \bbeta}
\bigg[
\widehat{\mu}_{\post} \big( Y_t - \tau(t \con \widehat{\bbeta}) \con \widehat{g} \big)
\Big\{
Y_t - \tau(t \con \widehat{\bbeta})
-
\widehat{\mu}_{\post} \big( Y_t - \tau(t \con \widehat{\bbeta}) \con \widehat{h}_0 \big)
\Big\}
\bigg] \ .
\end{align*}
Under regularity conditions, we establish that
\begin{align*}
\sup_{\widehat{h}_0 \in \widehat{\mathcal{H}}_0}
\sqrt{T_1}
\Big|
\widehat{r}_T(\widehat{h}_0) - {r}_T(\widehat{h}_0)
\Big| = o_P(1) \ ;
\end{align*} 
see Lemma 1 of \citet{Li2023} and Lemma 3.3 of \citet{Zhang2023} for details. Based on this result, we subtract $T_1^{1/2} \widehat{r}_T(\widehat{h}_0)$ in both hand sides of \eqref{eq-asymptotic rep}. We then obtain a de-biased estimator $\widehat{\bbeta}_{\text{db}}$ as
\begin{align*}
\widehat{\bbeta}_{\text{db}}
=
\widehat{\bbeta} 
-
V^{-1} (\widehat{\bbeta}, \widehat{h}_0)
\sqrt{T_1} \widehat{r}_T(\widehat{h}_0) \ ,
\end{align*}
which is asymptotically normal in that $T_{1}^{1/2} \big( \widehat{\bbeta}_{\text{db}} - \bbeta^* \big)$ converges in distribution to $
N \big( 0, S_1^* S_2^* S_1\sT \big)$ as $T \rightarrow \infty$ where $S_1^*$ and $S_2^*$ are given as follows:
\begin{align*}
&
S_1^*
=
\bigg[
\frac{\partial \EXP \big\{ {\Psi}_{\post} (\bO_t \con \bbeta^*, h_0) \big\} }{\partial \bbeta} 
\bigg]^{-1} 
\\
&
S_2^*
=
\VAR
\left[
{\Psi}_{\post} (\bO_t \con \bbeta^*, h_0)
-
\frac{\partial \tau(t \con \bbeta^*)}{\partial \bbeta} 
\EXP \big\{ g_{0,h_0} (\bW_{ t}) \cond \potY{t}{0} \big\} \big\{ \potY{t}{0} - h_0(\bW_{ t} ) \big\}
\right] \ ,
\end{align*}
Here, ${\Psi}_{\post} (\bO_t \con \bbeta,h )$ is defined in \eqref{eq-supp-postEE} for $t \in \{T_0+1,\ldots,T\}$. 
The ATT estimator is obtained from the plug-in formula $\widehat{\tau}_t = \tau(t \con \widehat{\bbeta}_{\text{db}})$. Consequently, inference of the ATT can be attained based on the standard delta-method applied to the asymptotic normal distribution of $\widehat{\bbeta}_{\text{db}}$.

\newpage

\section{Proof of Theorems}	\label{sec:supp:proof}

\subsection{Proof of Theorems \ref{thm:SC}, \ref{thm:ATT}, \ref{thm:Extension NP Cov}}

We first prove the most general case with a nonlinear bridge function $h^*$ and under the presence of covariates (i.e., Theorem \ref{thm:Extension NP Cov}). For the pre-treatment periods $t \in \{1,\ldots,T_0\}$, we establish
\begin{align*}
y 
=
\EXP \big\{ h^* (\bW_{ t}, \bX_{0t}, \bX_{ t}) \cond \potY{t}{0} = y, \bX_{0t}, \bX_{ t}  \big\} 
=
\EXP \big\{ h^* ( \bW_{ t}, \bX_{0t}, \bX_{ t} ) \cond Y_t=y, \bX_{0t}, \bX_{ t} \big\} \ .
\end{align*} 
The first equality holds from Assumption \ref{assumption:valid proxy Cov}. The second equality holds from Assumption \ref{assumption:consistency}.  

Furthermore, for any $t \in \{1,\ldots,T\}$, we establish
\begin{align*} 
\EXP \big\{ \potY{t}{0} \cond \bX_{0 t} , \bX_{ t} \big\} 
& = 
\EXP \big[
\EXP \big\{ h^*(\bW_{ t}, \bX_{0 t} , \bX_{ t}) \cond \potY{t}{0} , \bX_{0 t} , \bX_{ t} \big\}		
\cond \bX_{0 t} , \bX_{ t}
\big]
\nonumber 
\\
&
=		
\EXP \big\{ h^* (\bW_{ t}, \bX_{0 t} , \bX_{ t}) \cond \bX_{0 t} , \bX_{ t} \big\} \ .
\end{align*}	
The first equality holds from Assumption \ref{assumption:valid proxy Cov}, and the second equality holds from the law of iterated expectation. Therefore, we have 
\begin{align}
\label{eq-Identity1}
\EXP \big\{ \potY{t}{0} \big\} 
=
\EXP \big\{ h^* (\bW_{ t}, \bX_{0 t} , \bX_{ t}) \big\} \ .
\end{align}

Next, we prove the second result. For the post-treatment periods $t \in \{T_0+1,\ldots,T\}$, we have 
\begin{align*}
& \EXP \big\{ \potY{t}{1} - \potY{t}{0} \big\} = 
\EXP \big\{ Y_t - \potY{t}{0} \big\}
=
\EXP \big\{ Y_t - h^* (\bW_{ t}, \bX_{0 t} , \bX_{ t}) \big\}
\end{align*}
The first equality holds from Assumption \ref{assumption:consistency}. The second equality holds from \eqref{eq-Identity1}.

We remark that Theorems  \ref{thm:SC} and \ref{thm:ATT} can be shown in a similar manner. 

\subsection{Proof of Theorems \ref{thm:AN} and \ref{thm:AN Cov}}		\label{sec:supp:AN}

We denote the collection of parameters as $\btheta$. 
When there is no covariate as in Section \ref{sec:Estimation}, we have $\btheta = (\Beta, \bgamma, \bbeta)$; when there are covariates as in Section \ref{sec:Cov}, we have $\btheta = (\Beta, \bgamma, \bdelta, \bbeta)$. In what follows, we focus on the proof of Theorems \ref{thm:AN} (i.e., the case with no covariates), since the proof of Theorem \ref{thm:AN Cov} follows a similar approach.

\subsubsection{Notation}

Let the estimating function be $\Psi(\bO_t \con \btheta)$ where 
\begin{align}
\Psi (\bO_t \con \btheta)
&
=
\begin{bmatrix}
\Psi_{\pre} (\bO_t \con \Beta, \bgamma)
\\
\Psi_{\post} (\bO_t \con \bgamma , \bbeta)
\end{bmatrix}
\nonumber
\\
& 
=
\begin{bmatrix}
(1-A_t)
\bD_t
\big( Y_t - \bD_t\T \Beta \big)
\\
(1-A_t)
\bg(t,Y_t \con \Beta)
\big( Y_t - \bW_{ t} \T \bgamma \big)
\\
A_t 
\frac{ \partial \tau(t \con \bbeta) }{\partial \bbeta\T}
\big(
Y_t - \bW_{ t} \T \bgamma - \tau(t \con \bbeta)
\big)
\end{bmatrix} 
=
\begin{bmatrix}
(1-A_t)
\bD_t
\big( Y_t - \bD_t\T \Beta \big)
\\
(1-A_t)
\begin{bmatrix}
\bD_t \\ \bh(Y_t - \bD_t\T \Beta)
\end{bmatrix}
\big( Y_t - \bW_{ t} \T \bgamma \big)
\\
A_t 
\frac{ \partial \tau(t \con \bbeta) }{\partial \bbeta\T}
\big(
Y_t - \bW_{ t} \T \bgamma - \tau(t \con \bbeta)
\big)
\end{bmatrix} 
\label{eq-supp-EE-general}
\ .
\end{align}
The derivative of $\Psi$ is
\begin{align*}
&
\frac{\partial \Psi (\bO_t \con \Beta, \bgamma, \bbeta)}{\partial (\Beta, \bgamma, \bbeta)\T }
\\
&
=
-
\begin{bmatrix}
(1-A_t) \bD_t \bD_t\T
& 0_{d \times N}
& 0_{d \times b}
\\
0_{d \times d}
&
(1-A_t) \bD_t \bW_t\T 
&
0_{d \times b}
\\
(1-A_t) 
\bh'(Y_t - \bD_t \T \Beta)
(Y_t - \bW_t \T \bgamma) \bD_t\T
&
(1-A_t) \bh(Y_t - \bD_t\T \Beta) \bW_t\T 
&
0_{\dim(\bh) \times b}
\\
0_{b \times d}
& 
A_t  \tau'(t \con \bbeta)
& A_t
\big[
\tau''(t \con \bbeta) \tau(t \con \bbeta)
+
\tau'(t \con \bbeta)^{\otimes 2}
\big]
\end{bmatrix}
\\
&
\in \R^{(2d + \dim(\bh) + b) \times (d+N+b)}
\ ,
\end{align*}
where $\bh'(t) = \partial \bh(t) / \partial t \in \R^{\dim(\bh) \times 1}$, $\tau'(t \con \bbeta) = \partial \tau(t \con \bbeta)/\partial \bbeta\T \in \R^{b \times 1}$, $\tau''(t \con \bbeta) = \partial^2 \tau(t \con \bbeta)/(\partial \bbeta\partial \bbeta\T) \in \R^{b \times b}$.

We denote $ \mathcal{O}= \text{supp}(\bO_t)$, $\Theta = \text{supp}(\btheta)$, and the parameters of interest as $\btheta_{0}^* = (\Beta^*,\bgamma_0^*,\bdelta^*,\bbeta^*)$.
\begin{remark}
We will consider a simple case as an example to motivate the assumptions below. Specifically, suppose that $\bh(y)=y$ and $\tau(t \con \bbeta) = \bbeta$, i.e., constant treatment effect. Then, the estimating function is
\begin{align}		\label{eq-supp-simple}
\Psi (\bO_t \con \btheta)
=
\begin{bmatrix}
(1-A_t)
\bD_t
\big( Y_t - \bD_t\T \Beta \big)
\\
(1-A_t)
\bD_t
\big( Y_t - \bW_{ t} \T \bgamma \big)
\\
(1-A_t)
(Y_t - \bD_t\T \Beta)
\big( Y_t - \bW_{ t} \T \bgamma \big)
\\
A_t 
\big(
Y_t - \bW_{ t} \T \bgamma - \bbeta
\big)
\end{bmatrix}
\in \R^{2d+2}
\ .
\end{align}
Note that the derivative of \eqref{eq-supp-simple} is
\begin{align*}
\frac{\partial \Psi (\bO_t \con \Beta, \bgamma, \bbeta)}{\partial (\Beta, \bgamma, \bbeta)\T }
=
-
\begin{bmatrix}
(1-A_t) \bD_t \bD_t\T
& 0_{d \times N}
& 0_{d \times 1}
\\
0_{d \times d}
& (1-A_t) \bD_t \bW_t \T 
& 0_{d \times 1}
\\
(1-A_t) \bD_t\T (Y_t - \bW_t\T \bgamma)
&
(1-A_t) (Y_t - \bD_t\T \Beta) \bW_t\T
& 0
\\
0_{1 \times d}
& 
A_t \bW_{ t} \T & A_t
\end{bmatrix}
\in \R^{(2d+2) \times (d+N+1)}
\ .
\end{align*}
\end{remark}

\subsubsection{Assumptions}

We adapt the proof of Theorem S6 in \citet{Qiu2022} to our setting. First, we introduce regularity conditions that are applicable to general cases, without imposing strict requirements of strong stationarity and ergodicity.

\begin{GREG}[Sufficiently Long Pre- and Post-treatment Periods] 	\label{assumption-General-1}
As $T \rightarrow \infty$, $T_0, T_1 \rightarrow \infty$ and $T_a/T \rightarrow \pi_a \in (0,\infty)$ for $a=0,1$.
\end{GREG} 
Regularity Condition \ref{assumption-General-1} is reasonable if the pre- and post-treatment periods are of roughly the same size and sufficiently large.
\begin{GREG}[Compactness] 	\label{assumption-General-2}
The parameter space $\Theta$ is compact, and $\btheta_0 ^* \in \text{int} (\Theta)$; 
\end{GREG} 
Regularity Condition \ref{assumption-General-2} is standard in parametric estimation.
\begin{GREG}[Weighting Matrix] 	\label{assumption-General-3}
$\widehat{\Omega} = \text{diag} (I_{d \times d}, \widehat{\Omega}_{\bg}, \widehat{\Omega}_{\post})$ is a positive definite matrix, and converges to a non-random positive definite matrix $\Omega^* = \text{diag} (I_{d \times d}, {\Omega}_{\bg}^*, {\Omega}_{\post}^*)$ as $T \rightarrow \infty$.
\end{GREG}
Regularity Condition \ref{assumption-General-3} is easily satisfied if $\widehat{\Omega}$ is chosen as a fixed matrix such as the identity matrix.

\begin{GREG}[Population Moment Restriction \& Global Identification] 	\label{assumption-General-6}
Their exists unique $\btheta_0^* = (\Beta^*, \bgamma_0^*, \bbeta^*)$ where 
\begin{align*}
(i) 
& 
\quad 
\Beta^*
\text{ solves }
\lim_{T_0 \rightarrow \infty}
\frac{1}{T_0} \sum_{t=1}^{T_0}
\EXP \big[ \bm{D}_t \big\{ \potY{t}{0} - \bm{D}_t\T \Beta^* \big\} \big]
= 0 
\\
(ii) 
& 
\quad 
\bgamma_0^*
=
\argmin
\Bigg\{ \big\| \bgamma \big\|_2^2
\, \Bigg| \, 
\bgamma \text{ satisfies }
\lim_{T_0 \rightarrow \infty}
\frac{1}{T_0} \sum_{t=1}^{T_0}
\EXP
\big[
\bg(t, \potY{t}{0} \con \Beta^*)
\big\{ 
\potY{t}{0} -  \bW_t\T \bgamma
\big\} 
\big]
\Bigg\}
=
0
\\
(iii) 
& 
\quad 
\bbeta^*
\text{ solves }
\lim_{T_1 \rightarrow \infty}
\frac{1}{T_1} \sum_{t=T_0+1}^{T}
\EXP \bigg[ \frac{\partial \tau( t \con \bbeta^*) }{\partial \bbeta\T}
\big\{ \potY{t}{1} - \tau(t \con \bbeta^*) - \bW_t\T \bgamma_0^* \big\} \bigg]
= 0
\end{align*}
\end{GREG}
Regularity Condition \ref{assumption-General-6} (i) is satisfied if $[\lim_{T_0 \rightarrow \infty} T_0^{-1} \sum_{t=1}^{T_0} \bD_t \bD_t\T]$ is of full rank.  Regularity Condition \ref{assumption-General-6} (ii) is satisfied because the minimum norm solution is uniquely determined. Regularity Condition \ref{assumption-General-6} (iii) is satisfied if $\tau(t \con \bbeta^*)$ is specified based on an identifiable model.

\begin{GREG}[Regularity Conditions for $\Psi$]
\label{assumption-General-4}
The estimating function $\Psi(\bO_{t} \con \btheta) : \mathcal{O} \otimes \Theta \rightarrow \R^{\dim(\Psi)}$ satisfies
\begin{itemize}[itemsep=0cm]
\item[(i)] $ \lim_{T\rightarrow \infty} \big\{ T^{-1} \sum_{t=1}^{T} \Psi(\bO_{t} \con \btheta) \big\} $ is continuous on $\Theta$ for each $\bO_{t} \in \mathcal{O}$;
\item[(ii)] $ \lim_{T\rightarrow \infty} \big[ T^{-1} \sum_{t=1}^{T} \EXP \big\{ \Psi(\bO_{t} \con \btheta) \big\} \big] $ exists and is finite for any $\btheta \in \Theta$;
\item[(iii)] $ \lim_{T\rightarrow \infty} \big[ T^{-1} \sum_{t=1}^{T} \EXP \big\{ \Psi(\bO_{t} \con \btheta) \big\} \big] $ is continuous on $\Theta$.
\end{itemize}
\end{GREG}
Regularity Condition \ref{assumption-General-4} is satisfied for estimating equation \eqref{eq-supp-simple} if the following vectors/matrices are finite and well-defined:
\begin{align}
& 
\text{For $t \in \{1,\ldots,T_0\}$: }
&&
\EXP \big( \bD_t Y_t \big) \ , 
&&
\EXP \big( \bD_t \bD_t\T \big) \ , 
&& 
\EXP \big( \bD_t \bW_t\T \big) \ , 
&& 
\EXP \big( Y_t^2 \big) \ , 
&& \EXP \big( Y_t \bW_t \T \big) \ , 
\nonumber
\\
&
\text{For $t \in \{T_0+1,\ldots,T\}$: }
&&
\EXP \big( Y_t \big) \ , 
&&
\EXP \big( \bW_t\T\big) \ . 
\label{eq-finitevectormatrix}
\end{align}

\begin{GREG}[Regularity for $\partial \Psi( \bO_{t} \con \btheta) / \partial \btheta \T$ \& Local Identification] 	\label{assumption-General-5}
The function $\partial \Psi( \bO_{t} \con \btheta) / \partial \btheta \T \in \R^{\dim(\Psi) \times \dim(\btheta)}$ satisfies:
\begin{itemize}[itemsep=0cm]
\item[(i)] $ \lim_{T\rightarrow \infty} \big\{ T^{-1} \sum_{t=1}^{T} \partial \Psi( \bO_{t} \con \btheta) / \partial \btheta \T \big\} $ exists and is continuous on $\Theta$ for each $\bO_{t} \in \mathcal{O}$;
\item[(ii)] $ T^{-1} \sum_{t=1}^{T} \partial \Psi( \bO_{t} \con \btheta) / \partial \btheta \T $ is uniformly bounded for all $T \in \{1,2,\ldots\}$
\item[(iii)] $ \lim_{T\rightarrow \infty} \big[ T^{-1} \sum_{t=1}^{T} \EXP \big\{ \partial \Psi( \bO_{t} \con \btheta) / \partial \btheta \T \big\} \big]
\big|_{\btheta=\btheta_0^*} $ exists and is finite;
\item[(iv)] The column rank of $\lim_{T\rightarrow \infty} \big[ T^{-1} \sum_{t=1}^{T} \EXP \big\{ \partial \Psi( \bO_{t} \con \btheta) / \partial (\Beta, \bbeta) \T \big\} \big] \big|_{\btheta=\btheta_0^*}$ is $\dim(\Beta) + \dim(\bbeta) = d + b$.
\end{itemize}
\end{GREG}
Regularity Condition \ref{assumption-General-5} (i)-(iii) are satisfied for estimating equation \eqref{eq-supp-simple} if the following vectors/matrices are uniformly bounded:
\begin{align}
& 
\text{For $t \in \{1,\ldots,T_0\}$: }
&&
\bD_t Y_t \ , 
&&
\bD_t \bD_t\T  \ , 
&& 
\bD_t \bW_t\T \ , 
&& 
Y_t^2 \ , 
&& 
Y_t \bW_t \T \ , 
\nonumber
\\
&
\text{For $t \in \{T_0+1,\ldots,T\}$: }
&&
Y_t \ ,
&&
\bW_t\T \ .  
\nonumber
\end{align}
Regularity Condition \ref{assumption-General-5} (iv) accounts for both underspecified cases (i.e., $\dim (\Psi) < \dim(\btheta)$) and standard cases (i.e., $\dim(\Psi) \geq \dim(\btheta))$. 
We have {\small
\begin{align*}
&
\lim_{T \rightarrow \infty}
\frac{1}{T}
\sum_{t=1}^{T}
\EXP 
\bigg[ 
\frac{\partial \Psi (\bO_t \con \btheta)}{\partial (\Beta, \bbeta)\T }
\bigg] \bigg|_{\btheta=\btheta_0^*}
\\
&
=
-
\lim_{T \rightarrow \infty}
\frac{1}{T}
\sum_{t=1}^{T}
\EXP 
\left.
\begin{bmatrix}
(1-A_t) \bD_t \bD_t\T
& 0_{d \times b}
\\
0_{d \times d}
&
0_{d \times b}
\\
(1-A_t) 
\bh'(Y_t - \bD_t \T \Beta)
(Y_t - \bW_t \T \bgamma) \bD_t\T
&
0_{\dim(\bh) \times b}
\\
0_{b \times d}
& A_t 
\big[
\tau''(t \con \bbeta) \tau(t \con \bbeta)
+
\tau'(t \con \bbeta)^{\otimes 2}
\big]
\end{bmatrix}
\right|_{\btheta=\btheta_0^*}
\\
&
=
-
\begin{bmatrix}
\EXP_{\pre}^{\infty} \big( \bD_t \bD_t\T \big)
& 0_{d \times b}
\\
0_{d \times d}
&
0_{d \times b}
\\
\pi_0
\EXP_{\pre}^{\infty} \big\{
\bh'(Y_t - \bD_t \T \Beta_0^* )
(Y_t - \bW_t \T \bgamma_0^* ) \bD_t\T
\big\}
&
0_{\dim(\bh) \times b}
\\
0_{b \times d}
& 
\pi_1
\EXP_{\post}^{\infty}
\big[
\tau''(t \con \bbeta^*) \tau(t \con \bbeta^*)
+
\tau'(t \con \bbeta^*)^{\otimes 2}
\big]
\end{bmatrix} 
\\
&
=
-
\begin{bmatrix}
\pi_0 \EXP_{\pre}^{\infty}
\big( \bD_t \bD_t\T \big)
& 0_{d \times b}
\\
0_{d \times d}
&
0_{d \times b}
\\
0_{\dim(\bh) \times d}
&
0_{\dim(\bh) \times b}
\\
0_{b \times d}
& 
\pi_1
\EXP_{\post}^{\infty}
\big[
\tau''(t \con \bbeta^*) \tau(t \con \bbeta^*)
+
\tau'(t \con \bbeta^*)^{\otimes 2}
\big]
\end{bmatrix} 
\\
&
\in \R^{(2d + \dim(\bh) + b) \times (d+b)}
\ ,
\end{align*} }%
where $\EXP_{\pre}^{\infty}(f) =  \lim_{T_0 \rightarrow \infty} T_0^{-1} \sum_{t=1}^{T_0} \EXP\{ f(\bO_t) \}$ and $\EXP_{\post}^{\infty}(f) =  \lim_{T_1 \rightarrow \infty} T_1^{-1} \sum_{t=T_0+1}^{T} \EXP\{ f(\bO_t) \}$. Note that the last equality holds from
\begin{align*}
&
\EXP \big\{
\bh'(Y_t - \bD_t \T \Beta^*)
(Y_t - \bW_t \T \bgamma_0^*) \bD_t\T
\big\}
=
\EXP \big\{
\bh'(Y_t - \bD_t \T \Beta^*)
\EXP \big(Y_t - \bW_t \T \bgamma_0^* \cond Y_t \big)
\bD_t\T
\big\}
= 0 \ , \quad t \in \{1,\ldots,T_0\} \ .
\end{align*}
Therefore, Regularity Condition \ref{assumption-General-5} (iv) holds if $\EXP_{\pre}^{\infty}
\big( \bD_t \bD_t\T \big)$ and $\EXP_{\post}^{\infty}
\big[
\tau''(t \con \bbeta^*) \tau(t \con \bbeta^*)
+
\tau'(t \con \bbeta^*)^{\otimes 2}
\big]$ are of column full rank.

\begin{GREG}[Smoothness of $\bbeta$] 	\label{assumption-General-Continuous}
Let $\bbeta(\bgamma)$ be the solution to
\begin{align*}
\lim_{T_1 \rightarrow \infty}
\frac{1}{T_1} \sum_{t=T_0+1}^{T}
\EXP \bigg[ \frac{\partial \tau( t \con \bbeta) }{\partial \bbeta\T} \bigg|_{\bbeta = \bbeta(\bgamma)}
\big\{ \potY{t}{1} - \tau(t \con \bbeta(\bgamma)) - \bW_t\T \bgamma \big\} \bigg]
= 0 \ .
\end{align*}
Then, $\bbeta(\bgamma)$ is unique and uniformly bounded. Furthermore, its derivative with respect to $\bgamma$, i.e., $\partial{\bbeta(\bgamma)}/{\partial \bgamma\T}$, is continuous and uniformly bounded.
\end{GREG}

Regularity Condition \ref{assumption-General-Continuous} states that the parameter for the ATT is a smooth functional of the synthetic control weights $\bgamma$.

\begin{GREG}[Uniform Weak Law of Large Numbers for $\Psi$] 	\label{assumption-General-UWLLN}
\begin{align*}
\sup_{\btheta \in \Theta}
\bigg\|
\frac{1}{T} \sum_{t=1}^{T} \Psi(\bO_t \con \btheta) -
\lim_{T' \rightarrow \infty}
\frac{1}{T'} \sum_{t=1}^{T'} \EXP \big\{\Psi(\bO_t \con \btheta) \big\}
\bigg\|
=
o_P(1) \text{ as } T \rightarrow \infty \ .
\end{align*}
\end{GREG}

\begin{GREG}[Uniform Weak Law of Large Numbers for the Gradient of $\Psi$] 	\label{assumption-General-UWLLN2}
\begin{align*}
\sup_{\btheta \in \Theta}
\bigg\|
\frac{1}{T} \sum_{t=1}^{T} \frac{\partial}{\partial \btheta \T } \Psi(\bO_t \con \btheta) -
\lim_{T' \rightarrow \infty}
\frac{1}{T'} \sum_{t=1}^{T'} \EXP \bigg\{ \frac{\partial}{\partial \btheta \T } \Psi(\bO_t \con \btheta) \bigg\}
\bigg\|
=
o_P(1) \text{ as } T \rightarrow \infty \ .
\end{align*}
\end{GREG}
Regularity Conditions \ref{assumption-General-UWLLN} and \ref{assumption-General-UWLLN2} hold if the underlying process is strictly stationary, strongly mixing, or $\phi$-mixing processes; see \citet{Andrews1988}, \citet[Chapter 5]{PP1997} and \citet[Section S2]{Qiu2022} for details. 

\begin{GREG}[Asymptotic Normality]
\label{assumption-General-AN} 
As $T \rightarrow \infty$, we have
\begin{align*}
&
\frac{1}{\sqrt{T}} \sum_{t=1}^{T} \Psi(\bO_t \con \btheta^*)
\text{ converges in distribution to }
N (0, \Sigma_2^*)
\ , \\
&
\Sigma_2^* = \lim_{T \rightarrow \infty} \VAR \bigg\{ \frac{1}{\sqrt{T}} \sum_{t=1}^{T} \Psi (\bO_t \con \btheta^*) \bigg\} \ .
\end{align*}
Here, $\Sigma_2^*$ is a finite valued positive definite matrix. 
\end{GREG}
Assumption \ref{assumption-General-AN} directly assumes the asymptotic normality of the sample mean of the estimating function; see Section S2 of \citet{Qiu2022} for the plausibility of the assumption. We remark that Regularity Conditions \ref{assumption-General-UWLLN}--\ref{assumption-General-AN} are satisfied under standard assumptions for GMM; see Chapter 3 of \citet{Hall2004GMM} for details. 

\subsubsection{Proof}

Under Regularity Conditions \ref{assumption-General-1}--\ref{assumption-General-AN}, we establish the desired result. We simply denote $\widehat{\Psi}(\btheta) = T^{-1} \sum_{t=1}^T \Psi(\bO_t \con \btheta)$ and ${\Psi}(\btheta) = \lim_{T \rightarrow \infty} T^{-1} \sum_{t=1}^T \EXP \big\{ \Psi(\bO_t \con \btheta) \big\}$.  For a generic function $f$, we denote
\begin{align*}
&
\EXP_{\pre}^{\infty}(f) =  \lim_{T_0 \rightarrow \infty} \frac{1}{T_0} \sum_{t=1}^{T_0} \EXP\{ f(\bO_t) \}
&&
\AVER_{\pre}^{\infty}(f) =  \lim_{T_0 \rightarrow \infty} \frac{1}{T_0} \sum_{t=1}^{T_0} f(\bO_t)
\\
&
\EXP_{\post}^{\infty}(f) =  \lim_{T_1 \rightarrow \infty} \frac{1}{T_1} \sum_{t=T_0+1}^{T} \EXP\{ f(\bO_t) \}
&&
\AVER_{\post}^{\infty}(f) =  \lim_{T_1 \rightarrow \infty} \frac{1}{T_1} \sum_{t=T_0+1}^{T} f(\bO_t) \ .
\end{align*}

From the form of the estimating equation in \eqref{eq-supp-EE-general}, we find the following representations for $(\Beta^*,\bbeta^*)$:
\begin{align*}
&
\Beta^*
=
\big\{
\EXP_{\pre}^{\infty} ( \bD_t \bD_t\T )
\big\}^{-1}
\big\{
\EXP_{\pre}^{\infty} ( \bD_t Y_t )
\big\}^{-1}
\ , 
&&
\bbeta^* = \bbeta(\bgamma_0^*) 
\end{align*}
where $\bbeta( \cdot) $ is the function defined in Regularity Condition \ref{assumption-General-Continuous}. The synthetic control weight $\bgamma_0^*$ has the following representation. Let $ \bG_{YW}^{*} 
=
\EXP_{\pre}^{\infty} 
\big\{
\bg(t, Y_t \con \Beta)
\bW_t\T
\big\}$ and $
{\bG}_{YY}^*
=
\EXP_{\pre}^{\infty} 
\big\{ 
\bg(t, Y_t \con \Beta)
Y_t
\big\}$. Then, we find
\begin{align*}
&
\text{If $\bG_{YW}^{*}$ is of full row rank, }
&&
\text{then }
\bgamma_0^*
=
\big( \Omega_{\bg}^{*1/2} 
\bG_{YW}^* \big)^{+}
\big( 
\Omega_{\bg}^{*1/2} 
\bG_{YY}^*
\big)
\\
& \text{If $\bG_{YW}^{*}$ is of full column rank, }
&&
\text{then }
\bgamma_0^* 
=
(\bG_{YW}\sT \Omega_{\bg}^* \bG_{YW}^* )^{-1}
( \bG_{YW}\sT \Omega_{\bg}^* \bG_{YY}^* ) \ .
\end{align*}

Let $\btheta_{\rho}^*$ be the unique minimizer of 
\begin{align*}
& Q_\rho(\btheta)
\\
&
=
\bigg[
\lim_{T \rightarrow \infty}
\frac{1}{T} \sum_{t=1}^T \EXP \big\{ \Psi(\bO_t \con \btheta) \big\} \bigg]\T
\Omega^*
\bigg[
\lim_{T \rightarrow \infty}
\frac{1}{T} \sum_{t=1}^T \EXP \big\{ \Psi(\bO_t \con \btheta) \big\} \bigg]
+ \rho \big\| \bgamma \big\|_2^2 
\\
&
=
\left[ 
\begin{array}{l}
\pi_0^2
\big\| \EXP_{\pre}^{\infty}
\big\{ \bm{D}_t \big( Y_t - \bm{D}_t\T \Beta \big) \big\}
\big\|_2^2
\\
+
\pi_0^2
\big[ \EXP_{\pre}^{\infty}
\big\{ \bg(y, Y_t \con \Beta) \big( Y_t - \bW_t\T \bgamma \big) \big\} \big]\T \Omega_{\bg}^*
\big[ \EXP_{\pre}^{\infty}
\big\{ \bg(y, Y_t \con \Beta) \big( Y_t - \bW_t\T \bgamma \big) \big\} \big]
\\
+
\pi_1^2
\big[
\EXP_{\post}^{\infty} \big[ \tau' (t \con \bbeta)
\big\{ Y_t - \tau(t \con \bbeta) - \bW_t\T \bgamma \big\} \big]
\big]\T \Omega_{\post}^*
\big[
\EXP_{\post}^{\infty} \big[ \tau' (t \con \bbeta)
\big\{ Y_t - \tau(t \con \bbeta) - \bW_t\T \bgamma \big\} \big]
\big]
\end{array}
\right]     
+
\rho \big\| \bgamma \big\|_2^2 \ .
\end{align*}
From straightforward algebra, we find
\begin{align*}
&
\Beta_{\rho}^*
=
\big\{
\EXP_{\pre}^{\infty} ( \bD_t \bD_t\T )
\big\}^{-1}
\big\{
\EXP_{\pre}^{\infty} ( \bD_t Y_t )
\big\}^{-1}
\ , 
\\
&
\bgamma_{\rho}^* 
=
(\bG_{YW}\sT \Omega_{\bg}^* \bG_{YW}^* + \rho I_{N \times N} )^{-1}
( \bG_{YW}\sT \Omega_{\bg}^* \bG_{YY}^* ) 
\ , 
\\
&
\bbeta_{\rho}^* = \bbeta(\bgamma_\rho^*) \ .
\end{align*} 
As $\rho \downarrow 0$, we find
\begin{align*}
\Beta_{\rho}^*
&
\rightarrow
\Beta^*
\\
\bgamma_{\rho}^* 
&
=
(\bG_{YW}\sT \Omega_{\bg}^* \bG_{YW}^* + \rho I_{N \times N} )^{-1}
( \bG_{YW}\sT \Omega_{\bg}^* \bG_{YY}^* ) 
\\
&
\quad 
\rightarrow
\left\{
\begin{array}{ll}
\big( \Omega_{\bg}^{*1/2} 
\bG_{YW}^* \big)^{+}
\big( 
\Omega_{\bg}^{*1/2} 
\bG_{YY}^*
\big)
&
\text{ if $\bG_{YW}^*$ is of full row rank }
\\
(\bG_{YW}\sT \Omega_{\bg} \bG_{YW}^* )^{-1}
( \bG_{YW}\sT \Omega_{\bg} \bG_{YY}^* )
&
\text{ if $\bG_{YW}^*$ is of full column rank }
\end{array}
\right\}
=
\bgamma_0^*
\\
\bbeta_{\rho}^*
&
=
\bbeta(\bgamma_\rho^*) \rightarrow \bbeta(\bgamma_0^*) = \bbeta^* 
\end{align*}

First, we establish consistency, i.e., $\widehat{\btheta}_{\rho} = (\widehat{\Beta},
\widehat{\bgamma}_{\rho}, \widehat{\bbeta}) = (\Beta^*, \bgamma_0^*, \bbeta^*) + o_P(1) = \btheta_0^* + o_P(1)$. Under Regularity Conditions \ref{assumption-General-3} and \ref{assumption-General-UWLLN}, we obtain
\begin{align} \label{eq-UWLLN-GMM2}
&
\sup_{\btheta \in \Theta}
\left\|
\big\{
\widehat{\Psi}(\btheta)
\big\}\T 
\widehat{\Omega}
\big\{
\widehat{\Psi}(\btheta)
\big\} 
+ 
\rho \big\| \bgamma \big\|_2^2
-
\big\{
{\Psi}(\btheta)
\big\}\T 
\Omega^*
\big\{
{\Psi}(\btheta)
\big\}
-
\rho \big\| \bgamma \big\|_2^2
\right\|
\nonumber
\\
&
=
\sup_{\btheta \in \Theta}
\left\|
\big\{
\widehat{\Psi}(\btheta)
\big\}\T 
\widehat{\Omega}
\big\{
\widehat{\Psi}(\btheta)
\big\} 
-
\big\{
{\Psi}(\btheta)
\big\}\T 
\Omega^*
\big\{
{\Psi}(\btheta)
\big\}
\right\| 
\nonumber
\\
&
= o_P(1) \ .
\end{align}

Let $s > 0$ be an arbitrary positive constant. 
From \eqref{eq-UWLLN-GMM2} and the definition of $\widehat{\btheta}_{\rho}$, the following conditions hold with probability tending to one:
\begin{align*}
&
\left\|
\big\{
\widehat{\Psi}(\btheta_{\rho}^*)
\big\}\T 
\widehat{\Omega}
\big\{
\widehat{\Psi}(\btheta_{\rho}^*)
\big\} 
+
\rho \big\| \bgamma_{\rho}^* \big\|_2^2
-
\big\{
\Psi (\btheta_{\rho}^*)
\big\}\T 
\Omega^*
\big\{
\Psi (\btheta_{\rho}^*)
\big\}
-
\rho \big\| \bgamma_{\rho}^* \big\|_2^2
\right\| < s/2
\\
&
\left\|
\big\{
\widehat{\Psi}(\widehat{\btheta}_{\rho})
\big\}\T 
\widehat{\Omega}
\big\{
\widehat{\Psi}(\widehat{\btheta}_{\rho})
\big\} 
+
\rho \big\| \widehat{\bgamma}_{\rho} \big\|_2^2
-
\big\{
\Psi (\widehat{\btheta}_{\rho})
\big\}\T 
\Omega^*
\big\{
\Psi (\widehat{\btheta}_{\rho})
\big\}
-
\rho \big\| \widehat{\bgamma}_{\rho} \big\|_2^2
\right\| < s/2 
\\ 
&
\big\{
\widehat{\Psi} (\widehat{\btheta}_{\rho})
\big\}\T 
\widehat{\Omega}
\big\{
\widehat{\Psi} (\widehat{\btheta}_{\rho})
\big\}
+
\rho \big\| \widehat{\bgamma}_{\rho} \big\|_2^2
\leq
\big\{
\widehat{\Psi} (\btheta_{\rho}^*)
\big\}\T 
\widehat{\Omega}
\big\{
\widehat{\Psi} (\btheta_{\rho}^*)
\big\}
+
\rho \big\| \bgamma_{\rho}^* \big\|_2^2
\ .
\end{align*}
Note that the last inequality holds because  $\widehat{\btheta}_{\rho}$ is the minimizer of $\big\{
\widehat{\Psi}(\btheta)
\big\}\T 
\widehat{\Omega}
\big\{
\widehat{\Psi}(\btheta)
\big\} 
+ \rho \big\| \bgamma \big\|_2^2 $. 

These three inequalities imply that
\begin{align*}
\big\{
\Psi(\widehat{\btheta}_{\rho})
\big\}\T 
\Omega^*
\big\{
\Psi(\widehat{\btheta}_{\rho})
\big\} 
+
\rho \big\| \widehat{\bgamma}_{\rho} \big\|_2^2
<
\big\{
\Psi(\btheta_{\rho}^*)
\big\}\T 
\Omega^*
\big\{
\Psi(\btheta_{\rho}^*)
\big\} 
+
\rho \big\| {\bgamma}_{\rho}^* \big\|_2^2
+ s
=
m_\rho
+
s \ .
\end{align*} 
Here $m_\rho = \min_{\btheta} \big[ \big\{
\Psi(\btheta)
\big\}\T 
\Omega^*
\big\{
\Psi(\btheta)
\big\} 
+
\rho \big\| {\bgamma} \big\|_2^2 \big]$.

Let $\mathcal{N} \in \Theta$ be an arbitrary open set containing $\btheta_0^*$. Let us define the following quantity:
\begin{align*}
s_0
=
\inf_{\btheta \in \Theta \setminus \mathcal{N} }
\big[ 
\big\{ 
\Psi(\btheta)
\big\}\T
\Omega^*
\big\{ 
\Psi(\btheta)
\big\} +
\rho \big\| \bgamma \big\|_2^2
\big]
- m_{\rho}
\end{align*}
Note that $\Theta \setminus \mathcal{N}$ is compact under Regularity Condition \ref{assumption-General-2}. Also, for a fixed $\rho>0$, Regularity Conditions \ref{assumption-General-6} and \ref{assumption-General-4} imply that $s_0$ is positive. Therefore, by taking $s>s_0$, the event $\big\{ \big\{
\Psi(\widehat{\btheta}_{\rho})
\big\}\T 
\widehat{\Omega}
\big\{
\Psi(\widehat{\btheta}_{\rho})
\big\}
+
\rho \big\| \widehat{\bgamma}_{\rho} \big\|
< m_{\rho}+s_0 \big\}$ occurs with probability tending to one, which further implies that $\widehat{\btheta}_{\rho} \in \mathcal{N}$. Since $\mathcal{N}$ is arbitrary chosen, this establishes $\widehat{\btheta}_{\rho} = \btheta_{\rho}^* + o_P(1)$ as $T \rightarrow \infty$ for a fixed $\rho>0$. 

In addition, for any $T  \in \{ 1,2,\ldots\}$, we find $\widehat{\btheta}_{\rho} \rightarrow \widehat{\btheta}_0 $ as $\rho \downarrow 0$ where
\begin{align*} 
&
\widehat{\btheta}_0
=
\begin{bmatrix}
\widehat{\Beta}
\\
\widehat{\bgamma}_{0}
\\
\bbeta (  \widehat{\bgamma}_{0} )
\end{bmatrix}
\ , \quad
&&
\widehat{\bgamma}_0
=
\left\{
\begin{array}{ll}
\big( \widehat{\Omega}_{\bg}^{1/2}
\widehat{\bG}_{YW} \big)^{+}
\big( 
\widehat{\Omega}_{\bg}^{1/2}
\widehat{\bG}_{YY}
\big)
&
\text{if $\widehat{\bG}_{YW}$ is of full row rank}
\\
(\widehat{\bG}_{YW}\T \widehat{\Omega}_{\bg} \widehat{\bG}_{YW} )^{-1}
( \widehat{\bG}_{YW}\T \widehat{\Omega}_{\bg} \widehat{\bG}_{YY} )
&
\text{if $\widehat{\bG}_{YW}$ is of full column rank}
\end{array}
\right.\ .
\end{align*}
For $\rho,\rho'>0$, we get
\begin{align*}
&
\widehat{\btheta}_{\rho}
-
\widehat{\btheta}_{\rho'}=
\left[ 
\begin{array}{c}
0
\\
\widehat{\bgamma}_\rho
-
\widehat{\bgamma}_{\rho'}\\       
\bbeta(\widehat{\bgamma}_\rho)
-
\bbeta(\widehat{\bgamma}_{\rho'})
\end{array}
\right] \ .
\end{align*}
Consider a singular vector decomposition of $\widehat{\Omega}_{\bg}^{1/2} \widehat{\bG}_{YW} 
=
\widehat{\bm{\mathcal{U}}} \widehat{\bm{\mathcal{D}}} \widehat{\bm{\mathcal{V}}} \T$ where $\widehat{\bm{\mathcal{D}}} = \text{diag}(\widehat{d}_1,\ldots,d_{\dim(\bg)})$. Then, we find
\begin{align*}
\widehat{\bgamma}_{\rho} 
-
\widehat{\bgamma}_{\rho'} 
&
= \Big\{ \Big(
\widehat{\bG}_{YW}\T
\widehat{\Omega}_{\bg}
\widehat{\bG}_{YW}
+
\rho I_{N \times N}
\Big)^{-1}
-
\Big(
\widehat{\bG}_{YW}\T
\widehat{\Omega}_{\bg}
\widehat{\bG}_{YW}
+
\rho' I_{N \times N}
\Big)^{-1}
\Big\}
\Big( 
\widehat{\bG}_{YW}\T
\widehat{\Omega}_{\bg}
\widehat{\bG}_{YY}
\Big)
\\
&
=
\bigg\{
\widehat{\bm{\mathcal{V}}}
\text{diag} \bigg(
\frac{\widehat{d}_{j}}{\rho + \widehat{d}_{j}}
-
\frac{\widehat{d}_{j}}{\rho' + \widehat{d}_{j}}
\bigg) 
\widehat{\bm{\mathcal{U}}}\T
\bigg\} 
\Big( \widehat{\Omega}_{\bg}^{1/2} 
\widehat{\bG}_{YY}  
\Big)
\ .
\end{align*}
Therefore, $\big\| \widehat{\bgamma}_{\rho} 
-
\widehat{\bgamma}_{\rho'} \big\|
\leq 
C_1 \big\| \rho - \rho' \big\|$ where the constant $C_1$ does not depend on $T$ because $\widehat{\bG}_{YW}$ and $\widehat{\bG}_{YY}$ are uniformly bounded for any $T$ from Regularity Condition \ref{assumption-General-5} (ii).
Likewise, from the mean value theorem, there exists $\bgamma'$ satisfying 
\begin{align*}
\bbeta(\widehat{\bgamma}_\rho)
-
\bbeta(\widehat{\bgamma}_{\rho'})
=
\frac{ \partial \bbeta(\bgamma)}{\partial \bgamma\T}
\bigg|_{\bgamma=\bgamma'}
(\widehat{\bgamma}_{\rho} - \widehat{\bgamma}_{\rho'}) \ .
\end{align*}
Since $\partial \bbeta(\bgamma) / \partial \bgamma\T$ is uniformly bounded, we find the following result holds for any $T$:
\begin{align*}
\big\|
\bbeta(\widehat{\bgamma}_\rho)
-
\bbeta(\widehat{\bgamma}_{\rho'})
\big\|
\leq 
C_2 \big\| \widehat{\bgamma}_{\rho} 
-
\widehat{\bgamma}_{\rho'} \big\|
\leq 
C_1 C_2 \big\| \rho - \rho' \big\| \ .
\end{align*}
Therefore, for any $T$, we have
$\big\| \widehat{\btheta}_{\rho}
- \widehat{\btheta}_{\rho'}
\big\| \leq C \big\| \rho - \rho' \big\|$ for a constant $C$. This implies that the convergence $\widehat{\btheta}_{\rho} \rightarrow \widehat{\btheta}_0$ is uniform in $T$. Therefore, this implies the double in-probability limit of $\widehat{\btheta}_{\rho}$ is well-defined and converge to $\btheta_0^*$:
\begin{align*}
\plim_{\rho \downarrow 0}
\plim_{T \rightarrow \infty}
\widehat{\btheta}_{\rho}
=
\plim_{T \rightarrow \infty}
\plim_{\rho \downarrow 0}
\widehat{\btheta}_{\rho}
=
\plim_{T \rightarrow \infty,\rho \downarrow 0}
\widehat{\btheta}_{\rho}
=
\btheta_{0}^* \ .
\end{align*}
Therefore, taking $\rho = o(T^{-1/2})$, we find $\widehat{\btheta}_{\rho}=\btheta_0^*+o_P(1)$ as $T \rightarrow \infty$.

Next, we establish the asymptotic normality of $\widehat{\btheta}_{\rho}$. First, the following results hold from the assumptions and 
$\widehat{\btheta}_{\rho} = \btheta^*+o_P(1)$:
\begin{align} 
& 
\frac{\partial}{\partial \btheta\T} \widehat{\Psi}( \btheta) \bigg|_{\btheta=\widehat{\btheta}_\rho} 
= 
\frac{\partial}{\partial \btheta\T} \widehat{\Psi}( \btheta) \bigg|_{\btheta=\btheta^*} 
+
o_P(1) \ ,
\label{eq-proof-1}
\\
& 
\widehat{\Psi}( \widehat{\btheta}_{\rho})
=
\widehat{\Psi}( {\btheta}^*)
+
\bigg\{
\frac{\partial}{\partial \btheta\T} \widehat{\Psi}( \btheta) \bigg|_{\btheta=\btheta^*}
\bigg\}
(\widehat{\btheta}_{\rho}-\btheta^*)
+
o_P \Big( \big\| \widehat{\btheta}_{\rho} - \btheta^* \big\| \Big)
=
\widehat{\Psi}({\btheta}^*) 
+
o_P(1)
=
O_P(1)
\ . 
\label{eq-proof-2}
\end{align}
Specifically, \eqref{eq-proof-1} holds from Regularity Condition \ref{assumption-General-5}-(i) and the continuous mapping theorem. 
\eqref{eq-proof-2} holds from the first-order Taylor expansion, Regularity Condition \ref{assumption-General-AN}, and $\widehat{\btheta}_{\rho}-\btheta^* = o_P(1)$.

The first order condition of $\widehat{\btheta}_{\rho}$ along with Regularity Condition \ref{assumption-General-5} implies
\begin{align*}
0
& =
\frac{1}{2}
\frac{\partial}{\partial \btheta\T} 
\Big[
\big\{ \widehat{\Psi}( \btheta)
\big\}\T \widehat{\Omega} 
\big\{ \widehat{\Psi}( \btheta)
\big\}
+ \rho \big\| \bgamma \big\|_2^2
\Big] \Big|_{\btheta=\widehat{\btheta}_{\rho}}
\\
&
=
\bigg\{
\frac{\partial}{\partial \btheta\T} \widehat{\Psi}( \btheta) \bigg|_{\btheta=\widehat{\btheta}_\rho}
\bigg\}\T
\widehat{\Omega}
\big\{ \widehat{\Psi}( \widehat{\btheta}_{\rho})
\big\}
+ 
\rho 
\begin{bmatrix}
0_{\dim(\Beta)}
\\
\widehat{\bgamma}_{\rho}
\\
0_{\dim(\beta)}
\end{bmatrix}
\\
& =
\bigg\{
\frac{\partial}{\partial \btheta\T} \widehat{\Psi}( \btheta) \bigg|_{\btheta=\widehat{\btheta}_\rho}
\bigg\}\T
\widehat{\Omega}
\big\{ \widehat{\Psi}( \widehat{\btheta}_{\rho})
\big\} 
+ 
\rho 
\underbrace{
\begin{bmatrix}
0_{\dim(\Beta)} & 0_{\dim(\bgamma)} & 0_{\dim(\beta)}
\\
0_{\dim(\Beta)} & I_{\dim(\bgamma)} & 0_{\dim(\beta)}
\\
0_{\dim(\Beta)} & 0_{\dim(\bgamma)} & 0_{\dim(\beta)}
\end{bmatrix} }_{\mathcal{I}}
(\widehat{\btheta}_{\rho}
-
\btheta^*)
+
o_P(T^{-1/2})
\\
& =
\bigg\{
\frac{\partial}{\partial \btheta\T} \widehat{\Psi}( \btheta) \bigg|_{\btheta=\btheta^*}
\bigg\}\T
\widehat{\Omega}
\big\{ \widehat{\Psi}( \widehat{\btheta}_{\rho})
\big\} 
+
\rho \mathcal{I} (\widehat{\btheta}_{\rho}
-
\btheta^*)
+
o_P(T^{-1/2})
\\
&
=
\bigg\{
\frac{\partial}{\partial \btheta\T} \widehat{\Psi}( \btheta) \bigg|_{\btheta=\btheta^*}
\bigg\}\T
\widehat{\Omega}
\big\{ \widehat{\Psi}( \btheta^*)
\big\}
+
\Bigg[
\bigg\{
\frac{\partial}{\partial \btheta\T} \widehat{\Psi}( \btheta) \bigg|_{\btheta=\btheta^*}
\bigg\}\T
\widehat{\Omega}
\bigg\{
\frac{\partial}{\partial \btheta\T} \widehat{\Psi}( \btheta) \bigg|_{\btheta=\btheta^*}
\bigg\} 
+\rho \mathcal{I}
\Bigg]
(\widehat{\btheta}_{\rho}-\btheta^*)
\\
& 
\quad 
+
o_P
\Big(
\big\| \widehat{\btheta}_{\rho} - \btheta^* \big\|
+
T^{-1/2}
\Big) \ .
\end{align*} 
The third equality is from $T^{1/2} \rho \widehat{\bgamma}_{\rho}=o_P(1)$.
The fourth equality is from \eqref{eq-proof-1} and \eqref{eq-proof-2}. 
The last equality is from \eqref{eq-proof-2}.

By multiplying $T^{1/2}$, we get
\begin{align*}
0
&
=
\bigg\{
\frac{\partial}{\partial \btheta\T} \widehat{\Psi}( \btheta) \bigg|_{\btheta=\btheta^*}
\bigg\}\T
\widehat{\Omega}
\bigg\{
\frac{1}{T^{1/2}}
\sum_{t=1}^{T} \Psi(\bO_t \con \btheta^* )
\bigg\}
\\
&
\hspace*{0.3cm}
+
\Bigg[ 
\bigg\{
\frac{\partial}{\partial \btheta\T} \widehat{\Psi}( \btheta) \bigg|_{\btheta=\btheta^*}
\bigg\}\T
\widehat{\Omega}
\bigg\{
\frac{\partial}{\partial \btheta\T} \widehat{\Psi}( \btheta) \bigg|_{\btheta=\btheta^*}
\bigg\}
+
\rho 
\mathcal{I}
\Bigg]
T^{1/2}
(\widehat{\btheta}_{\rho} -\btheta^*)
+
o_P \Big( T^{1/2} \big\| \widehat{\btheta}_{\rho} - \btheta^* \big\|+ 1 \Big)
\\
&
=
\bigg\{
\underbrace{
\frac{\partial}{\partial \btheta\T} {\Psi}( \btheta) \bigg|_{\btheta=\btheta^*}
}_{=:G^*}
\bigg\}\T 
\Omega^*
\bigg\{
\lim_{T \rightarrow \infty}
\frac{1}{T^{1/2}}
\sum_{t=1}^{T} \Psi(\bO_t \con \btheta^* )
\bigg\}
\\
&
\hspace*{0.3cm}
+
\Bigg[
\bigg\{
\frac{\partial}{\partial \btheta\T} {\Psi}( \btheta) \bigg|_{\btheta=\btheta^*}
\bigg\}\T
\Omega^*
\bigg\{
\frac{\partial}{\partial \btheta\T} {\Psi}( \btheta) \bigg|_{\btheta=\btheta^*}
\bigg\} 
+
\rho 
\mathcal{I}
\Bigg]
T^{1/2}
(\widehat{\btheta}_{\rho}-\btheta^*)
+
o_P \Big( T^{1/2} \big\| \widehat{\btheta}_{\rho} - \btheta^* \big\| + 1 \Big) \ .
\end{align*}
The second equality holds from Regularity Condition \ref{assumption-General-UWLLN2}:
\begin{align*}
\frac{\partial}{\partial \btheta\T} \widehat{\Psi}( \btheta) \bigg|_{\btheta=\btheta^*}
=
\frac{\partial}{\partial \btheta\T} {\Psi}( \btheta) \bigg|_{\btheta=\btheta^*}
+
o_P(1)
\end{align*}
The last equality holds from Regularity Conditions \ref{assumption-General-3}, \ref{assumption-General-UWLLN}, and \ref{assumption-General-UWLLN2} and the consistency of $\widehat{\btheta}_{\rho}$. 

Therefore, we obtain
\begin{align*}
\Big\{ G\sT \Omega^* G^* + \rho \mathcal{I} + o_P(1) 
\Big\}
\Big\{
T^{1/2} \big( \widehat{\btheta}_{\rho} - \btheta^* \big)
\Big\}
=
-
G\sT \Omega^*
\bigg\{
\lim_{T \rightarrow \infty}
\frac{1}{T^{1/2}}
\sum_{t=1}^{T} \Psi(\bO_t \con \btheta^* )
\bigg\}
+ o_P(1) \ .
\end{align*}
This implies 
\begin{align*}
T^{1/2} \big( \widehat{\btheta}_{\rho} - \btheta^* \big)
=
\Big\{ G\sT \Omega^* G^* + \rho \mathcal{I} + o_P(1) 
\Big\}^{-1}
\Bigg[ 
G\sT \Omega^*
\bigg\{
\lim_{T \rightarrow \infty}
\frac{1}{T^{1/2}}
\sum_{t=1}^{T} \Psi(\bO_t \con \btheta^* )
\bigg\}
+ o_P(1)
\Bigg]  \ .
\end{align*}
Therefore, taking $\rho = o(T^{-1/2})$, we have 
\begin{align*}
\Big\{ 
G\sT \Omega^* G^* + \rho \mathcal{I}  
\Big\}^{-1}
G\sT \Omega^{*1/2}
\stackrel{\rho \downarrow 0}{\rightarrow}
\big( \Omega^{*1/2} G^*  \big)^{+} \ .
\end{align*}
As a side note, we have
\begin{align*}
\big( \Omega^{*1/2} G^*  \big)^{+}
\neq  
G^{*+} \Omega^{*-1/2} \ ,
\end{align*}
with a counterexample:
\begin{align*}
\Omega^{*1/2}
=
\begin{bmatrix}
2 & 1 \\ 1 & 1
\end{bmatrix} 
\ , \ 
G^{*+}
=
\begin{bmatrix}
1 \\ 0
\end{bmatrix} 
\quad \Rightarrow \quad 
\big( \Omega^{*1/2} G^*  \big)^{+}
=
\big[ 0.4 \ , \ 0.2 \big]
\neq 
\big[ 1 \ , \ -1 \big] 
=
G^{*+} \Omega^{*-1/2}
\ . 
\end{align*}

Consequently, we find
\begin{align*}
T^{1/2} \big( \widehat{\btheta}_{\rho} - \btheta^* \big)
& \text{ converges in distribution to }
N \big( 0, 
\big( \Omega^{*1/2} G^*  \big)^{+} \Omega^{*1/2}
\Sigma_2 
\Omega^{*1/2}
\big( \Omega^{*1/2} G^*  \big)^{+\intercal} \big)
\ . 
\end{align*}
This concludes the proof.

\newpage

\bibliographystyle{apa}
\bibliography{SPSC.bib}

\end{document}